\documentclass[prd,onecolumn,amsmath,amssymb,floatfix,superscriptaddress,notitlepage]{revtex4-1}

\usepackage{graphicx}
\usepackage{amssymb}
\usepackage{amsmath}
\usepackage{hyperref}
\usepackage{float}
\usepackage{multirow}

\usepackage{color}

\DeclareTextSymbol{\degre}{T1}{23}

\newcommand{\mnras}{Mon.~Not.~Roy.~Astron.~Soc.}
\newcommand{\physrep}{Phys.~Rep.}
\newcommand{\jcap}{JCAP}

\newcommand{\apjl}{\apj~Lett.}
\newcommand{\aap}{Astronomy \& Astrophysics}
\newcommand{\ssr}{Space Science Reviews}
\newcommand{\pasp}{Publications of the Astronomical Society of Pacific}
\newcommand{\procspie}{Proceedings of SPIE}

\newcommand{\beq} {\begin{equation}}
\newcommand{\eeq} {\end{equation}}
\newcommand{\bal} {\begin{aligned}}
\newcommand{\eal} {\end{aligned}}

\newcommand{\om}{\Omega^0_\text{m}}
\newcommand{\omb}{\Omega^0_\text{b}}
\newcommand{\sig}{\sigma_8}
\newcommand{\ns}{n_\text{s}}
\newcommand{\w}{w_0}
\newcommand{\wa}{w_\text{a}}

\begin{document}

\title{Looking through the same lens: shear calibration for LSST, Euclid \& WFIRST with stage 4 CMB lensing}

\author{Emmanuel Schaan}
\affiliation{Department of Astrophysical Sciences, Princeton University, Peyton Hall, Princeton NJ 08544, USA}
\author{Elisabeth Krause}
\affiliation{Kavli Institute for Particle Cosmology and Astrophysics, Stanford University, Stanford, CA 94305, USA}
\author{Tim Eifler}
\affiliation{Jet Propulsion Laboratory, California Institute of Technology, Pasadena, CA 91109, USA}
\affiliation{Department of Physics, California Institute of Technology, Pasadena, CA 91125, USA}
\author{Olivier Dor\'e}
\affiliation{Jet Propulsion Laboratory, California Institute of Technology, Pasadena, CA 91109, USA}
\affiliation{Department of Physics, California Institute of Technology, Pasadena, CA 91125, USA}
\author{Hironao Miyatake}
\affiliation{Jet Propulsion Laboratory, California Institute of Technology, Pasadena, CA 91109, USA}
\affiliation{Department of Physics, California Institute of Technology, Pasadena, CA 91125, USA}
\author{Jason Rhodes}
\affiliation{Jet Propulsion Laboratory, California Institute of Technology, Pasadena, CA 91109, USA}
\affiliation{Department of Physics, California Institute of Technology, Pasadena, CA 91125, USA}
\author{David N. Spergel}
\affiliation{Department of Astrophysical Sciences, Princeton University, Peyton Hall, Princeton NJ 08544, USA}

\begin{abstract} 
The next generation weak lensing surveys (i.e., LSST, Euclid and WFIRST) will require  exquisite control over systematic effects.
In this paper, we address shear calibration and present the most realistic forecast to date for LSST/Euclid/WFIRST and CMB lensing from a stage 4 CMB experiment (``CMB S4'').
We use the \textsc{CosmoLike} code to simulate a joint analysis of all the two-point functions of galaxy density, galaxy shear and CMB lensing convergence. 
We include the full Gaussian and non-Gaussian covariances and explore the resulting joint likelihood with Monte Carlo Markov Chains.
We constrain shear calibration biases while simultaneously varying cosmological parameters, galaxy biases and photometric redshift uncertainties.
We find that CMB lensing from CMB S4 enables the calibration of the shear biases down to $0.2\%-3\%$ in 10 tomographic bins for LSST (below the $\sim 0.5\%$ requirements in most tomographic bins), down to $0.4\%-2.4\%$ in 10 bins for Euclid and $0.6\%-3.2\%$ in 10 bins for WFIRST.
For a given lensing survey, the method works best at high redshift where shear calibration is otherwise most challenging.
This self-calibration is robust to Gaussian photometric redshift uncertainties and to a reasonable level of intrinsic alignment.  It is also robust to changes in the beam and the effectiveness of the component separation of the CMB experiment, and slowly dependent on its depth, making it possible with third generation CMB experiments such as AdvACT and SPT-3G, as well as the Simons Observatory.
\end{abstract}

\maketitle

\section{Introduction}
\label{sec:intro}

Understanding the physics of cosmic acceleration is the aim of many 
ongoing and 
upcoming imaging
surveys such as 
the Kilo-Degree Survey
(KiDS)\footnote{\url{http://www.astro-wise.org/projects/KIDS/}},
the Dark Energy Survey
(DES)\footnote{\url{http://www.darkenergysurvey.org}} \cite{DES},
the Subaru Hyper Suprime-Cam (HSC) survey\footnote{\url{http://www.naoj.org/Projects/HSC/index.html}}
\cite{Miyazakietal:06}, 
the Subaru Prime Focus Spectrograph
(PFS)\footnote{\url{http://sumire.ipmu.jp/en/2652}}\cite{Takadaetal:12},
the Dark Energy Spectroscopic Instrument (DESI)\footnote{\url{http://desi.lbl.gov}}, 
the Large Synoptic Survey Telescope (LSST)
\cite{LSSTScienceBook}, 
the ESA satellite mission Euclid \cite{EuclidDefinitionStudyReport},
and NASA's Wide-Field Infrared Survey Telescope (WFIRST) \cite{WFIRST}.
Through gravitational lensing, images of distant sources such as galaxies or the cosmic microwave background (CMB) are distorted by the presence of foreground mass. In the weak regime, lensing produces small distortions, arcminute deflections or $\sim1\%$ shear, coherent on degree scales, which are detected statistically.
Weak gravitational lensing is sensitive to the growth of structure and the geometry of the universe, making it a powerful probe of dark energy, modifications to General Relativity and the sum of the neutrino masses
(see \cite{2013PhR...530...87W} and references therein for a review).

Realizing the full potential of the stage 4 weak lensing surveys (i.e. LSST, Euclid and WFIRST) requires an exquisite understanding and control of systematics effects \cite{2006astro.ph..9591A}. 
In the case of LSST, the bias and scatter in photometric redshifts need to be controlled to better than a percent \cite{2006MNRAS.366..101H}, which may require more than $\sim 10^5$ galaxy spectra for calibration \cite{2008ApJ...682...39M}.
Interpreting cosmic shear, i.e. the power spectrum of the weak lensing of galaxies by the large-scale structure in the universe, requires knowledge of the matter power spectrum, down to scales where non-linear evolution and baryonic effects are important \cite{2008ApJ...672...19R, 2008PhRvD..77d3507Z, 2011MNRAS.418..536E, 2011MNRAS.415.3649V, 2014MNRAS.440.2997V, 2014MNRAS.442.2641V, 2015MNRAS.454.2451E, 2015ApJ...806..186O, 2016arXiv160303328H}.
Intrinsic alignments of galaxies, if not mitigated, could contaminate the cosmic shear signal by up to $1-10\%$ (see \cite{2015SSRv..193...67K, 2015SSRv..193..139K, 2015SSRv..193....1J, 2015PhR...558....1T} for a review).
%
Finally, estimating the shear from galaxy shapes may lead to additive and multiplicative biases, typically redshift dependent, which have to be controlled to a high accuracy \cite{2006MNRAS.366..101H, 2013MNRAS.429..661M}. The shear multiplicative bias is degenerate with the amplitude of the signal and its time evolution can hide the true evolution of the growth of structure, which probes dark energy and possible modifications to general relativity. \citet{2013MNRAS.429..661M} found that fully exploiting the statistical power of a stage 4 cosmic shear survey requires a shear multiplicative bias of $\lesssim 0.4\%$. 
The focus of this paper is to show how CMB lensing contributes to reaching this goal.

Many effects contribute to the shear multiplicative bias \cite{2013MNRAS.429..661M, 2016MNRAS.tmp..827J}, such as inaccuracies in the point-spread function (PSF) or detector effects (e.g. charge transfer inefficiency in CCDs or the brighter-fatter effect). Model biases may occur when estimating galaxy shapes with an inaccurate galaxy profile. Since lensing couples the short and long wavelength modes of a galaxy image, knowing the response of a galaxy image to shear requires knowing the galaxy image to a better resolution than the PSF, leading in practice to a ``noisy deconvolution'' shape bias. Furthermore, the galaxies used for shear estimation do not form a homogenous sample, and their detection signal-to-noise depends on their shape, leading to a shape selection bias. Correcting this bias perfectly would require knowledge of the galaxy population below the detection threshold
\cite{2013MNRAS.429..661M, 2016MNRAS.tmp..827J}.
State of the art shape algorithms calibrated on simulations (e.g. \cite{2016MNRAS.tmp..827J}) reach few percent accuracy on the shear multiplicative bias. They currently approach the requirements of stage 4 surveys (e.g. \cite{2015MNRAS.450.2963M}) in the absence of selection bias (i.e. when the galaxy population is known perfectly) and for images with slightly higher signal to noise.

Because the shear multiplicative bias is such a critical and difficult systematic for stage 4 weak lensing surveys, alternative methods to calibrate it provide a valuable redundancy. These consistency checks will be crucial in trusting the results, e.g. in the optimistic event where stage 4 weak lensing surveys would find deviations from $\Lambda$CDM.
 
The lensing convergence reconstructed from temperature and polarization of the cosmic microwave background (CMB) does not rely on galaxy shape estimation \cite{1999PhRvD..59l3507Z, 2002ApJ...574..566H, 2003PhRvD..67h3002O}. Although CMB lensing is most sensitive to the mater distribution at higher redshift, the lensing efficiencies for CMB and galaxy lensing overlap, thus potentially allowing to calibrate the shear multiplicative bias \cite{2012ApJ...759...32V, 2013ApJ...778..108V, 2013arXiv1311.2338D}. Using a Fisher forecast, \citet{2012ApJ...759...32V, 2013ApJ...778..108V, 2013arXiv1311.2338D} combine all the 2-point correlations of galaxy positions, galaxy shapes and CMB lensing convergence. They show that adding CMB lensing and galaxy spectroscopic data to a shape catalog allows to approach the requirements on shear calibration for stage 4 weak lensing surveys.
This method is appealing because it is a self-calibration: it relies on the data alone and not on image simulations of the whole galaxy sample. It is a practical example of real synergy between overlapping surveys, where combining probes leads to improved systematics and not just a marginally higher signal-to-noise. 
CMB lensing has been measured by the WMAP satellite \cite{2007PhRvD..76d3510S, 2008PhRvD..78d3520H}, the Atacama Cosmology Telescope (ACT) \cite{2011PhRvL.107b1301D, 2014JCAP...04..014D, 2015PhRvL.114o1302M, 2015ApJ...808....7V}, the South Pole Telescope (SPT) \cite{2012ApJ...756..142V, 2015ApJ...806..247B, 2015ApJ...810...50S}, POLARBEAR \cite{2014PhRvL.113b1301A} and the Planck satellite \cite{2014A&A...571A..17P, 2015arXiv150201591P}. In the future, Advanced ACT, SPT-3G and a stage 4 ground-based CMB experiment (CMB S4) \cite{2015APh....63...66A, 2015APh....63...55A}, as well as the Simons Observatory, will provide high fidelity maps of the CMB lensing convergence over a large fraction of the sky. 

Recent work has applied this method to existing stage 3 data. \citet{2016arXiv160105720L} correlated galaxy positions from CFHT with galaxy shapes from CFHT and CMB lensing from Planck. Assuming a fixed cosmology (WMAP or Planck parameters) and known photometric redshift errors, they constrain the shear bias in the CFHT shape catalog, finding a shear bias lower than unity at $2-4\sigma$. However, a shift in the cosmological parameters, a different redshift-evolution of the galaxy bias or uncharacterized photometric redshift errors might explain this tension.
Similarly, \citet{2016arXiv160207384B} correlated galaxy positions from DES with galaxy shapes from DES and CMB lensing from SPT, to constrain the shear bias and an overall additive photometric redshift bias. The constraints on shear bias are obtained by fixing cosmology and the photo-z bias to fiducial values.
Cross-correlations between galaxy shear and CMB lensing have been measured \cite{2015PhRvD..91f2001H, 2015PhRvD..92f3517L, 2016MNRAS.459...21K}. In \cite{2015MNRAS.449.2205K}, the combination of galaxy shear and CMB lensing is forecasted to improve dark energy and neutrino mass constraints.
In \cite{2016arXiv160505337M}, the CMB lensing from Planck and the galaxy lensing from CFHTLenS is measured around CMASS halos, yielding a $15\%$ measurement of a cosmographic distance ratio.
Finally, \cite{2016arXiv160608841S} measured all the two-point correlations of galaxy positions and shear from SDSS and CMB convergence from Planck, and constrained in turn galaxy bias to $2\%$ accuracy, cosmology, shear multiplicative bias to $15\%$ and distance ratio to $10\%$.
While these studies \cite{2016arXiv160105720L, 2016arXiv160207384B, 2016arXiv160608841S} currently constrain the shear bias to $\gtrsim10\%$, far from current state-of-the-art calibrations from simulations, they constitute very encouraging first steps in using CMB lensing and spectroscopic data to calibrate the shear bias.

In this paper, we address the following questions: can CMB lensing calibrate the shear bias down to the requirements of stage 4 surveys? Is this method competitive with image simulations? Is this possible without arbitrarily fixing cosmological and nuisance parameters? How robust is this calibration to photometric redshift uncertainties and intrinsic alignments?
To answer these questions in a realistic way, we simulate the full joint analysis of weak lensing, photometric clustering and CMB lensing. We compute the full covariances for all the two-point auto- and cross-correlations, including the non-Gaussian covariances which dominate on small scales. 
We analyze the full likelihood function with Monte Carlo Markov Chains (MCMC), allowing for a potentially non-Gaussian posterior distribution. This is particularly relevant when assessing non-linear degeneracies between parameters. We compare these MCMC forecasts with Fisher forecasts, thus assessing potential departures from Gaussianity of the posterior distributions.
We test the robustness of the method by simultaneously varying shear biases, cosmological parameters, galaxy biases, photometric redshift uncertainties (bias in each tomographic bin and overall scatter), and by contaminating the simulated data with intrinsic alignment.
We extend the \textsc{CosmoLike} \cite{2014MNRAS.440.1379E} framework to include CMB lensing
and produce the most realistic forecast to date for LSST/Euclid/WFIRST and CMB lensing from a stage 4 CMB experiment.
Given the importance of this result, and as an input for the design of CMB S4, we vary the depth, resolution and maximum multipole of the CMB experiment and show the robustness of this shear calibration method.

This paper is organized as follows. In Sec.~\ref{sec:method} we describe the observables considered, the survey specifications, the systematic effects included and the simulated likelihood. In Sec.~\ref{subsec:lsst_alone}, we revisit the LSST requirements and show that self-calibration of the shear biases with LSST alone is possible down to $1-2\%$. The calibration of shear multiplicative bias down to the LSST requirements by using stage 4 CMB lensing is presented in Sec.~\ref{subsec:lsst_cmbs4}, along with the impact of intrinsic alignments. We show the importance of sensitivity and resolution for CMB S4 in Sec.~\ref{subsec:varying_cmbs4}, the robustness to photometric redshift uncertainties in Sec.~\ref{subsec:photoz}, to non-linearities and baryonic effects in Sec.~\ref{subsec:varyinglmaxgks}, and present forecasts for Euclid and WFIRST instead of LSST in Sec.~\ref{subsec:euclid_wfirst}.

\section{Simulated joint analysis of LSST \& CMB S4: method}
\label{sec:method}

\subsection{Observables: $g$, $\kappa_\text{gal}$, $\kappa_\text{CMB}$}

We use the projected galaxy density field $g$, the convergence $\kappa_\text{gal}$ from galaxy shapes and $\kappa_\text{CMB}$ from CMB lensing reconstruction as probes of the matter density field. 
We consider two distinct galaxy samples for $g$ and $\kappa_\text{gal}$, with distinct redshift distributions and tomographic bins, as detailed in Sect.~\ref{sec:LSST specifications}.
Each observable $A \in \left\{ g, \kappa_\text{gal}, \kappa_\text{CMB} \right\}$ is a projection of the density contrast $\delta$, weighted by an efficiency kernel $W_A$:
\beq
A(\hat{n}) =  \int d\chi \; W_A(\chi) \; \delta(\chi \hat{n}, \chi)
\eeq
Thus the cross-spectrum $C_\ell^{AB}$ of observables $A,B$ is related to the matter power spectrum $P_m$ via
\beq
C_\ell^{AB} = \int \frac{d\chi}{\chi^2} \; W_A(\chi)W_B(\chi) \; P_m(k = \frac{\ell +1/2}{\chi}, \chi),
\eeq
in the Limber and flat sky approximations. Throughout, we assume a flat cosmology and therefore equate comoving radial and transverse distances.
%
For the projected density field $g_i$ in redshift bin $i$, the efficiency kernel is 
\beq
W_{g_i}(\chi) = b_g(z) \frac{1}{n_i} \frac{dn_i}{dz}  \frac{dz}{d\chi},
\text{ with }
n_i = \int dz \frac{dn_i}{dz},
\eeq
and $dn_i/dz$ is the redshift distribution of the galaxies in the $i$th bin.
%
For a source at comoving distance $\chi_S$, the lensing efficiency is 
\beq
W_\kappa (\chi, \chi_S) = \frac{3}{2} \left( \frac{H_0}{c} \right)^2 \Omega_m^0 \frac{\chi}{a(\chi)} \left( 1 - \chi/\chi_S \right)
\eeq
Thus the CMB lensing efficiency is simply $W_{\kappa_\text{CMB}} (\chi) = W_\kappa (\chi, \chi_\text{LSS})$, where $\chi_\text{LSS}$ is the comoving distance to the surface of last scattering at $z\sim 1100$ (see curve in Fig.~\ref{fig:specs_lsst}).
For the convergence $\kappa_{\text{gal},i}$ in the tomographic bin $i$, the efficiency kernel is obtained by integrating over the source distribution in the same bin:
\beq
W_{\kappa_{\text{gal},i}} (\chi) = \;\frac{1}{n_{\text{source}, i}} \int dz_S \frac{dn_{\text{source}, i}}{dz_S} \; W_\kappa(\chi, \chi(z_S)),
\eeq
In this simulated analysis, we compute all the cross and auto-spectra of $g$, $\kappa_\text{gal}$ and $\kappa_\text{CMB}$ in different tomographic redshift bins. The analysis therefore includes galaxy clustering ($C_\ell^{g_i g_j}$), galaxy-galaxy lensing ($C_\ell^{g_i \kappa_{\text{gal}, j}}$), galaxy-CMB lensing ($C_\ell^{g_i \kappa_\text{CMB}}$), cosmic shear tomography ($C_\ell^{\kappa_{\text{gal}, i} \kappa_{\text{gal}, j}}$), CMB lensing power spectrum ($C_\ell^{\kappa_\text{CMB} \kappa_\text{CMB}}$) and CMB lensing-galaxy lensing ($C_\ell^{\kappa_\text{CMB} \kappa_{\text{gal}, j}}$).
Our specific assumptions about CMB S4 and LSST are detailed in the next sections, as well as the treatment of the systematic effects.

\subsection{CMB S4 specifications}
\label{sec:specs_cmbs4}

\begin{figure}[h]
\begin{minipage}[c]{0.4\linewidth}
\centering
\renewcommand{\arraystretch}{1.5}
\begin{tabular*}{1\textwidth}{@{\extracolsep{\fill}}| c l |}
\hline
\multicolumn{2}{|c|}{\textbf{CMB S4 specifications}} \\
\hline
\hline
$\Omega_{\mathrm{s}}$ & 18,000 deg$^2$ ($f_\text{sky}=44\%$)\\
beam & FWHM$=1^\prime$\\
white noise & $1\mu K ^\prime$ for T; $1.4\mu K ^\prime$ for E,B\\
$\ell_\text{min}$ & 30 for T, E, B, $\kappa_\text{CMB}$\\
$\ell_\text{max}$ & 3000 for T; 5000 for E, B, $\kappa_\text{CMB}$\\
\hline
\end{tabular*}
\renewcommand{\arraystretch}{1.0}
\par\vspace{0pt}
\end{minipage}
\begin{minipage}[c]{0.57\linewidth}
\centering
\includegraphics[width=1\columnwidth]{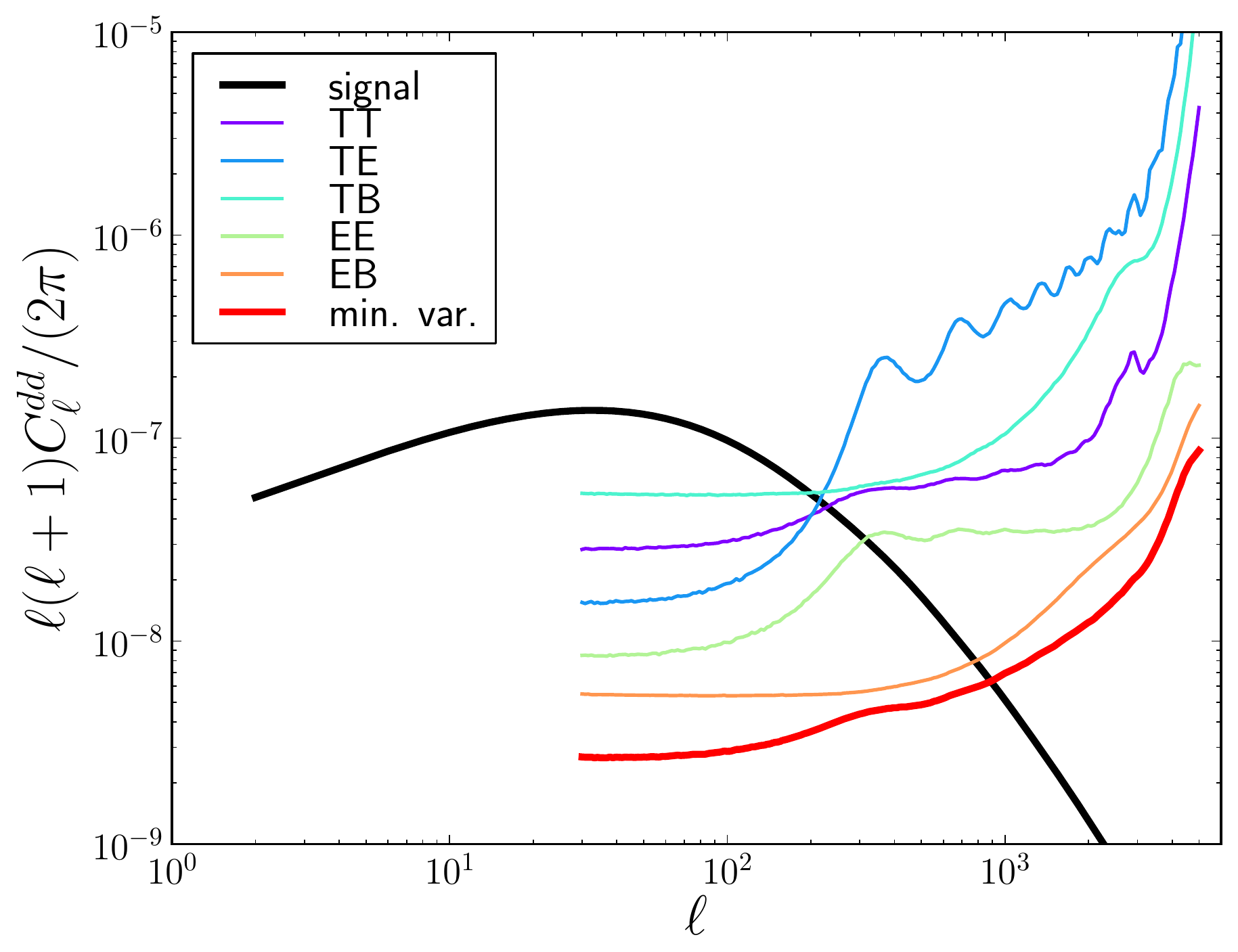}
\par\vspace{0pt}
\end{minipage}
\caption{
Assumed specifications for CMB S4. \\
\textbf{Left panel:} Assumed survey area, resolution, sensitivity and $\ell$-limits. The cutoff $\ell_\text{max}$ is determined by our ability to effectively remove foregrounds with multifrequency data, and the values assumed here represent reasonable assumptions. \\
\textbf{Left panel:} Corresponding noise per $\ell$-mode for the various quadratic estimators (colored lines) and for the minimum-variance quadratic estimator (solid black line), compared to the amplitude of the CMB lensing signal (dashed black line). The CMB lensing convergence is cosmic variance limited up to $\ell=1000$.
}
\label{fig:specs_cmbs4}
\end{figure}

We simulate a Stage 4 CMB experiment (CMB S4) \cite{2015APh....63...66A, 2015APh....63...55A}, with specifications presented in Fig.~\ref{fig:specs_cmbs4}. We assume full overlap with LSST, high resolution (beam FWHM$=1^\prime$) and sensitivity (white noise level $1\mu K^\prime$). 
We adopt reasonable $\ell$-cuts for the cleaned CMB temperature and polarization maps ($\ell_\text{min}=30$ for T, E, B; $\ell_\text{max}=3000$ for T; $\ell_\text{max}=5000$ for E,B). As a result, our forecast only uses the convergence $\kappa_\text{CMB}$ between $\ell=30$ and $\ell=5000$.
As an input for the design of CMB S4, we quantify the separate impacts of resolution, depth and effectiveness of component separation in Sec.~\ref{subsec:varying_cmbs4}.

Our likelihood analysis uses the reconstructed convergence $\kappa_\text{CMB}$ from CMB S4, and assumes the minimum variance quadratic estimator from \cite{1999PhRvD..59l3507Z, 2002ApJ...574..566H}. This minimum variance estimator is the optimal linear combination of the quadratic estimators from temperature and E and B polarizations. The corresponding reconstruction noise is shown in Fig.~\ref{fig:specs_cmbs4}: the reconstructed convergence is cosmic variance limited up to $\ell=1000$.
At the resolution and sensitivity considered, iterative techniques making use of the full likelihood function for the CMB convergence may improve the reconstruction noise by a factor of order unity, compared to the minimum variance quadratic estimator \cite{2003PhRvD..67d3001H, 2003PhRvD..68h3002H}. Using only the quadratic estimators gives a conservative forecast for CMB S4 lensing.

We do not include temperature and polarization power spectra from CMB S4, nor Planck priors on cosmological parameters: 
we wish to use the minimal number of probes in the shear calibration.
Furthermore, given the high statistical signal to noise, 
a consistent analysis might need to account for correlations between the CMB temperature and polarization and the large-scale structure.
We found that including Planck priors on cosmological parameters improves the shear calibration by several tens of percent.

\subsection{LSST specifications}
\label{sec:LSST specifications}

Following closely \cite{2016arXiv160105779K},
we simulate an LSST-like survey \cite{LSSTScienceBook} over a total area $\Omega_\text{s} = 18,000$ deg$^2$. The assumed specifications are presented in Fig.~\ref{fig:specs_lsst}.

We assume the source redshift distribution to follow
$d n_\text{source} / dz \propto z^\alpha e^{-(z/z_0)^\beta}$,
with $\alpha=1.27, \beta = 1.02, z_0=0.5$, 
with a total number density $n_\text{source}=26$ arcmin$^{-2}$ \cite{2013MNRAS.434.2121C}, and a shape noise $\sigma_\epsilon = 0.26$ in each ellipticity component. We split the source galaxies into 10 tomographic redshift bins.

The assumed galaxy lens sample is similar to the redMaGiC sample \cite{2015arXiv150705460R}, with a constant comoving volume density $\bar{n}_{\mathrm{lens}}(z) = 10^{-3} \left(h/\mathrm{Mpc}\right)^3$ and corresponding redshift distribution $dn_\text{lens}/dz \propto \chi(z)^2/H(z)$, giving a total number density of $n_\text{lens}=0.25$ arcmin$^{-2}$. We split these galaxies into 4 tomographic bins.
The lens sample is also used as the clustering sample: the same projected galaxy density field $g$ is used for clustering and lensing-tracer correlations. 
In both cases, we only use the $\ell$-modes with $\ell \geq 20$ and $2\pi \chi(z_\text{mean}) / \ell  > 10$ Mpc/h, corresponding to the smallest scale where we assume linear biasing to be valid. In practice, this corresponds to $\ell_\text{max} = 420, 714, 930$ and $1212$  respectively for the four redshift bins. 
We introduce the effective galaxy bias $b_\text{g}^i$ for each bin $i$, as 4 nuisance parameters.
Several comments are in order.
We have selected a clustering and lens sample with excellent photo-z accuracy (see next subsection), in order to get a robust shear calibration. 
For this reason, we have restricted the lens sample to $z<1$, beyond which photo-z accuracy is expected to degrade considerably. 
This limits the signal to noise in clustering and tracer-lensing correlations considerably.  As a result, our forecast for shear calibration from data combinations involving clustering and tracer-lensing should be considered very conservative.
We note however that the comoving volume density of this redMaGiG-like sample is not limiting: the galaxy shot noise in this sample is subdominant on almost all the scales we retain.

For the convergence field $\kappa_\text{gal}$, we use all the modes $20 \leq \ell \leq 5000$. This is the current baseline for lensing forecasts with LSST.
On the smallest scales, baryonic effects  in the matter power spectrum may constitute a source of systematic error. We discuss it in the next subsection.

\subsection{Systematics \& nuisance parameters}
\label{subsec:syst_nuisance_params}

In addition to the basic LSST survey parameters summarized above, this section specifies our detailed assumptions on systematic uncertainties affecting the LSST galaxy samples.

\subsubsection*{Photometric redshift uncertainties} We assume Gaussian photometric redshift uncertainties, with a bias $\Delta_z$ and scatter $\sigma_z$, such that
\beq
p\left( z_\text{ph} | z\right)
=
\frac{1}{\sqrt{2\pi \sigma_z^2}}
\exp
\left[
-\frac{\left( z_\text{ph} - z - \Delta_z \right)^2}{2 \sigma_z^2}
\right]
\eeq
is the probability of measuring the photo-z $z_\text{ph}$ given a galaxy with true redshift $z$. This approach neglects catastrophic photo-z failures \cite{2010MNRAS.401.1399B}. Realistic distributions for $p\left( z_\text{ph} | z\right)$ can be much complex, however a careful treatment of photo-z outliers is beyond the scope of this paper.
We split the lens sample into 4 tomographic bins $z\in (0.2-0.4), (0.4-0.6), (0.6-0.8), (0.8-1)$. We split the source sample into 10 tomographic bins with equal number of objects per bin.
These bins are defined by sharp photo-z cuts and therefore have overlapping true redshift distributions \cite{2006ApJ...636...21M}:
\beq
\frac{dn_{\text{lens/source}, i}}{dz} = \frac{dn_\text{lens/source}}{dz} 
\;
\int_{z_\text{ph} \in \text{ bin }i} dz_\text{ph} \;  p(z_\text{ph} | z).
\eeq
These redshift distributions are shown in Fig.~\ref{fig:specs_lsst}.
The large number of tomographic bins for the lens and source samples is justified \textit{a posteriori} by the good signal to noise ratio (SNR) and low correlation coefficient for the various $\ell$-bins and tomographic bins. 
For each lens or source bin, we introduce a bias $\Delta_{\text{z,lens/source}, i}$, as well as an overall scatter $\sigma_{z, \text{lens/source}} / (1+z)$, resulting in 16 nuisance parameters. We marginalize over these nuisance parameters with priors as indicated in Table~\ref{tab:params_mcmc}.

\subsubsection*{Multiplicative shear bias} We describe shear calibration uncertainties via an overall shear multiplicative bias $m_i$ for each source bin $i$, resulting in 10 nuisance parameters:
$\kappa_{\text{gal}, i} \longrightarrow \left( 1 + m_i \right)\; \kappa_{\text{gal}, i} $.
The goal of this analysis is to forecast the constraints on $m_i$, with and without priors.
Note that we do not account for potential scale-dependent errors in the calibration. Such effects can be marginalized over if their scale-dependences are known, as in \cite{2016arXiv160509130T}.

\subsubsection*{Intrinsic Alignments}

To assess the potential contamination from galaxy intrinsic alignments (IA), we include the IA contamination in the data vector, but do not account for the contamination in the analysis.
We model IA using the non-linear linear alignment (NLA) model \cite{ HirataSeljak04,BridleKing07,Singh15} for red galaxies, and neglect the plausible, but much weaker alignment of blue galaxies \cite{Catelan2001,Hirata07,Mandelbaum11}. We calculate the expected intrinsic alignment amplitude by averaging the observed redshift and luminosity dependence of the intrinsic alignment amplitude of red galaxies in the MegaZ-LRG sample \cite{Joachimi11} over the luminosity function of source galaxies, for which we extrapolate the r-band luminosity function measurements from the GAMA survey \cite{LFGAMA} to LSST depth \cite[see][for details]{keb16}. 

Unaccounted IA biases the shear estimated from galaxy shapes, and can therefore affect all cross-correlations involving at least one $\kappa_\text{gal}$. We calculate the contamination of $g\kappa_\text{gal}$ and $\kappa_\text{gal} \kappa_\text{gal}$ following \cite{JoachimiBridle10}, and the contamination of $\kappa_\text{gal}\kappa_\text{CMB}$ following \cite{TroxelIshak14, HallTaylor14}.

A detailed study of all existing IA models and mitigation techniques is beyond the scope of this paper; in the case of cosmic shear with LSST we refer the reader to \cite{keb16}.

\subsubsection*{Nonlinearities and baryonic effects}

We account for non-linearities in the covariance matrix, by including the non-linear trispectrum terms as well as the super-sample covariances \cite{2014PhRvD..89h3519L, 2014JCAP...05..048C, 2014PhRvD..90j3530L, 2014MNRAS.441.2456T, 2014PhRvD..90l3523S} as in \cite{2016arXiv160105779K}.

For the projected density field $g$, we discard the small scales (see previous subsection) where linear bias might no longer be valid. This limits the potential effect of baryons.
For the lensing convergence $\kappa_\text{gal}$, we do not include any uncertainty in the modeling of the non-linear power spectrum or the baryonic effects.
However, we show in Sec.~\ref{subsec:varyinglmaxgks} that the shear calibration is only degraded by $10-40\%$ when varying the maximum multipole $\ell_\text{max}$ from our fiducial value of $5,000$ down to $1,000$. 

\subsubsection*{Biases in the CMB lensing reconstruction}

The quadratic estimators for CMB lensing exploit the statistical isotropy of the primary CMB. 
Any component that breaks this statistical isotropy will therefore contribute to the reconstructed $\kappa_\text{CMB}$ map. This the case of the cosmic infrared background, as well as radio point sources and thermal Sunyaev-Zel'dovich clusters, whether they are resolved or not.
These sources may contaminate the convergence map and its power spectrum at the sub-percent to percent level, but mitigation techniques exist \cite{2014ApJ...786...13V, 2014JCAP...03..024O}. In this analysis, we do not take into account these potential biases.

\begin{figure}[h]
\begin{minipage}[c]{0.42\linewidth}
\centering
\renewcommand{\arraystretch}{1.3}
\begin{tabular*}{1\textwidth}{@{\extracolsep{\fill}}| c l |}
\hline
\multicolumn{2}{|c|}{\textbf{LSST specifications}} \\
\hline
\hline
$\Omega_{\mathrm{s}}$ & 18,000 deg$^2$\\
\multicolumn{2}{|c|}{} \\
source distribution & $dn_\text{source}/dz \propto z^\alpha e^{-(z/z_0)^\beta}$,\\ 
& $\alpha=1.27, \beta = 1.02, z_0=0.5$,\\
& $n_\text{source}= 26$ arcmin$^{-2}$\\
& 10 bins\\
& $\sigma_\epsilon= 0.26$\\
\multicolumn{2}{|c|}{} \\
lens distribution & $dn_\text{lens}/dz \propto \chi(z)^2/H(z),$\\
& $n_\text{lens}=0.25$ arcmin$^{-2}$\\
& 4 bins\\
\hline 
\end{tabular*}
\renewcommand{\arraystretch}{1.0}
\par\vspace{0pt}
\end{minipage}
\begin{minipage}[c]{0.57\linewidth}
\centering
\includegraphics[width=1\columnwidth]{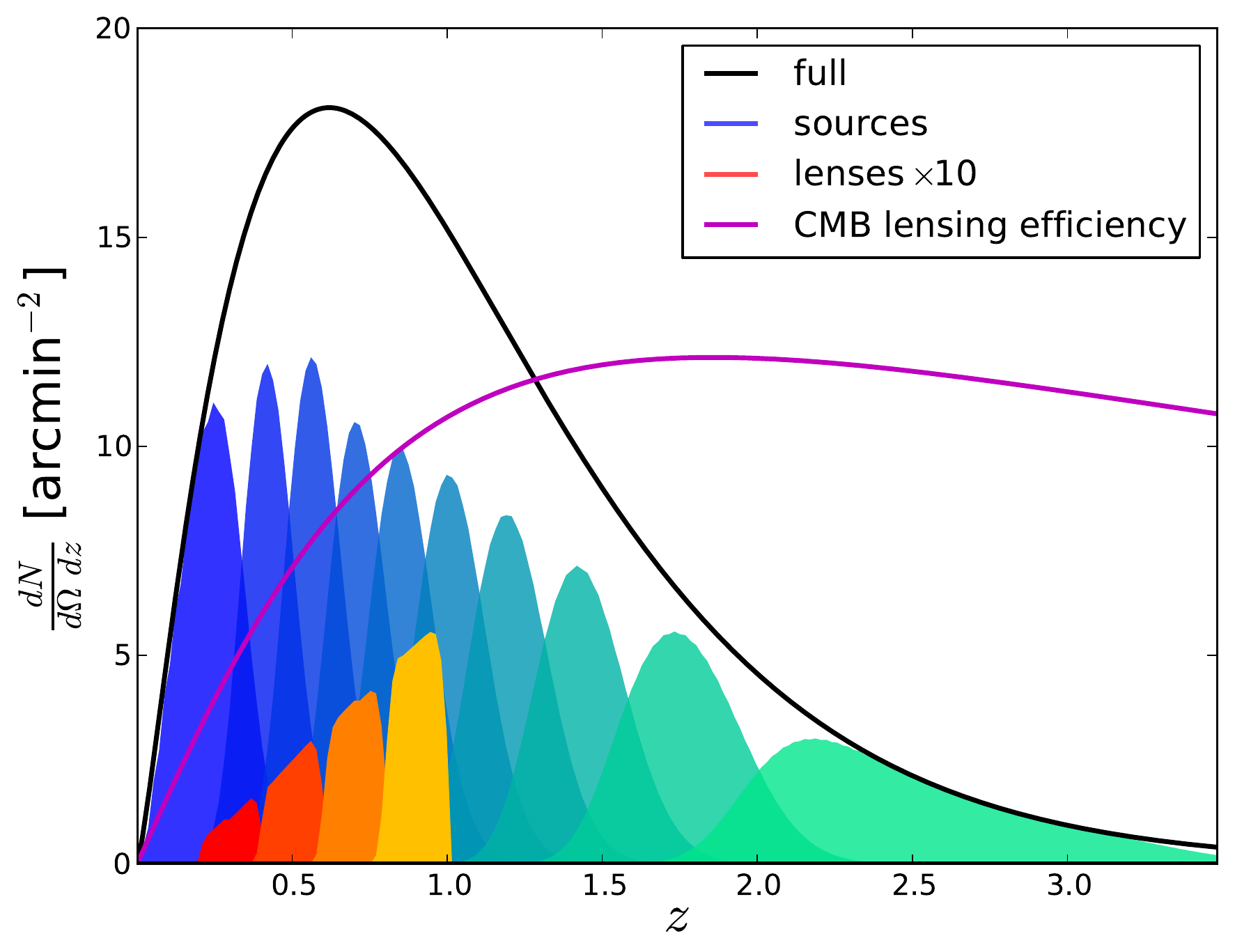}
\par\vspace{0pt}
\end{minipage}
\caption{
Assumed specifications for LSST \cite{LSSTScienceBook}, following \cite{2016arXiv160105779K}. The left panel shows the survey area, galaxy density, redshift distribution of the sources and shape noise. The right panel shows the full redshift distribution of the source galaxies (black curve), split into 10 tomographic bins (blue filled curves). The lens sample is redMaGiC-like \cite{2015arXiv150705460R}, split into 4 lens bins (red-yellow filled curves, multiplied by 10 to be visible on the same scale). Overlaid is the CMB lensing efficiency kernel (magenta line).
}
\label{fig:specs_lsst}
\end{figure}

\subsection{Likelihood analysis}

Our simulated data vector for the joint LSST \& CMB S4 analysis consists of all the auto and cross-spectra of galaxy projected density, galaxy convergence and CMB convergence for all the lens and source bins:
\beq
\mathbf{D}_\text{LSST \& CMB S4} = 
\left( 
\; \underbrace{ C_\ell^{g_i g_j} }_\text{\tiny clustering}, 
\; \underbrace{ C_\ell^{g_i \kappa_\text{CMB}} }_\text{\tiny galaxy-CMB lensing}, 
\; \underbrace{ C_\ell^{g_i \kappa_{\text{gal}, j}} }_\text{\tiny galaxy-galaxy lensing}, 
\; \underbrace{ C_\ell^{\kappa_\text{CMB} \kappa_\text{CMB}} }_\text{\tiny CMB lensing auto}, 
\; \underbrace{ C_\ell^{\kappa_\text{CMB} \kappa_{\text{gal}, j}} }_\text{\tiny CMB lensing-galaxy lensing}, 
\; \underbrace{ C_\ell^{\kappa_{\text{gal}, i} \kappa_{\text{gal}, j}} }_\text{\tiny shear tomography} 
\;  \right)
\label{eq:combi_full}
\eeq
Again, we note that our LSST \& CMB S4 analysis does not include temperature and polarization power spectra from CMB S4, only the convergence $\kappa_\text{CMB}$.
For comparison purposes, we also consider an LSST-only analysis:
\beq
\mathbf{D}_\text{LSST} = 
\left( 
\; \underbrace{ C_\ell^{g_i g_j} }_\text{\tiny clustering}, 
\; \underbrace{ C_\ell^{g_i \kappa_{\text{gal}, j}} }_\text{\tiny galaxy-galaxy lensing}, 
\; \underbrace{ C_\ell^{\kappa_{\text{gal}, i} \kappa_{\text{gal}, j}} }_\text{\tiny shear tomography} 
\;  \right)
\label{eq:combi_lsst}
\eeq
Finally, for the purpose of calibrating the shear bias, it is useful to compare the following two combinations:
\beq
\bal
\mathbf{D}_\text{Combi 1} &= 
\left( 
\; \underbrace{ C_\ell^{\kappa_\text{CMB} \kappa_\text{CMB}} }_\text{\tiny CMB lensing auto}, 
\; \underbrace{ C_\ell^{\kappa_\text{CMB} \kappa_{\text{gal}, j}} }_\text{\tiny CMB lensing-galaxy lensing}, 
\; \underbrace{ C_\ell^{\kappa_{\text{gal}, i} \kappa_{\text{gal}, j}} }_\text{\tiny shear tomography} 
\;  \right)
\\
\mathbf{D}_\text{Combi 2} &= 
\left( 
\; \underbrace{ C_\ell^{g_i g_j} }_\text{\tiny clustering}, 
\; \underbrace{ C_\ell^{g_i \kappa_\text{CMB}} }_\text{\tiny galaxy- CMB lensing}, 
\; \underbrace{ C_\ell^{g_i \kappa_{\text{gal}, j}} }_\text{\tiny galaxy-galaxy lensing}.
\;  \right)
\\
\eal
\label{eq:combi12}
\eeq

Combination 1 corresponds to the joint analysis of CMB lensing and LSST galaxy shapes through their auto- and cross-correlations. It uses only lensing-lensing correlations. This combination is natural since the CMB is distorted by the same foreground mass distribution as the galaxy shapes (although with a slightly different efficiency kernel), and so one would like to compare the two convergence maps directly, which can be done with the auto- and cross-spectra of $\kappa_\text{CMB}$ and $\kappa_\text{gal}$.

Combination 2 compares CMB lensing and galaxy shapes through cross-correlation with the lens sample, and adds clustering. It uses tracer-lensing and tracer-tracer correlations. This combination has the advantage of only using cross-correlations of the convergence fields, which are less prone to systematic effects than auto-correlations.
As we will show, this combination has slightly lower statistical SNR but is less affected by intrinsic alignment contamination (see Sec.~\ref{subsec:lsst_cmbs4}).
It is also less sensitive to uncertainties in the small-scale power spectrum: because of our $\ell$ cuts in the tracer population, the signal is not affected by comoving scales below $~10$ Mpc/h. This is not the case for lensing-lensing correlations, for which a fixed angular scale receives contributions from arbitrarily small comoving scales.

The various data vectors are computed assuming the survey parameters Fig.~\ref{fig:specs_cmbs4} and Fig.~\ref{fig:specs_lsst}, and the fiducial parameters in Table~\ref{tab:params_mcmc}.
We wish to constrain the parameters in Table~\ref{tab:params_mcmc} from the mock data vector $\mathbf{D}$. 
To give visual intuition on the effect of cosmological and nuisance parameters on the observables in $\mathbf{D}$, we show the logarithmic derivatives of $\mathbf{D}$ in App.~\ref{app:param_dpdce}. These give insight into the degeneracies between cosmological and nuisance parameters.
We assume a Gaussian likelihood for the data, and ignore the dependence of the covariance matrix on parameters
(see \cite{2009A&A...502..721E, 2016MNRAS.456L.132S, 2015JCAP...12..058W} for the impact of varying the covariance matrix and practical implementations):
\beq
\ln \mathcal{L}_{(\mathbf{D} | \Theta)} = -\frac{1}{2}\;  
\left(\mathbf{D} - \mathbf{M}_{(\Theta)} \right)^t \; 
C^{-1} \; 
\left(\mathbf{D} - \mathbf{M}_{(\Theta)} \right) \;
+
\text{ constant},
\eeq
where $\mathbf{M}_{(\Theta)}$ is the model for the data $\mathbf{D}$, computed in the same way as the data vector, but evaluated at the parameters $\Theta$ instead of their fiducial value.
%
The Gaussian approximation is most accurate at high $\ell$, where the $C_\ell$ estimator averages over a large number of modes and the central limit theorem applies. 
As shown in \cite{2013A&A...551A..88C}, varying the covariance matrix in the Gaussian likelihood may overestimate the information in the data, by including spurious information from the variance of the data vector. Keeping the covariance matrix fixed is therefore a conservative choice, as this includes only the information from the mean data vector.

We explore the posterior distribution with MCMC sampling, using the code \textsc{emcee} \cite{2013PASP..125..306F}. 
This method is appropriate for potentially non-Gaussian posterior distributions. This is relevant in the case of non-linear degeneracies. We show the convergence of the MCMC chains in App.~\ref{app:mcmc_fisher_convergence}.
We also perform Fisher forecasts, and validate them against the MCMC forecasts in App.~\ref{app:mcmc_fisher_convergence}. Confidence intervals from MCMC and Fisher agree to better than $5\%$, which is the result of both the convergence of the MCMC chains and the near-Gaussianity of the posterior for the shear biases.
As explained earlier, we do not include Planck priors on cosmological parameters.
\begin{table}[H]
\begin{center}
\begin{tabular*}{0.6\textwidth}{@{\extracolsep{\fill}}| c c l |}
\hline
Parameter & Fiducial & Prior \\  
\hline
\hline
\multicolumn{3}{|c|}{\textbf{Cosmology}} \\
$\om$ & 0.3156 & flat(0.1, 0.6)  \\ 
$\sig$ & 0.831 &  flat(0.6, 0.95) \\ 
$\ns$ & 0.9645 & flat(0.85, 1.06)  \\
$\w$ &  -1 & flat(-2, 0)   \\
$\wa$ &  0 & flat(-2.5, 2.5)  \\
$\omb$ &  0.0492 &  flat(0.04, 0.055) \\
$h_0$  & 0.6727 &  flat(0.6, 0.76) \\
\hline
\multicolumn{3}{|c|}{\textbf{Galaxy bias}} \\
$b_\text{g}^1$ & 1.35  & flat(0.8, 2.0) \\
$b_\text{g}^2$ & 1.5  &flat(0.8, 2.0) \\
$b_\text{g}^3$ & 1.65 & flat(0.8, 2.0) \\
$b_\text{g}^4$ & 1.8 & flat(0.8, 2.0) \\
\hline
\multicolumn{3}{|c|}{\textbf{Photo-z: lens sample}} \\
$\Delta_{z, \text{lens}, i} $ & 0 & Gauss(0, 0.0004) \\
$\sigma_{z, \text{lens}} / (1+z) $ & 0.01 & Gauss(0.01, 0.0006) \\
\hline
\multicolumn{3}{|c|}{\textbf{Photo-z: source sample}} \\
$\Delta_{z, \text{source}, i} $ & 0 & Gauss(0, 0.002) \\
$\sigma_{z, \text{source}} / (1+z)$ &0.05 & Gauss(0.05, 0.003) \\
\hline
\multicolumn{3}{|c|}{\textbf{Shear calibration}} \\
$m_i $ & 0 & Gauss(0, 0.004) or None\\
\hline
\end{tabular*}
\end{center}
\caption{We vary 37 parameters in the simulated likelihood analysis, including 7 cosmological parameters, 4 galaxy biases, 16 photo-z parameters and 10 shear biases. For each parameter, the fiducial value is shown, as well as the prior (either flat (min, max) or Gaussian ($\mu$, $\sigma$)). Priors for the nuisance parameters follow \cite{2016arXiv160105779K}. The fiducial values for the cosmological parameters follow \cite{2015arXiv150201589P}.
}
\label{tab:params_mcmc}
\end{table}

\begin{figure}[h]
\centering
\includegraphics[width=0.7\columnwidth]{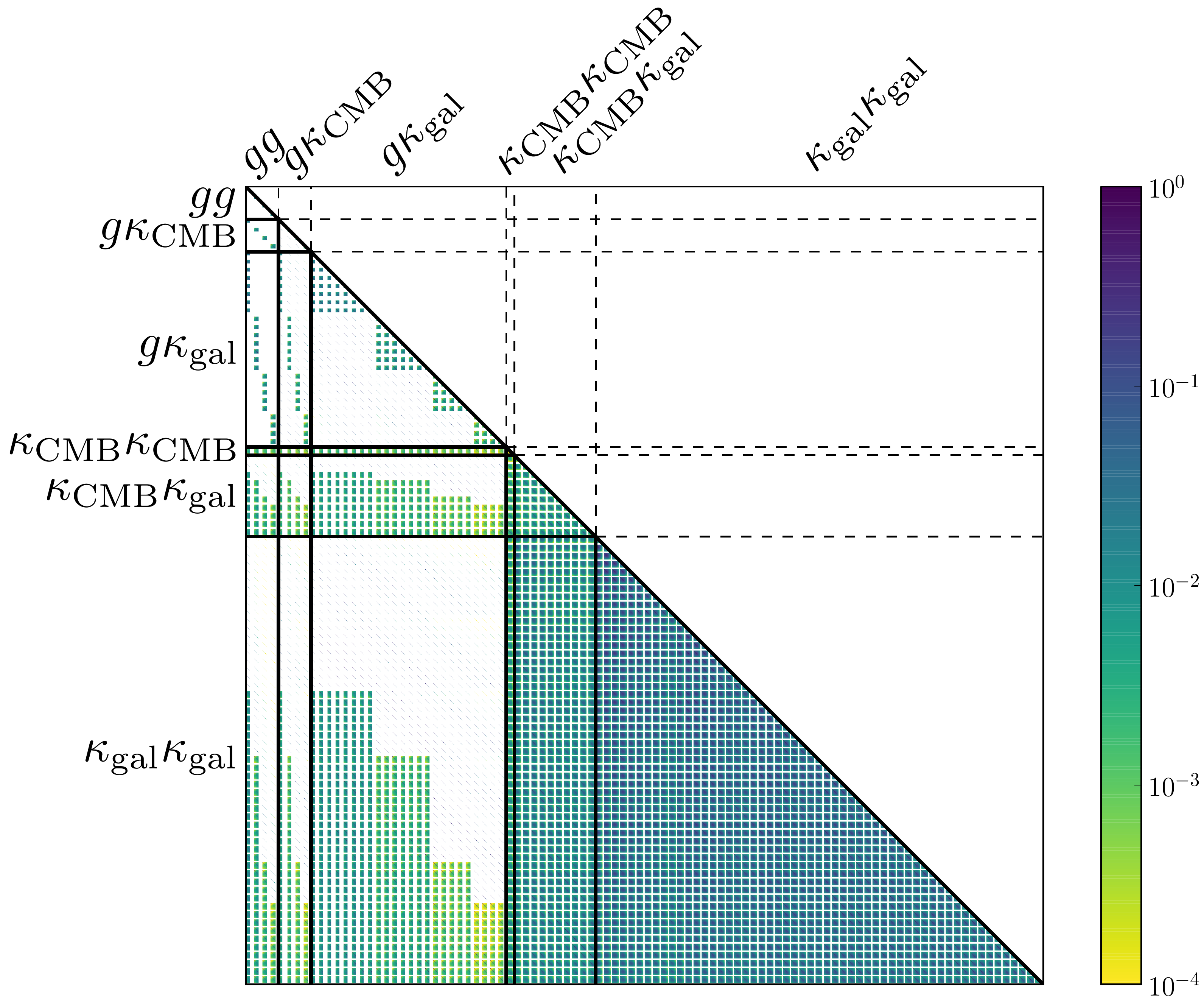}
\caption{
Correlation coefficient matrix for the full set of observables 
$\mathbf{D}_\text{LSST \& CMB S4}$
(i.e. clustering, 
galaxy-CMB lensing, 
galaxy-galaxy lensing, 
CMB lensing auto, 
CMB lensing-galaxy lensing, 
shear tomography).
The matrix is $2450\times 2450$ and includes galaxy clustering, galaxy-galaxy lensing, shear tomography, as well as CMB lensing, galaxy-CMB lensing, and galaxy lensing-CMB lensing. 
The details of the calculation are presented in the Appendix A of \cite{2016arXiv160105779K}.
Please zoom in to identify individual matrix elements.
}
\label{fig:lsst_cor}
\end{figure}

The covariance matrix is shown in Fig.~\ref{fig:lsst_cor}, and includes the Gaussian and non-Gaussian contributions (including the super-sample covariance \cite{2014PhRvD..89h3519L, 2014JCAP...05..048C, 2014PhRvD..90j3530L, 2014MNRAS.441.2456T, 2014PhRvD..90l3523S, 2016arXiv160105779K}). These covariances are computed analytically with \textsc{CosmoLike} \cite{2014MNRAS.440.1379E}, using Halofit for the non-linear power spectrum, and a halo model for the trispectrum, as in \cite{2016arXiv160105779K} (see their Appendix A). The most notable additional component here is the noise from CMB lensing reconstruction, described in Sec.~\ref{sec:specs_cmbs4}.

From the data vector and the covariance matrix, we compute the individual and combined signal-to-noise ratios (SNR) for the various probes. These are shown in Tab.~\ref{tab:snr_lsst}. 
Note that the statistical SNR is only a good figure of merit when predicting the constraint on a single parameter, the amplitude of the signal, in the absence of nuisance parameters. Instead, we present it here in order to give intuition about the relative statistical weight of each probe.
From Tab.~\ref{tab:snr_lsst}, we see that each probe will be measured with high significance, with SNR$\sim 100-1000$.

The SNR in cosmic shear is higher than in galaxy-galaxy lensing and clustering, which is due to our very conservative choice of tracer sample.
As explained earlier, in order to get a robust shear calibration and a conservative forecast, we restrict our tracer to $z<1$ where photo-z uncertainties are very well understood. We also discarding the small-scales, where linear bias breaks down and a more realistic halo occupation distribution model would be required. Again, this choice severely limits the signal to noise in clustering and tracer-lensing correlations. 

The total SNR for LSST \& CMB S4 lensing is only $\sim16\%$ higher than that of LSST alone.
However, this does not mean that adding CMB S4 lensing is pointless. Indeed, our goal is not to reduce the statistical error bars compared to LSST alone, but instead to constrain systematics by breaking degeneracies, due to the fact that CMB lensing is not affected by the same systematics as galaxy lensing. 
The SNR does not take into account nuisance parameters and their priors. AS we show later, CMB lensing from S4 basically replaces a prior on the shear biases.
We discuss this in more detail in the next section.

We also note that CMB lensing and galaxy lensing are relatively well matched in terms of SNR: the SNR in CMB lensing auto-correlation is $75\%$ of the SNR in cosmic shear. This justifies combining the two, and drives the calibration of the shear multiplicative biases.

\begin{table}[H]
\begin{center}
\begin{tabular*}{0.7\textwidth}{@{\extracolsep{\fill}}| l l |}
\hline
LSST \& CMB S4 & SNR \\  
\hline
\hline
\multicolumn{2}{|c|}{\textbf{Individual probes}} \\
clustering ($gg$) & 377 \\
\hline
galaxy-galaxy lensing: $g \kappa_\text{gal}$&276 \\
galaxy-CMB lensing: $g \kappa_\text{CMB}$&154 (56\% of $g \kappa_\text{gal}$) \\
\hline
shear tomography: $\kappa_\text{gal} \kappa_\text{gal}$ &532 \\
CMB lensing auto: $\kappa_\text{CMB} \kappa_\text{CMB}$&401  (75\% of $\kappa_\text{gal} \kappa_\text{gal}$) \\
galaxy lensing-CMB lensing: $\kappa_\text{gal} \kappa_\text{CMB}$&370 \\
\hline
\hline
\multicolumn{2}{|c|}{\textbf{Combinations}} \\
LSST: $gg, g\kappa_\text{gal}, \kappa_\text{gal}\kappa_\text{gal}$&620 \\
Combination 1: $\kappa_\text{CMB}\kappa_\text{CMB}, \kappa_\text{CMB}\kappa_\text{gal}, \kappa_\text{gal}\kappa_\text{gal}$&647 \\
Combination 2: $gg, g\kappa_\text{CMB}, g\kappa_\text{gal}$&403 \\
Full: $gg, g\kappa_\text{CMB}, g\kappa_\text{gal}, \kappa_\text{CMB}\kappa_\text{CMB}, \kappa_\text{CMB}\kappa_\text{gal}, \kappa_\text{gal}\kappa_\text{gal}$ & 718 (16\% more than LSST alone)\\
\hline
\end{tabular*}
\end{center}
\caption{
Individual and combined signal-to-noise ratios (SNR), giving insight on the statistical weight of each probe included in the joint analysis. All probes will be measured at high significance with LSST and CMB S4. 
The SNR in cosmic shear is higher than in galaxy-galaxy lensing and clustering, 
which is due to our very conservative choice of tracer sample.
%
CMB lensing from CMB S4 adds a small contribution to the total statistical significance, but will be important in breaking degeneracies and calibrating the shear multiplicative bias.
The SNR gives an idea of the relative statistical weight of the various observables. However, it doesn't take into account the presence of nuisance parameters and their priors. As we show later, CMB lensing from S4 basically replaces a prior on the shear multiplicative biases.}
\label{tab:snr_lsst}
\end{table}

\section{Shear calibration for LSST: requirements, self-calibration with and without CMB S4 lensing}
\label{sec:shear_reqs_calib}

\subsection{LSST requirements and self-calibration}
\label{subsec:lsst_alone}

In this subsection, we revisit the shear multiplicative bias requirements for LSST \cite{2013MNRAS.429..661M, Hutereretal:06}. We assess the degradation in cosmological parameters as a function of the prior on the shear biases $m_i$, while jointly fitting for cosmological parameters, galaxy biases and photo-z uncertainties. 

\begin{figure}[h]
\centering
\includegraphics[width=0.49\columnwidth]{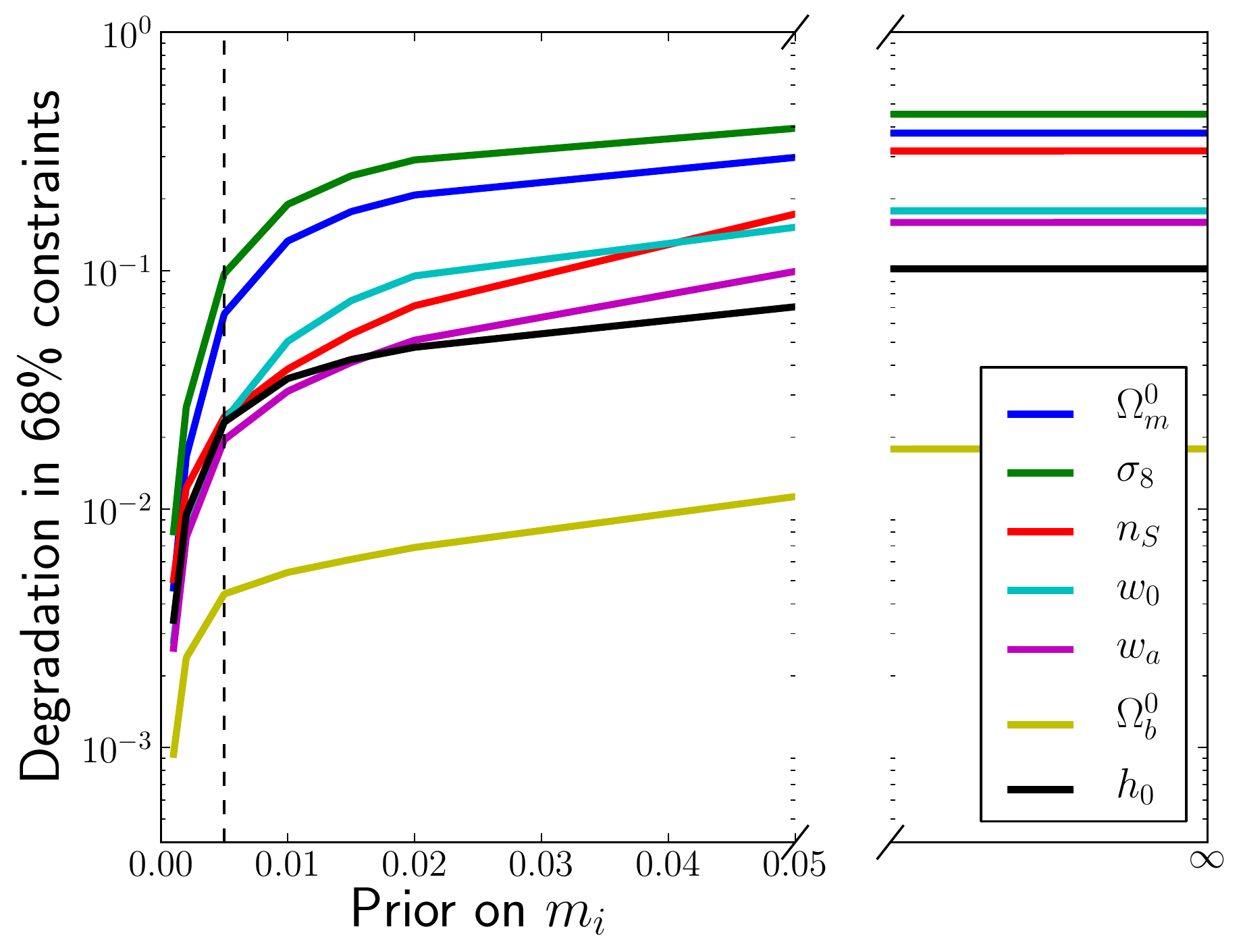}
\includegraphics[width=0.49\columnwidth]{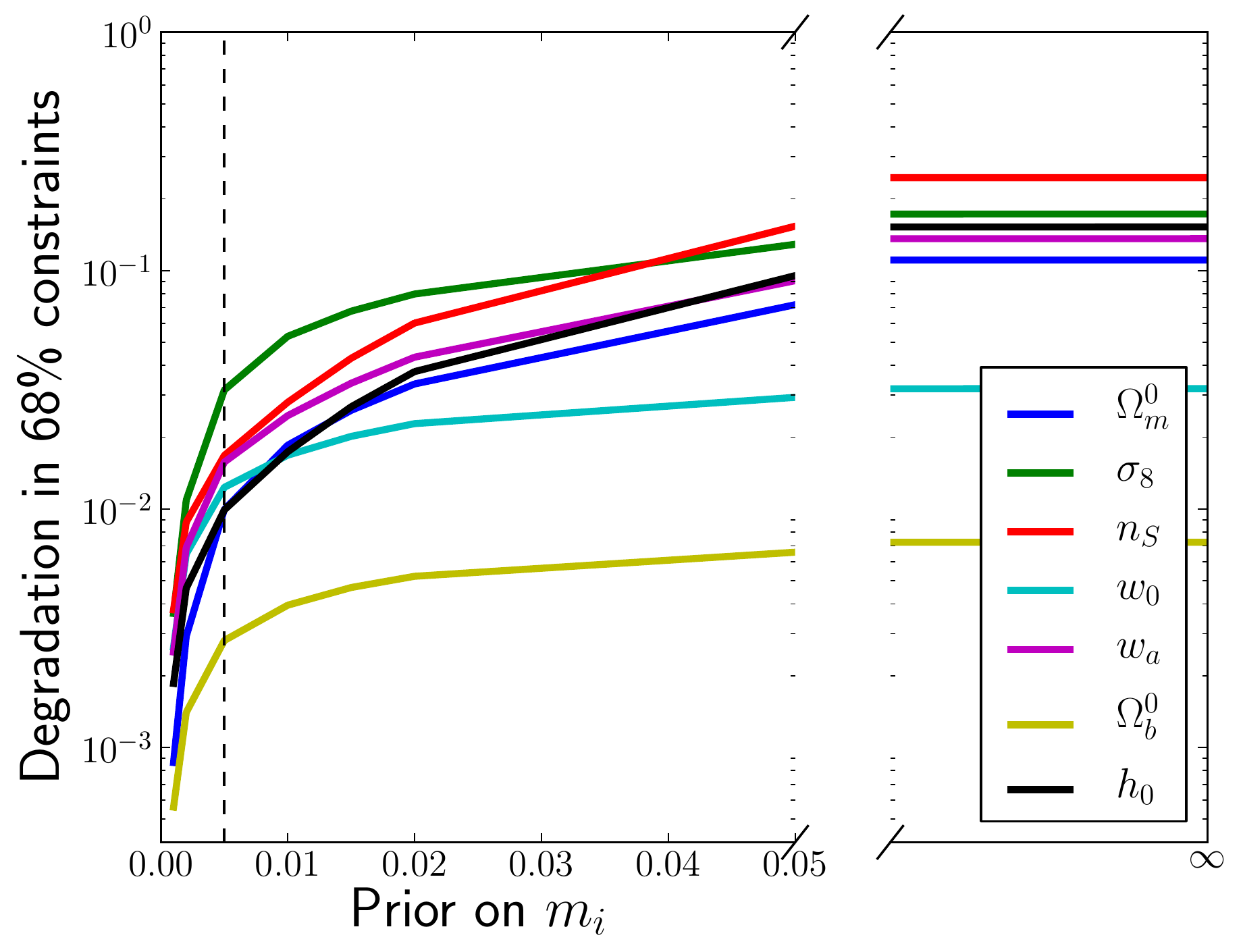}
\caption{
Degradation in cosmological parameter constraints, relative to a perfect shear calibration, as a function of shear bias prior.
A shear bias prior of $0$ corresponds to perfect shear calibration, and to $0$ degradation in cosmological constraints. A shear bias prior of $\infty$ amounts to completely relaxing any shear prior, i.e. self-calibrating the shear from the data. A degradation of $1$ corresponds to a $100\%$ increase in cosmological parameter error bars, compared to the case with perfect shear calibration.\\
\textbf{Left panel}: Degradation in cosmological parameter constraints for cosmic shear alone, from LSST.\\
\textbf{Right panel:} Degradation in cosmological parameter constraints for LSST, when combining cosmic shear, galaxy-galaxy lensing and clustering.
}
\label{fig:requirements}
\end{figure}

Fig.~\ref{fig:requirements} shows the degradation in cosmological parameters as a function of the prior on $m_i$. The left panel shows the case of cosmic shear alone, while the right panel shows the LSST combination (cosmic shear, galaxy-galaxy lensing, clustering). In both cases, we vary cosmological parameters as well as all the nuisance parameters in Tab.~\ref{tab:params_mcmc}.
For cosmic shear alone (left panel), we find that a prior on $m_i$ of 0.005 produces a $10\%$ degradation in cosmological parameters compared to a perfect shear calibration. This degradation is somewhat smaller than the $30-70\%$ found in \cite{2006MNRAS.366..101H} (in their Fig.~4), likely due to our marginalization of photo-z uncertainties.
Furthermore, \cite{2006MNRAS.366..101H} and \cite{2013MNRAS.429..661M} quote requirement for LSST shear calibration similar to our value of $0.5\%$, and we will therefore retain this value in the rest of the paper.
The left panel of Fig.~\ref{fig:summary_m} also shows that self-calibration of the shear is possible from cosmic shear alone.
This can be understood intuitively as follows. The dependence of cosmic shear in the $m_i$ is purely multiplicative on all scales, whereas the dependence in other cosmological and nuisance parameters changes with scale: for instance, the cosmic shear power spectrum scales as $\sig^2$ in the linear regime, and as $\sig^3$ in the non-linear regime (see Fig.~\ref{fig:param_dpdces} in App.~\ref{app:param_dpdce}). By including both large and small scales, the degeneracy with the shear bias can be broken.
We find a degradation of up to $50\%$ in cosmological parameters when relaxing completely the priors on shear calibration. This is again more optimistic compared to \cite{2006MNRAS.366..101H}, who found a factor of 2 degradation, and is again likely due to our marginalization of photo-z uncertainties.

We reproduce this analysis in the case of the full LSST (cosmic shear, galaxy-galaxy lensing, clustering), shown in the right panel of Fig.~\ref{fig:requirements}. Since clustering and galaxy-galaxy lensing have different dependences on the shear biases (if any), adding them allows for a better self-calibration of the shear, and a reduction in the degradation in cosmological parameters. Indeed, a $0.5\%$ prior on the $m_i$ only leads to at most a $3\%$ degradation in cosmological parameters. Completely relaxing shear bias priors only leads to a $25\%$ degradation at most.

In the absence of shear priors, the self-calibration of the shear is better than $2\%$ for cosmic shear alone, and better than $1.5\%$ for LSST, in most of the redshift range. This is shown in Fig.~\ref{fig:summary_m}, left panel.

Note that all these results assumed photo-z uncertainties as in Tab.~\ref{tab:params_mcmc}. We vary these photo-z assumptions in Sec.~\ref{subsec:photoz}.
Note also that this self-calibration from LSST alone relies on the comparison between large scales and small scales. It is therefore somewhat susceptible to non-linearities and baryonic effects in the matter power spectrum. 

\subsection{Approaching/surpassing the LSST requirements with CMB S4}
\label{subsec:lsst_cmbs4}

\begin{figure}[h]
\centering
\includegraphics[width=0.49\columnwidth]{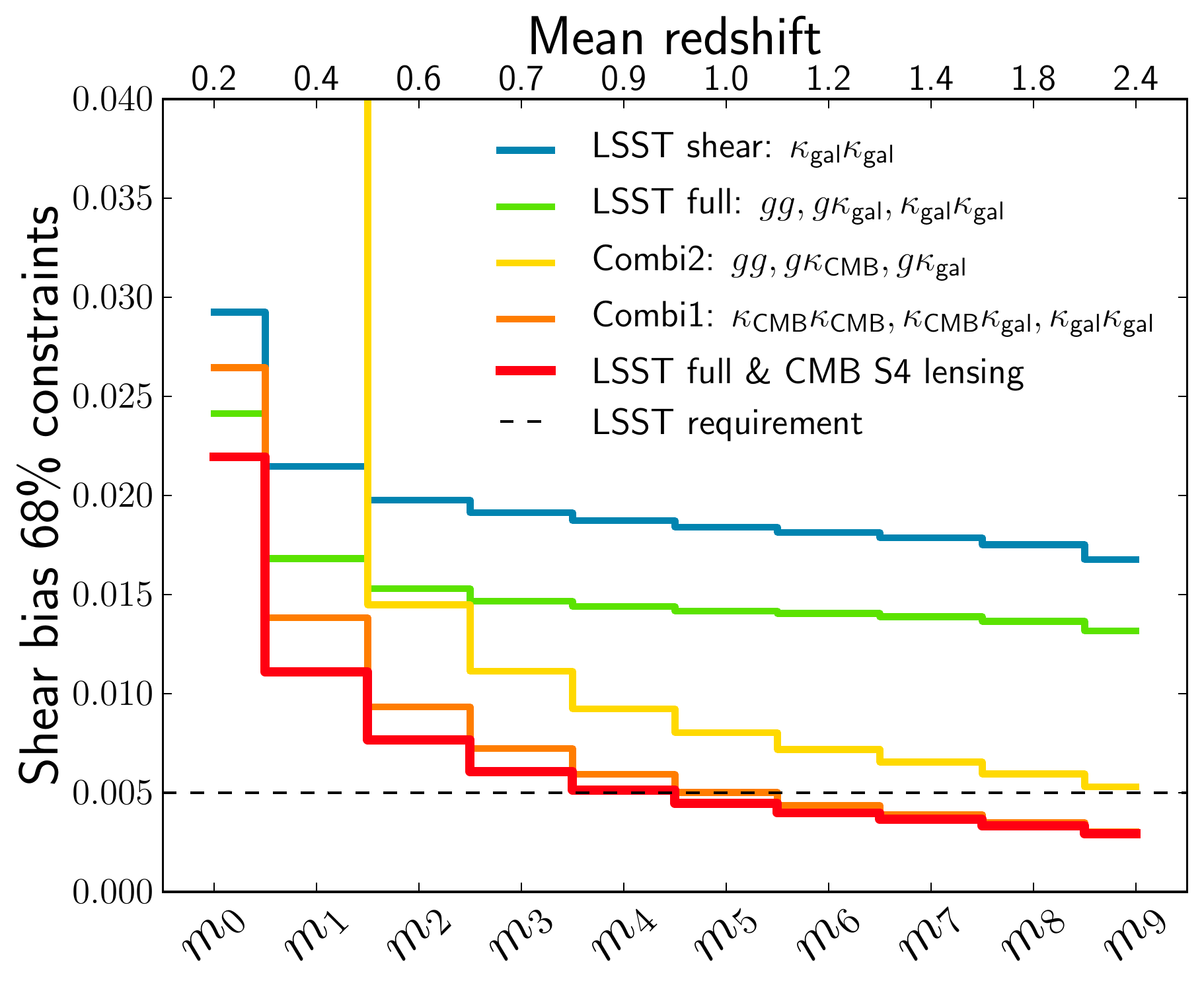}
\includegraphics[width=0.49\columnwidth]{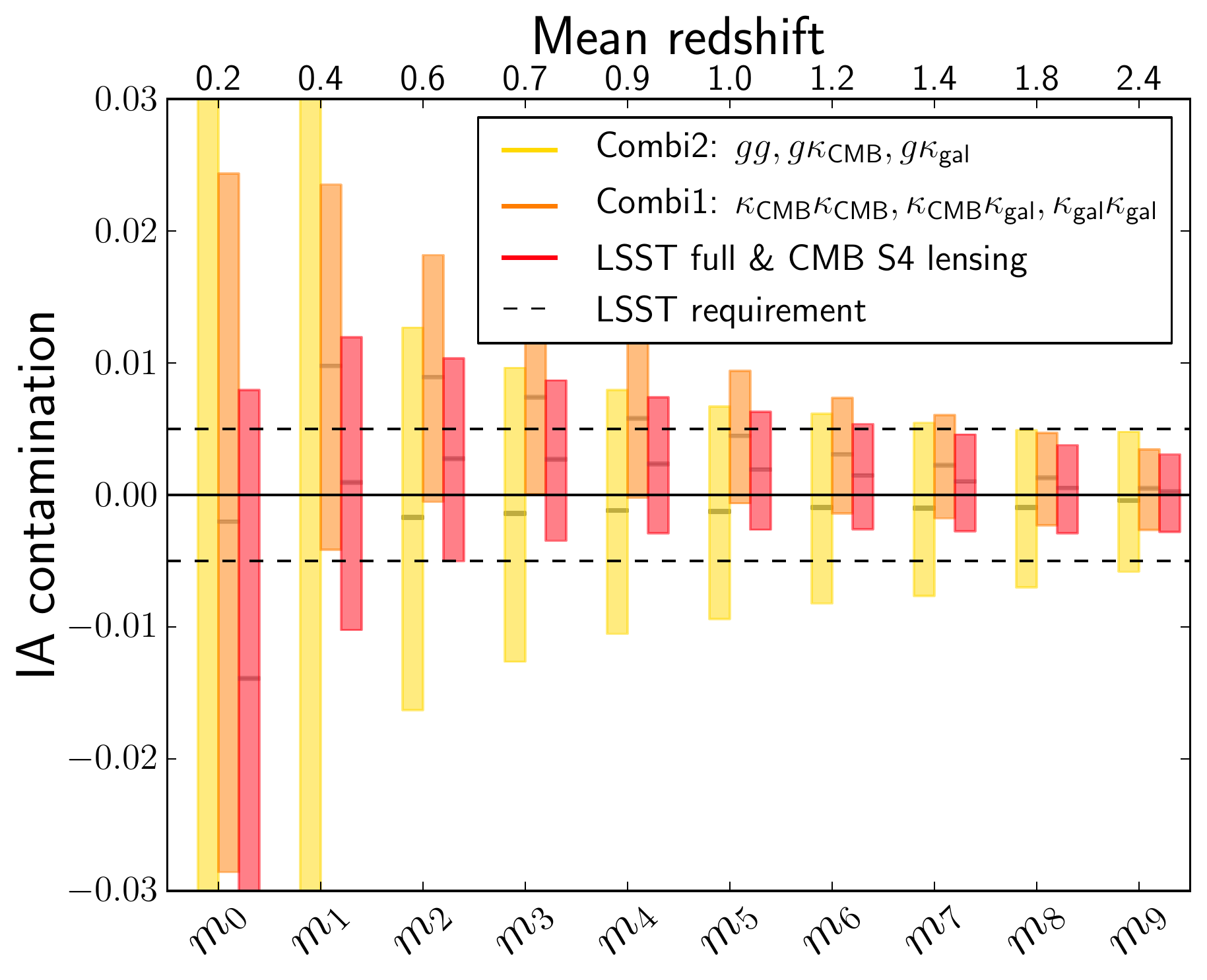}
\caption{
\textbf{Left panel:} $68\%$ confidence constraints on the shear biases $m_i$ for LSST, when self-calibrating them with LSST cosmic shear alone (blue), LSST full (i.e. clustering, galaxy-galaxy lensing and cosmic shear; green), combination 1 (orange), combination 2 (yellow) and the full LSST \& CMB S4 lensing (red). The self-calibration works down to the level of LSST requirements (dashed lines) for the highest redshift bins, where shear calibration is otherwise most difficult. We stress that all the solid lines correspond to self-calibration from the data alone, without relying on image simulations. Calibration from image simulations is expected to meet the LSST requirements, and CMB lensing will thus provide a valuable consistency check for building confidence in the results from LSST.\\
\textbf{Right panel:} Impact of unaccounted intrinsic alignments (see Sec.~\ref{subsec:syst_nuisance_params}). The lines show the bias in the self-calibrated value of $m_i$, and the colored bands show the $68\%$ confidence constraints, corresponding to the curves in the left panel. 
Intrinsic alignments produce a bias in the shear calibration, but not beyond the $68\%$ confidence region.
}
\label{fig:summary_m}
\end{figure}

Having examined the impact of shear calibration on cosmology for LSST alone, we now add CMB lensing data from CMB S4. We wish to identify the combination of observables that best constrains the shear biases $m_i$, taking into account statistical and systematic errors. To do so, we consider separately combinations 1 and 2 (Eq.~\ref{eq:combi12}), as well as all the two-point functions of LSST and CMB S4 lensing (``full LSST \& CMB S4 lensing'', Eq.~\ref{eq:combi_full}). We compare the resulting level of shear self-calibration to the case of LSST cosmic shear alone (``LSST shear''), and the combination of all two-point functions from LSST (``LSST full'', including clustering, galaxy-galaxy lensing and cosmic shear).
The shear bias constraints are shown in Fig.~\ref{fig:summary_m}.

As shown in Tab.~\ref{tab:snr_lsst}, the statistical SNR is higher for combination 1 than combination 2, which explains the slightly better constraints on the shear bias seen in Fig.~\ref{fig:summary_m}. 
As explained earlier, this is a consequence of our conservative tracer sample.
We also notice that the shear biases $m_0$ and $m_1$ of the first two tomographic bins are not constrained by combination 2. This is because we only include tracer-lensing correlations when the entire source bin is at higher redshift than the tracer bin.

In the case of the full LSST \& CMB S4 lensing, we find that the shear biases are constrained down to $0.3\%$ for the highest redshift bins and $2\%$ for the lowest redshift bins. 
For higher source bins, the lensing efficiency kernels for $\kappa_\text{gal}$ and $\kappa_\text{CMB}$ overlap more, leading to a larger signal, correlation coefficient and signal to noise (the SNR for $\kappa_\text{gal} \kappa_\text{CMB}$ goes from 15 for the lowest redshift bin to 330 for the highest redshift bin).
Thus CMB lensing from CMB S4 can approach and surpass the LSST requirements for shear calibration for most of the tomographic bins.

We assess the impact of intrinsic alignments by including our IA model into the data vector, and not accounting for it in the fit. This allows to estimate the size of the bias due to IA. Again, we do not account for catastrophic failures in the photo-z and restrict ourselves to Gaussian photo-z uncertainties.
The right panel of Fig.~\ref{fig:summary_m} shows that combination 2 is less affected by intrinsic alignments than combination 1.
This was expected, since combination 2 involves tracer-lensing correlations: it is completely exempt of IA in the case of perfect photo-z, and little affected in the case of Gaussian photo-z. Furthermore, we selected a very conservative tracer sample, for which excellent photo-z should be achievable. So combination 2 should be very robust to intrinsic alignment contamination.
In contrast, shear tomography is affected by the so-called GI-term for the inter-bin correlations, and by the II-term for the bin auto-correlations. Similarly, galaxy lensing-CMB lensing is affected by the GI-term \cite{2014MNRAS.443L.119H, 2015MNRAS.453..682C, 2014PhRvD..89f3528T}. This contamination is present even with perfect photo-z. However, IA only bias the shear calibration from combination 2 within the $68\%$ confidence region, for our reasonable IA model. Furthermore, no IA mitigation technique was applied here, suggesting that this forecast is certainly conservative.
This result was not obvious \textit{a priori}. Indeed, at low redshift, IA contaminate the power spectrum by up to $5-10\%$ for $\kappa_\text{gal}\kappa_\text{gal}$ and $\kappa_\text{gal}\kappa_\text{CMB}$, in agreement with \cite{2015MNRAS.453..682C, 2014MNRAS.443L.119H, 2015SSRv..193...67K, 2015SSRv..193..139K, 2015SSRv..193....1J, 2015PhR...558....1T}. However, this bias becomes much smaller at higher redshift (the IA signal is roughly constant with redshift, while the lensing signal increases), and the higher redshift power spectra are measured with the best signal to noise. As a result, the overall contamination due to IA is small.
Finally, when combining the full LSST \& CMB S4, the overall impact of intrinsic alignments is negligible for the shear bias constraints.

\begin{figure}[h]
\centering
\includegraphics[width=0.49\columnwidth]{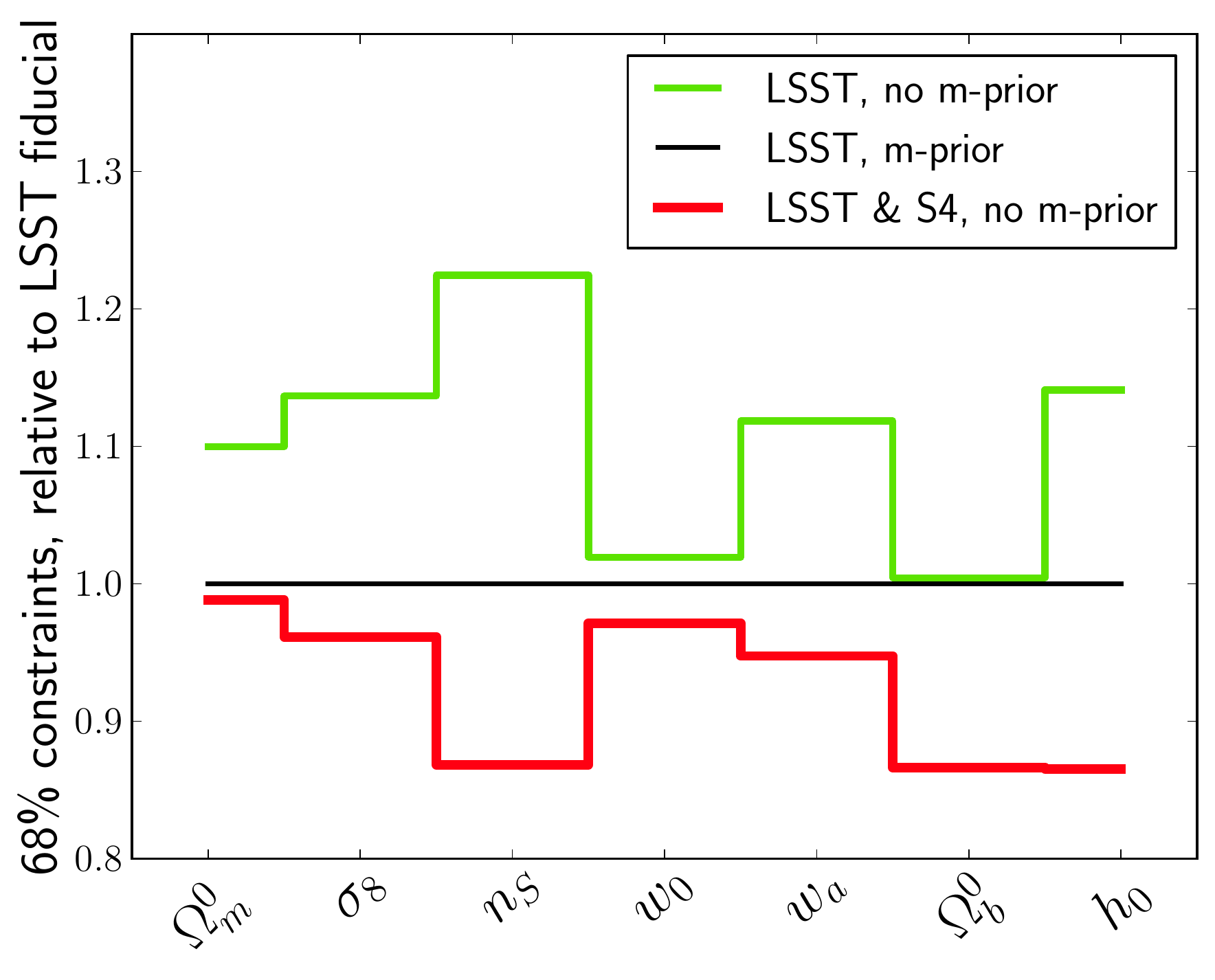}
\caption{
Radius of the $68\%$ confidence region for cosmological parameters, relative to the LSST fiducial forecast (with priors on the shear biases of $0.5\%$). The black line shows the fiducial LSST forecast, and the green line shows the degradation in cosmological constraints when relaxing the shear bias priors. The red line shows the forecast for LSST \& CMB S4 lensing, when relaxing the shear bias priors.
Thus adding CMB lensing from S4 successfully replaces a prior on the shear bias, and even surpasses the constraining power of LSST alone with shear priors at the level of the LSST requirements.
A contour plot of the cosmological constraints for LSST \& CMB S4 is shown in App.~\ref{app:param_dpdce}.
}
\label{fig:summary_cosmo}
\end{figure}

In conclusion, we find that CMB lensing from CMB S4 can constrain the shear multiplicative bias down to or beyond LSST requirements, while jointly fitting for cosmology, galaxy biases and photo-z uncertainties, and in the presence of reasonable intrinsic alignments. The method works best at higher redshift, where shear calibration is expected to be most difficult otherwise. This result is extremely encouraging, and is an example of synergy between Stage 4 surveys, where multi-probe analyses lead to a dramatic improvement in control over systematics.
In Fig.~\ref{fig:summary_cosmo}, we show that including CMB lensing from CMB S4 can successfully replace a prior on the shear biases $m_i$, and even improves the cosmological constraints over LSST alone with realistic priors on $m_i$. 
This forecast is conservative, as we include the non-Gaussian covariances, discard the small scales in clustering and galaxy-galaxy lensing and marginalize over galaxy bias, photo-z uncertainties and cosmological parameters. 
It is also robust to intrinsic alignments and the assumptions for the CMB S4 specifications (see Sec.~\ref{subsec:varying_cmbs4}) as well as photo-z priors (see Sec.~\ref{subsec:photoz}).


\section{Impact of CMB S4 design and lensing systematics; application to space-based lensing surveys}
\label{sec:cmbs4_syst_space}

\subsection{Importance of sensitivity and resolution for CMB S4}
\label{subsec:varying_cmbs4}

\begin{figure}[h]
\centering
\includegraphics[width=0.32\columnwidth]{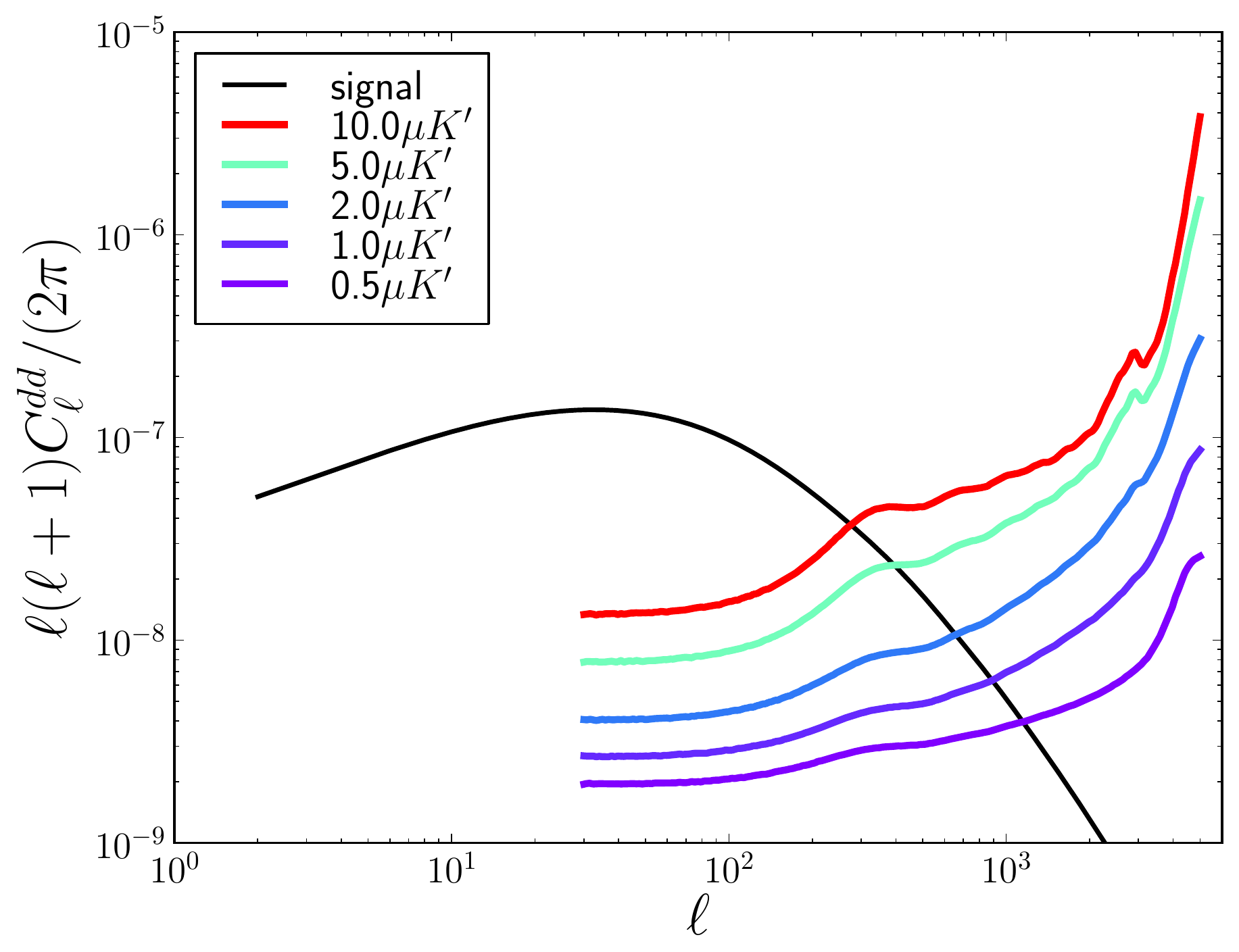}
\includegraphics[width=0.32\columnwidth]{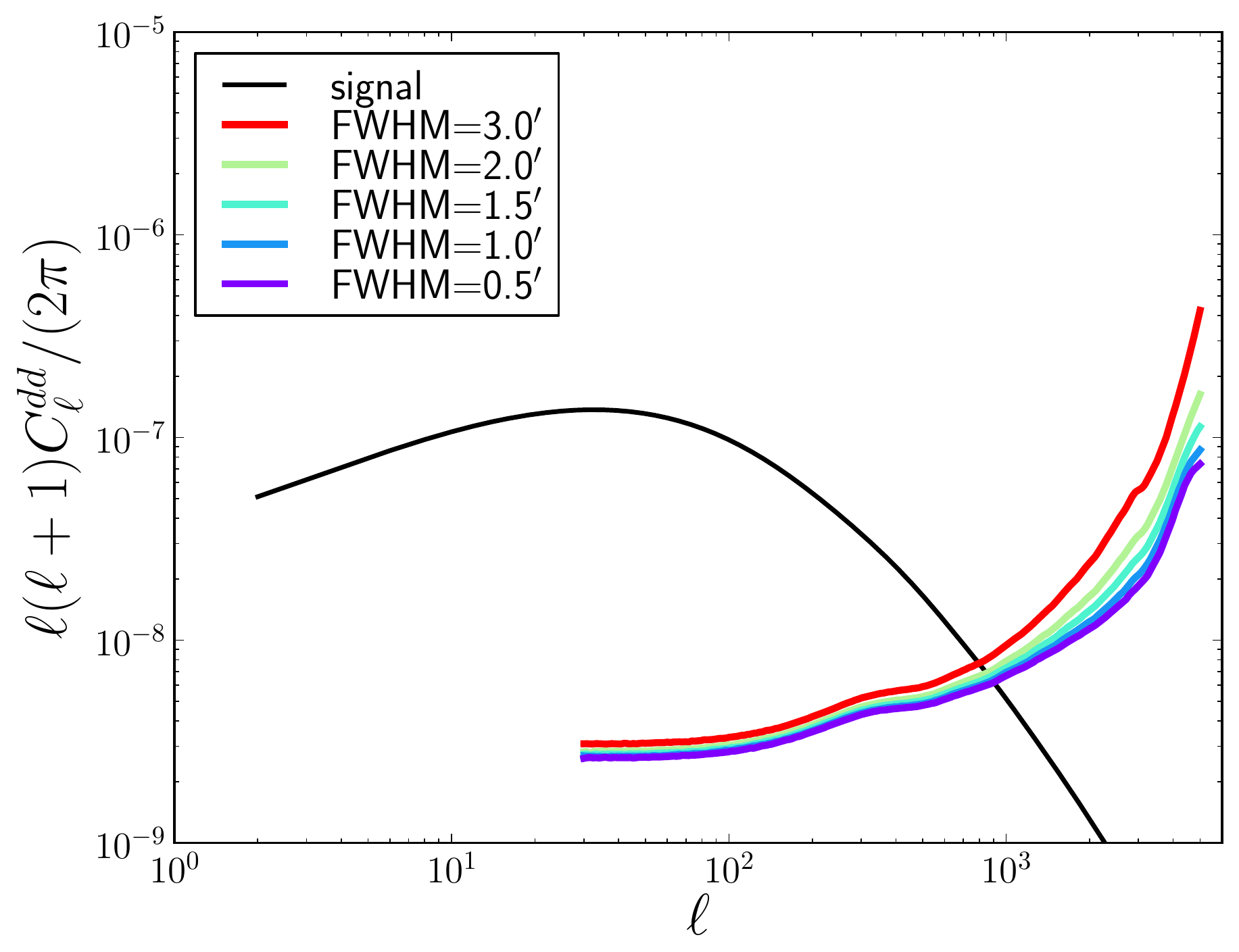}
\includegraphics[width=0.32\columnwidth]{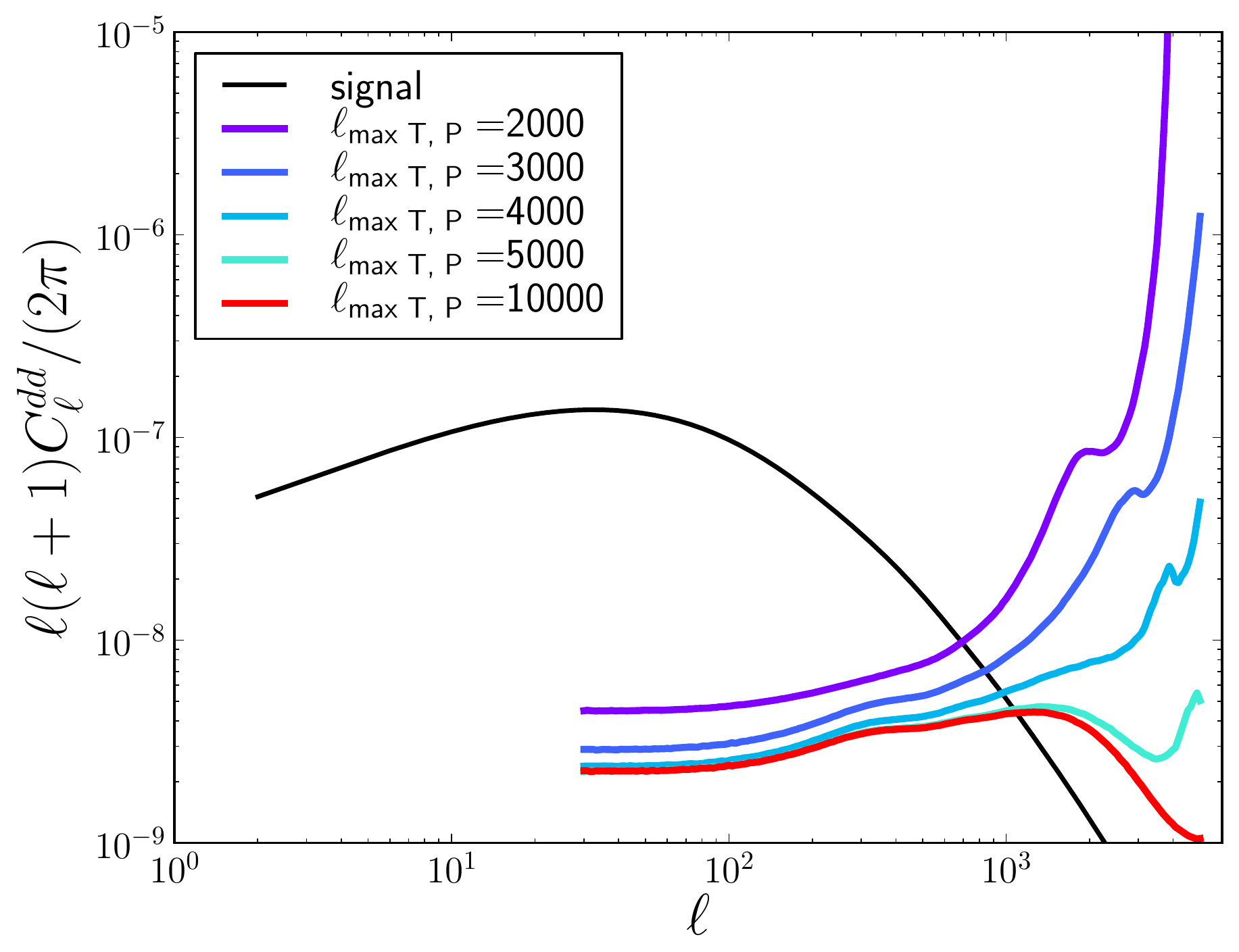}
\includegraphics[width=0.32\columnwidth]{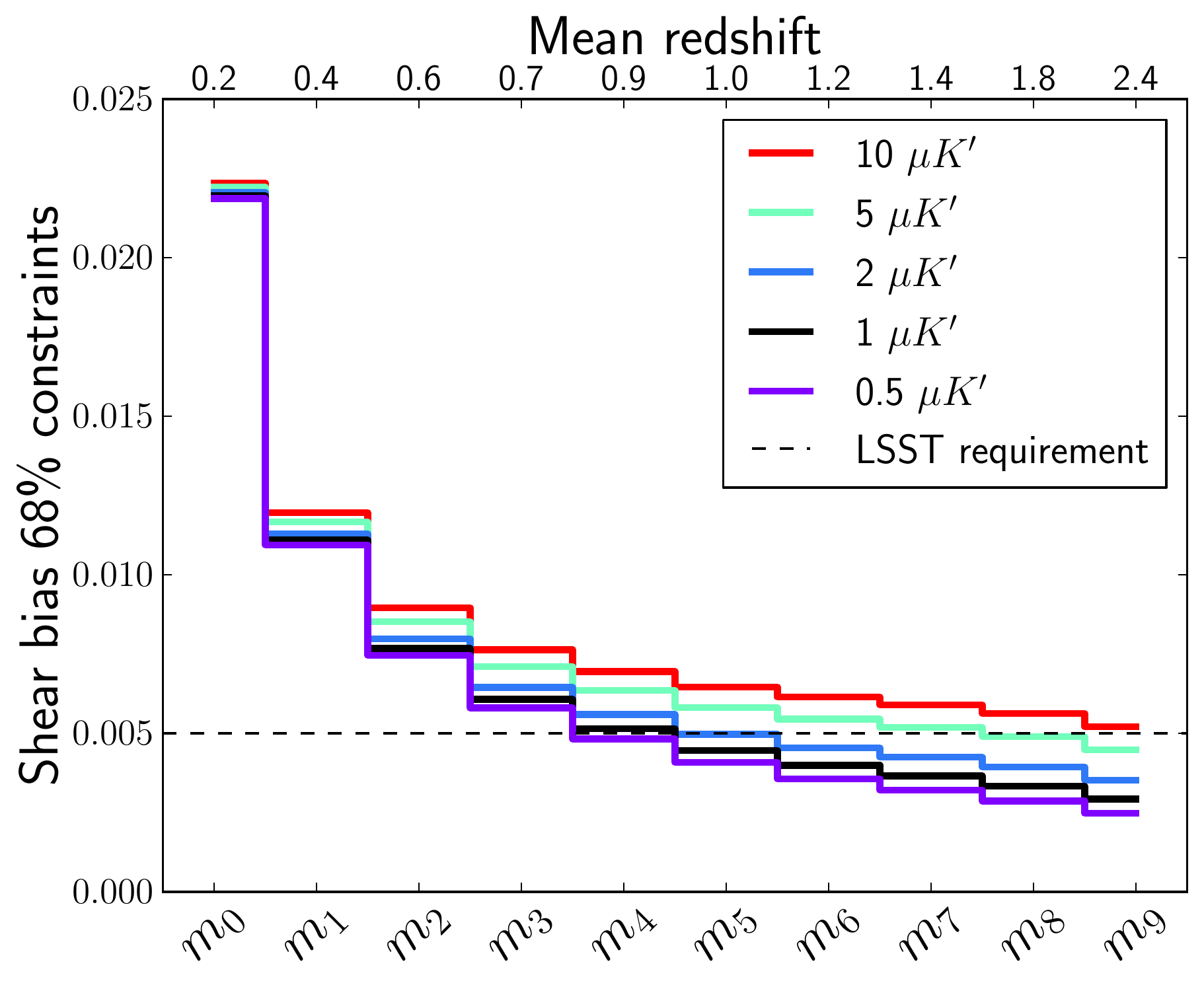}
\includegraphics[width=0.32\columnwidth]{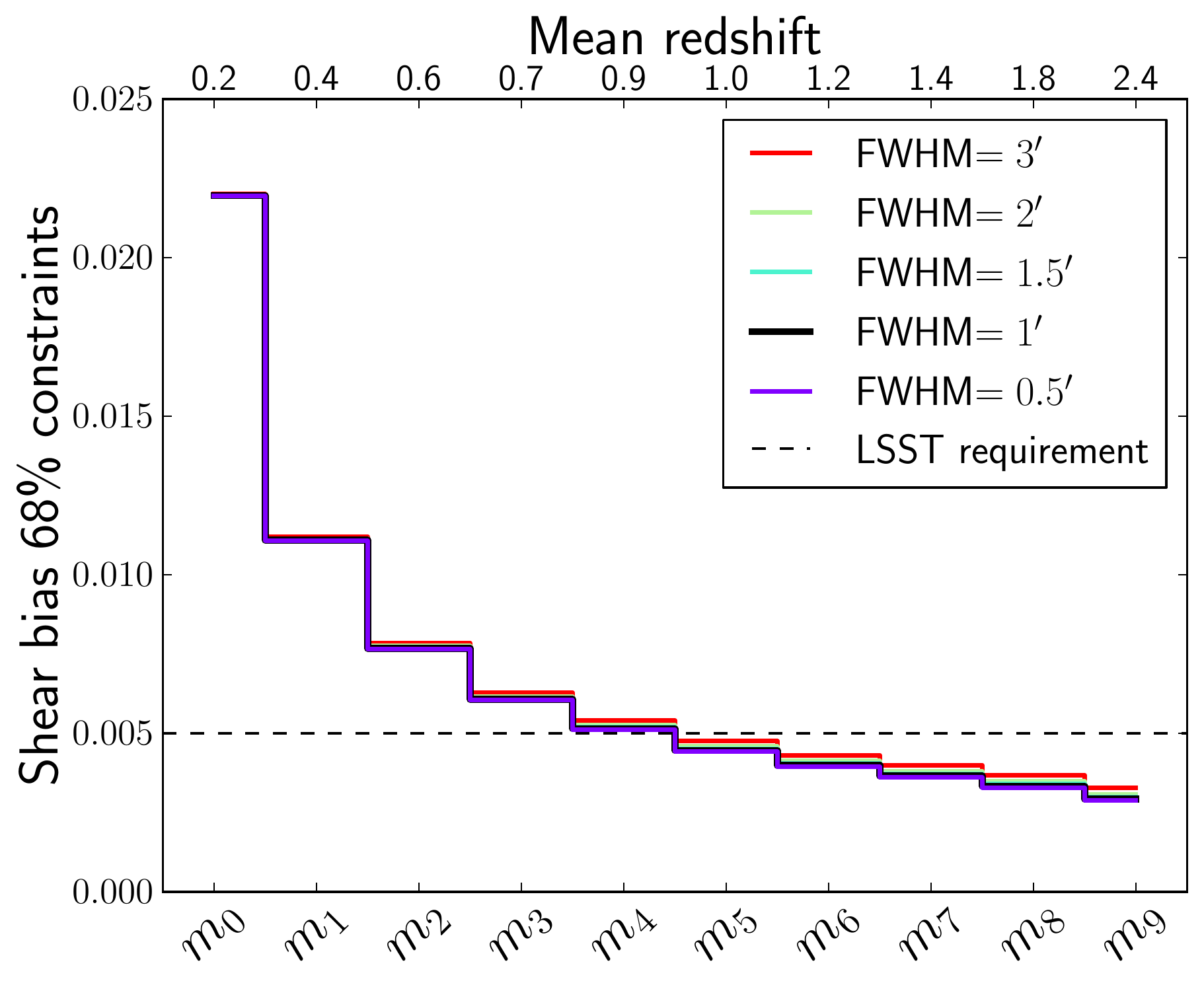}
\includegraphics[width=0.32\columnwidth]{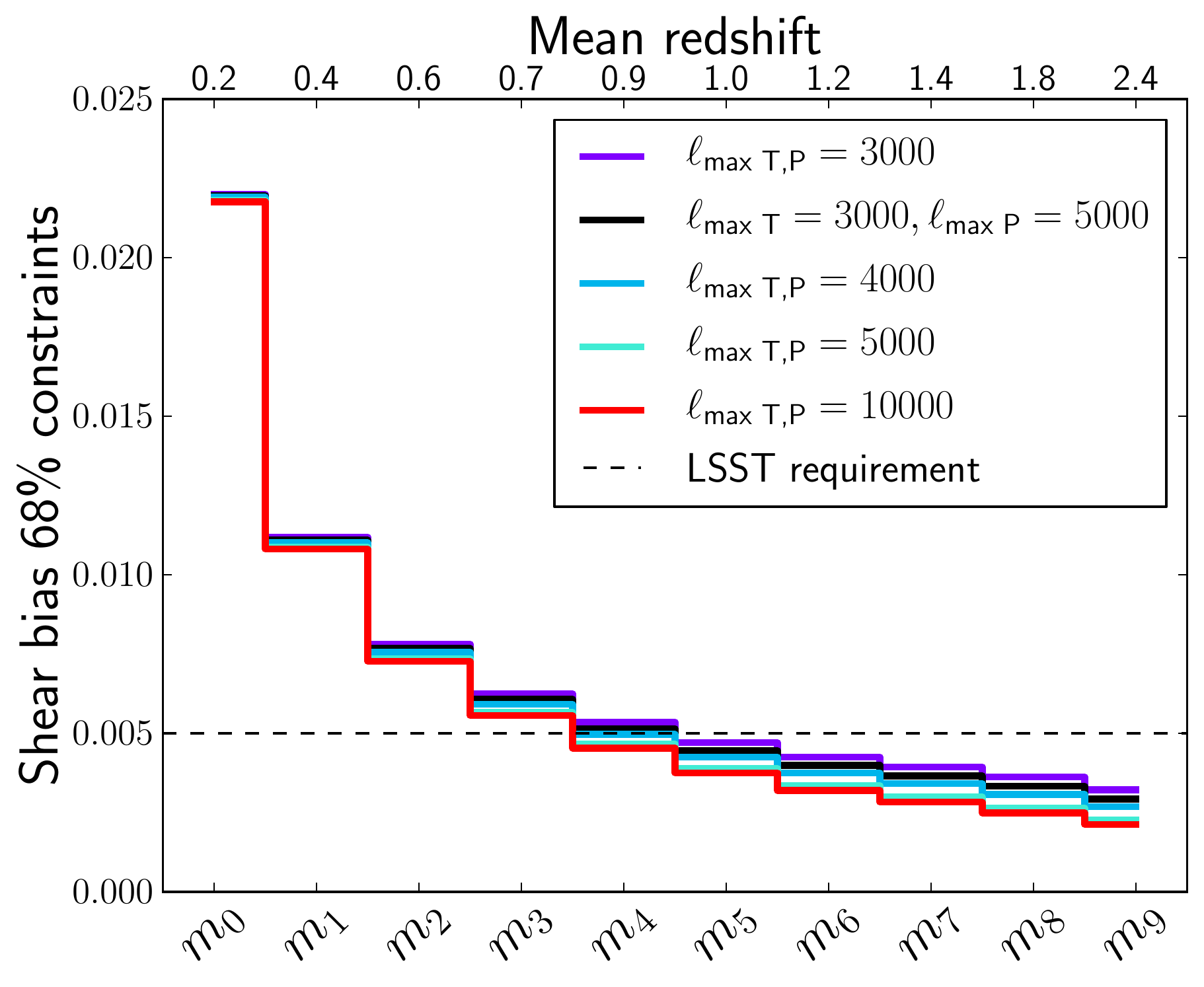}
\caption{
Impact of varying CMB S4 specifications on the CMB lensing reconstruction noise and the shear calibration for LSST. \\
\textbf{Top row:} the CMB lensing power spectrum (black line) is compared to the reconstruction noise per $\ell$-mode (colored lines) for different values of sensitivity (left), beam FWHM (center) and maximum multipole where a foreground-cleaned CMB map is available (right). When varying the sensitivity (left), we quote the white noise level in temperature, and use a $\sqrt{2}$ times larger value for $E$ and $B$ polarizations. When varying the maximum multipole (right), we assume $l_\text{max T} = l_\text{max P}$.\\
\textbf{Bottom row:} the shear calibration level for the fiducial CMBS4 (black solid line) is compared to the one obtained for each variation (solid colored lines). The LSST requirement is shown as the black dashed lines.\\
The shear calibration is affected by sensitivity, but is relatively insensitive to the beam and maximum multipole available. This is encouraging, and suggests that AdvACT \cite{2016JLTP..tmp..144H} and SPT-3G will already be useful for calibrating the shear from LSST.
}
\label{fig:cmbs4_vary_noise_beam_lmax}
\end{figure}

In this subsection, we quantify the impact of CMB S4 sensitivity, resolution and component separation on shear calibration. The top row of Fig.~\ref{fig:cmbs4_vary_noise_beam_lmax} shows the reconstruction noise on the CMB lensing convergence $\kappa_\text{CMB}$ as a function of sensitivity in temperature (assumed $\sqrt{2}$ times smaller than in polarization; left panel), beam FWHM (central panel) and maximum multipole included in the analysis $\ell_\text{max T,P}$ (parameterizing the effectiveness of component separation; right panel). When one parameter is varied, the others remain fixed to their fiducial values from Fig.~\ref{fig:specs_cmbs4}. Note that in all cases, the survey area is kept fixed at $18,000$ deg$^2$ ($f_\text{sky}=44\%$).
The bottom row of Fig.~\ref{fig:cmbs4_vary_noise_beam_lmax} shows the corresponding constraints on shear biases $m_i$ for each configuration. 

The shear calibration improves slowly with sensitivity, by a factor of $\sim 2$ when the noise varies from $10$ to $0.5 \mu K '$. 
This is understandable since the CMB lensing signal falls off quickly at high $\ell$, and therefore a significant reduction in reconstruction noise is needed to image higher $\ell$ lensing modes. For the same reason, we expect iterative lensing reconstruction methods \cite{2003PhRvD..67d3001H, 2003PhRvD..68h3002H} to only improve shear calibration by a few tens of percent.

For our choice of fiducial $\ell$-limits ($\ell_\text{max}=3000$ for T; $\ell_\text{max}=5000$ for E,B), set by foreground cleaning, varying the beam FWHM between $0.5^\prime$ and $3^\prime$ has basically no impact on the shear calibration: a higher resolution experiment can image higher $\ell$-modes, but we are discarding these small scales to avoid foreground contamination. 

More realistically, a higher resolution experiment might perform better at component separation and allow to use higher temperature and polarization multipoles. However, for our fiducial parameters, we find that varying $\ell_\text{max T, P}$ between $2,000$ and $10,000$ only changes the shear calibration by about $25\%$. 

This is encouraging and shows that upcoming third generation experiments such as Advanced ACT (AdvACT, $1.4^\prime$ resolution, $\sim 10\mu K^\prime$ sensitivity on half of the sky) \cite{2016JLTP..tmp..144H} and SPT-3G ($1^\prime$ resolution, $2.5 \mu K^\prime$ sensitivity on $2,500$ deg$^2$) \cite{2014SPIE.9153E..1PB} can already calibrate the shear from LSST. This calibration will be less precise than from CMB S4, but already at a useful level.
The amount of overlap of AdvACT and SPT-3G with LSST may evolve in the future, and will affect the shear calibration.

\subsection{Sensitivity to photometric redshift uncertainties}
\label{subsec:photoz}

\begin{figure}[h]
\centering
\includegraphics[width=0.49\columnwidth]{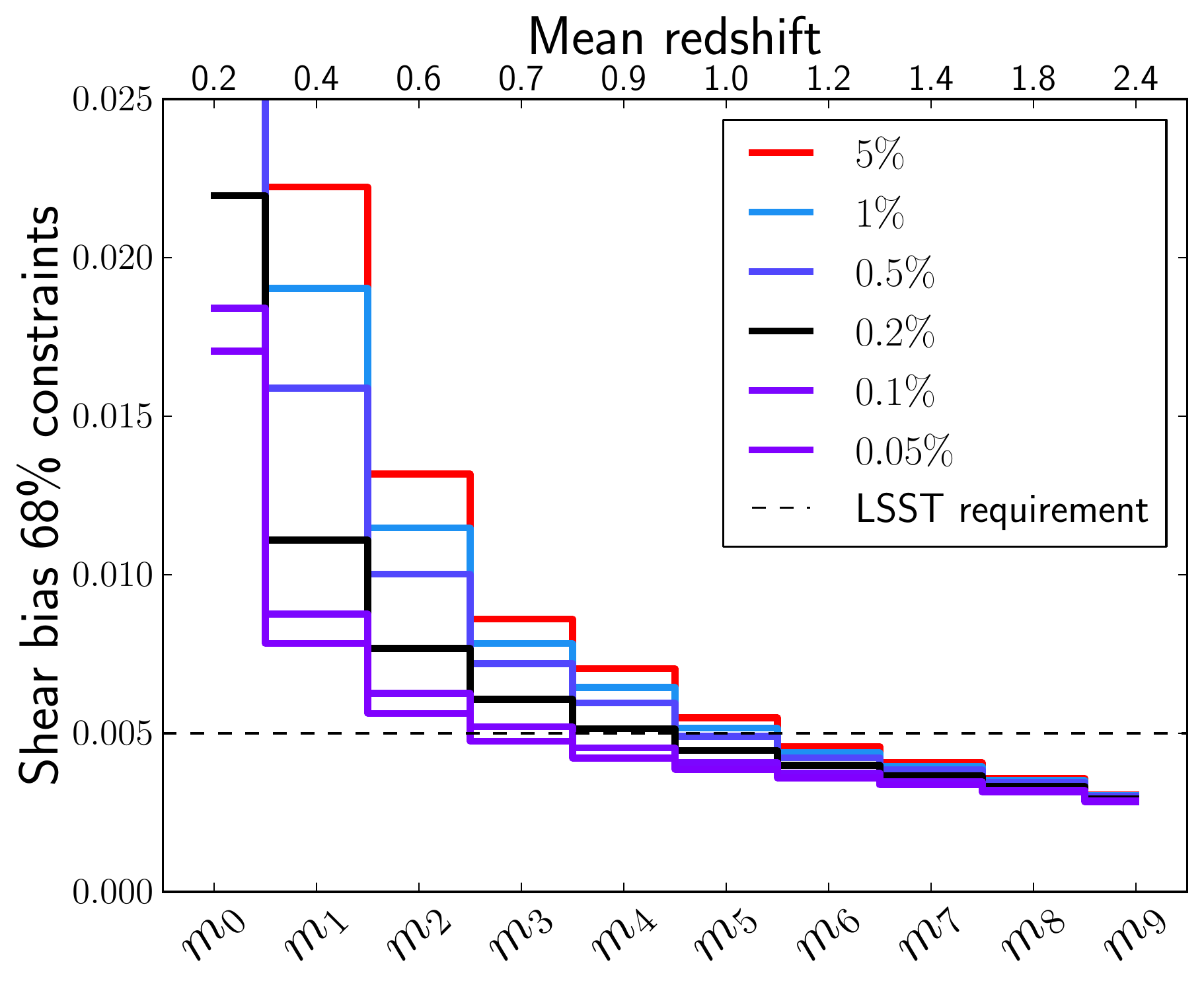}
\includegraphics[width=0.49\columnwidth]{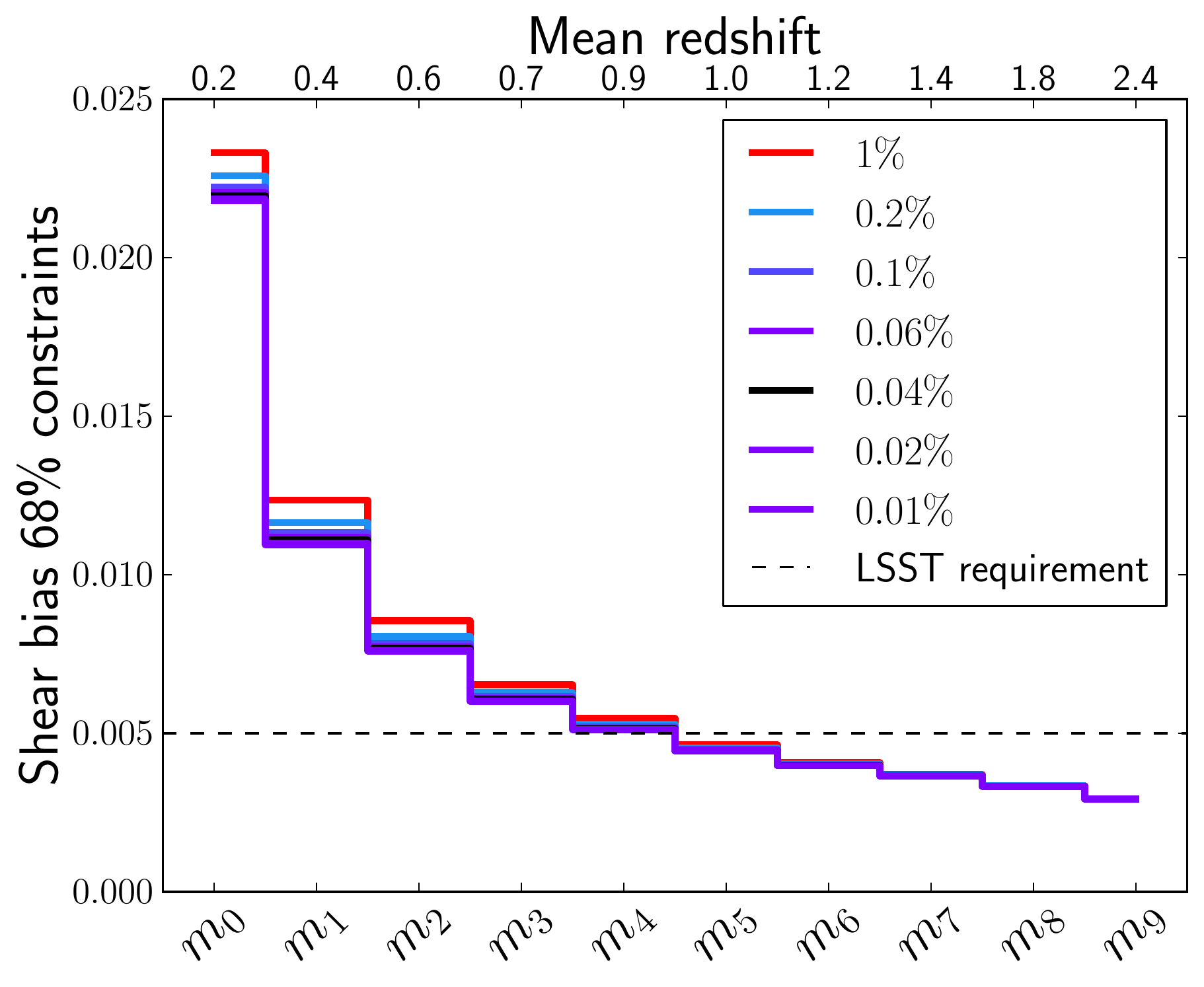}
\caption{
\textbf{Left panel:} level of shear calibration for LSST with CMB S4 lensing, when varying the source photo-z priors.
The black line corresponds to the fiducial priors, and the colored lines are labelled with ``$x\%$'' such that
the prior on $\Delta_{z, \text{source}, i}$ is Gauss(0, $x\%$)
and the prior on $\sigma_{z, \text{source}} / (1+z)$ is Gauss(0.05, $1.5\times x\%$).
The dependence is more important at low redshift, where a fixed absolute change in $z$ corresponds to a larger relative change in comoving distance.\\
\textbf{Right panel:} varying lens photo-z priors.
Similarly to the left panel, the black line corresponds to the fiducial priors, and the colored lines are labelled with ``$x\%$'' such that
the prior on $\Delta_{z, \text{lens}, i}$ is Gauss(0, $x\%$)
and the prior on $\sigma_{z, \text{lens}} / (1+z)$ is Gauss(0.01, $1.5\times x\%$).
The shear calibration is almost completely insensitive to the lens photo-z priors.
}
\label{fig:varying_sourcephotoz_m}
\end{figure}

In Sec.~\ref{subsec:lsst_cmbs4}, we showed that CMB S4 lensing can calibrate the shear from LSST, assuming that the photometric redshift uncertainties are under control. In this subsection, we ask how crucial this assumption is. We therefore vary the priors on source and lens photo-z uncertainties and re-run our forecast. The left panel of Fig.~\ref{fig:varying_sourcephotoz_m} shows that the shear calibration is mildly dependent on the source photo-z uncertainties. The dependence is higher at low redshift, where a fixed change in redshift corresponds to a larger relative change in comoving distance. The right panel of Fig.~\ref{fig:varying_sourcephotoz_m} shows that the shear calibration is very insensitive to the lens photo-z uncertainties. This is in large part because most of the constraining power comes from auto- and cross-correlations of the galaxy lensing and CMB lensing, which are not affected by the lens photo-z uncertainties.

These results are encouraging: they show that the shear calibration is robust to larger photometric redshift uncertainties.
Note however that  we have not taken into account catastrophic photo-z failures.

\subsection{Robustness to non-linearities and baryonic effects}
\label{subsec:varyinglmaxgks}

As pointed out earlier, our clustering and galaxy-galaxy lensing only uses large scales
$2\pi \chi(z_\text{mean})/l > 10$ Mpc/h, which correspond to $\ell_\text{max} = 420, 714, 930$ and 1212 respectively for the four lens bins. 
Thus, combination 2 (i.e. $gg$, $g\kappa_\text{gal}$, $g\kappa_\text{CMB}$) is only sensitive to scales greater than 10 Mpc/h.
This conservative cut makes linear biasing valid, and avoids systematic errors from halo occupation modeling, non-linearities and baryonic effects in the matter power spectrum.
 As a result, the shear calibration from combination 2 (see Fig.~\ref{fig:summary_m}) should be very robust to these effects.

\begin{figure}[h]
\centering
\includegraphics[width=0.55\columnwidth]{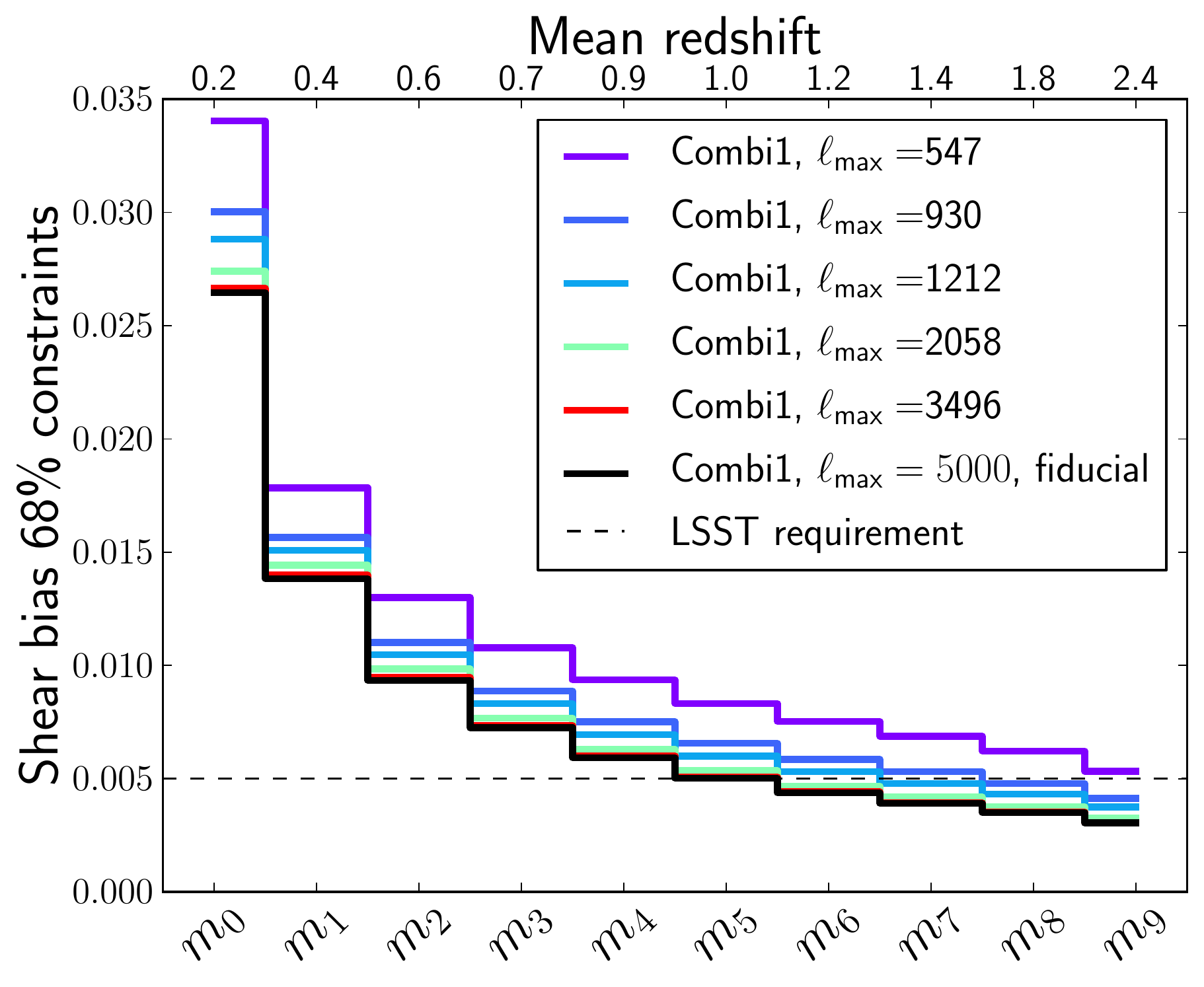}
\caption{
In this figure, we vary the maximum multipole included in the lensing-lensing correlations and compare the resulting shear calibrations from combination 1 (i.e. $\kappa_\text{gal}\kappa_\text{gal}$, $\kappa_\text{gal}\kappa_\text{CMB}$, $\kappa_\text{CMB}\kappa_\text{CMB}$). Between $\ell_\text{max}=5,000$ and $\ell_\text{max}= 930$, the shear calibration is only degraded by $10-40\%$.
Besides, the calibration from combination 2 (i.e. $gg$, $g\kappa_\text{gal}$, $g\kappa_\text{CMB}$; not shown in this figure) only uses lower multipoles ($\ell_\text{max}=420, 714, 9390, 1212$ for the four lens bins). As a result, the calibration from the full LSST \& CMB S4 lensing is rather insensitive to the maximum multipole included, beyond $\sim 1,000$. Therefore, it should be robust to uncertainties in non-linearities and baryonic effects in the matter power spectrum.
}
\label{fig:varyinglmaxgks}
\end{figure}

On the other hand, combination 1 (i.e. $\kappa_\text{gal}\kappa_\text{gal}$, $\kappa_\text{gal}\kappa_\text{CMB}$, $\kappa_\text{CMB}\kappa_\text{CMB}$) uses lensing-lensing correlations. Because the lensing kernels extend to very low redshift, a fixed angular scale receives contributions from arbitrarily small scales. Furthermore, our fiducial forecast includes the modes of $\kappa_\text{gal}$ and $\kappa_\text{CMB}$ up to $\ell_\text{max}=5000$. Assessing rigorously the contamination from uncertainties in the matter power spectrum is beyond the scope of this paper (see \cite{2015MNRAS.454.2451E} for details). However, this contamination is expected to be less important for small multipoles. We thus vary the maximum multipole $\ell_\text{max}$ and show the impact on shear calibration from combination 1 in Fig.~\ref{fig:varyinglmaxgks}. The shear calibration from combination 1 is only degraded by $10-40\%$ when reducing $\ell_\text{max}$ from 5,000 to 930. The shear calibration from the full LSST \& CMB S4 lensing is therefore even less affected. This is likely because at $\ell\gtrsim1,000$, both CMB and galaxy lensing stop being cosmic variance limited, and the relative uncertainty on the lensing-lensing power spectra starts growing.

In conclusion, the shear calibration is little affected by the maximum multipole included, beyond $\ell_\text{max}\sim 1,000$, which makes it robust to uncertainties in the modeling of non-linearities and baryonic effects. 

\subsection{Application to space-based lensing surveys: Euclid and WFIRST}
\label{subsec:euclid_wfirst}

\begin{figure}[h]
\begin{minipage}[c]{0.42\linewidth}
\renewcommand{\arraystretch}{1.3}
\begin{tabular*}{1\textwidth}{@{\extracolsep{\fill}}| c l |}
\hline
\multicolumn{2}{|c|}{\textbf{Euclid specifications}} \\
\hline
\hline
$\Omega_{\mathrm{s}}$ & 15,000 deg$^2$\\
\multicolumn{2}{|c|}{} \\
source distribution & $dn_\text{source}/dz \propto z^\alpha e^{-(z/z_0)^\beta}$,\\ 
& $\alpha=1.3, \beta = 1.5, z_0=0.65$,\\
& $n_\text{source}= 30$ arcmin$^{-2}$\\
& 10 bins\\
& $\sigma_\epsilon= 0.26$\\
\multicolumn{2}{|c|}{} \\
lens distribution & $dn_\text{lens}/dz \propto \chi(z)^2/H(z),$\\
& $n_\text{lens}=0.25$ arcmin$^{-2}$\\
& 4 bins\\
\hline 
\end{tabular*}
\renewcommand{\arraystretch}{1.0}
\par\vspace{0pt}
\end{minipage}
\begin{minipage}[c]{0.5\linewidth}
\centering
\includegraphics[width=1\columnwidth]{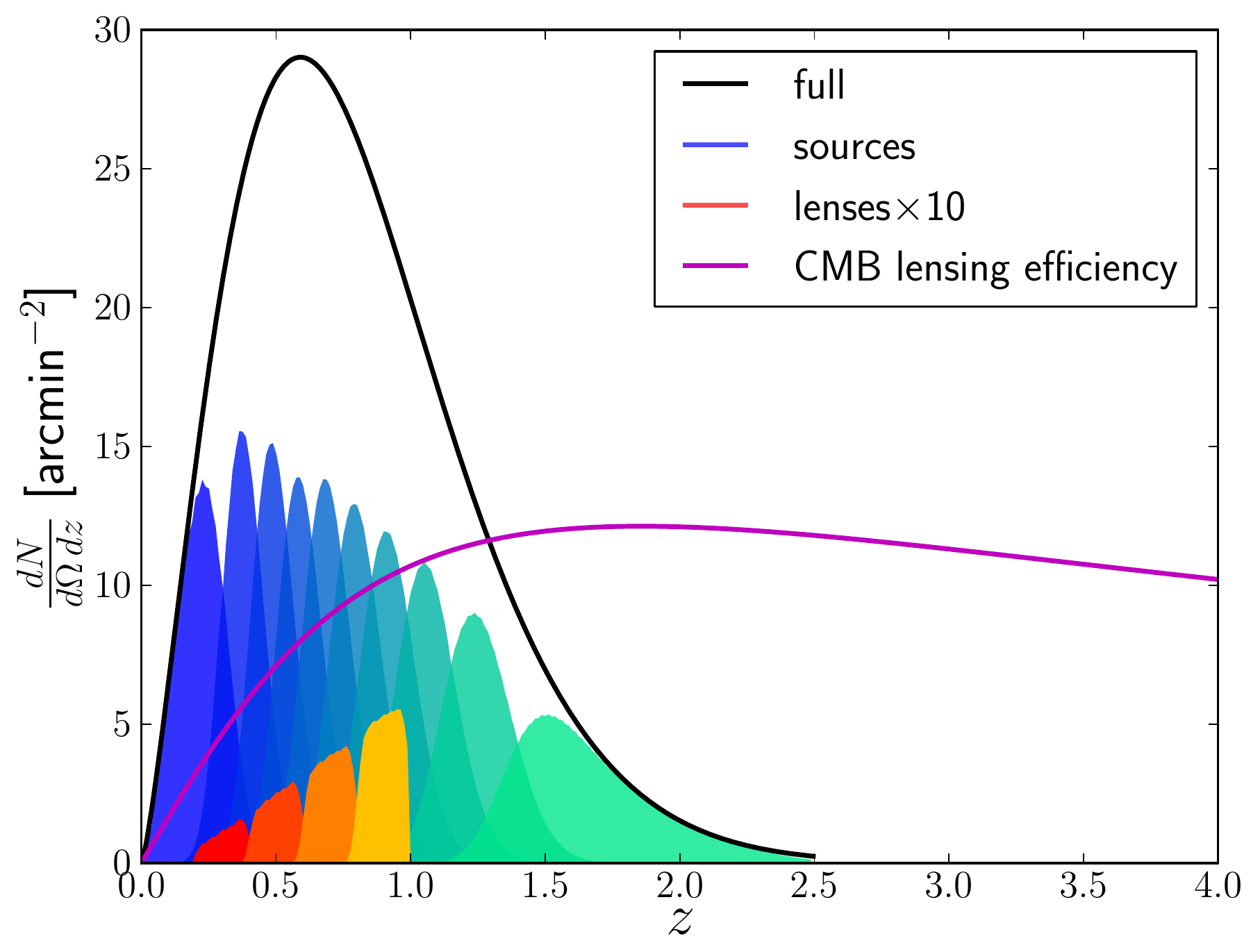}
\par\vspace{0pt}
\end{minipage}\\
\begin{minipage}[c]{0.49\linewidth}
\begin{tabular*}{1\textwidth}{@{\extracolsep{\fill}}| l l |}
\hline
Euclid \& CMB S4 & SNR \\  
\hline
\hline
\multicolumn{2}{|c|}{\textbf{Individual probes}} \\
clustering ($gg$) & 344 \\
\hline
galaxy-galaxy lensing: $g \kappa_\text{gal}$&230 \\
galaxy-CMB lensing: $g \kappa_\text{CMB}$&141 \\
\hline
shear tomography: $\kappa_\text{gal} \kappa_\text{gal}$ &375 \\
CMB lensing auto: $\kappa_\text{CMB} \kappa_\text{CMB}$&366 \\
galaxy lensing-CMB lensing: $\kappa_\text{gal} \kappa_\text{CMB}$&261 \\
\hline
\hline
\multicolumn{2}{|c|}{\textbf{Combinations}} \\
Euclid: $gg, g\kappa_\text{gal}, \kappa_\text{gal}\kappa_\text{gal}$&486 \\
Combination 1: $\kappa_\text{CMB}\kappa_\text{CMB}, \kappa_\text{CMB}\kappa_\text{gal}, \kappa_\text{gal}\kappa_\text{gal}$&510 \\
Combination 2: $gg, g\kappa_\text{CMB}, g\kappa_\text{gal}$&364 \\
Full: $gg, g\kappa_\text{CMB}, g\kappa_\text{gal}, \kappa_\text{CMB}\kappa_\text{CMB}, \kappa_\text{CMB}\kappa_\text{gal}, \kappa_\text{gal}\kappa_\text{gal}$ & 592 \\
\hline
\end{tabular*}
\end{minipage}
\begin{minipage}[c]{0.45\linewidth}
\centering
\includegraphics[width=1\columnwidth]{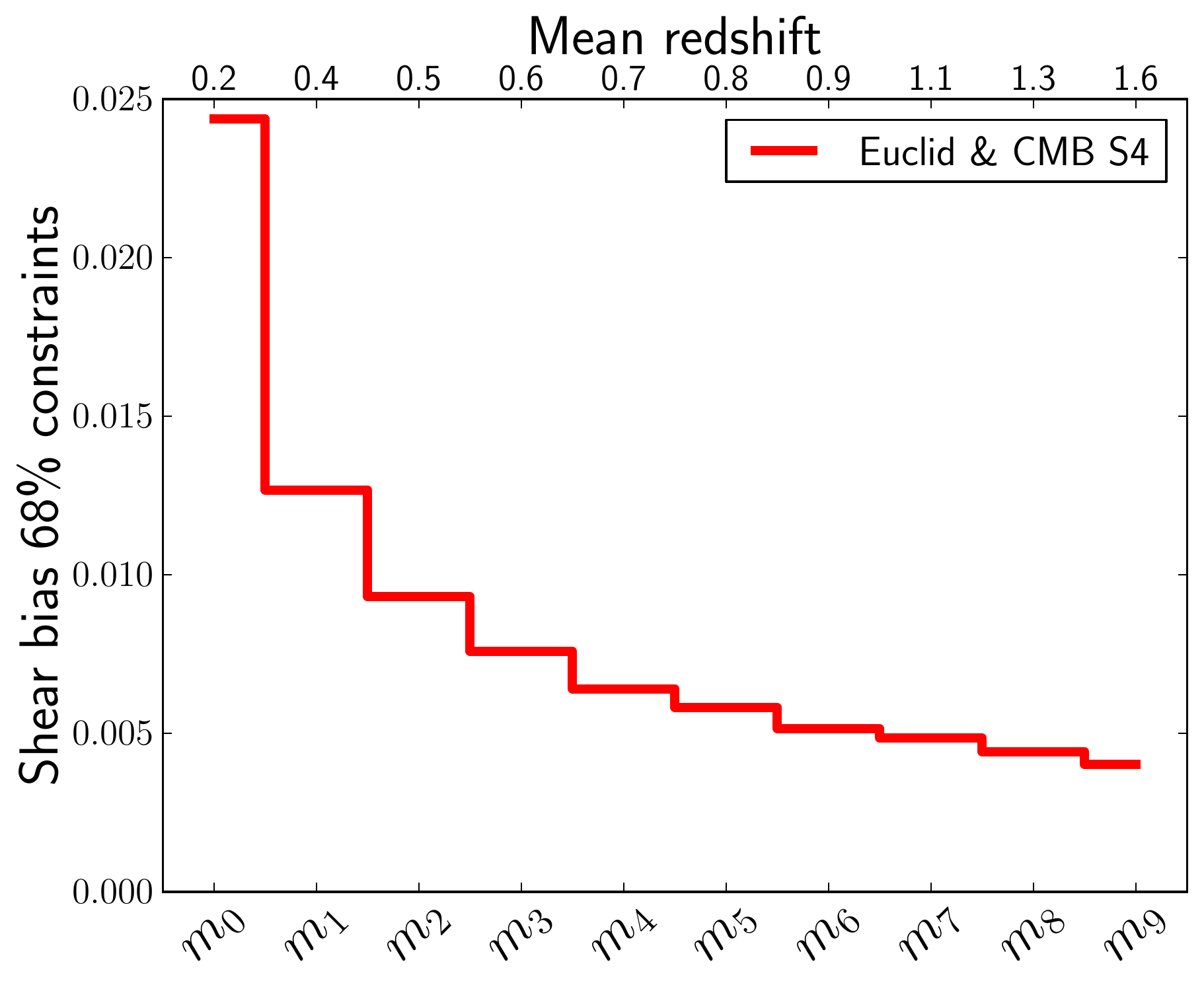}
\par\vspace{0pt}
\end{minipage}\\
\caption{
Summary of the forecast for Euclid and CMB S4 lensing. The top left panel summarizes our assumptions for the Euclid survey. The top right panel displays the photo-z bins assumed for the lens and source samples. The bottom left panel summarizes the signal to noise ratios for the various probes and combinations. The bottom right panel shows the level of shear calibration in Euclid's ten source bins.
}
\label{fig:euclid}
\end{figure}

\begin{figure}[h]
\begin{minipage}[c]{0.42\linewidth}
\renewcommand{\arraystretch}{1.3}
\begin{tabular*}{1\textwidth}{@{\extracolsep{\fill}}| c l |}
\hline
\multicolumn{2}{|c|}{\textbf{WFIRST specifications}} \\
\hline
\hline
$\Omega_{\mathrm{s}}$ & 2,200 deg$^2$\\
\multicolumn{2}{|c|}{} \\
source distribution & $dn_\text{source}/dz \propto z^\alpha e^{-(z/z_0)^\beta}$,\\ 
& $\alpha=1.27, \beta = 1.02, z_0=0.6$,\\
& $n_\text{source}= 45$ arcmin$^{-2}$\\
& 10 bins\\
& $\sigma_\epsilon= 0.26$\\
\multicolumn{2}{|c|}{} \\
lens distribution & $dn_\text{lens}/dz \propto \chi(z)^2/H(z),$\\
& $n_\text{lens}=0.25$ arcmin$^{-2}$\\
& 4 bins\\
\hline 
\end{tabular*}
\renewcommand{\arraystretch}{1.0}
\par\vspace{0pt}
\end{minipage}
\begin{minipage}[c]{0.5\linewidth}
\centering
\includegraphics[width=1\columnwidth]{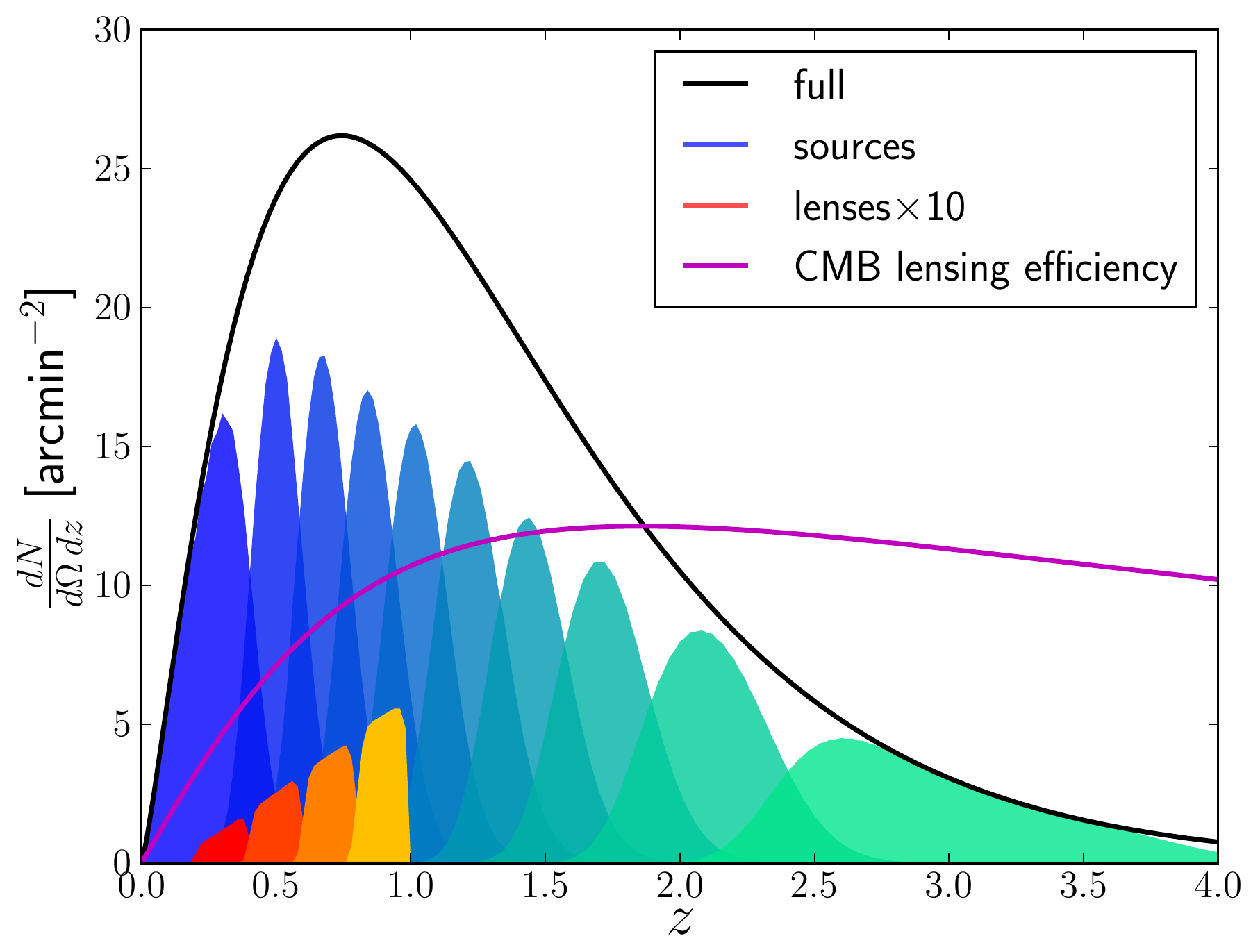}
\par\vspace{0pt}
\end{minipage}\\
\begin{minipage}[c]{0.49\linewidth}
\begin{tabular*}{1\textwidth}{@{\extracolsep{\fill}}| l l |}
\hline
WFIRST \& CMB S4 & SNR \\  
\hline
\hline
\multicolumn{2}{|c|}{\textbf{Individual probes}} \\
clustering ($gg$) & 131 \\
\hline
galaxy-galaxy lensing: $g \kappa_\text{gal}$&103 \\
galaxy-CMB lensing: $g \kappa_\text{CMB}$&54 \\
\hline
shear tomography: $\kappa_\text{gal} \kappa_\text{gal}$ &232 \\
CMB lensing auto: $\kappa_\text{CMB} \kappa_\text{CMB}$&140 \\
galaxy lensing-CMB lensing: $\kappa_\text{gal} \kappa_\text{CMB}$&154 \\
\hline
\hline
\multicolumn{2}{|c|}{\textbf{Combinations}} \\
WFIRST: $gg, g\kappa_\text{gal}, \kappa_\text{gal}\kappa_\text{gal}$&254 \\
Combination 1: $\kappa_\text{CMB}\kappa_\text{CMB}, \kappa_\text{CMB}\kappa_\text{gal}, \kappa_\text{gal}\kappa_\text{gal}$&262 \\
Combination 2: $gg, g\kappa_\text{CMB}, g\kappa_\text{gal}$&142 \\
Full: $gg, g\kappa_\text{CMB}, g\kappa_\text{gal}, \kappa_\text{CMB}\kappa_\text{CMB}, \kappa_\text{CMB}\kappa_\text{gal}, \kappa_\text{gal}\kappa_\text{gal}$ & 282\\
\hline
\end{tabular*}
\end{minipage}
\begin{minipage}[c]{0.45\linewidth}
\centering
\includegraphics[width=1\columnwidth]{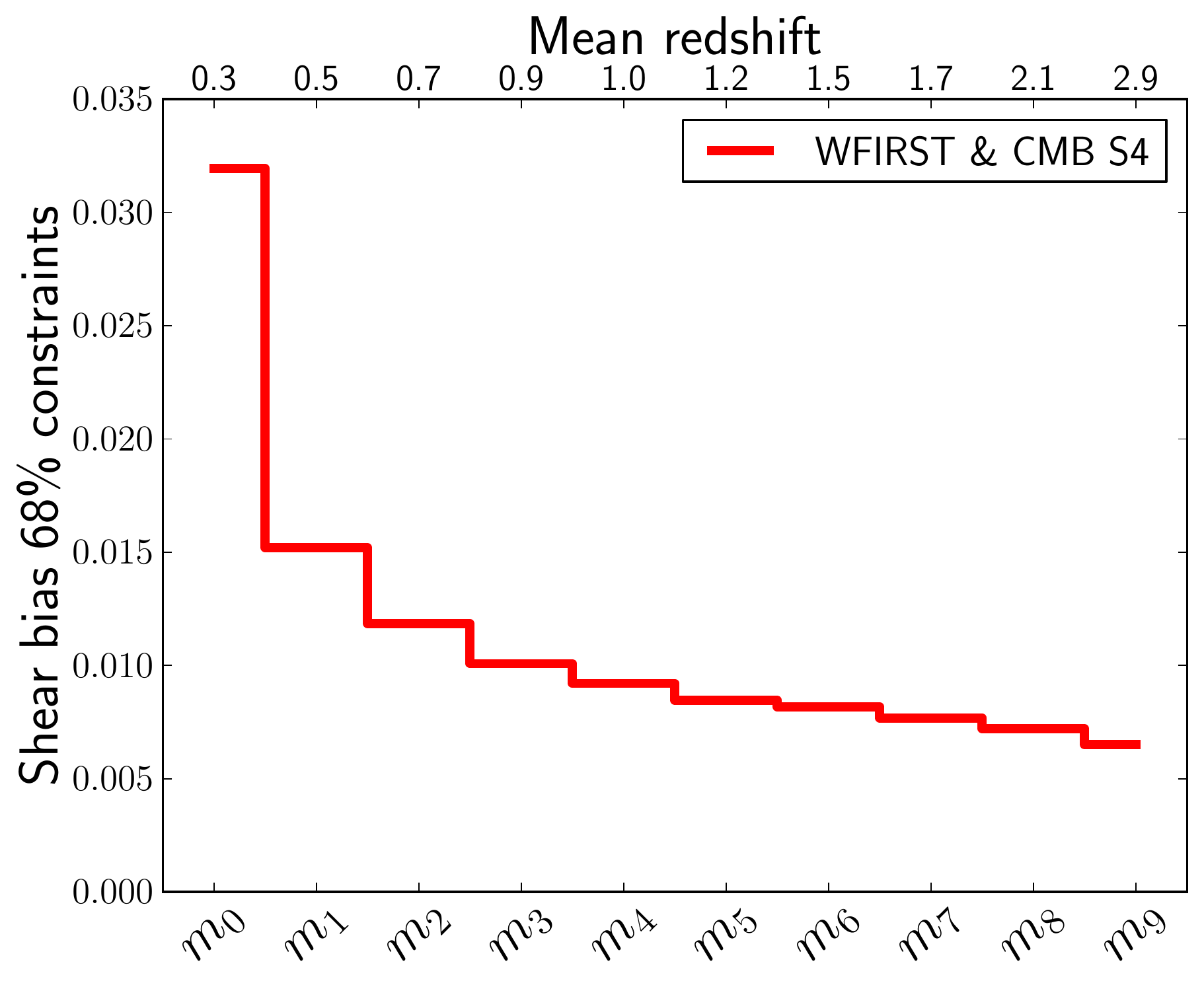}
\par\vspace{0pt}
\end{minipage}\\
\caption{
Summary of the forecast for WFIRST and CMB S4 lensing. The top left panel summarizes our assumptions for the WFIRST survey. The top right panel displays the photo-z bins assumed for the lens and source samples. The bottom left panel summarizes the signal to noise ratios for the various probes and combinations. The bottom right panel shows the level of shear calibration in WFIRST's ten source bins.
}
\label{fig:wfirst}
\end{figure}

In this subsection, we reproduce our main forecast on shear calibration for Euclid and WFIRST. 
Our assumptions and results for Euclid follow \cite{EuclidDefinitionStudyReport, 2013LRR....16....6A} and are summarized in Fig.~\ref{fig:euclid}. In particular, we assume a survey area of $15,000$ deg$^2$ with $30$ source galaxies per arcmin$^2$.
For WFIRST, we follow \cite{WFIRST} and present assumptions and results in Fig.~\ref{fig:wfirst}.
We assume a $2,200$ deg$^2$ survey area with $45$ sources/arcmin$^2$.
In both cases, we use 10 tomographic source bins, and the same redMaGiC-like lens sample as for LSST, with 4 lens bins.
For all cosmological and nuisance parameters, including photo-z uncertainties, we use the same priors as for LSST (see Tab.~\ref{tab:params_mcmc}).

As shown in Fig.~\ref{fig:euclid} and Fig.~\ref{fig:wfirst}, CMB lensing from S4 can calibrate the shear for the 10 Euclid source bins down to $0.4\%-2.4\%$, and for the 10 WFIRST source bins down to $0.6\%-3.2\%$. 
Note that the exact requirements for shear calibration for Euclid and WFIRST may differ from each other and from LSST.
Furthermore, the exact redshift distributions and survey parameters may evolve in the future, in particular for WFIRST.
Nevertheless, these results are highly encouraging.

\section{Conclusion}

In this study, we answer the following questions: can CMB lensing calibrate the shear bias down to a useful accuracy, competitive with image simulations and comparable with the LSST requirements? Is this possible while marginalizing over cosmological and nuisance parameters? How robust is this calibration to intrinsic alignments, photo-z uncertainties, non-linear and baryonic effects, and assumptions on the CMB S4 experiment?
To do so, we extend the \textsc{CosmoLike} framework to include CMB lensing. We jointly analyze all the two-point correlation functions of galaxy positions, shear and CMB lensing convergence. We include the non-Gaussian covariances and explore the posterior distribution with MCMC sampling and the Fisher approximation. Our forecasts simultaneously vary cosmological parameters, galaxy biases, photo-z uncertainties for each source and lens bin and shear calibration for each source bins. We make conservative choices of galaxy samples and scales. We therefore expect our forecast to be realistic and robust.

We show that CMB lensing from S4 can calibrate the shear multiplicative biases for LSST down to $0.3\%-2\%$ in 10 tomographic bins, surpassing the LSST requirements of $\sim 0.5\%$ in most of the redshift range. This method performs best in the highest redshift bins, where shear calibration is otherwise most challenging. We show a shear calibration of $0.4\%-2.4\%$ for Euclid's 10 tomographic source bins and $0.6\%-3.2\%$ for WFIRST's 10 bins.
For a reasonable level of intrinsic alignments and Gaussian photo-z uncertainties, the shear calibration from CMB S4 lensing is only biased at a fraction of the statistical uncertainty. 
This shear calibration is sensitive to the noise level in CMB S4 maps, but insensitive to the beam and maximum multipole at which component separation is performed, within sensible values. Thus stage 3 CMB surveys such as AdvACT and SPT-3G, as well as the Simons Observatory, will already provide a meaningful shear calibration.
It is mildly dependent on the photo-z priors for Gaussian photo-z errors, and on the maximum multipole included in the analysis, beyond $\ell_\text{max}\sim 1,000$.
We did not consider explicitly photo-z outliers \cite{2010MNRAS.401.1399B} or potential biases in the CMB lensing reconstruction \cite{2014ApJ...786...13V, 2014JCAP...03..024O}.

In conclusion, we find that shear calibration from CMB lensing will be possible at a level competitive with or even exceeding the LSST requirements.
This method is a powerful alternative to simulation-based calibration techniques, because it relies on the data directly.
In the systematics-limited era of stage 4 weak lensing surveys, this method will provide redundancy and serve as a cross check, in order to reliably measure  the properties of dark energy, the neutrino masses and possible modifications to general relativity.

\subsection*{Acknowledgements}

We thank 
Patricia Burchat, Scott Dodelson, Jo Dunkley, Simone Ferraro, Colin Hill, Gil Holder, Bhuvnesh Jain, Alexie Leauthaud, Jia Liu, Mathew Madhavacheril, Rachel Mandelbaum, Roland de Putter, Uro\v s Seljak, Blake Sherwin, Sukhdeep Singh and Martin White
 for useful discussion about shear calibration with CMB lensing.

We thank 
Bob Armstrong, Eric Huff and Peter Melchior
for useful discussion about shear systematics.

We thank
Elisa Chisari, Rachel Mandelbaum and Sukhdeep Singh
for useful discussion about intrinsic alignments.

We thank 
Scott Dodelson, Jo Dunkley, Simone Ferraro, Bhuvnesh Jain, Rachel Mandelbaum, Peter Melchior and Sukhdeep Singh
for feedback on an earlier version of this paper.

Numerical calculations in this work were carried out using computational resources 
supported by the Princeton Institute of Computational Science and Engineering.
We thank
Jim Stone
and
the Computational Science and Engineering Support
for access to these resources and invaluable help.

ES was supported, in part, by a JPL Strategic Universities Research Partnership grant.
JR, TE and HM were supported by JPL, which is operated by Caltech under a contract from NASA.

Part of the research described in this paper was carried out at the Jet Propulsion Laboratory, California Institute of Technology, under a contract with the National Aeronautics and Space Administration. Copyright 2016. All rights reserved.

\newpage
\bibliography{references}

\begin{thebibliography}{101}%
\makeatletter
\providecommand \@ifxundefined [1]{%
 \@ifx{#1\undefined}
}%
\providecommand \@ifnum [1]{%
 \ifnum #1\expandafter \@firstoftwo
 \else \expandafter \@secondoftwo
 \fi
}%
\providecommand \@ifx [1]{%
 \ifx #1\expandafter \@firstoftwo
 \else \expandafter \@secondoftwo
 \fi
}%
\providecommand \natexlab [1]{#1}%
\providecommand \enquote  [1]{``#1''}%
\providecommand \bibnamefont  [1]{#1}%
\providecommand \bibfnamefont [1]{#1}%
\providecommand \citenamefont [1]{#1}%
\providecommand \href@noop [0]{\@secondoftwo}%
\providecommand \href [0]{\begingroup \@sanitize@url \@href}%
\providecommand \@href[1]{\@@startlink{#1}\@@href}%
\providecommand \@@href[1]{\endgroup#1\@@endlink}%
\providecommand \@sanitize@url [0]{\catcode `\\12\catcode `\$12\catcode
  `\&12\catcode `\#12\catcode `\^12\catcode `\_12\catcode `\%12\relax}%
\providecommand \@@startlink[1]{}%
\providecommand \@@endlink[0]{}%
\providecommand \url  [0]{\begingroup\@sanitize@url \@url }%
\providecommand \@url [1]{\endgroup\@href {#1}{\urlprefix }}%
\providecommand \urlprefix  [0]{URL }%
\providecommand \Eprint [0]{\href }%
\providecommand \doibase [0]{http://dx.doi.org/}%
\providecommand \selectlanguage [0]{\@gobble}%
\providecommand \bibinfo  [0]{\@secondoftwo}%
\providecommand \bibfield  [0]{\@secondoftwo}%
\providecommand \translation [1]{[#1]}%
\providecommand \BibitemOpen [0]{}%
\providecommand \bibitemStop [0]{}%
\providecommand \bibitemNoStop [0]{.\EOS\space}%
\providecommand \EOS [0]{\spacefactor3000\relax}%
\providecommand \BibitemShut  [1]{\csname bibitem#1\endcsname}%
\let\auto@bib@innerbib\@empty
\bibitem [{Note1()}]{Note1}%
  \BibitemOpen
  \bibinfo {note} {\protect \url
  {http://www.astro-wise.org/projects/KIDS/}}\BibitemShut {NoStop}%
\bibitem [{Note2()}]{Note2}%
  \BibitemOpen
  \bibinfo {note} {\protect \url {http://www.darkenergysurvey.org}}\BibitemShut
  {NoStop}%
\bibitem [{\citenamefont {{The Dark Energy Survey Collaboration}}(2005)}]{DES}%
  \BibitemOpen
  \bibfield  {author} {\bibinfo {author} {\bibnamefont {{The Dark Energy Survey
  Collaboration}}},\ }\href@noop {} {\bibfield  {journal} {\bibinfo  {journal}
  {ArXiv Astrophysics e-prints}\ } (\bibinfo {year} {2005})},\ \Eprint
  {http://arxiv.org/abs/arXiv:astro-ph/0510346} {arXiv:astro-ph/0510346}
  \BibitemShut {NoStop}%
\bibitem [{Note3()}]{Note3}%
  \BibitemOpen
  \bibinfo {note} {\protect \url
  {http://www.naoj.org/Projects/HSC/index.html}}\BibitemShut {NoStop}%
\bibitem [{\citenamefont {{Miyazaki}}\ \emph {et~al.}(2006)\citenamefont
  {{Miyazaki}}, \citenamefont {{Komiyama}}, \citenamefont {{Nakaya}},
  \citenamefont {{Doi}}, \citenamefont {{Furusawa}}, \citenamefont
  {{Gillingham}}, \citenamefont {{Kamata}}, \citenamefont {{Takeshi}},\ and\
  \citenamefont {{Nariai}}}]{Miyazakietal:06}%
  \BibitemOpen
  \bibfield  {author} {\bibinfo {author} {\bibfnamefont {S.}~\bibnamefont
  {{Miyazaki}}}, \bibinfo {author} {\bibfnamefont {Y.}~\bibnamefont
  {{Komiyama}}}, \bibinfo {author} {\bibfnamefont {H.}~\bibnamefont
  {{Nakaya}}}, \bibinfo {author} {\bibfnamefont {Y.}~\bibnamefont {{Doi}}},
  \bibinfo {author} {\bibfnamefont {H.}~\bibnamefont {{Furusawa}}}, \bibinfo
  {author} {\bibfnamefont {P.}~\bibnamefont {{Gillingham}}}, \bibinfo {author}
  {\bibfnamefont {Y.}~\bibnamefont {{Kamata}}}, \bibinfo {author}
  {\bibfnamefont {K.}~\bibnamefont {{Takeshi}}}, \ and\ \bibinfo {author}
  {\bibfnamefont {K.}~\bibnamefont {{Nariai}}},\ }in\ \href {\doibase
  10.1117/12.672739} {\emph {\bibinfo {booktitle} {Society of Photo-Optical
  Instrumentation Engineers (SPIE) Conference Series}}},\ \bibinfo {series}
  {Society of Photo-Optical Instrumentation Engineers (SPIE) Conference
  Series}, Vol.\ \bibinfo {volume} {6269}\ (\bibinfo {year} {2006})\BibitemShut
  {NoStop}%
\bibitem [{Note4()}]{Note4}%
  \BibitemOpen
  \bibinfo {note} {\protect \url {http://sumire.ipmu.jp/en/2652}}\BibitemShut
  {NoStop}%
\bibitem [{\citenamefont {{Takada}}\ \emph {et~al.}(2012)\citenamefont
  {{Takada}}, \citenamefont {{Ellis}}, \citenamefont {{Chiba}}, \citenamefont
  {{Greene}}, \citenamefont {{Aihara}}, \citenamefont {{Arimoto}},
  \citenamefont {{Bundy}}, \citenamefont {{Cohen}}, \citenamefont {{Dor{\'e}}},
  \citenamefont {{Graves}}, \citenamefont {{Gunn}}, \citenamefont {{Heckman}},
  \citenamefont {{Hirata}}, \citenamefont {{Ho}}, \citenamefont {{Kneib}},
  \citenamefont {{Le F{\`e}vre}}, \citenamefont {{Lin}}, \citenamefont
  {{More}}, \citenamefont {{Murayama}}, \citenamefont {{Nagao}}, \citenamefont
  {{Ouchi}}, \citenamefont {{Seiffert}}, \citenamefont {{Silverman}},
  \citenamefont {{Sodr{\'e}}}, \citenamefont {{Spergel}}, \citenamefont
  {{Strauss}}, \citenamefont {{Sugai}}, \citenamefont {{Suto}}, \citenamefont
  {{Takami}},\ and\ \citenamefont {{Wyse}}}]{Takadaetal:12}%
  \BibitemOpen
  \bibfield  {author} {\bibinfo {author} {\bibfnamefont {M.}~\bibnamefont
  {{Takada}}}, \bibinfo {author} {\bibfnamefont {R.}~\bibnamefont {{Ellis}}},
  \bibinfo {author} {\bibfnamefont {M.}~\bibnamefont {{Chiba}}}, \bibinfo
  {author} {\bibfnamefont {J.~E.}\ \bibnamefont {{Greene}}}, \bibinfo {author}
  {\bibfnamefont {H.}~\bibnamefont {{Aihara}}}, \bibinfo {author}
  {\bibfnamefont {N.}~\bibnamefont {{Arimoto}}}, \bibinfo {author}
  {\bibfnamefont {K.}~\bibnamefont {{Bundy}}}, \bibinfo {author} {\bibfnamefont
  {J.}~\bibnamefont {{Cohen}}}, \bibinfo {author} {\bibfnamefont
  {O.}~\bibnamefont {{Dor{\'e}}}}, \bibinfo {author} {\bibfnamefont
  {G.}~\bibnamefont {{Graves}}}, \bibinfo {author} {\bibfnamefont {J.~E.}\
  \bibnamefont {{Gunn}}}, \bibinfo {author} {\bibfnamefont {T.}~\bibnamefont
  {{Heckman}}}, \bibinfo {author} {\bibfnamefont {C.}~\bibnamefont {{Hirata}}},
  \bibinfo {author} {\bibfnamefont {P.}~\bibnamefont {{Ho}}}, \bibinfo {author}
  {\bibfnamefont {J.-P.}\ \bibnamefont {{Kneib}}}, \bibinfo {author}
  {\bibfnamefont {O.}~\bibnamefont {{Le F{\`e}vre}}}, \bibinfo {author}
  {\bibfnamefont {L.}~\bibnamefont {{Lin}}}, \bibinfo {author} {\bibfnamefont
  {S.}~\bibnamefont {{More}}}, \bibinfo {author} {\bibfnamefont
  {H.}~\bibnamefont {{Murayama}}}, \bibinfo {author} {\bibfnamefont
  {T.}~\bibnamefont {{Nagao}}}, \bibinfo {author} {\bibfnamefont
  {M.}~\bibnamefont {{Ouchi}}}, \bibinfo {author} {\bibfnamefont
  {M.}~\bibnamefont {{Seiffert}}}, \bibinfo {author} {\bibfnamefont
  {J.}~\bibnamefont {{Silverman}}}, \bibinfo {author} {\bibfnamefont
  {L.}~\bibnamefont {{Sodr{\'e}}}, \bibfnamefont {Jr}}, \bibinfo {author}
  {\bibfnamefont {D.~N.}\ \bibnamefont {{Spergel}}}, \bibinfo {author}
  {\bibfnamefont {M.~A.}\ \bibnamefont {{Strauss}}}, \bibinfo {author}
  {\bibfnamefont {H.}~\bibnamefont {{Sugai}}}, \bibinfo {author} {\bibfnamefont
  {Y.}~\bibnamefont {{Suto}}}, \bibinfo {author} {\bibfnamefont
  {H.}~\bibnamefont {{Takami}}}, \ and\ \bibinfo {author} {\bibfnamefont
  {R.}~\bibnamefont {{Wyse}}},\ }\href@noop {} {\bibfield  {journal} {\bibinfo
  {journal} {ArXiv e-prints}\ } (\bibinfo {year} {2012})},\ \Eprint
  {http://arxiv.org/abs/1206.0737} {arXiv:1206.0737 [astro-ph.CO]} \BibitemShut
  {NoStop}%
\bibitem [{Note5()}]{Note5}%
  \BibitemOpen
  \bibinfo {note} {\protect \url {http://desi.lbl.gov}}\BibitemShut {NoStop}%
\bibitem [{\citenamefont {{LSST Science Collaboration}}\ \emph
  {et~al.}(2009)\citenamefont {{LSST Science Collaboration}}, \citenamefont
  {{Abell}}, \citenamefont {{Allison}}, \citenamefont {{Anderson}},
  \citenamefont {{Andrew}}, \citenamefont {{Angel}}, \citenamefont {{Armus}},
  \citenamefont {{Arnett}}, \citenamefont {{Asztalos}}, \citenamefont
  {{Axelrod}},\ and\ \citenamefont {et~al.}}]{LSSTScienceBook}%
  \BibitemOpen
  \bibfield  {author} {\bibinfo {author} {\bibnamefont {{LSST Science
  Collaboration}}}, \bibinfo {author} {\bibfnamefont {P.~A.}\ \bibnamefont
  {{Abell}}}, \bibinfo {author} {\bibfnamefont {J.}~\bibnamefont {{Allison}}},
  \bibinfo {author} {\bibfnamefont {S.~F.}\ \bibnamefont {{Anderson}}},
  \bibinfo {author} {\bibfnamefont {J.~R.}\ \bibnamefont {{Andrew}}}, \bibinfo
  {author} {\bibfnamefont {J.~R.~P.}\ \bibnamefont {{Angel}}}, \bibinfo
  {author} {\bibfnamefont {L.}~\bibnamefont {{Armus}}}, \bibinfo {author}
  {\bibfnamefont {D.}~\bibnamefont {{Arnett}}}, \bibinfo {author}
  {\bibfnamefont {S.~J.}\ \bibnamefont {{Asztalos}}}, \bibinfo {author}
  {\bibfnamefont {T.~S.}\ \bibnamefont {{Axelrod}}}, \ and\ \bibinfo {author}
  {\bibnamefont {et~al.}},\ }\href@noop {} {\bibfield  {journal} {\bibinfo
  {journal} {ArXiv e-prints}\ } (\bibinfo {year} {2009})},\ \Eprint
  {http://arxiv.org/abs/0912.0201} {arXiv:0912.0201 [astro-ph.IM]} \BibitemShut
  {NoStop}%
\bibitem [{\citenamefont {{Laureijs}}\ \emph {et~al.}(2011)\citenamefont
  {{Laureijs}}, \citenamefont {{Amiaux}}, \citenamefont {{Arduini}},
  \citenamefont {{Augu{\`e}res}}, \citenamefont {{Brinchmann}}, \citenamefont
  {{Cole}}, \citenamefont {{Cropper}}, \citenamefont {{Dabin}}, \citenamefont
  {{Duvet}}, \citenamefont {{Ealet}},\ and\ \citenamefont
  {et~al.}}]{EuclidDefinitionStudyReport}%
  \BibitemOpen
  \bibfield  {author} {\bibinfo {author} {\bibfnamefont {R.}~\bibnamefont
  {{Laureijs}}}, \bibinfo {author} {\bibfnamefont {J.}~\bibnamefont
  {{Amiaux}}}, \bibinfo {author} {\bibfnamefont {S.}~\bibnamefont {{Arduini}}},
  \bibinfo {author} {\bibfnamefont {J.~.}\ \bibnamefont {{Augu{\`e}res}}},
  \bibinfo {author} {\bibfnamefont {J.}~\bibnamefont {{Brinchmann}}}, \bibinfo
  {author} {\bibfnamefont {R.}~\bibnamefont {{Cole}}}, \bibinfo {author}
  {\bibfnamefont {M.}~\bibnamefont {{Cropper}}}, \bibinfo {author}
  {\bibfnamefont {C.}~\bibnamefont {{Dabin}}}, \bibinfo {author} {\bibfnamefont
  {L.}~\bibnamefont {{Duvet}}}, \bibinfo {author} {\bibfnamefont
  {A.}~\bibnamefont {{Ealet}}}, \ and\ \bibinfo {author} {\bibnamefont
  {et~al.}},\ }\href@noop {} {\bibfield  {journal} {\bibinfo  {journal} {ArXiv
  e-prints}\ } (\bibinfo {year} {2011})},\ \Eprint
  {http://arxiv.org/abs/1110.3193} {arXiv:1110.3193 [astro-ph.CO]} \BibitemShut
  {NoStop}%
\bibitem [{\citenamefont {{Spergel}}\ \emph {et~al.}(2013)\citenamefont
  {{Spergel}}, \citenamefont {{Gehrels}}, \citenamefont {{Breckinridge}},
  \citenamefont {{Donahue}}, \citenamefont {{Dressler}}, \citenamefont
  {{Gaudi}}, \citenamefont {{Greene}}, \citenamefont {{Guyon}}, \citenamefont
  {{Hirata}}, \citenamefont {{Kalirai}}, \citenamefont {{Kasdin}},
  \citenamefont {{Moos}}, \citenamefont {{Perlmutter}}, \citenamefont
  {{Postman}}, \citenamefont {{Rauscher}}, \citenamefont {{Rhodes}},
  \citenamefont {{Wang}}, \citenamefont {{Weinberg}}, \citenamefont
  {{Centrella}}, \citenamefont {{Traub}}, \citenamefont {{Baltay}},
  \citenamefont {{Colbert}}, \citenamefont {{Bennett}}, \citenamefont
  {{Kiessling}}, \citenamefont {{Macintosh}}, \citenamefont {{Merten}},
  \citenamefont {{Mortonson}}, \citenamefont {{Penny}}, \citenamefont {{Rozo}},
  \citenamefont {{Savransky}}, \citenamefont {{Stapelfeldt}}, \citenamefont
  {{Zu}}, \citenamefont {{Baker}}, \citenamefont {{Cheng}}, \citenamefont
  {{Content}}, \citenamefont {{Dooley}}, \citenamefont {{Foote}}, \citenamefont
  {{Goullioud}}, \citenamefont {{Grady}}, \citenamefont {{Jackson}},
  \citenamefont {{Kruk}}, \citenamefont {{Levine}}, \citenamefont {{Melton}},
  \citenamefont {{Peddie}}, \citenamefont {{Ruffa}},\ and\ \citenamefont
  {{Shaklan}}}]{WFIRST}%
  \BibitemOpen
  \bibfield  {author} {\bibinfo {author} {\bibfnamefont {D.}~\bibnamefont
  {{Spergel}}}, \bibinfo {author} {\bibfnamefont {N.}~\bibnamefont
  {{Gehrels}}}, \bibinfo {author} {\bibfnamefont {J.}~\bibnamefont
  {{Breckinridge}}}, \bibinfo {author} {\bibfnamefont {M.}~\bibnamefont
  {{Donahue}}}, \bibinfo {author} {\bibfnamefont {A.}~\bibnamefont
  {{Dressler}}}, \bibinfo {author} {\bibfnamefont {B.~S.}\ \bibnamefont
  {{Gaudi}}}, \bibinfo {author} {\bibfnamefont {T.}~\bibnamefont {{Greene}}},
  \bibinfo {author} {\bibfnamefont {O.}~\bibnamefont {{Guyon}}}, \bibinfo
  {author} {\bibfnamefont {C.}~\bibnamefont {{Hirata}}}, \bibinfo {author}
  {\bibfnamefont {J.}~\bibnamefont {{Kalirai}}}, \bibinfo {author}
  {\bibfnamefont {N.~J.}\ \bibnamefont {{Kasdin}}}, \bibinfo {author}
  {\bibfnamefont {W.}~\bibnamefont {{Moos}}}, \bibinfo {author} {\bibfnamefont
  {S.}~\bibnamefont {{Perlmutter}}}, \bibinfo {author} {\bibfnamefont
  {M.}~\bibnamefont {{Postman}}}, \bibinfo {author} {\bibfnamefont
  {B.}~\bibnamefont {{Rauscher}}}, \bibinfo {author} {\bibfnamefont
  {J.}~\bibnamefont {{Rhodes}}}, \bibinfo {author} {\bibfnamefont
  {Y.}~\bibnamefont {{Wang}}}, \bibinfo {author} {\bibfnamefont
  {D.}~\bibnamefont {{Weinberg}}}, \bibinfo {author} {\bibfnamefont
  {J.}~\bibnamefont {{Centrella}}}, \bibinfo {author} {\bibfnamefont
  {W.}~\bibnamefont {{Traub}}}, \bibinfo {author} {\bibfnamefont
  {C.}~\bibnamefont {{Baltay}}}, \bibinfo {author} {\bibfnamefont
  {J.}~\bibnamefont {{Colbert}}}, \bibinfo {author} {\bibfnamefont
  {D.}~\bibnamefont {{Bennett}}}, \bibinfo {author} {\bibfnamefont
  {A.}~\bibnamefont {{Kiessling}}}, \bibinfo {author} {\bibfnamefont
  {B.}~\bibnamefont {{Macintosh}}}, \bibinfo {author} {\bibfnamefont
  {J.}~\bibnamefont {{Merten}}}, \bibinfo {author} {\bibfnamefont
  {M.}~\bibnamefont {{Mortonson}}}, \bibinfo {author} {\bibfnamefont
  {M.}~\bibnamefont {{Penny}}}, \bibinfo {author} {\bibfnamefont
  {E.}~\bibnamefont {{Rozo}}}, \bibinfo {author} {\bibfnamefont
  {D.}~\bibnamefont {{Savransky}}}, \bibinfo {author} {\bibfnamefont
  {K.}~\bibnamefont {{Stapelfeldt}}}, \bibinfo {author} {\bibfnamefont
  {Y.}~\bibnamefont {{Zu}}}, \bibinfo {author} {\bibfnamefont {C.}~\bibnamefont
  {{Baker}}}, \bibinfo {author} {\bibfnamefont {E.}~\bibnamefont {{Cheng}}},
  \bibinfo {author} {\bibfnamefont {D.}~\bibnamefont {{Content}}}, \bibinfo
  {author} {\bibfnamefont {J.}~\bibnamefont {{Dooley}}}, \bibinfo {author}
  {\bibfnamefont {M.}~\bibnamefont {{Foote}}}, \bibinfo {author} {\bibfnamefont
  {R.}~\bibnamefont {{Goullioud}}}, \bibinfo {author} {\bibfnamefont
  {K.}~\bibnamefont {{Grady}}}, \bibinfo {author} {\bibfnamefont
  {C.}~\bibnamefont {{Jackson}}}, \bibinfo {author} {\bibfnamefont
  {J.}~\bibnamefont {{Kruk}}}, \bibinfo {author} {\bibfnamefont
  {M.}~\bibnamefont {{Levine}}}, \bibinfo {author} {\bibfnamefont
  {M.}~\bibnamefont {{Melton}}}, \bibinfo {author} {\bibfnamefont
  {C.}~\bibnamefont {{Peddie}}}, \bibinfo {author} {\bibfnamefont
  {J.}~\bibnamefont {{Ruffa}}}, \ and\ \bibinfo {author} {\bibfnamefont
  {S.}~\bibnamefont {{Shaklan}}},\ }\href@noop {} {\bibfield  {journal}
  {\bibinfo  {journal} {ArXiv e-prints}\ } (\bibinfo {year} {2013})},\ \Eprint
  {http://arxiv.org/abs/1305.5425} {arXiv:1305.5425 [astro-ph.IM]} \BibitemShut
  {NoStop}%
\bibitem [{\citenamefont {{Weinberg}}\ \emph {et~al.}(2013)\citenamefont
  {{Weinberg}}, \citenamefont {{Mortonson}}, \citenamefont {{Eisenstein}},
  \citenamefont {{Hirata}}, \citenamefont {{Riess}},\ and\ \citenamefont
  {{Rozo}}}]{2013PhR...530...87W}%
  \BibitemOpen
  \bibfield  {author} {\bibinfo {author} {\bibfnamefont {D.~H.}\ \bibnamefont
  {{Weinberg}}}, \bibinfo {author} {\bibfnamefont {M.~J.}\ \bibnamefont
  {{Mortonson}}}, \bibinfo {author} {\bibfnamefont {D.~J.}\ \bibnamefont
  {{Eisenstein}}}, \bibinfo {author} {\bibfnamefont {C.}~\bibnamefont
  {{Hirata}}}, \bibinfo {author} {\bibfnamefont {A.~G.}\ \bibnamefont
  {{Riess}}}, \ and\ \bibinfo {author} {\bibfnamefont {E.}~\bibnamefont
  {{Rozo}}},\ }\href {\doibase 10.1016/j.physrep.2013.05.001} {\bibfield
  {journal} {\bibinfo  {journal} {\physrep}\ }\textbf {\bibinfo {volume}
  {530}},\ \bibinfo {pages} {87} (\bibinfo {year} {2013})},\ \Eprint
  {http://arxiv.org/abs/1201.2434} {arXiv:1201.2434} \BibitemShut {NoStop}%
\bibitem [{\citenamefont {{Albrecht}}\ \emph {et~al.}(2006)\citenamefont
  {{Albrecht}}, \citenamefont {{Bernstein}}, \citenamefont {{Cahn}},
  \citenamefont {{Freedman}}, \citenamefont {{Hewitt}}, \citenamefont {{Hu}},
  \citenamefont {{Huth}}, \citenamefont {{Kamionkowski}}, \citenamefont
  {{Kolb}}, \citenamefont {{Knox}}, \citenamefont {{Mather}}, \citenamefont
  {{Staggs}},\ and\ \citenamefont {{Suntzeff}}}]{2006astro.ph..9591A}%
  \BibitemOpen
  \bibfield  {author} {\bibinfo {author} {\bibfnamefont {A.}~\bibnamefont
  {{Albrecht}}}, \bibinfo {author} {\bibfnamefont {G.}~\bibnamefont
  {{Bernstein}}}, \bibinfo {author} {\bibfnamefont {R.}~\bibnamefont {{Cahn}}},
  \bibinfo {author} {\bibfnamefont {W.~L.}\ \bibnamefont {{Freedman}}},
  \bibinfo {author} {\bibfnamefont {J.}~\bibnamefont {{Hewitt}}}, \bibinfo
  {author} {\bibfnamefont {W.}~\bibnamefont {{Hu}}}, \bibinfo {author}
  {\bibfnamefont {J.}~\bibnamefont {{Huth}}}, \bibinfo {author} {\bibfnamefont
  {M.}~\bibnamefont {{Kamionkowski}}}, \bibinfo {author} {\bibfnamefont
  {E.~W.}\ \bibnamefont {{Kolb}}}, \bibinfo {author} {\bibfnamefont
  {L.}~\bibnamefont {{Knox}}}, \bibinfo {author} {\bibfnamefont {J.~C.}\
  \bibnamefont {{Mather}}}, \bibinfo {author} {\bibfnamefont {S.}~\bibnamefont
  {{Staggs}}}, \ and\ \bibinfo {author} {\bibfnamefont {N.~B.}\ \bibnamefont
  {{Suntzeff}}},\ }\href@noop {} {\bibfield  {journal} {\bibinfo  {journal}
  {ArXiv Astrophysics e-prints}\ } (\bibinfo {year} {2006})},\ \Eprint
  {http://arxiv.org/abs/astro-ph/0609591} {astro-ph/0609591} \BibitemShut
  {NoStop}%
\bibitem [{\citenamefont {{Huterer}}\ \emph
  {et~al.}(2006{\natexlab{a}})\citenamefont {{Huterer}}, \citenamefont
  {{Takada}}, \citenamefont {{Bernstein}},\ and\ \citenamefont
  {{Jain}}}]{2006MNRAS.366..101H}%
  \BibitemOpen
  \bibfield  {author} {\bibinfo {author} {\bibfnamefont {D.}~\bibnamefont
  {{Huterer}}}, \bibinfo {author} {\bibfnamefont {M.}~\bibnamefont {{Takada}}},
  \bibinfo {author} {\bibfnamefont {G.}~\bibnamefont {{Bernstein}}}, \ and\
  \bibinfo {author} {\bibfnamefont {B.}~\bibnamefont {{Jain}}},\ }\href
  {\doibase 10.1111/j.1365-2966.2005.09782.x} {\bibfield  {journal} {\bibinfo
  {journal} {\mnras}\ }\textbf {\bibinfo {volume} {366}},\ \bibinfo {pages}
  {101} (\bibinfo {year} {2006}{\natexlab{a}})},\ \Eprint
  {http://arxiv.org/abs/astro-ph/0506030} {astro-ph/0506030} \BibitemShut
  {NoStop}%
\bibitem [{\citenamefont {{Ma}}\ and\ \citenamefont
  {{Bernstein}}(2008)}]{2008ApJ...682...39M}%
  \BibitemOpen
  \bibfield  {author} {\bibinfo {author} {\bibfnamefont {Z.}~\bibnamefont
  {{Ma}}}\ and\ \bibinfo {author} {\bibfnamefont {G.}~\bibnamefont
  {{Bernstein}}},\ }\href {\doibase 10.1086/588214} {\bibfield  {journal}
  {\bibinfo  {journal} {\apj}\ }\textbf {\bibinfo {volume} {682}},\ \bibinfo
  {eid} {39-48} (\bibinfo {year} {2008})},\ \Eprint
  {http://arxiv.org/abs/0712.1562} {arXiv:0712.1562} \BibitemShut {NoStop}%
\bibitem [{\citenamefont {{Rudd}}\ \emph {et~al.}(2008)\citenamefont {{Rudd}},
  \citenamefont {{Zentner}},\ and\ \citenamefont
  {{Kravtsov}}}]{2008ApJ...672...19R}%
  \BibitemOpen
  \bibfield  {author} {\bibinfo {author} {\bibfnamefont {D.~H.}\ \bibnamefont
  {{Rudd}}}, \bibinfo {author} {\bibfnamefont {A.~R.}\ \bibnamefont
  {{Zentner}}}, \ and\ \bibinfo {author} {\bibfnamefont {A.~V.}\ \bibnamefont
  {{Kravtsov}}},\ }\href {\doibase 10.1086/523836} {\bibfield  {journal}
  {\bibinfo  {journal} {\apj}\ }\textbf {\bibinfo {volume} {672}},\ \bibinfo
  {eid} {19-32} (\bibinfo {year} {2008})},\ \Eprint
  {http://arxiv.org/abs/astro-ph/0703741} {astro-ph/0703741} \BibitemShut
  {NoStop}%
\bibitem [{\citenamefont {{Zentner}}\ \emph {et~al.}(2008)\citenamefont
  {{Zentner}}, \citenamefont {{Rudd}},\ and\ \citenamefont
  {{Hu}}}]{2008PhRvD..77d3507Z}%
  \BibitemOpen
  \bibfield  {author} {\bibinfo {author} {\bibfnamefont {A.~R.}\ \bibnamefont
  {{Zentner}}}, \bibinfo {author} {\bibfnamefont {D.~H.}\ \bibnamefont
  {{Rudd}}}, \ and\ \bibinfo {author} {\bibfnamefont {W.}~\bibnamefont
  {{Hu}}},\ }\href {\doibase 10.1103/PhysRevD.77.043507} {\bibfield  {journal}
  {\bibinfo  {journal} {\prd}\ }\textbf {\bibinfo {volume} {77}},\ \bibinfo
  {eid} {043507} (\bibinfo {year} {2008})},\ \Eprint
  {http://arxiv.org/abs/0709.4029} {arXiv:0709.4029} \BibitemShut {NoStop}%
\bibitem [{\citenamefont {{Eifler}}(2011)}]{2011MNRAS.418..536E}%
  \BibitemOpen
  \bibfield  {author} {\bibinfo {author} {\bibfnamefont {T.}~\bibnamefont
  {{Eifler}}},\ }\href {\doibase 10.1111/j.1365-2966.2011.19502.x} {\bibfield
  {journal} {\bibinfo  {journal} {\mnras}\ }\textbf {\bibinfo {volume} {418}},\
  \bibinfo {pages} {536} (\bibinfo {year} {2011})},\ \Eprint
  {http://arxiv.org/abs/1012.2978} {arXiv:1012.2978} \BibitemShut {NoStop}%
\bibitem [{\citenamefont {{van Daalen}}\ \emph {et~al.}(2011)\citenamefont
  {{van Daalen}}, \citenamefont {{Schaye}}, \citenamefont {{Booth}},\ and\
  \citenamefont {{Dalla Vecchia}}}]{2011MNRAS.415.3649V}%
  \BibitemOpen
  \bibfield  {author} {\bibinfo {author} {\bibfnamefont {M.~P.}\ \bibnamefont
  {{van Daalen}}}, \bibinfo {author} {\bibfnamefont {J.}~\bibnamefont
  {{Schaye}}}, \bibinfo {author} {\bibfnamefont {C.~M.}\ \bibnamefont
  {{Booth}}}, \ and\ \bibinfo {author} {\bibfnamefont {C.}~\bibnamefont {{Dalla
  Vecchia}}},\ }\href {\doibase 10.1111/j.1365-2966.2011.18981.x} {\bibfield
  {journal} {\bibinfo  {journal} {\mnras}\ }\textbf {\bibinfo {volume} {415}},\
  \bibinfo {pages} {3649} (\bibinfo {year} {2011})},\ \Eprint
  {http://arxiv.org/abs/1104.1174} {arXiv:1104.1174 [astro-ph.CO]} \BibitemShut
  {NoStop}%
\bibitem [{\citenamefont {{van Daalen}}\ \emph {et~al.}(2014)\citenamefont
  {{van Daalen}}, \citenamefont {{Schaye}}, \citenamefont {{McCarthy}},
  \citenamefont {{Booth}},\ and\ \citenamefont {{Dalla
  Vecchia}}}]{2014MNRAS.440.2997V}%
  \BibitemOpen
  \bibfield  {author} {\bibinfo {author} {\bibfnamefont {M.~P.}\ \bibnamefont
  {{van Daalen}}}, \bibinfo {author} {\bibfnamefont {J.}~\bibnamefont
  {{Schaye}}}, \bibinfo {author} {\bibfnamefont {I.~G.}\ \bibnamefont
  {{McCarthy}}}, \bibinfo {author} {\bibfnamefont {C.~M.}\ \bibnamefont
  {{Booth}}}, \ and\ \bibinfo {author} {\bibfnamefont {C.}~\bibnamefont {{Dalla
  Vecchia}}},\ }\href {\doibase 10.1093/mnras/stu482} {\bibfield  {journal}
  {\bibinfo  {journal} {\mnras}\ }\textbf {\bibinfo {volume} {440}},\ \bibinfo
  {pages} {2997} (\bibinfo {year} {2014})},\ \Eprint
  {http://arxiv.org/abs/1310.7571} {arXiv:1310.7571} \BibitemShut {NoStop}%
\bibitem [{\citenamefont {{Velliscig}}\ \emph {et~al.}(2014)\citenamefont
  {{Velliscig}}, \citenamefont {{van Daalen}}, \citenamefont {{Schaye}},
  \citenamefont {{McCarthy}}, \citenamefont {{Cacciato}}, \citenamefont {{Le
  Brun}},\ and\ \citenamefont {{Dalla Vecchia}}}]{2014MNRAS.442.2641V}%
  \BibitemOpen
  \bibfield  {author} {\bibinfo {author} {\bibfnamefont {M.}~\bibnamefont
  {{Velliscig}}}, \bibinfo {author} {\bibfnamefont {M.~P.}\ \bibnamefont {{van
  Daalen}}}, \bibinfo {author} {\bibfnamefont {J.}~\bibnamefont {{Schaye}}},
  \bibinfo {author} {\bibfnamefont {I.~G.}\ \bibnamefont {{McCarthy}}},
  \bibinfo {author} {\bibfnamefont {M.}~\bibnamefont {{Cacciato}}}, \bibinfo
  {author} {\bibfnamefont {A.~M.~C.}\ \bibnamefont {{Le Brun}}}, \ and\
  \bibinfo {author} {\bibfnamefont {C.}~\bibnamefont {{Dalla Vecchia}}},\
  }\href {\doibase 10.1093/mnras/stu1044} {\bibfield  {journal} {\bibinfo
  {journal} {\mnras}\ }\textbf {\bibinfo {volume} {442}},\ \bibinfo {pages}
  {2641} (\bibinfo {year} {2014})},\ \Eprint {http://arxiv.org/abs/1402.4461}
  {arXiv:1402.4461} \BibitemShut {NoStop}%
\bibitem [{\citenamefont {{Eifler}}\ \emph {et~al.}(2015)\citenamefont
  {{Eifler}}, \citenamefont {{Krause}}, \citenamefont {{Dodelson}},
  \citenamefont {{Zentner}}, \citenamefont {{Hearin}},\ and\ \citenamefont
  {{Gnedin}}}]{2015MNRAS.454.2451E}%
  \BibitemOpen
  \bibfield  {author} {\bibinfo {author} {\bibfnamefont {T.}~\bibnamefont
  {{Eifler}}}, \bibinfo {author} {\bibfnamefont {E.}~\bibnamefont {{Krause}}},
  \bibinfo {author} {\bibfnamefont {S.}~\bibnamefont {{Dodelson}}}, \bibinfo
  {author} {\bibfnamefont {A.~R.}\ \bibnamefont {{Zentner}}}, \bibinfo {author}
  {\bibfnamefont {A.~P.}\ \bibnamefont {{Hearin}}}, \ and\ \bibinfo {author}
  {\bibfnamefont {N.~Y.}\ \bibnamefont {{Gnedin}}},\ }\href {\doibase
  10.1093/mnras/stv2000} {\bibfield  {journal} {\bibinfo  {journal} {\mnras}\
  }\textbf {\bibinfo {volume} {454}},\ \bibinfo {pages} {2451} (\bibinfo {year}
  {2015})},\ \Eprint {http://arxiv.org/abs/1405.7423} {arXiv:1405.7423}
  \BibitemShut {NoStop}%
\bibitem [{\citenamefont {{Osato}}\ \emph {et~al.}(2015)\citenamefont
  {{Osato}}, \citenamefont {{Shirasaki}},\ and\ \citenamefont
  {{Yoshida}}}]{2015ApJ...806..186O}%
  \BibitemOpen
  \bibfield  {author} {\bibinfo {author} {\bibfnamefont {K.}~\bibnamefont
  {{Osato}}}, \bibinfo {author} {\bibfnamefont {M.}~\bibnamefont
  {{Shirasaki}}}, \ and\ \bibinfo {author} {\bibfnamefont {N.}~\bibnamefont
  {{Yoshida}}},\ }\href {\doibase 10.1088/0004-637X/806/2/186} {\bibfield
  {journal} {\bibinfo  {journal} {\apj}\ }\textbf {\bibinfo {volume} {806}},\
  \bibinfo {eid} {186} (\bibinfo {year} {2015})},\ \Eprint
  {http://arxiv.org/abs/1501.02055} {arXiv:1501.02055} \BibitemShut {NoStop}%
\bibitem [{\citenamefont {{Hellwing}}\ \emph {et~al.}(2016)\citenamefont
  {{Hellwing}}, \citenamefont {{Schaller}}, \citenamefont {{Frenk}},
  \citenamefont {{Theuns}}, \citenamefont {{Schaye}}, \citenamefont {{Bower}},\
  and\ \citenamefont {{Crain}}}]{2016arXiv160303328H}%
  \BibitemOpen
  \bibfield  {author} {\bibinfo {author} {\bibfnamefont {W.~A.}\ \bibnamefont
  {{Hellwing}}}, \bibinfo {author} {\bibfnamefont {M.}~\bibnamefont
  {{Schaller}}}, \bibinfo {author} {\bibfnamefont {C.~S.}\ \bibnamefont
  {{Frenk}}}, \bibinfo {author} {\bibfnamefont {T.}~\bibnamefont {{Theuns}}},
  \bibinfo {author} {\bibfnamefont {J.}~\bibnamefont {{Schaye}}}, \bibinfo
  {author} {\bibfnamefont {R.~G.}\ \bibnamefont {{Bower}}}, \ and\ \bibinfo
  {author} {\bibfnamefont {R.~A.}\ \bibnamefont {{Crain}}},\ }\href@noop {}
  {\bibfield  {journal} {\bibinfo  {journal} {ArXiv e-prints}\ } (\bibinfo
  {year} {2016})},\ \Eprint {http://arxiv.org/abs/1603.03328}
  {arXiv:1603.03328} \BibitemShut {NoStop}%
\bibitem [{\citenamefont {{Kiessling}}\ \emph {et~al.}(2015)\citenamefont
  {{Kiessling}}, \citenamefont {{Cacciato}}, \citenamefont {{Joachimi}},
  \citenamefont {{Kirk}}, \citenamefont {{Kitching}}, \citenamefont
  {{Leonard}}, \citenamefont {{Mandelbaum}}, \citenamefont {{Sch{\"a}fer}},
  \citenamefont {{Sif{\'o}n}}, \citenamefont {{Brown}},\ and\ \citenamefont
  {{Rassat}}}]{2015SSRv..193...67K}%
  \BibitemOpen
  \bibfield  {author} {\bibinfo {author} {\bibfnamefont {A.}~\bibnamefont
  {{Kiessling}}}, \bibinfo {author} {\bibfnamefont {M.}~\bibnamefont
  {{Cacciato}}}, \bibinfo {author} {\bibfnamefont {B.}~\bibnamefont
  {{Joachimi}}}, \bibinfo {author} {\bibfnamefont {D.}~\bibnamefont {{Kirk}}},
  \bibinfo {author} {\bibfnamefont {T.~D.}\ \bibnamefont {{Kitching}}},
  \bibinfo {author} {\bibfnamefont {A.}~\bibnamefont {{Leonard}}}, \bibinfo
  {author} {\bibfnamefont {R.}~\bibnamefont {{Mandelbaum}}}, \bibinfo {author}
  {\bibfnamefont {B.~M.}\ \bibnamefont {{Sch{\"a}fer}}}, \bibinfo {author}
  {\bibfnamefont {C.}~\bibnamefont {{Sif{\'o}n}}}, \bibinfo {author}
  {\bibfnamefont {M.~L.}\ \bibnamefont {{Brown}}}, \ and\ \bibinfo {author}
  {\bibfnamefont {A.}~\bibnamefont {{Rassat}}},\ }\href {\doibase
  10.1007/s11214-015-0203-6} {\bibfield  {journal} {\bibinfo  {journal} {\ssr}\
  }\textbf {\bibinfo {volume} {193}},\ \bibinfo {pages} {67} (\bibinfo {year}
  {2015})},\ \Eprint {http://arxiv.org/abs/1504.05546} {arXiv:1504.05546}
  \BibitemShut {NoStop}%
\bibitem [{\citenamefont {{Kirk}}\ \emph {et~al.}(2015)\citenamefont {{Kirk}},
  \citenamefont {{Brown}}, \citenamefont {{Hoekstra}}, \citenamefont
  {{Joachimi}}, \citenamefont {{Kitching}}, \citenamefont {{Mandelbaum}},
  \citenamefont {{Sif{\'o}n}}, \citenamefont {{Cacciato}}, \citenamefont
  {{Choi}}, \citenamefont {{Kiessling}}, \citenamefont {{Leonard}},
  \citenamefont {{Rassat}},\ and\ \citenamefont
  {{Sch{\"a}fer}}}]{2015SSRv..193..139K}%
  \BibitemOpen
  \bibfield  {author} {\bibinfo {author} {\bibfnamefont {D.}~\bibnamefont
  {{Kirk}}}, \bibinfo {author} {\bibfnamefont {M.~L.}\ \bibnamefont {{Brown}}},
  \bibinfo {author} {\bibfnamefont {H.}~\bibnamefont {{Hoekstra}}}, \bibinfo
  {author} {\bibfnamefont {B.}~\bibnamefont {{Joachimi}}}, \bibinfo {author}
  {\bibfnamefont {T.~D.}\ \bibnamefont {{Kitching}}}, \bibinfo {author}
  {\bibfnamefont {R.}~\bibnamefont {{Mandelbaum}}}, \bibinfo {author}
  {\bibfnamefont {C.}~\bibnamefont {{Sif{\'o}n}}}, \bibinfo {author}
  {\bibfnamefont {M.}~\bibnamefont {{Cacciato}}}, \bibinfo {author}
  {\bibfnamefont {A.}~\bibnamefont {{Choi}}}, \bibinfo {author} {\bibfnamefont
  {A.}~\bibnamefont {{Kiessling}}}, \bibinfo {author} {\bibfnamefont
  {A.}~\bibnamefont {{Leonard}}}, \bibinfo {author} {\bibfnamefont
  {A.}~\bibnamefont {{Rassat}}}, \ and\ \bibinfo {author} {\bibfnamefont
  {B.~M.}\ \bibnamefont {{Sch{\"a}fer}}},\ }\href {\doibase
  10.1007/s11214-015-0213-4} {\bibfield  {journal} {\bibinfo  {journal} {\ssr}\
  }\textbf {\bibinfo {volume} {193}},\ \bibinfo {pages} {139} (\bibinfo {year}
  {2015})},\ \Eprint {http://arxiv.org/abs/1504.05465} {arXiv:1504.05465}
  \BibitemShut {NoStop}%
\bibitem [{\citenamefont {{Joachimi}}\ \emph {et~al.}(2015)\citenamefont
  {{Joachimi}}, \citenamefont {{Cacciato}}, \citenamefont {{Kitching}},
  \citenamefont {{Leonard}}, \citenamefont {{Mandelbaum}}, \citenamefont
  {{Sch{\"a}fer}}, \citenamefont {{Sif{\'o}n}}, \citenamefont {{Hoekstra}},
  \citenamefont {{Kiessling}}, \citenamefont {{Kirk}},\ and\ \citenamefont
  {{Rassat}}}]{2015SSRv..193....1J}%
  \BibitemOpen
  \bibfield  {author} {\bibinfo {author} {\bibfnamefont {B.}~\bibnamefont
  {{Joachimi}}}, \bibinfo {author} {\bibfnamefont {M.}~\bibnamefont
  {{Cacciato}}}, \bibinfo {author} {\bibfnamefont {T.~D.}\ \bibnamefont
  {{Kitching}}}, \bibinfo {author} {\bibfnamefont {A.}~\bibnamefont
  {{Leonard}}}, \bibinfo {author} {\bibfnamefont {R.}~\bibnamefont
  {{Mandelbaum}}}, \bibinfo {author} {\bibfnamefont {B.~M.}\ \bibnamefont
  {{Sch{\"a}fer}}}, \bibinfo {author} {\bibfnamefont {C.}~\bibnamefont
  {{Sif{\'o}n}}}, \bibinfo {author} {\bibfnamefont {H.}~\bibnamefont
  {{Hoekstra}}}, \bibinfo {author} {\bibfnamefont {A.}~\bibnamefont
  {{Kiessling}}}, \bibinfo {author} {\bibfnamefont {D.}~\bibnamefont {{Kirk}}},
  \ and\ \bibinfo {author} {\bibfnamefont {A.}~\bibnamefont {{Rassat}}},\
  }\href {\doibase 10.1007/s11214-015-0177-4} {\bibfield  {journal} {\bibinfo
  {journal} {\ssr}\ }\textbf {\bibinfo {volume} {193}},\ \bibinfo {pages} {1}
  (\bibinfo {year} {2015})},\ \Eprint {http://arxiv.org/abs/1504.05456}
  {arXiv:1504.05456} \BibitemShut {NoStop}%
\bibitem [{\citenamefont {{Troxel}}\ and\ \citenamefont
  {{Ishak}}(2015)}]{2015PhR...558....1T}%
  \BibitemOpen
  \bibfield  {author} {\bibinfo {author} {\bibfnamefont {M.~A.}\ \bibnamefont
  {{Troxel}}}\ and\ \bibinfo {author} {\bibfnamefont {M.}~\bibnamefont
  {{Ishak}}},\ }\href {\doibase 10.1016/j.physrep.2014.11.001} {\bibfield
  {journal} {\bibinfo  {journal} {\physrep}\ }\textbf {\bibinfo {volume}
  {558}},\ \bibinfo {pages} {1} (\bibinfo {year} {2015})},\ \Eprint
  {http://arxiv.org/abs/1407.6990} {arXiv:1407.6990} \BibitemShut {NoStop}%
\bibitem [{\citenamefont {{Massey}}\ \emph {et~al.}(2013)\citenamefont
  {{Massey}}, \citenamefont {{Hoekstra}}, \citenamefont {{Kitching}},
  \citenamefont {{Rhodes}}, \citenamefont {{Cropper}}, \citenamefont
  {{Amiaux}}, \citenamefont {{Harvey}}, \citenamefont {{Mellier}},
  \citenamefont {{Meneghetti}}, \citenamefont {{Miller}}, \citenamefont
  {{Paulin-Henriksson}}, \citenamefont {{Pires}}, \citenamefont
  {{Scaramella}},\ and\ \citenamefont {{Schrabback}}}]{2013MNRAS.429..661M}%
  \BibitemOpen
  \bibfield  {author} {\bibinfo {author} {\bibfnamefont {R.}~\bibnamefont
  {{Massey}}}, \bibinfo {author} {\bibfnamefont {H.}~\bibnamefont
  {{Hoekstra}}}, \bibinfo {author} {\bibfnamefont {T.}~\bibnamefont
  {{Kitching}}}, \bibinfo {author} {\bibfnamefont {J.}~\bibnamefont
  {{Rhodes}}}, \bibinfo {author} {\bibfnamefont {M.}~\bibnamefont {{Cropper}}},
  \bibinfo {author} {\bibfnamefont {J.}~\bibnamefont {{Amiaux}}}, \bibinfo
  {author} {\bibfnamefont {D.}~\bibnamefont {{Harvey}}}, \bibinfo {author}
  {\bibfnamefont {Y.}~\bibnamefont {{Mellier}}}, \bibinfo {author}
  {\bibfnamefont {M.}~\bibnamefont {{Meneghetti}}}, \bibinfo {author}
  {\bibfnamefont {L.}~\bibnamefont {{Miller}}}, \bibinfo {author}
  {\bibfnamefont {S.}~\bibnamefont {{Paulin-Henriksson}}}, \bibinfo {author}
  {\bibfnamefont {S.}~\bibnamefont {{Pires}}}, \bibinfo {author} {\bibfnamefont
  {R.}~\bibnamefont {{Scaramella}}}, \ and\ \bibinfo {author} {\bibfnamefont
  {T.}~\bibnamefont {{Schrabback}}},\ }\href {\doibase 10.1093/mnras/sts371}
  {\bibfield  {journal} {\bibinfo  {journal} {\mnras}\ }\textbf {\bibinfo
  {volume} {429}},\ \bibinfo {pages} {661} (\bibinfo {year} {2013})},\ \Eprint
  {http://arxiv.org/abs/1210.7690} {arXiv:1210.7690} \BibitemShut {NoStop}%
\bibitem [{\citenamefont {{Jarvis}}\ \emph {et~al.}(2016)\citenamefont
  {{Jarvis}}, \citenamefont {{Sheldon}}, \citenamefont {{Zuntz}}, \citenamefont
  {{Kacprzak}}, \citenamefont {{Bridle}}, \citenamefont {{Amara}},
  \citenamefont {{Armstrong}}, \citenamefont {{Becker}}, \citenamefont
  {{Bernstein}}, \citenamefont {{Bonnett}}, \citenamefont {{Chang}},
  \citenamefont {{Das}}, \citenamefont {{Dietrich}}, \citenamefont
  {{Drlica-Wagner}}, \citenamefont {{Eifler}}, \citenamefont {{Gangkofner}},
  \citenamefont {{Gruen}}, \citenamefont {{Hirsch}}, \citenamefont {{Huff}},
  \citenamefont {{Jain}}, \citenamefont {{Kent}}, \citenamefont {{Kirk}},
  \citenamefont {{MacCrann}}, \citenamefont {{Melchior}}, \citenamefont
  {{Plazas}}, \citenamefont {{Refregier}}, \citenamefont {{Rowe}},
  \citenamefont {{Rykoff}}, \citenamefont {{Samuroff}}, \citenamefont
  {{S{\'a}nchez}}, \citenamefont {{Suchyta}}, \citenamefont {{Troxel}},
  \citenamefont {{Vikram}}, \citenamefont {{Abbott}}, \citenamefont
  {{Abdalla}}, \citenamefont {{Allam}}, \citenamefont {{Annis}}, \citenamefont
  {{Benoit-L{\'e}vy}}, \citenamefont {{Bertin}}, \citenamefont {{Brooks}},
  \citenamefont {{Buckley-Geer}}, \citenamefont {{Burke}}, \citenamefont
  {{Capozzi}}, \citenamefont {{Rosell}}, \citenamefont {{Kind}}, \citenamefont
  {{Carretero}}, \citenamefont {{Castander}}, \citenamefont {{Clampitt}},
  \citenamefont {{Crocce}}, \citenamefont {{Cunha}}, \citenamefont
  {{D'Andrea}}, \citenamefont {{da Costa}}, \citenamefont {{DePoy}},
  \citenamefont {{Desai}}, \citenamefont {{Diehl}}, \citenamefont {{Doel}},
  \citenamefont {{Neto}}, \citenamefont {{Flaugher}}, \citenamefont
  {{Fosalba}}, \citenamefont {{Frieman}}, \citenamefont {{Gaztanaga}},
  \citenamefont {{Gerdes}}, \citenamefont {{Gruendl}}, \citenamefont
  {{Gutierrez}}, \citenamefont {{Honscheid}}, \citenamefont {{James}},
  \citenamefont {{Kuehn}}, \citenamefont {{Kuropatkin}}, \citenamefont
  {{Lahav}}, \citenamefont {{Li}}, \citenamefont {{Lima}}, \citenamefont
  {{March}}, \citenamefont {{Martini}}, \citenamefont {{Miquel}}, \citenamefont
  {{Mohr}}, \citenamefont {{Neilsen}}, \citenamefont {{Nord}}, \citenamefont
  {{Ogando}}, \citenamefont {{Reil}}, \citenamefont {{Romer}}, \citenamefont
  {{Roodman}}, \citenamefont {{Sako}}, \citenamefont {{Sanchez}}, \citenamefont
  {{Scarpine}}, \citenamefont {{Schubnell}}, \citenamefont {{Sevilla-Noarbe}},
  \citenamefont {{Smith}}, \citenamefont {{Soares-Santos}}, \citenamefont
  {{Sobreira}}, \citenamefont {{Swanson}}, \citenamefont {{Tarle}},
  \citenamefont {{Thaler}}, \citenamefont {{Thomas}}, \citenamefont
  {{Walker}},\ and\ \citenamefont {{Wechsler}}}]{2016MNRAS.tmp..827J}%
  \BibitemOpen
  \bibfield  {author} {\bibinfo {author} {\bibfnamefont {M.}~\bibnamefont
  {{Jarvis}}}, \bibinfo {author} {\bibfnamefont {E.}~\bibnamefont {{Sheldon}}},
  \bibinfo {author} {\bibfnamefont {J.}~\bibnamefont {{Zuntz}}}, \bibinfo
  {author} {\bibfnamefont {T.}~\bibnamefont {{Kacprzak}}}, \bibinfo {author}
  {\bibfnamefont {S.~L.}\ \bibnamefont {{Bridle}}}, \bibinfo {author}
  {\bibfnamefont {A.}~\bibnamefont {{Amara}}}, \bibinfo {author} {\bibfnamefont
  {R.}~\bibnamefont {{Armstrong}}}, \bibinfo {author} {\bibfnamefont {M.~R.}\
  \bibnamefont {{Becker}}}, \bibinfo {author} {\bibfnamefont {G.~M.}\
  \bibnamefont {{Bernstein}}}, \bibinfo {author} {\bibfnamefont
  {C.}~\bibnamefont {{Bonnett}}}, \bibinfo {author} {\bibfnamefont
  {C.}~\bibnamefont {{Chang}}}, \bibinfo {author} {\bibfnamefont
  {R.}~\bibnamefont {{Das}}}, \bibinfo {author} {\bibfnamefont {J.~P.}\
  \bibnamefont {{Dietrich}}}, \bibinfo {author} {\bibfnamefont
  {A.}~\bibnamefont {{Drlica-Wagner}}}, \bibinfo {author} {\bibfnamefont
  {T.~F.}\ \bibnamefont {{Eifler}}}, \bibinfo {author} {\bibfnamefont
  {C.}~\bibnamefont {{Gangkofner}}}, \bibinfo {author} {\bibfnamefont
  {D.}~\bibnamefont {{Gruen}}}, \bibinfo {author} {\bibfnamefont
  {M.}~\bibnamefont {{Hirsch}}}, \bibinfo {author} {\bibfnamefont {E.~M.}\
  \bibnamefont {{Huff}}}, \bibinfo {author} {\bibfnamefont {B.}~\bibnamefont
  {{Jain}}}, \bibinfo {author} {\bibfnamefont {S.}~\bibnamefont {{Kent}}},
  \bibinfo {author} {\bibfnamefont {D.}~\bibnamefont {{Kirk}}}, \bibinfo
  {author} {\bibfnamefont {N.}~\bibnamefont {{MacCrann}}}, \bibinfo {author}
  {\bibfnamefont {P.}~\bibnamefont {{Melchior}}}, \bibinfo {author}
  {\bibfnamefont {A.~A.}\ \bibnamefont {{Plazas}}}, \bibinfo {author}
  {\bibfnamefont {A.}~\bibnamefont {{Refregier}}}, \bibinfo {author}
  {\bibfnamefont {B.}~\bibnamefont {{Rowe}}}, \bibinfo {author} {\bibfnamefont
  {E.~S.}\ \bibnamefont {{Rykoff}}}, \bibinfo {author} {\bibfnamefont
  {S.}~\bibnamefont {{Samuroff}}}, \bibinfo {author} {\bibfnamefont
  {C.}~\bibnamefont {{S{\'a}nchez}}}, \bibinfo {author} {\bibfnamefont
  {E.}~\bibnamefont {{Suchyta}}}, \bibinfo {author} {\bibfnamefont {M.~A.}\
  \bibnamefont {{Troxel}}}, \bibinfo {author} {\bibfnamefont {V.}~\bibnamefont
  {{Vikram}}}, \bibinfo {author} {\bibfnamefont {T.}~\bibnamefont {{Abbott}}},
  \bibinfo {author} {\bibfnamefont {F.~B.}\ \bibnamefont {{Abdalla}}}, \bibinfo
  {author} {\bibfnamefont {S.}~\bibnamefont {{Allam}}}, \bibinfo {author}
  {\bibfnamefont {J.}~\bibnamefont {{Annis}}}, \bibinfo {author} {\bibfnamefont
  {A.}~\bibnamefont {{Benoit-L{\'e}vy}}}, \bibinfo {author} {\bibfnamefont
  {E.}~\bibnamefont {{Bertin}}}, \bibinfo {author} {\bibfnamefont
  {D.}~\bibnamefont {{Brooks}}}, \bibinfo {author} {\bibfnamefont
  {E.}~\bibnamefont {{Buckley-Geer}}}, \bibinfo {author} {\bibfnamefont
  {D.~L.}\ \bibnamefont {{Burke}}}, \bibinfo {author} {\bibfnamefont
  {D.}~\bibnamefont {{Capozzi}}}, \bibinfo {author} {\bibfnamefont {A.~C.}\
  \bibnamefont {{Rosell}}}, \bibinfo {author} {\bibfnamefont {M.~C.}\
  \bibnamefont {{Kind}}}, \bibinfo {author} {\bibfnamefont {J.}~\bibnamefont
  {{Carretero}}}, \bibinfo {author} {\bibfnamefont {F.~J.}\ \bibnamefont
  {{Castander}}}, \bibinfo {author} {\bibfnamefont {J.}~\bibnamefont
  {{Clampitt}}}, \bibinfo {author} {\bibfnamefont {M.}~\bibnamefont
  {{Crocce}}}, \bibinfo {author} {\bibfnamefont {C.~E.}\ \bibnamefont
  {{Cunha}}}, \bibinfo {author} {\bibfnamefont {C.~B.}\ \bibnamefont
  {{D'Andrea}}}, \bibinfo {author} {\bibfnamefont {L.~N.}\ \bibnamefont {{da
  Costa}}}, \bibinfo {author} {\bibfnamefont {D.~L.}\ \bibnamefont {{DePoy}}},
  \bibinfo {author} {\bibfnamefont {S.}~\bibnamefont {{Desai}}}, \bibinfo
  {author} {\bibfnamefont {H.~T.}\ \bibnamefont {{Diehl}}}, \bibinfo {author}
  {\bibfnamefont {P.}~\bibnamefont {{Doel}}}, \bibinfo {author} {\bibfnamefont
  {A.~F.}\ \bibnamefont {{Neto}}}, \bibinfo {author} {\bibfnamefont
  {B.}~\bibnamefont {{Flaugher}}}, \bibinfo {author} {\bibfnamefont
  {P.}~\bibnamefont {{Fosalba}}}, \bibinfo {author} {\bibfnamefont
  {J.}~\bibnamefont {{Frieman}}}, \bibinfo {author} {\bibfnamefont
  {E.}~\bibnamefont {{Gaztanaga}}}, \bibinfo {author} {\bibfnamefont {D.~W.}\
  \bibnamefont {{Gerdes}}}, \bibinfo {author} {\bibfnamefont {R.~A.}\
  \bibnamefont {{Gruendl}}}, \bibinfo {author} {\bibfnamefont {G.}~\bibnamefont
  {{Gutierrez}}}, \bibinfo {author} {\bibfnamefont {K.}~\bibnamefont
  {{Honscheid}}}, \bibinfo {author} {\bibfnamefont {D.~J.}\ \bibnamefont
  {{James}}}, \bibinfo {author} {\bibfnamefont {K.}~\bibnamefont {{Kuehn}}},
  \bibinfo {author} {\bibfnamefont {N.}~\bibnamefont {{Kuropatkin}}}, \bibinfo
  {author} {\bibfnamefont {O.}~\bibnamefont {{Lahav}}}, \bibinfo {author}
  {\bibfnamefont {T.~S.}\ \bibnamefont {{Li}}}, \bibinfo {author}
  {\bibfnamefont {M.}~\bibnamefont {{Lima}}}, \bibinfo {author} {\bibfnamefont
  {M.}~\bibnamefont {{March}}}, \bibinfo {author} {\bibfnamefont
  {P.}~\bibnamefont {{Martini}}}, \bibinfo {author} {\bibfnamefont
  {R.}~\bibnamefont {{Miquel}}}, \bibinfo {author} {\bibfnamefont {J.~J.}\
  \bibnamefont {{Mohr}}}, \bibinfo {author} {\bibfnamefont {E.}~\bibnamefont
  {{Neilsen}}}, \bibinfo {author} {\bibfnamefont {B.}~\bibnamefont {{Nord}}},
  \bibinfo {author} {\bibfnamefont {R.}~\bibnamefont {{Ogando}}}, \bibinfo
  {author} {\bibfnamefont {K.}~\bibnamefont {{Reil}}}, \bibinfo {author}
  {\bibfnamefont {A.~K.}\ \bibnamefont {{Romer}}}, \bibinfo {author}
  {\bibfnamefont {A.}~\bibnamefont {{Roodman}}}, \bibinfo {author}
  {\bibfnamefont {M.}~\bibnamefont {{Sako}}}, \bibinfo {author} {\bibfnamefont
  {E.}~\bibnamefont {{Sanchez}}}, \bibinfo {author} {\bibfnamefont
  {V.}~\bibnamefont {{Scarpine}}}, \bibinfo {author} {\bibfnamefont
  {M.}~\bibnamefont {{Schubnell}}}, \bibinfo {author} {\bibfnamefont
  {I.}~\bibnamefont {{Sevilla-Noarbe}}}, \bibinfo {author} {\bibfnamefont
  {R.~C.}\ \bibnamefont {{Smith}}}, \bibinfo {author} {\bibfnamefont
  {M.}~\bibnamefont {{Soares-Santos}}}, \bibinfo {author} {\bibfnamefont
  {F.}~\bibnamefont {{Sobreira}}}, \bibinfo {author} {\bibfnamefont {M.~E.~C.}\
  \bibnamefont {{Swanson}}}, \bibinfo {author} {\bibfnamefont {G.}~\bibnamefont
  {{Tarle}}}, \bibinfo {author} {\bibfnamefont {J.}~\bibnamefont {{Thaler}}},
  \bibinfo {author} {\bibfnamefont {D.}~\bibnamefont {{Thomas}}}, \bibinfo
  {author} {\bibfnamefont {A.~R.}\ \bibnamefont {{Walker}}}, \ and\ \bibinfo
  {author} {\bibfnamefont {R.~H.}\ \bibnamefont {{Wechsler}}},\ }\href
  {\doibase 10.1093/mnras/stw990} {\bibfield  {journal} {\bibinfo  {journal}
  {\mnras}\ } (\bibinfo {year} {2016}),\ 10.1093/mnras/stw990}\BibitemShut
  {NoStop}%
\bibitem [{\citenamefont {{Mandelbaum}}\ \emph {et~al.}(2015)\citenamefont
  {{Mandelbaum}}, \citenamefont {{Rowe}}, \citenamefont {{Armstrong}},
  \citenamefont {{Bard}}, \citenamefont {{Bertin}}, \citenamefont {{Bosch}},
  \citenamefont {{Boutigny}}, \citenamefont {{Courbin}}, \citenamefont
  {{Dawson}}, \citenamefont {{Donnarumma}}, \citenamefont {{Fenech Conti}},
  \citenamefont {{Gavazzi}}, \citenamefont {{Gentile}}, \citenamefont {{Gill}},
  \citenamefont {{Hogg}}, \citenamefont {{Huff}}, \citenamefont {{Jee}},
  \citenamefont {{Kacprzak}}, \citenamefont {{Kilbinger}}, \citenamefont
  {{Kuntzer}}, \citenamefont {{Lang}}, \citenamefont {{Luo}}, \citenamefont
  {{March}}, \citenamefont {{Marshall}}, \citenamefont {{Meyers}},
  \citenamefont {{Miller}}, \citenamefont {{Miyatake}}, \citenamefont
  {{Nakajima}}, \citenamefont {{Ngol{\'e} Mboula}}, \citenamefont {{Nurbaeva}},
  \citenamefont {{Okura}}, \citenamefont {{Paulin-Henriksson}}, \citenamefont
  {{Rhodes}}, \citenamefont {{Schneider}}, \citenamefont {{Shan}},
  \citenamefont {{Sheldon}}, \citenamefont {{Simet}}, \citenamefont {{Starck}},
  \citenamefont {{Sureau}}, \citenamefont {{Tewes}}, \citenamefont {{Zarb
  Adami}}, \citenamefont {{Zhang}},\ and\ \citenamefont
  {{Zuntz}}}]{2015MNRAS.450.2963M}%
  \BibitemOpen
  \bibfield  {author} {\bibinfo {author} {\bibfnamefont {R.}~\bibnamefont
  {{Mandelbaum}}}, \bibinfo {author} {\bibfnamefont {B.}~\bibnamefont
  {{Rowe}}}, \bibinfo {author} {\bibfnamefont {R.}~\bibnamefont {{Armstrong}}},
  \bibinfo {author} {\bibfnamefont {D.}~\bibnamefont {{Bard}}}, \bibinfo
  {author} {\bibfnamefont {E.}~\bibnamefont {{Bertin}}}, \bibinfo {author}
  {\bibfnamefont {J.}~\bibnamefont {{Bosch}}}, \bibinfo {author} {\bibfnamefont
  {D.}~\bibnamefont {{Boutigny}}}, \bibinfo {author} {\bibfnamefont
  {F.}~\bibnamefont {{Courbin}}}, \bibinfo {author} {\bibfnamefont {W.~A.}\
  \bibnamefont {{Dawson}}}, \bibinfo {author} {\bibfnamefont {A.}~\bibnamefont
  {{Donnarumma}}}, \bibinfo {author} {\bibfnamefont {I.}~\bibnamefont {{Fenech
  Conti}}}, \bibinfo {author} {\bibfnamefont {R.}~\bibnamefont {{Gavazzi}}},
  \bibinfo {author} {\bibfnamefont {M.}~\bibnamefont {{Gentile}}}, \bibinfo
  {author} {\bibfnamefont {M.~S.~S.}\ \bibnamefont {{Gill}}}, \bibinfo {author}
  {\bibfnamefont {D.~W.}\ \bibnamefont {{Hogg}}}, \bibinfo {author}
  {\bibfnamefont {E.~M.}\ \bibnamefont {{Huff}}}, \bibinfo {author}
  {\bibfnamefont {M.~J.}\ \bibnamefont {{Jee}}}, \bibinfo {author}
  {\bibfnamefont {T.}~\bibnamefont {{Kacprzak}}}, \bibinfo {author}
  {\bibfnamefont {M.}~\bibnamefont {{Kilbinger}}}, \bibinfo {author}
  {\bibfnamefont {T.}~\bibnamefont {{Kuntzer}}}, \bibinfo {author}
  {\bibfnamefont {D.}~\bibnamefont {{Lang}}}, \bibinfo {author} {\bibfnamefont
  {W.}~\bibnamefont {{Luo}}}, \bibinfo {author} {\bibfnamefont {M.~C.}\
  \bibnamefont {{March}}}, \bibinfo {author} {\bibfnamefont {P.~J.}\
  \bibnamefont {{Marshall}}}, \bibinfo {author} {\bibfnamefont {J.~E.}\
  \bibnamefont {{Meyers}}}, \bibinfo {author} {\bibfnamefont {L.}~\bibnamefont
  {{Miller}}}, \bibinfo {author} {\bibfnamefont {H.}~\bibnamefont
  {{Miyatake}}}, \bibinfo {author} {\bibfnamefont {R.}~\bibnamefont
  {{Nakajima}}}, \bibinfo {author} {\bibfnamefont {F.~M.}\ \bibnamefont
  {{Ngol{\'e} Mboula}}}, \bibinfo {author} {\bibfnamefont {G.}~\bibnamefont
  {{Nurbaeva}}}, \bibinfo {author} {\bibfnamefont {Y.}~\bibnamefont {{Okura}}},
  \bibinfo {author} {\bibfnamefont {S.}~\bibnamefont {{Paulin-Henriksson}}},
  \bibinfo {author} {\bibfnamefont {J.}~\bibnamefont {{Rhodes}}}, \bibinfo
  {author} {\bibfnamefont {M.~D.}\ \bibnamefont {{Schneider}}}, \bibinfo
  {author} {\bibfnamefont {H.}~\bibnamefont {{Shan}}}, \bibinfo {author}
  {\bibfnamefont {E.~S.}\ \bibnamefont {{Sheldon}}}, \bibinfo {author}
  {\bibfnamefont {M.}~\bibnamefont {{Simet}}}, \bibinfo {author} {\bibfnamefont
  {J.-L.}\ \bibnamefont {{Starck}}}, \bibinfo {author} {\bibfnamefont
  {F.}~\bibnamefont {{Sureau}}}, \bibinfo {author} {\bibfnamefont
  {M.}~\bibnamefont {{Tewes}}}, \bibinfo {author} {\bibfnamefont
  {K.}~\bibnamefont {{Zarb Adami}}}, \bibinfo {author} {\bibfnamefont
  {J.}~\bibnamefont {{Zhang}}}, \ and\ \bibinfo {author} {\bibfnamefont
  {J.}~\bibnamefont {{Zuntz}}},\ }\href {\doibase 10.1093/mnras/stv781}
  {\bibfield  {journal} {\bibinfo  {journal} {\mnras}\ }\textbf {\bibinfo
  {volume} {450}},\ \bibinfo {pages} {2963} (\bibinfo {year} {2015})},\ \Eprint
  {http://arxiv.org/abs/1412.1825} {arXiv:1412.1825} \BibitemShut {NoStop}%
\bibitem [{\citenamefont {{Zaldarriaga}}\ and\ \citenamefont
  {{Seljak}}(1999)}]{1999PhRvD..59l3507Z}%
  \BibitemOpen
  \bibfield  {author} {\bibinfo {author} {\bibfnamefont {M.}~\bibnamefont
  {{Zaldarriaga}}}\ and\ \bibinfo {author} {\bibfnamefont {U.}~\bibnamefont
  {{Seljak}}},\ }\href {\doibase 10.1103/PhysRevD.59.123507} {\bibfield
  {journal} {\bibinfo  {journal} {\prd}\ }\textbf {\bibinfo {volume} {59}},\
  \bibinfo {eid} {123507} (\bibinfo {year} {1999})},\ \Eprint
  {http://arxiv.org/abs/astro-ph/9810257} {astro-ph/9810257} \BibitemShut
  {NoStop}%
\bibitem [{\citenamefont {{Hu}}\ and\ \citenamefont
  {{Okamoto}}(2002)}]{2002ApJ...574..566H}%
  \BibitemOpen
  \bibfield  {author} {\bibinfo {author} {\bibfnamefont {W.}~\bibnamefont
  {{Hu}}}\ and\ \bibinfo {author} {\bibfnamefont {T.}~\bibnamefont
  {{Okamoto}}},\ }\href {\doibase 10.1086/341110} {\bibfield  {journal}
  {\bibinfo  {journal} {\apj}\ }\textbf {\bibinfo {volume} {574}},\ \bibinfo
  {pages} {566} (\bibinfo {year} {2002})},\ \Eprint
  {http://arxiv.org/abs/astro-ph/0111606} {astro-ph/0111606} \BibitemShut
  {NoStop}%
\bibitem [{\citenamefont {{Okamoto}}\ and\ \citenamefont
  {{Hu}}(2003)}]{2003PhRvD..67h3002O}%
  \BibitemOpen
  \bibfield  {author} {\bibinfo {author} {\bibfnamefont {T.}~\bibnamefont
  {{Okamoto}}}\ and\ \bibinfo {author} {\bibfnamefont {W.}~\bibnamefont
  {{Hu}}},\ }\href {\doibase 10.1103/PhysRevD.67.083002} {\bibfield  {journal}
  {\bibinfo  {journal} {\prd}\ }\textbf {\bibinfo {volume} {67}},\ \bibinfo
  {eid} {083002} (\bibinfo {year} {2003})},\ \Eprint
  {http://arxiv.org/abs/astro-ph/0301031} {astro-ph/0301031} \BibitemShut
  {NoStop}%
\bibitem [{\citenamefont {{Vallinotto}}(2012)}]{2012ApJ...759...32V}%
  \BibitemOpen
  \bibfield  {author} {\bibinfo {author} {\bibfnamefont {A.}~\bibnamefont
  {{Vallinotto}}},\ }\href {\doibase 10.1088/0004-637X/759/1/32} {\bibfield
  {journal} {\bibinfo  {journal} {\apj}\ }\textbf {\bibinfo {volume} {759}},\
  \bibinfo {eid} {32} (\bibinfo {year} {2012})},\ \Eprint
  {http://arxiv.org/abs/1110.5339} {arXiv:1110.5339 [astro-ph.CO]} \BibitemShut
  {NoStop}%
\bibitem [{\citenamefont {{Vallinotto}}(2013)}]{2013ApJ...778..108V}%
  \BibitemOpen
  \bibfield  {author} {\bibinfo {author} {\bibfnamefont {A.}~\bibnamefont
  {{Vallinotto}}},\ }\href {\doibase 10.1088/0004-637X/778/2/108} {\bibfield
  {journal} {\bibinfo  {journal} {\apj}\ }\textbf {\bibinfo {volume} {778}},\
  \bibinfo {eid} {108} (\bibinfo {year} {2013})},\ \Eprint
  {http://arxiv.org/abs/1304.3474} {arXiv:1304.3474 [astro-ph.CO]} \BibitemShut
  {NoStop}%
\bibitem [{\citenamefont {{Das}}\ \emph {et~al.}(2013)\citenamefont {{Das}},
  \citenamefont {{Errard}},\ and\ \citenamefont
  {{Spergel}}}]{2013arXiv1311.2338D}%
  \BibitemOpen
  \bibfield  {author} {\bibinfo {author} {\bibfnamefont {S.}~\bibnamefont
  {{Das}}}, \bibinfo {author} {\bibfnamefont {J.}~\bibnamefont {{Errard}}}, \
  and\ \bibinfo {author} {\bibfnamefont {D.}~\bibnamefont {{Spergel}}},\
  }\href@noop {} {\bibfield  {journal} {\bibinfo  {journal} {ArXiv e-prints}\ }
  (\bibinfo {year} {2013})},\ \Eprint {http://arxiv.org/abs/1311.2338}
  {arXiv:1311.2338 [astro-ph.CO]} \BibitemShut {NoStop}%
\bibitem [{\citenamefont {{Smith}}\ \emph {et~al.}(2007)\citenamefont
  {{Smith}}, \citenamefont {{Zahn}},\ and\ \citenamefont
  {{Dor{\'e}}}}]{2007PhRvD..76d3510S}%
  \BibitemOpen
  \bibfield  {author} {\bibinfo {author} {\bibfnamefont {K.~M.}\ \bibnamefont
  {{Smith}}}, \bibinfo {author} {\bibfnamefont {O.}~\bibnamefont {{Zahn}}}, \
  and\ \bibinfo {author} {\bibfnamefont {O.}~\bibnamefont {{Dor{\'e}}}},\
  }\href {\doibase 10.1103/PhysRevD.76.043510} {\bibfield  {journal} {\bibinfo
  {journal} {\prd}\ }\textbf {\bibinfo {volume} {76}},\ \bibinfo {eid} {043510}
  (\bibinfo {year} {2007})},\ \Eprint {http://arxiv.org/abs/0705.3980}
  {arXiv:0705.3980} \BibitemShut {NoStop}%
\bibitem [{\citenamefont {{Hirata}}\ \emph {et~al.}(2008)\citenamefont
  {{Hirata}}, \citenamefont {{Ho}}, \citenamefont {{Padmanabhan}},
  \citenamefont {{Seljak}},\ and\ \citenamefont
  {{Bahcall}}}]{2008PhRvD..78d3520H}%
  \BibitemOpen
  \bibfield  {author} {\bibinfo {author} {\bibfnamefont {C.~M.}\ \bibnamefont
  {{Hirata}}}, \bibinfo {author} {\bibfnamefont {S.}~\bibnamefont {{Ho}}},
  \bibinfo {author} {\bibfnamefont {N.}~\bibnamefont {{Padmanabhan}}}, \bibinfo
  {author} {\bibfnamefont {U.}~\bibnamefont {{Seljak}}}, \ and\ \bibinfo
  {author} {\bibfnamefont {N.~A.}\ \bibnamefont {{Bahcall}}},\ }\href {\doibase
  10.1103/PhysRevD.78.043520} {\bibfield  {journal} {\bibinfo  {journal}
  {\prd}\ }\textbf {\bibinfo {volume} {78}},\ \bibinfo {eid} {043520} (\bibinfo
  {year} {2008})},\ \Eprint {http://arxiv.org/abs/0801.0644} {arXiv:0801.0644}
  \BibitemShut {NoStop}%
\bibitem [{\citenamefont {{Das}}\ \emph {et~al.}(2011)\citenamefont {{Das}},
  \citenamefont {{Sherwin}}, \citenamefont {{Aguirre}}, \citenamefont
  {{Appel}}, \citenamefont {{Bond}}, \citenamefont {{Carvalho}}, \citenamefont
  {{Devlin}}, \citenamefont {{Dunkley}}, \citenamefont {{D{\"u}nner}},
  \citenamefont {{Essinger-Hileman}}, \citenamefont {{Fowler}}, \citenamefont
  {{Hajian}}, \citenamefont {{Halpern}}, \citenamefont {{Hasselfield}},
  \citenamefont {{Hincks}}, \citenamefont {{Hlozek}}, \citenamefont
  {{Huffenberger}}, \citenamefont {{Hughes}}, \citenamefont {{Irwin}},
  \citenamefont {{Klein}}, \citenamefont {{Kosowsky}}, \citenamefont
  {{Lupton}}, \citenamefont {{Marriage}}, \citenamefont {{Marsden}},
  \citenamefont {{Menanteau}}, \citenamefont {{Moodley}}, \citenamefont
  {{Niemack}}, \citenamefont {{Nolta}}, \citenamefont {{Page}}, \citenamefont
  {{Parker}}, \citenamefont {{Reese}}, \citenamefont {{Schmitt}}, \citenamefont
  {{Sehgal}}, \citenamefont {{Sievers}}, \citenamefont {{Spergel}},
  \citenamefont {{Staggs}}, \citenamefont {{Swetz}}, \citenamefont {{Switzer}},
  \citenamefont {{Thornton}}, \citenamefont {{Visnjic}},\ and\ \citenamefont
  {{Wollack}}}]{2011PhRvL.107b1301D}%
  \BibitemOpen
  \bibfield  {author} {\bibinfo {author} {\bibfnamefont {S.}~\bibnamefont
  {{Das}}}, \bibinfo {author} {\bibfnamefont {B.~D.}\ \bibnamefont
  {{Sherwin}}}, \bibinfo {author} {\bibfnamefont {P.}~\bibnamefont
  {{Aguirre}}}, \bibinfo {author} {\bibfnamefont {J.~W.}\ \bibnamefont
  {{Appel}}}, \bibinfo {author} {\bibfnamefont {J.~R.}\ \bibnamefont {{Bond}}},
  \bibinfo {author} {\bibfnamefont {C.~S.}\ \bibnamefont {{Carvalho}}},
  \bibinfo {author} {\bibfnamefont {M.~J.}\ \bibnamefont {{Devlin}}}, \bibinfo
  {author} {\bibfnamefont {J.}~\bibnamefont {{Dunkley}}}, \bibinfo {author}
  {\bibfnamefont {R.}~\bibnamefont {{D{\"u}nner}}}, \bibinfo {author}
  {\bibfnamefont {T.}~\bibnamefont {{Essinger-Hileman}}}, \bibinfo {author}
  {\bibfnamefont {J.~W.}\ \bibnamefont {{Fowler}}}, \bibinfo {author}
  {\bibfnamefont {A.}~\bibnamefont {{Hajian}}}, \bibinfo {author}
  {\bibfnamefont {M.}~\bibnamefont {{Halpern}}}, \bibinfo {author}
  {\bibfnamefont {M.}~\bibnamefont {{Hasselfield}}}, \bibinfo {author}
  {\bibfnamefont {A.~D.}\ \bibnamefont {{Hincks}}}, \bibinfo {author}
  {\bibfnamefont {R.}~\bibnamefont {{Hlozek}}}, \bibinfo {author}
  {\bibfnamefont {K.~M.}\ \bibnamefont {{Huffenberger}}}, \bibinfo {author}
  {\bibfnamefont {J.~P.}\ \bibnamefont {{Hughes}}}, \bibinfo {author}
  {\bibfnamefont {K.~D.}\ \bibnamefont {{Irwin}}}, \bibinfo {author}
  {\bibfnamefont {J.}~\bibnamefont {{Klein}}}, \bibinfo {author} {\bibfnamefont
  {A.}~\bibnamefont {{Kosowsky}}}, \bibinfo {author} {\bibfnamefont {R.~H.}\
  \bibnamefont {{Lupton}}}, \bibinfo {author} {\bibfnamefont {T.~A.}\
  \bibnamefont {{Marriage}}}, \bibinfo {author} {\bibfnamefont
  {D.}~\bibnamefont {{Marsden}}}, \bibinfo {author} {\bibfnamefont
  {F.}~\bibnamefont {{Menanteau}}}, \bibinfo {author} {\bibfnamefont
  {K.}~\bibnamefont {{Moodley}}}, \bibinfo {author} {\bibfnamefont {M.~D.}\
  \bibnamefont {{Niemack}}}, \bibinfo {author} {\bibfnamefont {M.~R.}\
  \bibnamefont {{Nolta}}}, \bibinfo {author} {\bibfnamefont {L.~A.}\
  \bibnamefont {{Page}}}, \bibinfo {author} {\bibfnamefont {L.}~\bibnamefont
  {{Parker}}}, \bibinfo {author} {\bibfnamefont {E.~D.}\ \bibnamefont
  {{Reese}}}, \bibinfo {author} {\bibfnamefont {B.~L.}\ \bibnamefont
  {{Schmitt}}}, \bibinfo {author} {\bibfnamefont {N.}~\bibnamefont {{Sehgal}}},
  \bibinfo {author} {\bibfnamefont {J.}~\bibnamefont {{Sievers}}}, \bibinfo
  {author} {\bibfnamefont {D.~N.}\ \bibnamefont {{Spergel}}}, \bibinfo {author}
  {\bibfnamefont {S.~T.}\ \bibnamefont {{Staggs}}}, \bibinfo {author}
  {\bibfnamefont {D.~S.}\ \bibnamefont {{Swetz}}}, \bibinfo {author}
  {\bibfnamefont {E.~R.}\ \bibnamefont {{Switzer}}}, \bibinfo {author}
  {\bibfnamefont {R.}~\bibnamefont {{Thornton}}}, \bibinfo {author}
  {\bibfnamefont {K.}~\bibnamefont {{Visnjic}}}, \ and\ \bibinfo {author}
  {\bibfnamefont {E.}~\bibnamefont {{Wollack}}},\ }\href {\doibase
  10.1103/PhysRevLett.107.021301} {\bibfield  {journal} {\bibinfo  {journal}
  {Physical Review Letters}\ }\textbf {\bibinfo {volume} {107}},\ \bibinfo
  {eid} {021301} (\bibinfo {year} {2011})},\ \Eprint
  {http://arxiv.org/abs/1103.2124} {arXiv:1103.2124} \BibitemShut {NoStop}%
\bibitem [{\citenamefont {{Das}}\ \emph {et~al.}(2014)\citenamefont {{Das}},
  \citenamefont {{Louis}}, \citenamefont {{Nolta}}, \citenamefont {{Addison}},
  \citenamefont {{Battistelli}}, \citenamefont {{Bond}}, \citenamefont
  {{Calabrese}}, \citenamefont {{Crichton}}, \citenamefont {{Devlin}},
  \citenamefont {{Dicker}}, \citenamefont {{Dunkley}}, \citenamefont
  {{D{\"u}nner}}, \citenamefont {{Fowler}}, \citenamefont {{Gralla}},
  \citenamefont {{Hajian}}, \citenamefont {{Halpern}}, \citenamefont
  {{Hasselfield}}, \citenamefont {{Hilton}}, \citenamefont {{Hincks}},
  \citenamefont {{Hlozek}}, \citenamefont {{Huffenberger}}, \citenamefont
  {{Hughes}}, \citenamefont {{Irwin}}, \citenamefont {{Kosowsky}},
  \citenamefont {{Lupton}}, \citenamefont {{Marriage}}, \citenamefont
  {{Marsden}}, \citenamefont {{Menanteau}}, \citenamefont {{Moodley}},
  \citenamefont {{Niemack}}, \citenamefont {{Page}}, \citenamefont
  {{Partridge}}, \citenamefont {{Reese}}, \citenamefont {{Schmitt}},
  \citenamefont {{Sehgal}}, \citenamefont {{Sherwin}}, \citenamefont
  {{Sievers}}, \citenamefont {{Spergel}}, \citenamefont {{Staggs}},
  \citenamefont {{Swetz}}, \citenamefont {{Switzer}}, \citenamefont
  {{Thornton}}, \citenamefont {{Trac}},\ and\ \citenamefont
  {{Wollack}}}]{2014JCAP...04..014D}%
  \BibitemOpen
  \bibfield  {author} {\bibinfo {author} {\bibfnamefont {S.}~\bibnamefont
  {{Das}}}, \bibinfo {author} {\bibfnamefont {T.}~\bibnamefont {{Louis}}},
  \bibinfo {author} {\bibfnamefont {M.~R.}\ \bibnamefont {{Nolta}}}, \bibinfo
  {author} {\bibfnamefont {G.~E.}\ \bibnamefont {{Addison}}}, \bibinfo {author}
  {\bibfnamefont {E.~S.}\ \bibnamefont {{Battistelli}}}, \bibinfo {author}
  {\bibfnamefont {J.~R.}\ \bibnamefont {{Bond}}}, \bibinfo {author}
  {\bibfnamefont {E.}~\bibnamefont {{Calabrese}}}, \bibinfo {author}
  {\bibfnamefont {D.}~\bibnamefont {{Crichton}}}, \bibinfo {author}
  {\bibfnamefont {M.~J.}\ \bibnamefont {{Devlin}}}, \bibinfo {author}
  {\bibfnamefont {S.}~\bibnamefont {{Dicker}}}, \bibinfo {author}
  {\bibfnamefont {J.}~\bibnamefont {{Dunkley}}}, \bibinfo {author}
  {\bibfnamefont {R.}~\bibnamefont {{D{\"u}nner}}}, \bibinfo {author}
  {\bibfnamefont {J.~W.}\ \bibnamefont {{Fowler}}}, \bibinfo {author}
  {\bibfnamefont {M.}~\bibnamefont {{Gralla}}}, \bibinfo {author}
  {\bibfnamefont {A.}~\bibnamefont {{Hajian}}}, \bibinfo {author}
  {\bibfnamefont {M.}~\bibnamefont {{Halpern}}}, \bibinfo {author}
  {\bibfnamefont {M.}~\bibnamefont {{Hasselfield}}}, \bibinfo {author}
  {\bibfnamefont {M.}~\bibnamefont {{Hilton}}}, \bibinfo {author}
  {\bibfnamefont {A.~D.}\ \bibnamefont {{Hincks}}}, \bibinfo {author}
  {\bibfnamefont {R.}~\bibnamefont {{Hlozek}}}, \bibinfo {author}
  {\bibfnamefont {K.~M.}\ \bibnamefont {{Huffenberger}}}, \bibinfo {author}
  {\bibfnamefont {J.~P.}\ \bibnamefont {{Hughes}}}, \bibinfo {author}
  {\bibfnamefont {K.~D.}\ \bibnamefont {{Irwin}}}, \bibinfo {author}
  {\bibfnamefont {A.}~\bibnamefont {{Kosowsky}}}, \bibinfo {author}
  {\bibfnamefont {R.~H.}\ \bibnamefont {{Lupton}}}, \bibinfo {author}
  {\bibfnamefont {T.~A.}\ \bibnamefont {{Marriage}}}, \bibinfo {author}
  {\bibfnamefont {D.}~\bibnamefont {{Marsden}}}, \bibinfo {author}
  {\bibfnamefont {F.}~\bibnamefont {{Menanteau}}}, \bibinfo {author}
  {\bibfnamefont {K.}~\bibnamefont {{Moodley}}}, \bibinfo {author}
  {\bibfnamefont {M.~D.}\ \bibnamefont {{Niemack}}}, \bibinfo {author}
  {\bibfnamefont {L.~A.}\ \bibnamefont {{Page}}}, \bibinfo {author}
  {\bibfnamefont {B.}~\bibnamefont {{Partridge}}}, \bibinfo {author}
  {\bibfnamefont {E.~D.}\ \bibnamefont {{Reese}}}, \bibinfo {author}
  {\bibfnamefont {B.~L.}\ \bibnamefont {{Schmitt}}}, \bibinfo {author}
  {\bibfnamefont {N.}~\bibnamefont {{Sehgal}}}, \bibinfo {author}
  {\bibfnamefont {B.~D.}\ \bibnamefont {{Sherwin}}}, \bibinfo {author}
  {\bibfnamefont {J.~L.}\ \bibnamefont {{Sievers}}}, \bibinfo {author}
  {\bibfnamefont {D.~N.}\ \bibnamefont {{Spergel}}}, \bibinfo {author}
  {\bibfnamefont {S.~T.}\ \bibnamefont {{Staggs}}}, \bibinfo {author}
  {\bibfnamefont {D.~S.}\ \bibnamefont {{Swetz}}}, \bibinfo {author}
  {\bibfnamefont {E.~R.}\ \bibnamefont {{Switzer}}}, \bibinfo {author}
  {\bibfnamefont {R.}~\bibnamefont {{Thornton}}}, \bibinfo {author}
  {\bibfnamefont {H.}~\bibnamefont {{Trac}}}, \ and\ \bibinfo {author}
  {\bibfnamefont {E.}~\bibnamefont {{Wollack}}},\ }\href {\doibase
  10.1088/1475-7516/2014/04/014} {\bibfield  {journal} {\bibinfo  {journal}
  {\jcap}\ }\textbf {\bibinfo {volume} {4}},\ \bibinfo {eid} {014} (\bibinfo
  {year} {2014})},\ \Eprint {http://arxiv.org/abs/1301.1037} {arXiv:1301.1037}
  \BibitemShut {NoStop}%
\bibitem [{\citenamefont {{Madhavacheril}}\ \emph {et~al.}(2015)\citenamefont
  {{Madhavacheril}}, \citenamefont {{Sehgal}}, \citenamefont {{Allison}},
  \citenamefont {{Battaglia}}, \citenamefont {{Bond}}, \citenamefont
  {{Calabrese}}, \citenamefont {{Caliguiri}}, \citenamefont {{Coughlin}},
  \citenamefont {{Crichton}}, \citenamefont {{Datta}}, \citenamefont
  {{Devlin}}, \citenamefont {{Dunkley}}, \citenamefont {{D{\"u}nner}},
  \citenamefont {{Fogarty}}, \citenamefont {{Grace}}, \citenamefont {{Hajian}},
  \citenamefont {{Hasselfield}}, \citenamefont {{Hill}}, \citenamefont
  {{Hilton}}, \citenamefont {{Hincks}}, \citenamefont {{Hlozek}}, \citenamefont
  {{Hughes}}, \citenamefont {{Kosowsky}}, \citenamefont {{Louis}},
  \citenamefont {{Lungu}}, \citenamefont {{McMahon}}, \citenamefont
  {{Moodley}}, \citenamefont {{Munson}}, \citenamefont {{Naess}}, \citenamefont
  {{Nati}}, \citenamefont {{Newburgh}}, \citenamefont {{Niemack}},
  \citenamefont {{Page}}, \citenamefont {{Partridge}}, \citenamefont
  {{Schmitt}}, \citenamefont {{Sherwin}}, \citenamefont {{Sievers}},
  \citenamefont {{Spergel}}, \citenamefont {{Staggs}}, \citenamefont
  {{Thornton}}, \citenamefont {{Van Engelen}}, \citenamefont {{Ward}},
  \citenamefont {{Wollack}},\ and\ \citenamefont {{Atacama Cosmology Telescope
  Collaboration}}}]{2015PhRvL.114o1302M}%
  \BibitemOpen
  \bibfield  {author} {\bibinfo {author} {\bibfnamefont {M.}~\bibnamefont
  {{Madhavacheril}}}, \bibinfo {author} {\bibfnamefont {N.}~\bibnamefont
  {{Sehgal}}}, \bibinfo {author} {\bibfnamefont {R.}~\bibnamefont {{Allison}}},
  \bibinfo {author} {\bibfnamefont {N.}~\bibnamefont {{Battaglia}}}, \bibinfo
  {author} {\bibfnamefont {J.~R.}\ \bibnamefont {{Bond}}}, \bibinfo {author}
  {\bibfnamefont {E.}~\bibnamefont {{Calabrese}}}, \bibinfo {author}
  {\bibfnamefont {J.}~\bibnamefont {{Caliguiri}}}, \bibinfo {author}
  {\bibfnamefont {K.}~\bibnamefont {{Coughlin}}}, \bibinfo {author}
  {\bibfnamefont {D.}~\bibnamefont {{Crichton}}}, \bibinfo {author}
  {\bibfnamefont {R.}~\bibnamefont {{Datta}}}, \bibinfo {author} {\bibfnamefont
  {M.~J.}\ \bibnamefont {{Devlin}}}, \bibinfo {author} {\bibfnamefont
  {J.}~\bibnamefont {{Dunkley}}}, \bibinfo {author} {\bibfnamefont
  {R.}~\bibnamefont {{D{\"u}nner}}}, \bibinfo {author} {\bibfnamefont
  {K.}~\bibnamefont {{Fogarty}}}, \bibinfo {author} {\bibfnamefont
  {E.}~\bibnamefont {{Grace}}}, \bibinfo {author} {\bibfnamefont
  {A.}~\bibnamefont {{Hajian}}}, \bibinfo {author} {\bibfnamefont
  {M.}~\bibnamefont {{Hasselfield}}}, \bibinfo {author} {\bibfnamefont {J.~C.}\
  \bibnamefont {{Hill}}}, \bibinfo {author} {\bibfnamefont {M.}~\bibnamefont
  {{Hilton}}}, \bibinfo {author} {\bibfnamefont {A.~D.}\ \bibnamefont
  {{Hincks}}}, \bibinfo {author} {\bibfnamefont {R.}~\bibnamefont {{Hlozek}}},
  \bibinfo {author} {\bibfnamefont {J.~P.}\ \bibnamefont {{Hughes}}}, \bibinfo
  {author} {\bibfnamefont {A.}~\bibnamefont {{Kosowsky}}}, \bibinfo {author}
  {\bibfnamefont {T.}~\bibnamefont {{Louis}}}, \bibinfo {author} {\bibfnamefont
  {M.}~\bibnamefont {{Lungu}}}, \bibinfo {author} {\bibfnamefont
  {J.}~\bibnamefont {{McMahon}}}, \bibinfo {author} {\bibfnamefont
  {K.}~\bibnamefont {{Moodley}}}, \bibinfo {author} {\bibfnamefont
  {C.}~\bibnamefont {{Munson}}}, \bibinfo {author} {\bibfnamefont
  {S.}~\bibnamefont {{Naess}}}, \bibinfo {author} {\bibfnamefont
  {F.}~\bibnamefont {{Nati}}}, \bibinfo {author} {\bibfnamefont
  {L.}~\bibnamefont {{Newburgh}}}, \bibinfo {author} {\bibfnamefont {M.~D.}\
  \bibnamefont {{Niemack}}}, \bibinfo {author} {\bibfnamefont {L.~A.}\
  \bibnamefont {{Page}}}, \bibinfo {author} {\bibfnamefont {B.}~\bibnamefont
  {{Partridge}}}, \bibinfo {author} {\bibfnamefont {B.}~\bibnamefont
  {{Schmitt}}}, \bibinfo {author} {\bibfnamefont {B.~D.}\ \bibnamefont
  {{Sherwin}}}, \bibinfo {author} {\bibfnamefont {J.}~\bibnamefont
  {{Sievers}}}, \bibinfo {author} {\bibfnamefont {D.~N.}\ \bibnamefont
  {{Spergel}}}, \bibinfo {author} {\bibfnamefont {S.~T.}\ \bibnamefont
  {{Staggs}}}, \bibinfo {author} {\bibfnamefont {R.}~\bibnamefont
  {{Thornton}}}, \bibinfo {author} {\bibfnamefont {A.}~\bibnamefont {{Van
  Engelen}}}, \bibinfo {author} {\bibfnamefont {J.~T.}\ \bibnamefont {{Ward}}},
  \bibinfo {author} {\bibfnamefont {E.~J.}\ \bibnamefont {{Wollack}}}, \ and\
  \bibinfo {author} {\bibnamefont {{Atacama Cosmology Telescope
  Collaboration}}},\ }\href {\doibase 10.1103/PhysRevLett.114.151302}
  {\bibfield  {journal} {\bibinfo  {journal} {Physical Review Letters}\
  }\textbf {\bibinfo {volume} {114}},\ \bibinfo {eid} {151302} (\bibinfo {year}
  {2015})},\ \Eprint {http://arxiv.org/abs/1411.7999} {arXiv:1411.7999}
  \BibitemShut {NoStop}%
\bibitem [{\citenamefont {{van Engelen}}\ \emph {et~al.}(2015)\citenamefont
  {{van Engelen}}, \citenamefont {{Sherwin}}, \citenamefont {{Sehgal}},
  \citenamefont {{Addison}}, \citenamefont {{Allison}}, \citenamefont
  {{Battaglia}}, \citenamefont {{de Bernardis}}, \citenamefont {{Bond}},
  \citenamefont {{Calabrese}}, \citenamefont {{Coughlin}}, \citenamefont
  {{Crichton}}, \citenamefont {{Datta}}, \citenamefont {{Devlin}},
  \citenamefont {{Dunkley}}, \citenamefont {{D{\"u}nner}}, \citenamefont
  {{Gallardo}}, \citenamefont {{Grace}}, \citenamefont {{Gralla}},
  \citenamefont {{Hajian}}, \citenamefont {{Hasselfield}}, \citenamefont
  {{Henderson}}, \citenamefont {{Hill}}, \citenamefont {{Hilton}},
  \citenamefont {{Hincks}}, \citenamefont {{Hlozek}}, \citenamefont
  {{Huffenberger}}, \citenamefont {{Hughes}}, \citenamefont {{Koopman}},
  \citenamefont {{Kosowsky}}, \citenamefont {{Louis}}, \citenamefont {{Lungu}},
  \citenamefont {{Madhavacheril}}, \citenamefont {{Maurin}}, \citenamefont
  {{McMahon}}, \citenamefont {{Moodley}}, \citenamefont {{Munson}},
  \citenamefont {{Naess}}, \citenamefont {{Nati}}, \citenamefont {{Newburgh}},
  \citenamefont {{Niemack}}, \citenamefont {{Nolta}}, \citenamefont {{Page}},
  \citenamefont {{Pappas}}, \citenamefont {{Partridge}}, \citenamefont
  {{Schmitt}}, \citenamefont {{Sievers}}, \citenamefont {{Simon}},
  \citenamefont {{Spergel}}, \citenamefont {{Staggs}}, \citenamefont
  {{Switzer}}, \citenamefont {{Ward}},\ and\ \citenamefont
  {{Wollack}}}]{2015ApJ...808....7V}%
  \BibitemOpen
  \bibfield  {author} {\bibinfo {author} {\bibfnamefont {A.}~\bibnamefont {{van
  Engelen}}}, \bibinfo {author} {\bibfnamefont {B.~D.}\ \bibnamefont
  {{Sherwin}}}, \bibinfo {author} {\bibfnamefont {N.}~\bibnamefont {{Sehgal}}},
  \bibinfo {author} {\bibfnamefont {G.~E.}\ \bibnamefont {{Addison}}}, \bibinfo
  {author} {\bibfnamefont {R.}~\bibnamefont {{Allison}}}, \bibinfo {author}
  {\bibfnamefont {N.}~\bibnamefont {{Battaglia}}}, \bibinfo {author}
  {\bibfnamefont {F.}~\bibnamefont {{de Bernardis}}}, \bibinfo {author}
  {\bibfnamefont {J.~R.}\ \bibnamefont {{Bond}}}, \bibinfo {author}
  {\bibfnamefont {E.}~\bibnamefont {{Calabrese}}}, \bibinfo {author}
  {\bibfnamefont {K.}~\bibnamefont {{Coughlin}}}, \bibinfo {author}
  {\bibfnamefont {D.}~\bibnamefont {{Crichton}}}, \bibinfo {author}
  {\bibfnamefont {R.}~\bibnamefont {{Datta}}}, \bibinfo {author} {\bibfnamefont
  {M.~J.}\ \bibnamefont {{Devlin}}}, \bibinfo {author} {\bibfnamefont
  {J.}~\bibnamefont {{Dunkley}}}, \bibinfo {author} {\bibfnamefont
  {R.}~\bibnamefont {{D{\"u}nner}}}, \bibinfo {author} {\bibfnamefont
  {P.}~\bibnamefont {{Gallardo}}}, \bibinfo {author} {\bibfnamefont
  {E.}~\bibnamefont {{Grace}}}, \bibinfo {author} {\bibfnamefont
  {M.}~\bibnamefont {{Gralla}}}, \bibinfo {author} {\bibfnamefont
  {A.}~\bibnamefont {{Hajian}}}, \bibinfo {author} {\bibfnamefont
  {M.}~\bibnamefont {{Hasselfield}}}, \bibinfo {author} {\bibfnamefont
  {S.}~\bibnamefont {{Henderson}}}, \bibinfo {author} {\bibfnamefont {J.~C.}\
  \bibnamefont {{Hill}}}, \bibinfo {author} {\bibfnamefont {M.}~\bibnamefont
  {{Hilton}}}, \bibinfo {author} {\bibfnamefont {A.~D.}\ \bibnamefont
  {{Hincks}}}, \bibinfo {author} {\bibfnamefont {R.}~\bibnamefont {{Hlozek}}},
  \bibinfo {author} {\bibfnamefont {K.~M.}\ \bibnamefont {{Huffenberger}}},
  \bibinfo {author} {\bibfnamefont {J.~P.}\ \bibnamefont {{Hughes}}}, \bibinfo
  {author} {\bibfnamefont {B.}~\bibnamefont {{Koopman}}}, \bibinfo {author}
  {\bibfnamefont {A.}~\bibnamefont {{Kosowsky}}}, \bibinfo {author}
  {\bibfnamefont {T.}~\bibnamefont {{Louis}}}, \bibinfo {author} {\bibfnamefont
  {M.}~\bibnamefont {{Lungu}}}, \bibinfo {author} {\bibfnamefont
  {M.}~\bibnamefont {{Madhavacheril}}}, \bibinfo {author} {\bibfnamefont
  {L.}~\bibnamefont {{Maurin}}}, \bibinfo {author} {\bibfnamefont
  {J.}~\bibnamefont {{McMahon}}}, \bibinfo {author} {\bibfnamefont
  {K.}~\bibnamefont {{Moodley}}}, \bibinfo {author} {\bibfnamefont
  {C.}~\bibnamefont {{Munson}}}, \bibinfo {author} {\bibfnamefont
  {S.}~\bibnamefont {{Naess}}}, \bibinfo {author} {\bibfnamefont
  {F.}~\bibnamefont {{Nati}}}, \bibinfo {author} {\bibfnamefont
  {L.}~\bibnamefont {{Newburgh}}}, \bibinfo {author} {\bibfnamefont {M.~D.}\
  \bibnamefont {{Niemack}}}, \bibinfo {author} {\bibfnamefont {M.~R.}\
  \bibnamefont {{Nolta}}}, \bibinfo {author} {\bibfnamefont {L.~A.}\
  \bibnamefont {{Page}}}, \bibinfo {author} {\bibfnamefont {C.}~\bibnamefont
  {{Pappas}}}, \bibinfo {author} {\bibfnamefont {B.}~\bibnamefont
  {{Partridge}}}, \bibinfo {author} {\bibfnamefont {B.~L.}\ \bibnamefont
  {{Schmitt}}}, \bibinfo {author} {\bibfnamefont {J.~L.}\ \bibnamefont
  {{Sievers}}}, \bibinfo {author} {\bibfnamefont {S.}~\bibnamefont {{Simon}}},
  \bibinfo {author} {\bibfnamefont {D.~N.}\ \bibnamefont {{Spergel}}}, \bibinfo
  {author} {\bibfnamefont {S.~T.}\ \bibnamefont {{Staggs}}}, \bibinfo {author}
  {\bibfnamefont {E.~R.}\ \bibnamefont {{Switzer}}}, \bibinfo {author}
  {\bibfnamefont {J.~T.}\ \bibnamefont {{Ward}}}, \ and\ \bibinfo {author}
  {\bibfnamefont {E.~J.}\ \bibnamefont {{Wollack}}},\ }\href {\doibase
  10.1088/0004-637X/808/1/7} {\bibfield  {journal} {\bibinfo  {journal} {\apj}\
  }\textbf {\bibinfo {volume} {808}},\ \bibinfo {eid} {7} (\bibinfo {year}
  {2015})},\ \Eprint {http://arxiv.org/abs/1412.0626} {arXiv:1412.0626}
  \BibitemShut {NoStop}%
\bibitem [{\citenamefont {{van Engelen}}\ \emph {et~al.}(2012)\citenamefont
  {{van Engelen}}, \citenamefont {{Keisler}}, \citenamefont {{Zahn}},
  \citenamefont {{Aird}}, \citenamefont {{Benson}}, \citenamefont {{Bleem}},
  \citenamefont {{Carlstrom}}, \citenamefont {{Chang}}, \citenamefont {{Cho}},
  \citenamefont {{Crawford}}, \citenamefont {{Crites}}, \citenamefont {{de
  Haan}}, \citenamefont {{Dobbs}}, \citenamefont {{Dudley}}, \citenamefont
  {{George}}, \citenamefont {{Halverson}}, \citenamefont {{Holder}},
  \citenamefont {{Holzapfel}}, \citenamefont {{Hoover}}, \citenamefont {{Hou}},
  \citenamefont {{Hrubes}}, \citenamefont {{Joy}}, \citenamefont {{Knox}},
  \citenamefont {{Lee}}, \citenamefont {{Leitch}}, \citenamefont {{Lueker}},
  \citenamefont {{Luong-Van}}, \citenamefont {{McMahon}}, \citenamefont
  {{Mehl}}, \citenamefont {{Meyer}}, \citenamefont {{Millea}}, \citenamefont
  {{Mohr}}, \citenamefont {{Montroy}}, \citenamefont {{Natoli}}, \citenamefont
  {{Padin}}, \citenamefont {{Plagge}}, \citenamefont {{Pryke}}, \citenamefont
  {{Reichardt}}, \citenamefont {{Ruhl}}, \citenamefont {{Sayre}}, \citenamefont
  {{Schaffer}}, \citenamefont {{Shaw}}, \citenamefont {{Shirokoff}},
  \citenamefont {{Spieler}}, \citenamefont {{Staniszewski}}, \citenamefont
  {{Stark}}, \citenamefont {{Story}}, \citenamefont {{Vanderlinde}},
  \citenamefont {{Vieira}},\ and\ \citenamefont
  {{Williamson}}}]{2012ApJ...756..142V}%
  \BibitemOpen
  \bibfield  {author} {\bibinfo {author} {\bibfnamefont {A.}~\bibnamefont {{van
  Engelen}}}, \bibinfo {author} {\bibfnamefont {R.}~\bibnamefont {{Keisler}}},
  \bibinfo {author} {\bibfnamefont {O.}~\bibnamefont {{Zahn}}}, \bibinfo
  {author} {\bibfnamefont {K.~A.}\ \bibnamefont {{Aird}}}, \bibinfo {author}
  {\bibfnamefont {B.~A.}\ \bibnamefont {{Benson}}}, \bibinfo {author}
  {\bibfnamefont {L.~E.}\ \bibnamefont {{Bleem}}}, \bibinfo {author}
  {\bibfnamefont {J.~E.}\ \bibnamefont {{Carlstrom}}}, \bibinfo {author}
  {\bibfnamefont {C.~L.}\ \bibnamefont {{Chang}}}, \bibinfo {author}
  {\bibfnamefont {H.~M.}\ \bibnamefont {{Cho}}}, \bibinfo {author}
  {\bibfnamefont {T.~M.}\ \bibnamefont {{Crawford}}}, \bibinfo {author}
  {\bibfnamefont {A.~T.}\ \bibnamefont {{Crites}}}, \bibinfo {author}
  {\bibfnamefont {T.}~\bibnamefont {{de Haan}}}, \bibinfo {author}
  {\bibfnamefont {M.~A.}\ \bibnamefont {{Dobbs}}}, \bibinfo {author}
  {\bibfnamefont {J.}~\bibnamefont {{Dudley}}}, \bibinfo {author}
  {\bibfnamefont {E.~M.}\ \bibnamefont {{George}}}, \bibinfo {author}
  {\bibfnamefont {N.~W.}\ \bibnamefont {{Halverson}}}, \bibinfo {author}
  {\bibfnamefont {G.~P.}\ \bibnamefont {{Holder}}}, \bibinfo {author}
  {\bibfnamefont {W.~L.}\ \bibnamefont {{Holzapfel}}}, \bibinfo {author}
  {\bibfnamefont {S.}~\bibnamefont {{Hoover}}}, \bibinfo {author}
  {\bibfnamefont {Z.}~\bibnamefont {{Hou}}}, \bibinfo {author} {\bibfnamefont
  {J.~D.}\ \bibnamefont {{Hrubes}}}, \bibinfo {author} {\bibfnamefont
  {M.}~\bibnamefont {{Joy}}}, \bibinfo {author} {\bibfnamefont
  {L.}~\bibnamefont {{Knox}}}, \bibinfo {author} {\bibfnamefont {A.~T.}\
  \bibnamefont {{Lee}}}, \bibinfo {author} {\bibfnamefont {E.~M.}\ \bibnamefont
  {{Leitch}}}, \bibinfo {author} {\bibfnamefont {M.}~\bibnamefont {{Lueker}}},
  \bibinfo {author} {\bibfnamefont {D.}~\bibnamefont {{Luong-Van}}}, \bibinfo
  {author} {\bibfnamefont {J.~J.}\ \bibnamefont {{McMahon}}}, \bibinfo {author}
  {\bibfnamefont {J.}~\bibnamefont {{Mehl}}}, \bibinfo {author} {\bibfnamefont
  {S.~S.}\ \bibnamefont {{Meyer}}}, \bibinfo {author} {\bibfnamefont
  {M.}~\bibnamefont {{Millea}}}, \bibinfo {author} {\bibfnamefont {J.~J.}\
  \bibnamefont {{Mohr}}}, \bibinfo {author} {\bibfnamefont {T.~E.}\
  \bibnamefont {{Montroy}}}, \bibinfo {author} {\bibfnamefont {T.}~\bibnamefont
  {{Natoli}}}, \bibinfo {author} {\bibfnamefont {S.}~\bibnamefont {{Padin}}},
  \bibinfo {author} {\bibfnamefont {T.}~\bibnamefont {{Plagge}}}, \bibinfo
  {author} {\bibfnamefont {C.}~\bibnamefont {{Pryke}}}, \bibinfo {author}
  {\bibfnamefont {C.~L.}\ \bibnamefont {{Reichardt}}}, \bibinfo {author}
  {\bibfnamefont {J.~E.}\ \bibnamefont {{Ruhl}}}, \bibinfo {author}
  {\bibfnamefont {J.~T.}\ \bibnamefont {{Sayre}}}, \bibinfo {author}
  {\bibfnamefont {K.~K.}\ \bibnamefont {{Schaffer}}}, \bibinfo {author}
  {\bibfnamefont {L.}~\bibnamefont {{Shaw}}}, \bibinfo {author} {\bibfnamefont
  {E.}~\bibnamefont {{Shirokoff}}}, \bibinfo {author} {\bibfnamefont {H.~G.}\
  \bibnamefont {{Spieler}}}, \bibinfo {author} {\bibfnamefont {Z.}~\bibnamefont
  {{Staniszewski}}}, \bibinfo {author} {\bibfnamefont {A.~A.}\ \bibnamefont
  {{Stark}}}, \bibinfo {author} {\bibfnamefont {K.}~\bibnamefont {{Story}}},
  \bibinfo {author} {\bibfnamefont {K.}~\bibnamefont {{Vanderlinde}}}, \bibinfo
  {author} {\bibfnamefont {J.~D.}\ \bibnamefont {{Vieira}}}, \ and\ \bibinfo
  {author} {\bibfnamefont {R.}~\bibnamefont {{Williamson}}},\ }\href {\doibase
  10.1088/0004-637X/756/2/142} {\bibfield  {journal} {\bibinfo  {journal}
  {\apj}\ }\textbf {\bibinfo {volume} {756}},\ \bibinfo {eid} {142} (\bibinfo
  {year} {2012})},\ \Eprint {http://arxiv.org/abs/1202.0546} {arXiv:1202.0546}
  \BibitemShut {NoStop}%
\bibitem [{\citenamefont {{Baxter}}\ \emph {et~al.}(2015)\citenamefont
  {{Baxter}}, \citenamefont {{Keisler}}, \citenamefont {{Dodelson}},
  \citenamefont {{Aird}}, \citenamefont {{Allen}}, \citenamefont {{Ashby}},
  \citenamefont {{Bautz}}, \citenamefont {{Bayliss}}, \citenamefont {{Benson}},
  \citenamefont {{Bleem}}, \citenamefont {{Bocquet}}, \citenamefont
  {{Brodwin}}, \citenamefont {{Carlstrom}}, \citenamefont {{Chang}},
  \citenamefont {{Chiu}}, \citenamefont {{Cho}}, \citenamefont {{Clocchiatti}},
  \citenamefont {{Crawford}}, \citenamefont {{Crites}}, \citenamefont
  {{Desai}}, \citenamefont {{Dietrich}}, \citenamefont {{de Haan}},
  \citenamefont {{Dobbs}}, \citenamefont {{Foley}}, \citenamefont {{Forman}},
  \citenamefont {{George}}, \citenamefont {{Gladders}}, \citenamefont
  {{Gonzalez}}, \citenamefont {{Halverson}}, \citenamefont {{Harrington}},
  \citenamefont {{Hennig}}, \citenamefont {{Hoekstra}}, \citenamefont
  {{Holder}}, \citenamefont {{Holzapfel}}, \citenamefont {{Hou}}, \citenamefont
  {{Hrubes}}, \citenamefont {{Jones}}, \citenamefont {{Knox}}, \citenamefont
  {{Lee}}, \citenamefont {{Leitch}}, \citenamefont {{Liu}}, \citenamefont
  {{Lueker}}, \citenamefont {{Luong-Van}}, \citenamefont {{Mantz}},
  \citenamefont {{Marrone}}, \citenamefont {{McDonald}}, \citenamefont
  {{McMahon}}, \citenamefont {{Meyer}}, \citenamefont {{Millea}}, \citenamefont
  {{Mocanu}}, \citenamefont {{Murray}}, \citenamefont {{Padin}}, \citenamefont
  {{Pryke}}, \citenamefont {{Reichardt}}, \citenamefont {{Rest}}, \citenamefont
  {{Ruhl}}, \citenamefont {{Saliwanchik}}, \citenamefont {{Saro}},
  \citenamefont {{Sayre}}, \citenamefont {{Schaffer}}, \citenamefont
  {{Shirokoff}}, \citenamefont {{Song}}, \citenamefont {{Spieler}},
  \citenamefont {{Stalder}}, \citenamefont {{Stanford}}, \citenamefont
  {{Staniszewski}}, \citenamefont {{Stark}}, \citenamefont {{Story}},
  \citenamefont {{van Engelen}}, \citenamefont {{Vanderlinde}}, \citenamefont
  {{Vieira}}, \citenamefont {{Vikhlinin}}, \citenamefont {{Williamson}},
  \citenamefont {{Zahn}},\ and\ \citenamefont
  {{Zenteno}}}]{2015ApJ...806..247B}%
  \BibitemOpen
  \bibfield  {author} {\bibinfo {author} {\bibfnamefont {E.~J.}\ \bibnamefont
  {{Baxter}}}, \bibinfo {author} {\bibfnamefont {R.}~\bibnamefont {{Keisler}}},
  \bibinfo {author} {\bibfnamefont {S.}~\bibnamefont {{Dodelson}}}, \bibinfo
  {author} {\bibfnamefont {K.~A.}\ \bibnamefont {{Aird}}}, \bibinfo {author}
  {\bibfnamefont {S.~W.}\ \bibnamefont {{Allen}}}, \bibinfo {author}
  {\bibfnamefont {M.~L.~N.}\ \bibnamefont {{Ashby}}}, \bibinfo {author}
  {\bibfnamefont {M.}~\bibnamefont {{Bautz}}}, \bibinfo {author} {\bibfnamefont
  {M.}~\bibnamefont {{Bayliss}}}, \bibinfo {author} {\bibfnamefont {B.~A.}\
  \bibnamefont {{Benson}}}, \bibinfo {author} {\bibfnamefont {L.~E.}\
  \bibnamefont {{Bleem}}}, \bibinfo {author} {\bibfnamefont {S.}~\bibnamefont
  {{Bocquet}}}, \bibinfo {author} {\bibfnamefont {M.}~\bibnamefont
  {{Brodwin}}}, \bibinfo {author} {\bibfnamefont {J.~E.}\ \bibnamefont
  {{Carlstrom}}}, \bibinfo {author} {\bibfnamefont {C.~L.}\ \bibnamefont
  {{Chang}}}, \bibinfo {author} {\bibfnamefont {I.}~\bibnamefont {{Chiu}}},
  \bibinfo {author} {\bibfnamefont {H.-M.}\ \bibnamefont {{Cho}}}, \bibinfo
  {author} {\bibfnamefont {A.}~\bibnamefont {{Clocchiatti}}}, \bibinfo {author}
  {\bibfnamefont {T.~M.}\ \bibnamefont {{Crawford}}}, \bibinfo {author}
  {\bibfnamefont {A.~T.}\ \bibnamefont {{Crites}}}, \bibinfo {author}
  {\bibfnamefont {S.}~\bibnamefont {{Desai}}}, \bibinfo {author} {\bibfnamefont
  {J.~P.}\ \bibnamefont {{Dietrich}}}, \bibinfo {author} {\bibfnamefont
  {T.}~\bibnamefont {{de Haan}}}, \bibinfo {author} {\bibfnamefont {M.~A.}\
  \bibnamefont {{Dobbs}}}, \bibinfo {author} {\bibfnamefont {R.~J.}\
  \bibnamefont {{Foley}}}, \bibinfo {author} {\bibfnamefont {W.~R.}\
  \bibnamefont {{Forman}}}, \bibinfo {author} {\bibfnamefont {E.~M.}\
  \bibnamefont {{George}}}, \bibinfo {author} {\bibfnamefont {M.~D.}\
  \bibnamefont {{Gladders}}}, \bibinfo {author} {\bibfnamefont {A.~H.}\
  \bibnamefont {{Gonzalez}}}, \bibinfo {author} {\bibfnamefont {N.~W.}\
  \bibnamefont {{Halverson}}}, \bibinfo {author} {\bibfnamefont {N.~L.}\
  \bibnamefont {{Harrington}}}, \bibinfo {author} {\bibfnamefont
  {C.}~\bibnamefont {{Hennig}}}, \bibinfo {author} {\bibfnamefont
  {H.}~\bibnamefont {{Hoekstra}}}, \bibinfo {author} {\bibfnamefont {G.~P.}\
  \bibnamefont {{Holder}}}, \bibinfo {author} {\bibfnamefont {W.~L.}\
  \bibnamefont {{Holzapfel}}}, \bibinfo {author} {\bibfnamefont
  {Z.}~\bibnamefont {{Hou}}}, \bibinfo {author} {\bibfnamefont {J.~D.}\
  \bibnamefont {{Hrubes}}}, \bibinfo {author} {\bibfnamefont {C.}~\bibnamefont
  {{Jones}}}, \bibinfo {author} {\bibfnamefont {L.}~\bibnamefont {{Knox}}},
  \bibinfo {author} {\bibfnamefont {A.~T.}\ \bibnamefont {{Lee}}}, \bibinfo
  {author} {\bibfnamefont {E.~M.}\ \bibnamefont {{Leitch}}}, \bibinfo {author}
  {\bibfnamefont {J.}~\bibnamefont {{Liu}}}, \bibinfo {author} {\bibfnamefont
  {M.}~\bibnamefont {{Lueker}}}, \bibinfo {author} {\bibfnamefont
  {D.}~\bibnamefont {{Luong-Van}}}, \bibinfo {author} {\bibfnamefont
  {A.}~\bibnamefont {{Mantz}}}, \bibinfo {author} {\bibfnamefont {D.~P.}\
  \bibnamefont {{Marrone}}}, \bibinfo {author} {\bibfnamefont {M.}~\bibnamefont
  {{McDonald}}}, \bibinfo {author} {\bibfnamefont {J.~J.}\ \bibnamefont
  {{McMahon}}}, \bibinfo {author} {\bibfnamefont {S.~S.}\ \bibnamefont
  {{Meyer}}}, \bibinfo {author} {\bibfnamefont {M.}~\bibnamefont {{Millea}}},
  \bibinfo {author} {\bibfnamefont {L.~M.}\ \bibnamefont {{Mocanu}}}, \bibinfo
  {author} {\bibfnamefont {S.~S.}\ \bibnamefont {{Murray}}}, \bibinfo {author}
  {\bibfnamefont {S.}~\bibnamefont {{Padin}}}, \bibinfo {author} {\bibfnamefont
  {C.}~\bibnamefont {{Pryke}}}, \bibinfo {author} {\bibfnamefont {C.~L.}\
  \bibnamefont {{Reichardt}}}, \bibinfo {author} {\bibfnamefont
  {A.}~\bibnamefont {{Rest}}}, \bibinfo {author} {\bibfnamefont {J.~E.}\
  \bibnamefont {{Ruhl}}}, \bibinfo {author} {\bibfnamefont {B.~R.}\
  \bibnamefont {{Saliwanchik}}}, \bibinfo {author} {\bibfnamefont
  {A.}~\bibnamefont {{Saro}}}, \bibinfo {author} {\bibfnamefont {J.~T.}\
  \bibnamefont {{Sayre}}}, \bibinfo {author} {\bibfnamefont {K.~K.}\
  \bibnamefont {{Schaffer}}}, \bibinfo {author} {\bibfnamefont
  {E.}~\bibnamefont {{Shirokoff}}}, \bibinfo {author} {\bibfnamefont
  {J.}~\bibnamefont {{Song}}}, \bibinfo {author} {\bibfnamefont {H.~G.}\
  \bibnamefont {{Spieler}}}, \bibinfo {author} {\bibfnamefont {B.}~\bibnamefont
  {{Stalder}}}, \bibinfo {author} {\bibfnamefont {S.~A.}\ \bibnamefont
  {{Stanford}}}, \bibinfo {author} {\bibfnamefont {Z.}~\bibnamefont
  {{Staniszewski}}}, \bibinfo {author} {\bibfnamefont {A.~A.}\ \bibnamefont
  {{Stark}}}, \bibinfo {author} {\bibfnamefont {K.~T.}\ \bibnamefont
  {{Story}}}, \bibinfo {author} {\bibfnamefont {A.}~\bibnamefont {{van
  Engelen}}}, \bibinfo {author} {\bibfnamefont {K.}~\bibnamefont
  {{Vanderlinde}}}, \bibinfo {author} {\bibfnamefont {J.~D.}\ \bibnamefont
  {{Vieira}}}, \bibinfo {author} {\bibfnamefont {A.}~\bibnamefont
  {{Vikhlinin}}}, \bibinfo {author} {\bibfnamefont {R.}~\bibnamefont
  {{Williamson}}}, \bibinfo {author} {\bibfnamefont {O.}~\bibnamefont
  {{Zahn}}}, \ and\ \bibinfo {author} {\bibfnamefont {A.}~\bibnamefont
  {{Zenteno}}},\ }\href {\doibase 10.1088/0004-637X/806/2/247} {\bibfield
  {journal} {\bibinfo  {journal} {\apj}\ }\textbf {\bibinfo {volume} {806}},\
  \bibinfo {eid} {247} (\bibinfo {year} {2015})},\ \Eprint
  {http://arxiv.org/abs/1412.7521} {arXiv:1412.7521} \BibitemShut {NoStop}%
\bibitem [{\citenamefont {{Story}}\ \emph {et~al.}(2015)\citenamefont
  {{Story}}, \citenamefont {{Hanson}}, \citenamefont {{Ade}}, \citenamefont
  {{Aird}}, \citenamefont {{Austermann}}, \citenamefont {{Beall}},
  \citenamefont {{Bender}}, \citenamefont {{Benson}}, \citenamefont {{Bleem}},
  \citenamefont {{Carlstrom}}, \citenamefont {{Chang}}, \citenamefont
  {{Chiang}}, \citenamefont {{Cho}}, \citenamefont {{Citron}}, \citenamefont
  {{Crawford}}, \citenamefont {{Crites}}, \citenamefont {{de Haan}},
  \citenamefont {{Dobbs}}, \citenamefont {{Everett}}, \citenamefont
  {{Gallicchio}}, \citenamefont {{Gao}}, \citenamefont {{George}},
  \citenamefont {{Gilbert}}, \citenamefont {{Halverson}}, \citenamefont
  {{Harrington}}, \citenamefont {{Henning}}, \citenamefont {{Hilton}},
  \citenamefont {{Holder}}, \citenamefont {{Holzapfel}}, \citenamefont
  {{Hoover}}, \citenamefont {{Hou}}, \citenamefont {{Hrubes}}, \citenamefont
  {{Huang}}, \citenamefont {{Hubmayr}}, \citenamefont {{Irwin}}, \citenamefont
  {{Keisler}}, \citenamefont {{Knox}}, \citenamefont {{Lee}}, \citenamefont
  {{Leitch}}, \citenamefont {{Li}}, \citenamefont {{Liang}}, \citenamefont
  {{Luong-Van}}, \citenamefont {{McMahon}}, \citenamefont {{Mehl}},
  \citenamefont {{Meyer}}, \citenamefont {{Mocanu}}, \citenamefont {{Montroy}},
  \citenamefont {{Natoli}}, \citenamefont {{Nibarger}}, \citenamefont
  {{Novosad}}, \citenamefont {{Padin}}, \citenamefont {{Pryke}}, \citenamefont
  {{Reichardt}}, \citenamefont {{Ruhl}}, \citenamefont {{Saliwanchik}},
  \citenamefont {{Sayre}}, \citenamefont {{Schaffer}}, \citenamefont
  {{Smecher}}, \citenamefont {{Stark}}, \citenamefont {{Tucker}}, \citenamefont
  {{Vanderlinde}}, \citenamefont {{Vieira}}, \citenamefont {{Wang}},
  \citenamefont {{Whitehorn}}, \citenamefont {{Yefremenko}},\ and\
  \citenamefont {{Zahn}}}]{2015ApJ...810...50S}%
  \BibitemOpen
  \bibfield  {author} {\bibinfo {author} {\bibfnamefont {K.~T.}\ \bibnamefont
  {{Story}}}, \bibinfo {author} {\bibfnamefont {D.}~\bibnamefont {{Hanson}}},
  \bibinfo {author} {\bibfnamefont {P.~A.~R.}\ \bibnamefont {{Ade}}}, \bibinfo
  {author} {\bibfnamefont {K.~A.}\ \bibnamefont {{Aird}}}, \bibinfo {author}
  {\bibfnamefont {J.~E.}\ \bibnamefont {{Austermann}}}, \bibinfo {author}
  {\bibfnamefont {J.~A.}\ \bibnamefont {{Beall}}}, \bibinfo {author}
  {\bibfnamefont {A.~N.}\ \bibnamefont {{Bender}}}, \bibinfo {author}
  {\bibfnamefont {B.~A.}\ \bibnamefont {{Benson}}}, \bibinfo {author}
  {\bibfnamefont {L.~E.}\ \bibnamefont {{Bleem}}}, \bibinfo {author}
  {\bibfnamefont {J.~E.}\ \bibnamefont {{Carlstrom}}}, \bibinfo {author}
  {\bibfnamefont {C.~L.}\ \bibnamefont {{Chang}}}, \bibinfo {author}
  {\bibfnamefont {H.~C.}\ \bibnamefont {{Chiang}}}, \bibinfo {author}
  {\bibfnamefont {H.-M.}\ \bibnamefont {{Cho}}}, \bibinfo {author}
  {\bibfnamefont {R.}~\bibnamefont {{Citron}}}, \bibinfo {author}
  {\bibfnamefont {T.~M.}\ \bibnamefont {{Crawford}}}, \bibinfo {author}
  {\bibfnamefont {A.~T.}\ \bibnamefont {{Crites}}}, \bibinfo {author}
  {\bibfnamefont {T.}~\bibnamefont {{de Haan}}}, \bibinfo {author}
  {\bibfnamefont {M.~A.}\ \bibnamefont {{Dobbs}}}, \bibinfo {author}
  {\bibfnamefont {W.}~\bibnamefont {{Everett}}}, \bibinfo {author}
  {\bibfnamefont {J.}~\bibnamefont {{Gallicchio}}}, \bibinfo {author}
  {\bibfnamefont {J.}~\bibnamefont {{Gao}}}, \bibinfo {author} {\bibfnamefont
  {E.~M.}\ \bibnamefont {{George}}}, \bibinfo {author} {\bibfnamefont
  {A.}~\bibnamefont {{Gilbert}}}, \bibinfo {author} {\bibfnamefont {N.~W.}\
  \bibnamefont {{Halverson}}}, \bibinfo {author} {\bibfnamefont
  {N.}~\bibnamefont {{Harrington}}}, \bibinfo {author} {\bibfnamefont {J.~W.}\
  \bibnamefont {{Henning}}}, \bibinfo {author} {\bibfnamefont {G.~C.}\
  \bibnamefont {{Hilton}}}, \bibinfo {author} {\bibfnamefont {G.~P.}\
  \bibnamefont {{Holder}}}, \bibinfo {author} {\bibfnamefont {W.~L.}\
  \bibnamefont {{Holzapfel}}}, \bibinfo {author} {\bibfnamefont
  {S.}~\bibnamefont {{Hoover}}}, \bibinfo {author} {\bibfnamefont
  {Z.}~\bibnamefont {{Hou}}}, \bibinfo {author} {\bibfnamefont {J.~D.}\
  \bibnamefont {{Hrubes}}}, \bibinfo {author} {\bibfnamefont {N.}~\bibnamefont
  {{Huang}}}, \bibinfo {author} {\bibfnamefont {J.}~\bibnamefont {{Hubmayr}}},
  \bibinfo {author} {\bibfnamefont {K.~D.}\ \bibnamefont {{Irwin}}}, \bibinfo
  {author} {\bibfnamefont {R.}~\bibnamefont {{Keisler}}}, \bibinfo {author}
  {\bibfnamefont {L.}~\bibnamefont {{Knox}}}, \bibinfo {author} {\bibfnamefont
  {A.~T.}\ \bibnamefont {{Lee}}}, \bibinfo {author} {\bibfnamefont {E.~M.}\
  \bibnamefont {{Leitch}}}, \bibinfo {author} {\bibfnamefont {D.}~\bibnamefont
  {{Li}}}, \bibinfo {author} {\bibfnamefont {C.}~\bibnamefont {{Liang}}},
  \bibinfo {author} {\bibfnamefont {D.}~\bibnamefont {{Luong-Van}}}, \bibinfo
  {author} {\bibfnamefont {J.~J.}\ \bibnamefont {{McMahon}}}, \bibinfo {author}
  {\bibfnamefont {J.}~\bibnamefont {{Mehl}}}, \bibinfo {author} {\bibfnamefont
  {S.~S.}\ \bibnamefont {{Meyer}}}, \bibinfo {author} {\bibfnamefont
  {L.}~\bibnamefont {{Mocanu}}}, \bibinfo {author} {\bibfnamefont {T.~E.}\
  \bibnamefont {{Montroy}}}, \bibinfo {author} {\bibfnamefont {T.}~\bibnamefont
  {{Natoli}}}, \bibinfo {author} {\bibfnamefont {J.~P.}\ \bibnamefont
  {{Nibarger}}}, \bibinfo {author} {\bibfnamefont {V.}~\bibnamefont
  {{Novosad}}}, \bibinfo {author} {\bibfnamefont {S.}~\bibnamefont {{Padin}}},
  \bibinfo {author} {\bibfnamefont {C.}~\bibnamefont {{Pryke}}}, \bibinfo
  {author} {\bibfnamefont {C.~L.}\ \bibnamefont {{Reichardt}}}, \bibinfo
  {author} {\bibfnamefont {J.~E.}\ \bibnamefont {{Ruhl}}}, \bibinfo {author}
  {\bibfnamefont {B.~R.}\ \bibnamefont {{Saliwanchik}}}, \bibinfo {author}
  {\bibfnamefont {J.~T.}\ \bibnamefont {{Sayre}}}, \bibinfo {author}
  {\bibfnamefont {K.~K.}\ \bibnamefont {{Schaffer}}}, \bibinfo {author}
  {\bibfnamefont {G.}~\bibnamefont {{Smecher}}}, \bibinfo {author}
  {\bibfnamefont {A.~A.}\ \bibnamefont {{Stark}}}, \bibinfo {author}
  {\bibfnamefont {C.}~\bibnamefont {{Tucker}}}, \bibinfo {author}
  {\bibfnamefont {K.}~\bibnamefont {{Vanderlinde}}}, \bibinfo {author}
  {\bibfnamefont {J.~D.}\ \bibnamefont {{Vieira}}}, \bibinfo {author}
  {\bibfnamefont {G.}~\bibnamefont {{Wang}}}, \bibinfo {author} {\bibfnamefont
  {N.}~\bibnamefont {{Whitehorn}}}, \bibinfo {author} {\bibfnamefont
  {V.}~\bibnamefont {{Yefremenko}}}, \ and\ \bibinfo {author} {\bibfnamefont
  {O.}~\bibnamefont {{Zahn}}},\ }\href {\doibase 10.1088/0004-637X/810/1/50}
  {\bibfield  {journal} {\bibinfo  {journal} {\apj}\ }\textbf {\bibinfo
  {volume} {810}},\ \bibinfo {eid} {50} (\bibinfo {year} {2015})},\ \Eprint
  {http://arxiv.org/abs/1412.4760} {arXiv:1412.4760} \BibitemShut {NoStop}%
\bibitem [{\citenamefont {{Ade}}\ \emph {et~al.}(2014)\citenamefont {{Ade}},
  \citenamefont {{Akiba}}, \citenamefont {{Anthony}}, \citenamefont {{Arnold}},
  \citenamefont {{Atlas}}, \citenamefont {{Barron}}, \citenamefont
  {{Boettger}}, \citenamefont {{Borrill}}, \citenamefont {{Chapman}},
  \citenamefont {{Chinone}}, \citenamefont {{Dobbs}}, \citenamefont
  {{Elleflot}}, \citenamefont {{Errard}}, \citenamefont {{Fabbian}},
  \citenamefont {{Feng}}, \citenamefont {{Flanigan}}, \citenamefont
  {{Gilbert}}, \citenamefont {{Grainger}}, \citenamefont {{Halverson}},
  \citenamefont {{Hasegawa}}, \citenamefont {{Hattori}}, \citenamefont
  {{Hazumi}}, \citenamefont {{Holzapfel}}, \citenamefont {{Hori}},
  \citenamefont {{Howard}}, \citenamefont {{Hyland}}, \citenamefont {{Inoue}},
  \citenamefont {{Jaehnig}}, \citenamefont {{Jaffe}}, \citenamefont
  {{Keating}}, \citenamefont {{Kermish}}, \citenamefont {{Keskitalo}},
  \citenamefont {{Kisner}}, \citenamefont {{Le Jeune}}, \citenamefont {{Lee}},
  \citenamefont {{Linder}}, \citenamefont {{Leitch}}, \citenamefont {{Lungu}},
  \citenamefont {{Matsuda}}, \citenamefont {{Matsumura}}, \citenamefont
  {{Meng}}, \citenamefont {{Miller}}, \citenamefont {{Morii}}, \citenamefont
  {{Moyerman}}, \citenamefont {{Myers}}, \citenamefont {{Navaroli}},
  \citenamefont {{Nishino}}, \citenamefont {{Paar}}, \citenamefont {{Peloton}},
  \citenamefont {{Quealy}}, \citenamefont {{Rebeiz}}, \citenamefont
  {{Reichardt}}, \citenamefont {{Richards}}, \citenamefont {{Ross}},
  \citenamefont {{Schanning}}, \citenamefont {{Schenck}}, \citenamefont
  {{Sherwin}}, \citenamefont {{Shimizu}}, \citenamefont {{Shimmin}},
  \citenamefont {{Shimon}}, \citenamefont {{Siritanasak}}, \citenamefont
  {{Smecher}}, \citenamefont {{Spieler}}, \citenamefont {{Stebor}},
  \citenamefont {{Steinbach}}, \citenamefont {{Stompor}}, \citenamefont
  {{Suzuki}}, \citenamefont {{Takakura}}, \citenamefont {{Tomaru}},
  \citenamefont {{Wilson}}, \citenamefont {{Yadav}}, \citenamefont {{Zahn}},\
  and\ \citenamefont {{Polarbear Collaboration}}}]{2014PhRvL.113b1301A}%
  \BibitemOpen
  \bibfield  {author} {\bibinfo {author} {\bibfnamefont {P.~A.~R.}\
  \bibnamefont {{Ade}}}, \bibinfo {author} {\bibfnamefont {Y.}~\bibnamefont
  {{Akiba}}}, \bibinfo {author} {\bibfnamefont {A.~E.}\ \bibnamefont
  {{Anthony}}}, \bibinfo {author} {\bibfnamefont {K.}~\bibnamefont {{Arnold}}},
  \bibinfo {author} {\bibfnamefont {M.}~\bibnamefont {{Atlas}}}, \bibinfo
  {author} {\bibfnamefont {D.}~\bibnamefont {{Barron}}}, \bibinfo {author}
  {\bibfnamefont {D.}~\bibnamefont {{Boettger}}}, \bibinfo {author}
  {\bibfnamefont {J.}~\bibnamefont {{Borrill}}}, \bibinfo {author}
  {\bibfnamefont {S.}~\bibnamefont {{Chapman}}}, \bibinfo {author}
  {\bibfnamefont {Y.}~\bibnamefont {{Chinone}}}, \bibinfo {author}
  {\bibfnamefont {M.}~\bibnamefont {{Dobbs}}}, \bibinfo {author} {\bibfnamefont
  {T.}~\bibnamefont {{Elleflot}}}, \bibinfo {author} {\bibfnamefont
  {J.}~\bibnamefont {{Errard}}}, \bibinfo {author} {\bibfnamefont
  {G.}~\bibnamefont {{Fabbian}}}, \bibinfo {author} {\bibfnamefont
  {C.}~\bibnamefont {{Feng}}}, \bibinfo {author} {\bibfnamefont
  {D.}~\bibnamefont {{Flanigan}}}, \bibinfo {author} {\bibfnamefont
  {A.}~\bibnamefont {{Gilbert}}}, \bibinfo {author} {\bibfnamefont
  {W.}~\bibnamefont {{Grainger}}}, \bibinfo {author} {\bibfnamefont {N.~W.}\
  \bibnamefont {{Halverson}}}, \bibinfo {author} {\bibfnamefont
  {M.}~\bibnamefont {{Hasegawa}}}, \bibinfo {author} {\bibfnamefont
  {K.}~\bibnamefont {{Hattori}}}, \bibinfo {author} {\bibfnamefont
  {M.}~\bibnamefont {{Hazumi}}}, \bibinfo {author} {\bibfnamefont {W.~L.}\
  \bibnamefont {{Holzapfel}}}, \bibinfo {author} {\bibfnamefont
  {Y.}~\bibnamefont {{Hori}}}, \bibinfo {author} {\bibfnamefont
  {J.}~\bibnamefont {{Howard}}}, \bibinfo {author} {\bibfnamefont
  {P.}~\bibnamefont {{Hyland}}}, \bibinfo {author} {\bibfnamefont
  {Y.}~\bibnamefont {{Inoue}}}, \bibinfo {author} {\bibfnamefont {G.~C.}\
  \bibnamefont {{Jaehnig}}}, \bibinfo {author} {\bibfnamefont {A.}~\bibnamefont
  {{Jaffe}}}, \bibinfo {author} {\bibfnamefont {B.}~\bibnamefont {{Keating}}},
  \bibinfo {author} {\bibfnamefont {Z.}~\bibnamefont {{Kermish}}}, \bibinfo
  {author} {\bibfnamefont {R.}~\bibnamefont {{Keskitalo}}}, \bibinfo {author}
  {\bibfnamefont {T.}~\bibnamefont {{Kisner}}}, \bibinfo {author}
  {\bibfnamefont {M.}~\bibnamefont {{Le Jeune}}}, \bibinfo {author}
  {\bibfnamefont {A.~T.}\ \bibnamefont {{Lee}}}, \bibinfo {author}
  {\bibfnamefont {E.}~\bibnamefont {{Linder}}}, \bibinfo {author}
  {\bibfnamefont {E.~M.}\ \bibnamefont {{Leitch}}}, \bibinfo {author}
  {\bibfnamefont {M.}~\bibnamefont {{Lungu}}}, \bibinfo {author} {\bibfnamefont
  {F.}~\bibnamefont {{Matsuda}}}, \bibinfo {author} {\bibfnamefont
  {T.}~\bibnamefont {{Matsumura}}}, \bibinfo {author} {\bibfnamefont
  {X.}~\bibnamefont {{Meng}}}, \bibinfo {author} {\bibfnamefont {N.~J.}\
  \bibnamefont {{Miller}}}, \bibinfo {author} {\bibfnamefont {H.}~\bibnamefont
  {{Morii}}}, \bibinfo {author} {\bibfnamefont {S.}~\bibnamefont {{Moyerman}}},
  \bibinfo {author} {\bibfnamefont {M.~J.}\ \bibnamefont {{Myers}}}, \bibinfo
  {author} {\bibfnamefont {M.}~\bibnamefont {{Navaroli}}}, \bibinfo {author}
  {\bibfnamefont {H.}~\bibnamefont {{Nishino}}}, \bibinfo {author}
  {\bibfnamefont {H.}~\bibnamefont {{Paar}}}, \bibinfo {author} {\bibfnamefont
  {J.}~\bibnamefont {{Peloton}}}, \bibinfo {author} {\bibfnamefont
  {E.}~\bibnamefont {{Quealy}}}, \bibinfo {author} {\bibfnamefont
  {G.}~\bibnamefont {{Rebeiz}}}, \bibinfo {author} {\bibfnamefont {C.~L.}\
  \bibnamefont {{Reichardt}}}, \bibinfo {author} {\bibfnamefont {P.~L.}\
  \bibnamefont {{Richards}}}, \bibinfo {author} {\bibfnamefont
  {C.}~\bibnamefont {{Ross}}}, \bibinfo {author} {\bibfnamefont
  {I.}~\bibnamefont {{Schanning}}}, \bibinfo {author} {\bibfnamefont {D.~E.}\
  \bibnamefont {{Schenck}}}, \bibinfo {author} {\bibfnamefont {B.}~\bibnamefont
  {{Sherwin}}}, \bibinfo {author} {\bibfnamefont {A.}~\bibnamefont
  {{Shimizu}}}, \bibinfo {author} {\bibfnamefont {C.}~\bibnamefont
  {{Shimmin}}}, \bibinfo {author} {\bibfnamefont {M.}~\bibnamefont {{Shimon}}},
  \bibinfo {author} {\bibfnamefont {P.}~\bibnamefont {{Siritanasak}}}, \bibinfo
  {author} {\bibfnamefont {G.}~\bibnamefont {{Smecher}}}, \bibinfo {author}
  {\bibfnamefont {H.}~\bibnamefont {{Spieler}}}, \bibinfo {author}
  {\bibfnamefont {N.}~\bibnamefont {{Stebor}}}, \bibinfo {author}
  {\bibfnamefont {B.}~\bibnamefont {{Steinbach}}}, \bibinfo {author}
  {\bibfnamefont {R.}~\bibnamefont {{Stompor}}}, \bibinfo {author}
  {\bibfnamefont {A.}~\bibnamefont {{Suzuki}}}, \bibinfo {author}
  {\bibfnamefont {S.}~\bibnamefont {{Takakura}}}, \bibinfo {author}
  {\bibfnamefont {T.}~\bibnamefont {{Tomaru}}}, \bibinfo {author}
  {\bibfnamefont {B.}~\bibnamefont {{Wilson}}}, \bibinfo {author}
  {\bibfnamefont {A.}~\bibnamefont {{Yadav}}}, \bibinfo {author} {\bibfnamefont
  {O.}~\bibnamefont {{Zahn}}}, \ and\ \bibinfo {author} {\bibnamefont
  {{Polarbear Collaboration}}},\ }\href {\doibase
  10.1103/PhysRevLett.113.021301} {\bibfield  {journal} {\bibinfo  {journal}
  {Physical Review Letters}\ }\textbf {\bibinfo {volume} {113}},\ \bibinfo
  {eid} {021301} (\bibinfo {year} {2014})},\ \Eprint
  {http://arxiv.org/abs/1312.6646} {arXiv:1312.6646} \BibitemShut {NoStop}%
\bibitem [{\citenamefont {{Planck Collaboration}}\ \emph
  {et~al.}(2014)\citenamefont {{Planck Collaboration}}, \citenamefont {{Ade}},
  \citenamefont {{Aghanim}}, \citenamefont {{Armitage-Caplan}}, \citenamefont
  {{Arnaud}}, \citenamefont {{Ashdown}}, \citenamefont {{Atrio-Barandela}},
  \citenamefont {{Aumont}}, \citenamefont {{Baccigalupi}}, \citenamefont
  {{Banday}},\ and\ \citenamefont {et~al.}}]{2014A&A...571A..17P}%
  \BibitemOpen
  \bibfield  {author} {\bibinfo {author} {\bibnamefont {{Planck
  Collaboration}}}, \bibinfo {author} {\bibfnamefont {P.~A.~R.}\ \bibnamefont
  {{Ade}}}, \bibinfo {author} {\bibfnamefont {N.}~\bibnamefont {{Aghanim}}},
  \bibinfo {author} {\bibfnamefont {C.}~\bibnamefont {{Armitage-Caplan}}},
  \bibinfo {author} {\bibfnamefont {M.}~\bibnamefont {{Arnaud}}}, \bibinfo
  {author} {\bibfnamefont {M.}~\bibnamefont {{Ashdown}}}, \bibinfo {author}
  {\bibfnamefont {F.}~\bibnamefont {{Atrio-Barandela}}}, \bibinfo {author}
  {\bibfnamefont {J.}~\bibnamefont {{Aumont}}}, \bibinfo {author}
  {\bibfnamefont {C.}~\bibnamefont {{Baccigalupi}}}, \bibinfo {author}
  {\bibfnamefont {A.~J.}\ \bibnamefont {{Banday}}}, \ and\ \bibinfo {author}
  {\bibnamefont {et~al.}},\ }\href {\doibase 10.1051/0004-6361/201321543}
  {\bibfield  {journal} {\bibinfo  {journal} {\aap}\ }\textbf {\bibinfo
  {volume} {571}},\ \bibinfo {eid} {A17} (\bibinfo {year} {2014})},\ \Eprint
  {http://arxiv.org/abs/1303.5077} {arXiv:1303.5077} \BibitemShut {NoStop}%
\bibitem [{\citenamefont {{Planck Collaboration}}\ \emph
  {et~al.}(2015{\natexlab{a}})\citenamefont {{Planck Collaboration}},
  \citenamefont {{Ade}}, \citenamefont {{Aghanim}}, \citenamefont {{Arnaud}},
  \citenamefont {{Ashdown}}, \citenamefont {{Aumont}}, \citenamefont
  {{Baccigalupi}}, \citenamefont {{Banday}}, \citenamefont {{Barreiro}},
  \citenamefont {{Bartlett}},\ and\ \citenamefont
  {et~al.}}]{2015arXiv150201591P}%
  \BibitemOpen
  \bibfield  {author} {\bibinfo {author} {\bibnamefont {{Planck
  Collaboration}}}, \bibinfo {author} {\bibfnamefont {P.~A.~R.}\ \bibnamefont
  {{Ade}}}, \bibinfo {author} {\bibfnamefont {N.}~\bibnamefont {{Aghanim}}},
  \bibinfo {author} {\bibfnamefont {M.}~\bibnamefont {{Arnaud}}}, \bibinfo
  {author} {\bibfnamefont {M.}~\bibnamefont {{Ashdown}}}, \bibinfo {author}
  {\bibfnamefont {J.}~\bibnamefont {{Aumont}}}, \bibinfo {author}
  {\bibfnamefont {C.}~\bibnamefont {{Baccigalupi}}}, \bibinfo {author}
  {\bibfnamefont {A.~J.}\ \bibnamefont {{Banday}}}, \bibinfo {author}
  {\bibfnamefont {R.~B.}\ \bibnamefont {{Barreiro}}}, \bibinfo {author}
  {\bibfnamefont {J.~G.}\ \bibnamefont {{Bartlett}}}, \ and\ \bibinfo {author}
  {\bibnamefont {et~al.}},\ }\href@noop {} {\bibfield  {journal} {\bibinfo
  {journal} {ArXiv e-prints}\ } (\bibinfo {year} {2015}{\natexlab{a}})},\
  \Eprint {http://arxiv.org/abs/1502.01591} {arXiv:1502.01591} \BibitemShut
  {NoStop}%
\bibitem [{\citenamefont {{Abazajian}}\ \emph
  {et~al.}(2015{\natexlab{a}})\citenamefont {{Abazajian}}, \citenamefont
  {{Arnold}}, \citenamefont {{Austermann}}, \citenamefont {{Benson}},
  \citenamefont {{Bischoff}}, \citenamefont {{Bock}}, \citenamefont {{Bond}},
  \citenamefont {{Borrill}}, \citenamefont {{Calabrese}}, \citenamefont
  {{Carlstrom}}, \citenamefont {{Carvalho}}, \citenamefont {{Chang}},
  \citenamefont {{Chiang}}, \citenamefont {{Church}}, \citenamefont {{Cooray}},
  \citenamefont {{Crawford}}, \citenamefont {{Dawson}}, \citenamefont {{Das}},
  \citenamefont {{Devlin}}, \citenamefont {{Dobbs}}, \citenamefont
  {{Dodelson}}, \citenamefont {{Dor{\'e}}}, \citenamefont {{Dunkley}},
  \citenamefont {{Errard}}, \citenamefont {{Fraisse}}, \citenamefont
  {{Gallicchio}}, \citenamefont {{Halverson}}, \citenamefont {{Hanany}},
  \citenamefont {{Hildebrandt}}, \citenamefont {{Hincks}}, \citenamefont
  {{Hlozek}}, \citenamefont {{Holder}}, \citenamefont {{Holzapfel}},
  \citenamefont {{Honscheid}}, \citenamefont {{Hu}}, \citenamefont {{Hubmayr}},
  \citenamefont {{Irwin}}, \citenamefont {{Jones}}, \citenamefont
  {{Kamionkowski}}, \citenamefont {{Keating}}, \citenamefont {{Keisler}},
  \citenamefont {{Knox}}, \citenamefont {{Komatsu}}, \citenamefont {{Kovac}},
  \citenamefont {{Kuo}}, \citenamefont {{Lawrence}}, \citenamefont {{Lee}},
  \citenamefont {{Leitch}}, \citenamefont {{Linder}}, \citenamefont {{Lubin}},
  \citenamefont {{McMahon}}, \citenamefont {{Miller}}, \citenamefont
  {{Newburgh}}, \citenamefont {{Niemack}}, \citenamefont {{Nguyen}},
  \citenamefont {{Nguyen}}, \citenamefont {{Page}}, \citenamefont {{Pryke}},
  \citenamefont {{Reichardt}}, \citenamefont {{Ruhl}}, \citenamefont
  {{Sehgal}}, \citenamefont {{Seljak}}, \citenamefont {{Sievers}},
  \citenamefont {{Silverstein}}, \citenamefont {{Slosar}}, \citenamefont
  {{Smith}}, \citenamefont {{Spergel}}, \citenamefont {{Staggs}}, \citenamefont
  {{Stark}}, \citenamefont {{Stompor}}, \citenamefont {{Vieregg}},
  \citenamefont {{Wang}}, \citenamefont {{Watson}}, \citenamefont {{Wollack}},
  \citenamefont {{Wu}}, \citenamefont {{Yoon}},\ and\ \citenamefont
  {{Zahn}}}]{2015APh....63...66A}%
  \BibitemOpen
  \bibfield  {author} {\bibinfo {author} {\bibfnamefont {K.~N.}\ \bibnamefont
  {{Abazajian}}}, \bibinfo {author} {\bibfnamefont {K.}~\bibnamefont
  {{Arnold}}}, \bibinfo {author} {\bibfnamefont {J.}~\bibnamefont
  {{Austermann}}}, \bibinfo {author} {\bibfnamefont {B.~A.}\ \bibnamefont
  {{Benson}}}, \bibinfo {author} {\bibfnamefont {C.}~\bibnamefont
  {{Bischoff}}}, \bibinfo {author} {\bibfnamefont {J.}~\bibnamefont {{Bock}}},
  \bibinfo {author} {\bibfnamefont {J.~R.}\ \bibnamefont {{Bond}}}, \bibinfo
  {author} {\bibfnamefont {J.}~\bibnamefont {{Borrill}}}, \bibinfo {author}
  {\bibfnamefont {E.}~\bibnamefont {{Calabrese}}}, \bibinfo {author}
  {\bibfnamefont {J.~E.}\ \bibnamefont {{Carlstrom}}}, \bibinfo {author}
  {\bibfnamefont {C.~S.}\ \bibnamefont {{Carvalho}}}, \bibinfo {author}
  {\bibfnamefont {C.~L.}\ \bibnamefont {{Chang}}}, \bibinfo {author}
  {\bibfnamefont {H.~C.}\ \bibnamefont {{Chiang}}}, \bibinfo {author}
  {\bibfnamefont {S.}~\bibnamefont {{Church}}}, \bibinfo {author}
  {\bibfnamefont {A.}~\bibnamefont {{Cooray}}}, \bibinfo {author}
  {\bibfnamefont {T.~M.}\ \bibnamefont {{Crawford}}}, \bibinfo {author}
  {\bibfnamefont {K.~S.}\ \bibnamefont {{Dawson}}}, \bibinfo {author}
  {\bibfnamefont {S.}~\bibnamefont {{Das}}}, \bibinfo {author} {\bibfnamefont
  {M.~J.}\ \bibnamefont {{Devlin}}}, \bibinfo {author} {\bibfnamefont
  {M.}~\bibnamefont {{Dobbs}}}, \bibinfo {author} {\bibfnamefont
  {S.}~\bibnamefont {{Dodelson}}}, \bibinfo {author} {\bibfnamefont
  {O.}~\bibnamefont {{Dor{\'e}}}}, \bibinfo {author} {\bibfnamefont
  {J.}~\bibnamefont {{Dunkley}}}, \bibinfo {author} {\bibfnamefont
  {J.}~\bibnamefont {{Errard}}}, \bibinfo {author} {\bibfnamefont
  {A.}~\bibnamefont {{Fraisse}}}, \bibinfo {author} {\bibfnamefont
  {J.}~\bibnamefont {{Gallicchio}}}, \bibinfo {author} {\bibfnamefont {N.~W.}\
  \bibnamefont {{Halverson}}}, \bibinfo {author} {\bibfnamefont
  {S.}~\bibnamefont {{Hanany}}}, \bibinfo {author} {\bibfnamefont {S.~R.}\
  \bibnamefont {{Hildebrandt}}}, \bibinfo {author} {\bibfnamefont
  {A.}~\bibnamefont {{Hincks}}}, \bibinfo {author} {\bibfnamefont
  {R.}~\bibnamefont {{Hlozek}}}, \bibinfo {author} {\bibfnamefont
  {G.}~\bibnamefont {{Holder}}}, \bibinfo {author} {\bibfnamefont {W.~L.}\
  \bibnamefont {{Holzapfel}}}, \bibinfo {author} {\bibfnamefont
  {K.}~\bibnamefont {{Honscheid}}}, \bibinfo {author} {\bibfnamefont
  {W.}~\bibnamefont {{Hu}}}, \bibinfo {author} {\bibfnamefont {J.}~\bibnamefont
  {{Hubmayr}}}, \bibinfo {author} {\bibfnamefont {K.}~\bibnamefont {{Irwin}}},
  \bibinfo {author} {\bibfnamefont {W.~C.}\ \bibnamefont {{Jones}}}, \bibinfo
  {author} {\bibfnamefont {M.}~\bibnamefont {{Kamionkowski}}}, \bibinfo
  {author} {\bibfnamefont {B.}~\bibnamefont {{Keating}}}, \bibinfo {author}
  {\bibfnamefont {R.}~\bibnamefont {{Keisler}}}, \bibinfo {author}
  {\bibfnamefont {L.}~\bibnamefont {{Knox}}}, \bibinfo {author} {\bibfnamefont
  {E.}~\bibnamefont {{Komatsu}}}, \bibinfo {author} {\bibfnamefont
  {J.}~\bibnamefont {{Kovac}}}, \bibinfo {author} {\bibfnamefont {C.-L.}\
  \bibnamefont {{Kuo}}}, \bibinfo {author} {\bibfnamefont {C.}~\bibnamefont
  {{Lawrence}}}, \bibinfo {author} {\bibfnamefont {A.~T.}\ \bibnamefont
  {{Lee}}}, \bibinfo {author} {\bibfnamefont {E.}~\bibnamefont {{Leitch}}},
  \bibinfo {author} {\bibfnamefont {E.}~\bibnamefont {{Linder}}}, \bibinfo
  {author} {\bibfnamefont {P.}~\bibnamefont {{Lubin}}}, \bibinfo {author}
  {\bibfnamefont {J.}~\bibnamefont {{McMahon}}}, \bibinfo {author}
  {\bibfnamefont {A.}~\bibnamefont {{Miller}}}, \bibinfo {author}
  {\bibfnamefont {L.}~\bibnamefont {{Newburgh}}}, \bibinfo {author}
  {\bibfnamefont {M.~D.}\ \bibnamefont {{Niemack}}}, \bibinfo {author}
  {\bibfnamefont {H.}~\bibnamefont {{Nguyen}}}, \bibinfo {author}
  {\bibfnamefont {H.~T.}\ \bibnamefont {{Nguyen}}}, \bibinfo {author}
  {\bibfnamefont {L.}~\bibnamefont {{Page}}}, \bibinfo {author} {\bibfnamefont
  {C.}~\bibnamefont {{Pryke}}}, \bibinfo {author} {\bibfnamefont {C.~L.}\
  \bibnamefont {{Reichardt}}}, \bibinfo {author} {\bibfnamefont {J.~E.}\
  \bibnamefont {{Ruhl}}}, \bibinfo {author} {\bibfnamefont {N.}~\bibnamefont
  {{Sehgal}}}, \bibinfo {author} {\bibfnamefont {U.}~\bibnamefont {{Seljak}}},
  \bibinfo {author} {\bibfnamefont {J.}~\bibnamefont {{Sievers}}}, \bibinfo
  {author} {\bibfnamefont {E.}~\bibnamefont {{Silverstein}}}, \bibinfo {author}
  {\bibfnamefont {A.}~\bibnamefont {{Slosar}}}, \bibinfo {author}
  {\bibfnamefont {K.~M.}\ \bibnamefont {{Smith}}}, \bibinfo {author}
  {\bibfnamefont {D.}~\bibnamefont {{Spergel}}}, \bibinfo {author}
  {\bibfnamefont {S.~T.}\ \bibnamefont {{Staggs}}}, \bibinfo {author}
  {\bibfnamefont {A.}~\bibnamefont {{Stark}}}, \bibinfo {author} {\bibfnamefont
  {R.}~\bibnamefont {{Stompor}}}, \bibinfo {author} {\bibfnamefont {A.~G.}\
  \bibnamefont {{Vieregg}}}, \bibinfo {author} {\bibfnamefont {G.}~\bibnamefont
  {{Wang}}}, \bibinfo {author} {\bibfnamefont {S.}~\bibnamefont {{Watson}}},
  \bibinfo {author} {\bibfnamefont {E.~J.}\ \bibnamefont {{Wollack}}}, \bibinfo
  {author} {\bibfnamefont {W.~L.~K.}\ \bibnamefont {{Wu}}}, \bibinfo {author}
  {\bibfnamefont {K.~W.}\ \bibnamefont {{Yoon}}}, \ and\ \bibinfo {author}
  {\bibfnamefont {O.}~\bibnamefont {{Zahn}}},\ }\href {\doibase
  10.1016/j.astropartphys.2014.05.014} {\bibfield  {journal} {\bibinfo
  {journal} {Astroparticle Physics}\ }\textbf {\bibinfo {volume} {63}},\
  \bibinfo {pages} {66} (\bibinfo {year} {2015}{\natexlab{a}})},\ \Eprint
  {http://arxiv.org/abs/1309.5383} {arXiv:1309.5383} \BibitemShut {NoStop}%
\bibitem [{\citenamefont {{Abazajian}}\ \emph
  {et~al.}(2015{\natexlab{b}})\citenamefont {{Abazajian}}, \citenamefont
  {{Arnold}}, \citenamefont {{Austermann}}, \citenamefont {{Benson}},
  \citenamefont {{Bischoff}}, \citenamefont {{Bock}}, \citenamefont {{Bond}},
  \citenamefont {{Borrill}}, \citenamefont {{Buder}}, \citenamefont {{Burke}},
  \citenamefont {{Calabrese}}, \citenamefont {{Carlstrom}}, \citenamefont
  {{Carvalho}}, \citenamefont {{Chang}}, \citenamefont {{Chiang}},
  \citenamefont {{Church}}, \citenamefont {{Cooray}}, \citenamefont
  {{Crawford}}, \citenamefont {{Crill}}, \citenamefont {{Dawson}},
  \citenamefont {{Das}}, \citenamefont {{Devlin}}, \citenamefont {{Dobbs}},
  \citenamefont {{Dodelson}}, \citenamefont {{Dor{\'e}}}, \citenamefont
  {{Dunkley}}, \citenamefont {{Feng}}, \citenamefont {{Fraisse}}, \citenamefont
  {{Gallicchio}}, \citenamefont {{Giddings}}, \citenamefont {{Green}},
  \citenamefont {{Halverson}}, \citenamefont {{Hanany}}, \citenamefont
  {{Hanson}}, \citenamefont {{Hildebrandt}}, \citenamefont {{Hincks}},
  \citenamefont {{Hlozek}}, \citenamefont {{Holder}}, \citenamefont
  {{Holzapfel}}, \citenamefont {{Honscheid}}, \citenamefont {{Horowitz}},
  \citenamefont {{Hu}}, \citenamefont {{Hubmayr}}, \citenamefont {{Irwin}},
  \citenamefont {{Jackson}}, \citenamefont {{Jones}}, \citenamefont
  {{Kallosh}}, \citenamefont {{Kamionkowski}}, \citenamefont {{Keating}},
  \citenamefont {{Keisler}}, \citenamefont {{Kinney}}, \citenamefont {{Knox}},
  \citenamefont {{Komatsu}}, \citenamefont {{Kovac}}, \citenamefont {{Kuo}},
  \citenamefont {{Kusaka}}, \citenamefont {{Lawrence}}, \citenamefont {{Lee}},
  \citenamefont {{Leitch}}, \citenamefont {{Linde}}, \citenamefont {{Linder}},
  \citenamefont {{Lubin}}, \citenamefont {{Maldacena}}, \citenamefont
  {{Martinec}}, \citenamefont {{McMahon}}, \citenamefont {{Miller}},
  \citenamefont {{Mukhanov}}, \citenamefont {{Newburgh}}, \citenamefont
  {{Niemack}}, \citenamefont {{Nguyen}}, \citenamefont {{Nguyen}},
  \citenamefont {{Page}}, \citenamefont {{Pryke}}, \citenamefont {{Reichardt}},
  \citenamefont {{Ruhl}}, \citenamefont {{Sehgal}}, \citenamefont {{Seljak}},
  \citenamefont {{Senatore}}, \citenamefont {{Sievers}}, \citenamefont
  {{Silverstein}}, \citenamefont {{Slosar}}, \citenamefont {{Smith}},
  \citenamefont {{Spergel}}, \citenamefont {{Staggs}}, \citenamefont {{Stark}},
  \citenamefont {{Stompor}}, \citenamefont {{Vieregg}}, \citenamefont {{Wang}},
  \citenamefont {{Watson}}, \citenamefont {{Wollack}}, \citenamefont {{Wu}},
  \citenamefont {{Yoon}}, \citenamefont {{Zahn}},\ and\ \citenamefont
  {{Zaldarriaga}}}]{2015APh....63...55A}%
  \BibitemOpen
  \bibfield  {author} {\bibinfo {author} {\bibfnamefont {K.~N.}\ \bibnamefont
  {{Abazajian}}}, \bibinfo {author} {\bibfnamefont {K.}~\bibnamefont
  {{Arnold}}}, \bibinfo {author} {\bibfnamefont {J.}~\bibnamefont
  {{Austermann}}}, \bibinfo {author} {\bibfnamefont {B.~A.}\ \bibnamefont
  {{Benson}}}, \bibinfo {author} {\bibfnamefont {C.}~\bibnamefont
  {{Bischoff}}}, \bibinfo {author} {\bibfnamefont {J.}~\bibnamefont {{Bock}}},
  \bibinfo {author} {\bibfnamefont {J.~R.}\ \bibnamefont {{Bond}}}, \bibinfo
  {author} {\bibfnamefont {J.}~\bibnamefont {{Borrill}}}, \bibinfo {author}
  {\bibfnamefont {I.}~\bibnamefont {{Buder}}}, \bibinfo {author} {\bibfnamefont
  {D.~L.}\ \bibnamefont {{Burke}}}, \bibinfo {author} {\bibfnamefont
  {E.}~\bibnamefont {{Calabrese}}}, \bibinfo {author} {\bibfnamefont {J.~E.}\
  \bibnamefont {{Carlstrom}}}, \bibinfo {author} {\bibfnamefont {C.~S.}\
  \bibnamefont {{Carvalho}}}, \bibinfo {author} {\bibfnamefont {C.~L.}\
  \bibnamefont {{Chang}}}, \bibinfo {author} {\bibfnamefont {H.~C.}\
  \bibnamefont {{Chiang}}}, \bibinfo {author} {\bibfnamefont {S.}~\bibnamefont
  {{Church}}}, \bibinfo {author} {\bibfnamefont {A.}~\bibnamefont {{Cooray}}},
  \bibinfo {author} {\bibfnamefont {T.~M.}\ \bibnamefont {{Crawford}}},
  \bibinfo {author} {\bibfnamefont {B.~P.}\ \bibnamefont {{Crill}}}, \bibinfo
  {author} {\bibfnamefont {K.~S.}\ \bibnamefont {{Dawson}}}, \bibinfo {author}
  {\bibfnamefont {S.}~\bibnamefont {{Das}}}, \bibinfo {author} {\bibfnamefont
  {M.~J.}\ \bibnamefont {{Devlin}}}, \bibinfo {author} {\bibfnamefont
  {M.}~\bibnamefont {{Dobbs}}}, \bibinfo {author} {\bibfnamefont
  {S.}~\bibnamefont {{Dodelson}}}, \bibinfo {author} {\bibfnamefont
  {O.}~\bibnamefont {{Dor{\'e}}}}, \bibinfo {author} {\bibfnamefont
  {J.}~\bibnamefont {{Dunkley}}}, \bibinfo {author} {\bibfnamefont {J.~L.}\
  \bibnamefont {{Feng}}}, \bibinfo {author} {\bibfnamefont {A.}~\bibnamefont
  {{Fraisse}}}, \bibinfo {author} {\bibfnamefont {J.}~\bibnamefont
  {{Gallicchio}}}, \bibinfo {author} {\bibfnamefont {S.~B.}\ \bibnamefont
  {{Giddings}}}, \bibinfo {author} {\bibfnamefont {D.}~\bibnamefont {{Green}}},
  \bibinfo {author} {\bibfnamefont {N.~W.}\ \bibnamefont {{Halverson}}},
  \bibinfo {author} {\bibfnamefont {S.}~\bibnamefont {{Hanany}}}, \bibinfo
  {author} {\bibfnamefont {D.}~\bibnamefont {{Hanson}}}, \bibinfo {author}
  {\bibfnamefont {S.~R.}\ \bibnamefont {{Hildebrandt}}}, \bibinfo {author}
  {\bibfnamefont {A.}~\bibnamefont {{Hincks}}}, \bibinfo {author}
  {\bibfnamefont {R.}~\bibnamefont {{Hlozek}}}, \bibinfo {author}
  {\bibfnamefont {G.}~\bibnamefont {{Holder}}}, \bibinfo {author}
  {\bibfnamefont {W.~L.}\ \bibnamefont {{Holzapfel}}}, \bibinfo {author}
  {\bibfnamefont {K.}~\bibnamefont {{Honscheid}}}, \bibinfo {author}
  {\bibfnamefont {G.}~\bibnamefont {{Horowitz}}}, \bibinfo {author}
  {\bibfnamefont {W.}~\bibnamefont {{Hu}}}, \bibinfo {author} {\bibfnamefont
  {J.}~\bibnamefont {{Hubmayr}}}, \bibinfo {author} {\bibfnamefont
  {K.}~\bibnamefont {{Irwin}}}, \bibinfo {author} {\bibfnamefont
  {M.}~\bibnamefont {{Jackson}}}, \bibinfo {author} {\bibfnamefont {W.~C.}\
  \bibnamefont {{Jones}}}, \bibinfo {author} {\bibfnamefont {R.}~\bibnamefont
  {{Kallosh}}}, \bibinfo {author} {\bibfnamefont {M.}~\bibnamefont
  {{Kamionkowski}}}, \bibinfo {author} {\bibfnamefont {B.}~\bibnamefont
  {{Keating}}}, \bibinfo {author} {\bibfnamefont {R.}~\bibnamefont
  {{Keisler}}}, \bibinfo {author} {\bibfnamefont {W.}~\bibnamefont {{Kinney}}},
  \bibinfo {author} {\bibfnamefont {L.}~\bibnamefont {{Knox}}}, \bibinfo
  {author} {\bibfnamefont {E.}~\bibnamefont {{Komatsu}}}, \bibinfo {author}
  {\bibfnamefont {J.}~\bibnamefont {{Kovac}}}, \bibinfo {author} {\bibfnamefont
  {C.-L.}\ \bibnamefont {{Kuo}}}, \bibinfo {author} {\bibfnamefont
  {A.}~\bibnamefont {{Kusaka}}}, \bibinfo {author} {\bibfnamefont
  {C.}~\bibnamefont {{Lawrence}}}, \bibinfo {author} {\bibfnamefont {A.~T.}\
  \bibnamefont {{Lee}}}, \bibinfo {author} {\bibfnamefont {E.}~\bibnamefont
  {{Leitch}}}, \bibinfo {author} {\bibfnamefont {A.}~\bibnamefont {{Linde}}},
  \bibinfo {author} {\bibfnamefont {E.}~\bibnamefont {{Linder}}}, \bibinfo
  {author} {\bibfnamefont {P.}~\bibnamefont {{Lubin}}}, \bibinfo {author}
  {\bibfnamefont {J.}~\bibnamefont {{Maldacena}}}, \bibinfo {author}
  {\bibfnamefont {E.}~\bibnamefont {{Martinec}}}, \bibinfo {author}
  {\bibfnamefont {J.}~\bibnamefont {{McMahon}}}, \bibinfo {author}
  {\bibfnamefont {A.}~\bibnamefont {{Miller}}}, \bibinfo {author}
  {\bibfnamefont {V.}~\bibnamefont {{Mukhanov}}}, \bibinfo {author}
  {\bibfnamefont {L.}~\bibnamefont {{Newburgh}}}, \bibinfo {author}
  {\bibfnamefont {M.~D.}\ \bibnamefont {{Niemack}}}, \bibinfo {author}
  {\bibfnamefont {H.}~\bibnamefont {{Nguyen}}}, \bibinfo {author}
  {\bibfnamefont {H.~T.}\ \bibnamefont {{Nguyen}}}, \bibinfo {author}
  {\bibfnamefont {L.}~\bibnamefont {{Page}}}, \bibinfo {author} {\bibfnamefont
  {C.}~\bibnamefont {{Pryke}}}, \bibinfo {author} {\bibfnamefont {C.~L.}\
  \bibnamefont {{Reichardt}}}, \bibinfo {author} {\bibfnamefont {J.~E.}\
  \bibnamefont {{Ruhl}}}, \bibinfo {author} {\bibfnamefont {N.}~\bibnamefont
  {{Sehgal}}}, \bibinfo {author} {\bibfnamefont {U.}~\bibnamefont {{Seljak}}},
  \bibinfo {author} {\bibfnamefont {L.}~\bibnamefont {{Senatore}}}, \bibinfo
  {author} {\bibfnamefont {J.}~\bibnamefont {{Sievers}}}, \bibinfo {author}
  {\bibfnamefont {E.}~\bibnamefont {{Silverstein}}}, \bibinfo {author}
  {\bibfnamefont {A.}~\bibnamefont {{Slosar}}}, \bibinfo {author}
  {\bibfnamefont {K.~M.}\ \bibnamefont {{Smith}}}, \bibinfo {author}
  {\bibfnamefont {D.}~\bibnamefont {{Spergel}}}, \bibinfo {author}
  {\bibfnamefont {S.~T.}\ \bibnamefont {{Staggs}}}, \bibinfo {author}
  {\bibfnamefont {A.}~\bibnamefont {{Stark}}}, \bibinfo {author} {\bibfnamefont
  {R.}~\bibnamefont {{Stompor}}}, \bibinfo {author} {\bibfnamefont {A.~G.}\
  \bibnamefont {{Vieregg}}}, \bibinfo {author} {\bibfnamefont {G.}~\bibnamefont
  {{Wang}}}, \bibinfo {author} {\bibfnamefont {S.}~\bibnamefont {{Watson}}},
  \bibinfo {author} {\bibfnamefont {E.~J.}\ \bibnamefont {{Wollack}}}, \bibinfo
  {author} {\bibfnamefont {W.~L.~K.}\ \bibnamefont {{Wu}}}, \bibinfo {author}
  {\bibfnamefont {K.~W.}\ \bibnamefont {{Yoon}}}, \bibinfo {author}
  {\bibfnamefont {O.}~\bibnamefont {{Zahn}}}, \ and\ \bibinfo {author}
  {\bibfnamefont {M.}~\bibnamefont {{Zaldarriaga}}},\ }\href {\doibase
  10.1016/j.astropartphys.2014.05.013} {\bibfield  {journal} {\bibinfo
  {journal} {Astroparticle Physics}\ }\textbf {\bibinfo {volume} {63}},\
  \bibinfo {pages} {55} (\bibinfo {year} {2015}{\natexlab{b}})},\ \Eprint
  {http://arxiv.org/abs/1309.5381} {arXiv:1309.5381} \BibitemShut {NoStop}%
\bibitem [{\citenamefont {{Liu}}\ \emph {et~al.}(2016)\citenamefont {{Liu}},
  \citenamefont {{Ortiz-Vazquez}},\ and\ \citenamefont
  {{Hill}}}]{2016arXiv160105720L}%
  \BibitemOpen
  \bibfield  {author} {\bibinfo {author} {\bibfnamefont {J.}~\bibnamefont
  {{Liu}}}, \bibinfo {author} {\bibfnamefont {A.}~\bibnamefont
  {{Ortiz-Vazquez}}}, \ and\ \bibinfo {author} {\bibfnamefont {J.~C.}\
  \bibnamefont {{Hill}}},\ }\href@noop {} {\bibfield  {journal} {\bibinfo
  {journal} {ArXiv e-prints}\ } (\bibinfo {year} {2016})},\ \Eprint
  {http://arxiv.org/abs/1601.05720} {arXiv:1601.05720} \BibitemShut {NoStop}%
\bibitem [{\citenamefont {{Baxter}}\ \emph {et~al.}(2016)\citenamefont
  {{Baxter}}, \citenamefont {{Clampitt}}, \citenamefont {{Giannantonio}},
  \citenamefont {{Dodelson}}, \citenamefont {{Jain}}, \citenamefont
  {{Huterer}}, \citenamefont {{Bleem}}, \citenamefont {{Crawford}},
  \citenamefont {{Efstathiou}}, \citenamefont {{Fosalba}}, \citenamefont
  {{Kirk}}, \citenamefont {{Kwan}}, \citenamefont {{S{\'a}nchez}},
  \citenamefont {{Story}}, \citenamefont {{Troxel}}, \citenamefont {{Abbott}},
  \citenamefont {{Abdalla}}, \citenamefont {{Armstrong}}, \citenamefont
  {{Benoit-L{\'e}vy}}, \citenamefont {{Benson}}, \citenamefont {{Bernstein}},
  \citenamefont {{Bernstein}}, \citenamefont {{Bertin}}, \citenamefont
  {{Brooks}}, \citenamefont {{Carlstrom}}, \citenamefont {{Carnero Rosell}},
  \citenamefont {{Carrasco Kind}}, \citenamefont {{Carretero}}, \citenamefont
  {{Chown}}, \citenamefont {{Crocce}}, \citenamefont {{Cunha}}, \citenamefont
  {{D'Andrea}}, \citenamefont {{da Costa}}, \citenamefont {{Desai}},
  \citenamefont {{Diehl}}, \citenamefont {{Dietrich}}, \citenamefont {{Doel}},
  \citenamefont {{Evrard}}, \citenamefont {{Fausti Neto}}, \citenamefont
  {{Flaugher}}, \citenamefont {{Frieman}}, \citenamefont {{Gruen}},
  \citenamefont {{Gruendl}}, \citenamefont {{Gutierrez}}, \citenamefont {{de
  Haan}}, \citenamefont {{Holder}}, \citenamefont {{Honscheid}}, \citenamefont
  {{Hou}}, \citenamefont {{James}}, \citenamefont {{Kuehn}}, \citenamefont
  {{Kuropatkin}}, \citenamefont {{Lima}}, \citenamefont {{March}},
  \citenamefont {{Marshall}}, \citenamefont {{Martini}}, \citenamefont
  {{Melchior}}, \citenamefont {{Miller}}, \citenamefont {{Miquel}},
  \citenamefont {{Mohr}}, \citenamefont {{Nord}}, \citenamefont {{Omori}},
  \citenamefont {{Plazas}}, \citenamefont {{Reichardt}}, \citenamefont
  {{Romer}}, \citenamefont {{Rykoff}}, \citenamefont {{Sanchez}}, \citenamefont
  {{Sevilla-Noarbe}}, \citenamefont {{Sheldon}}, \citenamefont {{Smith}},
  \citenamefont {{Soares-Santos}}, \citenamefont {{Sobreira}}, \citenamefont
  {{Suchyta}}, \citenamefont {{Stark}}, \citenamefont {{Swanson}},
  \citenamefont {{Tarle}}, \citenamefont {{Thomas}}, \citenamefont {{Walker}},\
  and\ \citenamefont {{Wechsler}}}]{2016arXiv160207384B}%
  \BibitemOpen
  \bibfield  {author} {\bibinfo {author} {\bibfnamefont {E.~J.}\ \bibnamefont
  {{Baxter}}}, \bibinfo {author} {\bibfnamefont {J.}~\bibnamefont
  {{Clampitt}}}, \bibinfo {author} {\bibfnamefont {T.}~\bibnamefont
  {{Giannantonio}}}, \bibinfo {author} {\bibfnamefont {S.}~\bibnamefont
  {{Dodelson}}}, \bibinfo {author} {\bibfnamefont {B.}~\bibnamefont {{Jain}}},
  \bibinfo {author} {\bibfnamefont {D.}~\bibnamefont {{Huterer}}}, \bibinfo
  {author} {\bibfnamefont {L.~E.}\ \bibnamefont {{Bleem}}}, \bibinfo {author}
  {\bibfnamefont {T.~M.}\ \bibnamefont {{Crawford}}}, \bibinfo {author}
  {\bibfnamefont {G.}~\bibnamefont {{Efstathiou}}}, \bibinfo {author}
  {\bibfnamefont {P.}~\bibnamefont {{Fosalba}}}, \bibinfo {author}
  {\bibfnamefont {D.}~\bibnamefont {{Kirk}}}, \bibinfo {author} {\bibfnamefont
  {J.}~\bibnamefont {{Kwan}}}, \bibinfo {author} {\bibfnamefont
  {C.}~\bibnamefont {{S{\'a}nchez}}}, \bibinfo {author} {\bibfnamefont {K.~T.}\
  \bibnamefont {{Story}}}, \bibinfo {author} {\bibfnamefont {M.~A.}\
  \bibnamefont {{Troxel}}}, \bibinfo {author} {\bibfnamefont {T.~M.~C.}\
  \bibnamefont {{Abbott}}}, \bibinfo {author} {\bibfnamefont {F.~B.}\
  \bibnamefont {{Abdalla}}}, \bibinfo {author} {\bibfnamefont {R.}~\bibnamefont
  {{Armstrong}}}, \bibinfo {author} {\bibfnamefont {A.}~\bibnamefont
  {{Benoit-L{\'e}vy}}}, \bibinfo {author} {\bibfnamefont {B.~A.}\ \bibnamefont
  {{Benson}}}, \bibinfo {author} {\bibfnamefont {G.~M.}\ \bibnamefont
  {{Bernstein}}}, \bibinfo {author} {\bibfnamefont {R.~A.}\ \bibnamefont
  {{Bernstein}}}, \bibinfo {author} {\bibfnamefont {E.}~\bibnamefont
  {{Bertin}}}, \bibinfo {author} {\bibfnamefont {D.}~\bibnamefont {{Brooks}}},
  \bibinfo {author} {\bibfnamefont {J.~E.}\ \bibnamefont {{Carlstrom}}},
  \bibinfo {author} {\bibfnamefont {A.}~\bibnamefont {{Carnero Rosell}}},
  \bibinfo {author} {\bibfnamefont {M.}~\bibnamefont {{Carrasco Kind}}},
  \bibinfo {author} {\bibfnamefont {J.}~\bibnamefont {{Carretero}}}, \bibinfo
  {author} {\bibfnamefont {R.}~\bibnamefont {{Chown}}}, \bibinfo {author}
  {\bibfnamefont {M.}~\bibnamefont {{Crocce}}}, \bibinfo {author}
  {\bibfnamefont {C.~E.}\ \bibnamefont {{Cunha}}}, \bibinfo {author}
  {\bibfnamefont {C.~B.}\ \bibnamefont {{D'Andrea}}}, \bibinfo {author}
  {\bibfnamefont {L.~N.}\ \bibnamefont {{da Costa}}}, \bibinfo {author}
  {\bibfnamefont {S.}~\bibnamefont {{Desai}}}, \bibinfo {author} {\bibfnamefont
  {H.~T.}\ \bibnamefont {{Diehl}}}, \bibinfo {author} {\bibfnamefont {J.~P.}\
  \bibnamefont {{Dietrich}}}, \bibinfo {author} {\bibfnamefont
  {P.}~\bibnamefont {{Doel}}}, \bibinfo {author} {\bibfnamefont {A.~E.}\
  \bibnamefont {{Evrard}}}, \bibinfo {author} {\bibfnamefont {A.}~\bibnamefont
  {{Fausti Neto}}}, \bibinfo {author} {\bibfnamefont {B.}~\bibnamefont
  {{Flaugher}}}, \bibinfo {author} {\bibfnamefont {J.}~\bibnamefont
  {{Frieman}}}, \bibinfo {author} {\bibfnamefont {D.}~\bibnamefont {{Gruen}}},
  \bibinfo {author} {\bibfnamefont {R.~A.}\ \bibnamefont {{Gruendl}}}, \bibinfo
  {author} {\bibfnamefont {G.}~\bibnamefont {{Gutierrez}}}, \bibinfo {author}
  {\bibfnamefont {T.}~\bibnamefont {{de Haan}}}, \bibinfo {author}
  {\bibfnamefont {G.~P.}\ \bibnamefont {{Holder}}}, \bibinfo {author}
  {\bibfnamefont {K.}~\bibnamefont {{Honscheid}}}, \bibinfo {author}
  {\bibfnamefont {Z.}~\bibnamefont {{Hou}}}, \bibinfo {author} {\bibfnamefont
  {D.~J.}\ \bibnamefont {{James}}}, \bibinfo {author} {\bibfnamefont
  {K.}~\bibnamefont {{Kuehn}}}, \bibinfo {author} {\bibfnamefont
  {N.}~\bibnamefont {{Kuropatkin}}}, \bibinfo {author} {\bibfnamefont
  {M.}~\bibnamefont {{Lima}}}, \bibinfo {author} {\bibfnamefont
  {M.}~\bibnamefont {{March}}}, \bibinfo {author} {\bibfnamefont {J.~L.}\
  \bibnamefont {{Marshall}}}, \bibinfo {author} {\bibfnamefont
  {P.}~\bibnamefont {{Martini}}}, \bibinfo {author} {\bibfnamefont
  {P.}~\bibnamefont {{Melchior}}}, \bibinfo {author} {\bibfnamefont {C.~J.}\
  \bibnamefont {{Miller}}}, \bibinfo {author} {\bibfnamefont {R.}~\bibnamefont
  {{Miquel}}}, \bibinfo {author} {\bibfnamefont {J.~J.}\ \bibnamefont
  {{Mohr}}}, \bibinfo {author} {\bibfnamefont {B.}~\bibnamefont {{Nord}}},
  \bibinfo {author} {\bibfnamefont {Y.}~\bibnamefont {{Omori}}}, \bibinfo
  {author} {\bibfnamefont {A.~A.}\ \bibnamefont {{Plazas}}}, \bibinfo {author}
  {\bibfnamefont {C.~L.}\ \bibnamefont {{Reichardt}}}, \bibinfo {author}
  {\bibfnamefont {A.~K.}\ \bibnamefont {{Romer}}}, \bibinfo {author}
  {\bibfnamefont {E.~S.}\ \bibnamefont {{Rykoff}}}, \bibinfo {author}
  {\bibfnamefont {E.}~\bibnamefont {{Sanchez}}}, \bibinfo {author}
  {\bibfnamefont {I.}~\bibnamefont {{Sevilla-Noarbe}}}, \bibinfo {author}
  {\bibfnamefont {E.}~\bibnamefont {{Sheldon}}}, \bibinfo {author}
  {\bibfnamefont {R.~C.}\ \bibnamefont {{Smith}}}, \bibinfo {author}
  {\bibfnamefont {M.}~\bibnamefont {{Soares-Santos}}}, \bibinfo {author}
  {\bibfnamefont {F.}~\bibnamefont {{Sobreira}}}, \bibinfo {author}
  {\bibfnamefont {E.}~\bibnamefont {{Suchyta}}}, \bibinfo {author}
  {\bibfnamefont {A.~A.}\ \bibnamefont {{Stark}}}, \bibinfo {author}
  {\bibfnamefont {M.~E.~C.}\ \bibnamefont {{Swanson}}}, \bibinfo {author}
  {\bibfnamefont {G.}~\bibnamefont {{Tarle}}}, \bibinfo {author} {\bibfnamefont
  {D.}~\bibnamefont {{Thomas}}}, \bibinfo {author} {\bibfnamefont {A.~R.}\
  \bibnamefont {{Walker}}}, \ and\ \bibinfo {author} {\bibfnamefont {R.~H.}\
  \bibnamefont {{Wechsler}}},\ }\href@noop {} {\bibfield  {journal} {\bibinfo
  {journal} {ArXiv e-prints}\ } (\bibinfo {year} {2016})},\ \Eprint
  {http://arxiv.org/abs/1602.07384} {arXiv:1602.07384} \BibitemShut {NoStop}%
\bibitem [{\citenamefont {{Hand}}\ \emph {et~al.}(2015)\citenamefont {{Hand}},
  \citenamefont {{Leauthaud}}, \citenamefont {{Das}}, \citenamefont
  {{Sherwin}}, \citenamefont {{Addison}}, \citenamefont {{Bond}}, \citenamefont
  {{Calabrese}}, \citenamefont {{Charbonnier}}, \citenamefont {{Devlin}},
  \citenamefont {{Dunkley}}, \citenamefont {{Erben}}, \citenamefont {{Hajian}},
  \citenamefont {{Halpern}}, \citenamefont {{Harnois-D{\'e}raps}},
  \citenamefont {{Heymans}}, \citenamefont {{Hildebrandt}}, \citenamefont
  {{Hincks}}, \citenamefont {{Kneib}}, \citenamefont {{Kosowsky}},
  \citenamefont {{Makler}}, \citenamefont {{Miller}}, \citenamefont
  {{Moodley}}, \citenamefont {{Moraes}}, \citenamefont {{Niemack}},
  \citenamefont {{Page}}, \citenamefont {{Partridge}}, \citenamefont
  {{Sehgal}}, \citenamefont {{Shan}}, \citenamefont {{Sievers}}, \citenamefont
  {{Spergel}}, \citenamefont {{Staggs}}, \citenamefont {{Switzer}},
  \citenamefont {{Taylor}}, \citenamefont {{Van Waerbeke}}, \citenamefont
  {{Welker}},\ and\ \citenamefont {{Wollack}}}]{2015PhRvD..91f2001H}%
  \BibitemOpen
  \bibfield  {author} {\bibinfo {author} {\bibfnamefont {N.}~\bibnamefont
  {{Hand}}}, \bibinfo {author} {\bibfnamefont {A.}~\bibnamefont {{Leauthaud}}},
  \bibinfo {author} {\bibfnamefont {S.}~\bibnamefont {{Das}}}, \bibinfo
  {author} {\bibfnamefont {B.~D.}\ \bibnamefont {{Sherwin}}}, \bibinfo {author}
  {\bibfnamefont {G.~E.}\ \bibnamefont {{Addison}}}, \bibinfo {author}
  {\bibfnamefont {J.~R.}\ \bibnamefont {{Bond}}}, \bibinfo {author}
  {\bibfnamefont {E.}~\bibnamefont {{Calabrese}}}, \bibinfo {author}
  {\bibfnamefont {A.}~\bibnamefont {{Charbonnier}}}, \bibinfo {author}
  {\bibfnamefont {M.~J.}\ \bibnamefont {{Devlin}}}, \bibinfo {author}
  {\bibfnamefont {J.}~\bibnamefont {{Dunkley}}}, \bibinfo {author}
  {\bibfnamefont {T.}~\bibnamefont {{Erben}}}, \bibinfo {author} {\bibfnamefont
  {A.}~\bibnamefont {{Hajian}}}, \bibinfo {author} {\bibfnamefont
  {M.}~\bibnamefont {{Halpern}}}, \bibinfo {author} {\bibfnamefont
  {J.}~\bibnamefont {{Harnois-D{\'e}raps}}}, \bibinfo {author} {\bibfnamefont
  {C.}~\bibnamefont {{Heymans}}}, \bibinfo {author} {\bibfnamefont
  {H.}~\bibnamefont {{Hildebrandt}}}, \bibinfo {author} {\bibfnamefont {A.~D.}\
  \bibnamefont {{Hincks}}}, \bibinfo {author} {\bibfnamefont {J.-P.}\
  \bibnamefont {{Kneib}}}, \bibinfo {author} {\bibfnamefont {A.}~\bibnamefont
  {{Kosowsky}}}, \bibinfo {author} {\bibfnamefont {M.}~\bibnamefont
  {{Makler}}}, \bibinfo {author} {\bibfnamefont {L.}~\bibnamefont {{Miller}}},
  \bibinfo {author} {\bibfnamefont {K.}~\bibnamefont {{Moodley}}}, \bibinfo
  {author} {\bibfnamefont {B.}~\bibnamefont {{Moraes}}}, \bibinfo {author}
  {\bibfnamefont {M.~D.}\ \bibnamefont {{Niemack}}}, \bibinfo {author}
  {\bibfnamefont {L.~A.}\ \bibnamefont {{Page}}}, \bibinfo {author}
  {\bibfnamefont {B.}~\bibnamefont {{Partridge}}}, \bibinfo {author}
  {\bibfnamefont {N.}~\bibnamefont {{Sehgal}}}, \bibinfo {author}
  {\bibfnamefont {H.}~\bibnamefont {{Shan}}}, \bibinfo {author} {\bibfnamefont
  {J.~L.}\ \bibnamefont {{Sievers}}}, \bibinfo {author} {\bibfnamefont {D.~N.}\
  \bibnamefont {{Spergel}}}, \bibinfo {author} {\bibfnamefont {S.~T.}\
  \bibnamefont {{Staggs}}}, \bibinfo {author} {\bibfnamefont {E.~R.}\
  \bibnamefont {{Switzer}}}, \bibinfo {author} {\bibfnamefont {J.~E.}\
  \bibnamefont {{Taylor}}}, \bibinfo {author} {\bibfnamefont {L.}~\bibnamefont
  {{Van Waerbeke}}}, \bibinfo {author} {\bibfnamefont {C.}~\bibnamefont
  {{Welker}}}, \ and\ \bibinfo {author} {\bibfnamefont {E.~J.}\ \bibnamefont
  {{Wollack}}},\ }\href {\doibase 10.1103/PhysRevD.91.062001} {\bibfield
  {journal} {\bibinfo  {journal} {\prd}\ }\textbf {\bibinfo {volume} {91}},\
  \bibinfo {eid} {062001} (\bibinfo {year} {2015})},\ \Eprint
  {http://arxiv.org/abs/1311.6200} {arXiv:1311.6200} \BibitemShut {NoStop}%
\bibitem [{\citenamefont {{Liu}}\ and\ \citenamefont
  {{Hill}}(2015)}]{2015PhRvD..92f3517L}%
  \BibitemOpen
  \bibfield  {author} {\bibinfo {author} {\bibfnamefont {J.}~\bibnamefont
  {{Liu}}}\ and\ \bibinfo {author} {\bibfnamefont {J.~C.}\ \bibnamefont
  {{Hill}}},\ }\href {\doibase 10.1103/PhysRevD.92.063517} {\bibfield
  {journal} {\bibinfo  {journal} {\prd}\ }\textbf {\bibinfo {volume} {92}},\
  \bibinfo {eid} {063517} (\bibinfo {year} {2015})},\ \Eprint
  {http://arxiv.org/abs/1504.05598} {arXiv:1504.05598} \BibitemShut {NoStop}%
\bibitem [{\citenamefont {{Kirk}}\ \emph {et~al.}(2016)\citenamefont {{Kirk}},
  \citenamefont {{Omori}}, \citenamefont {{Benoit-L{\'e}vy}}, \citenamefont
  {{Cawthon}}, \citenamefont {{Chang}}, \citenamefont {{Larsen}}, \citenamefont
  {{Amara}}, \citenamefont {{Bacon}}, \citenamefont {{Crawford}}, \citenamefont
  {{Dodelson}}, \citenamefont {{Fosalba}}, \citenamefont {{Giannantonio}},
  \citenamefont {{Holder}}, \citenamefont {{Jain}}, \citenamefont {{Kacprzak}},
  \citenamefont {{Lahav}}, \citenamefont {{MacCrann}}, \citenamefont
  {{Nicola}}, \citenamefont {{Refregier}}, \citenamefont {{Sheldon}},
  \citenamefont {{Story}}, \citenamefont {{Troxel}}, \citenamefont {{Vieira}},
  \citenamefont {{Vikram}}, \citenamefont {{Zuntz}}, \citenamefont {{Abbott}},
  \citenamefont {{Abdalla}}, \citenamefont {{Becker}}, \citenamefont
  {{Benson}}, \citenamefont {{Bernstein}}, \citenamefont {{Bernstein}},
  \citenamefont {{Bleem}}, \citenamefont {{Bonnett}}, \citenamefont {{Bridle}},
  \citenamefont {{Brooks}}, \citenamefont {{Buckley-Geer}}, \citenamefont
  {{Burke}}, \citenamefont {{Capozzi}}, \citenamefont {{Carlstrom}},
  \citenamefont {{Rosell}}, \citenamefont {{Kind}}, \citenamefont
  {{Carretero}}, \citenamefont {{Crocce}}, \citenamefont {{Cunha}},
  \citenamefont {{D'Andrea}}, \citenamefont {{da Costa}}, \citenamefont
  {{Desai}}, \citenamefont {{Diehl}}, \citenamefont {{Dietrich}}, \citenamefont
  {{Doel}}, \citenamefont {{Eifler}}, \citenamefont {{Evrard}}, \citenamefont
  {{Flaugher}}, \citenamefont {{Frieman}}, \citenamefont {{Gerdes}},
  \citenamefont {{Goldstein}}, \citenamefont {{Gruen}}, \citenamefont
  {{Gruendl}}, \citenamefont {{Honscheid}}, \citenamefont {{James}},
  \citenamefont {{Jarvis}}, \citenamefont {{Kent}}, \citenamefont {{Kuehn}},
  \citenamefont {{Kuropatkin}}, \citenamefont {{Lima}}, \citenamefont
  {{March}}, \citenamefont {{Martini}}, \citenamefont {{Melchior}},
  \citenamefont {{Miller}}, \citenamefont {{Miquel}}, \citenamefont {{Nichol}},
  \citenamefont {{Ogando}}, \citenamefont {{Plazas}}, \citenamefont
  {{Reichardt}}, \citenamefont {{Roodman}}, \citenamefont {{Rozo}},
  \citenamefont {{Rykoff}}, \citenamefont {{Sako}}, \citenamefont {{Sanchez}},
  \citenamefont {{Scarpine}}, \citenamefont {{Schubnell}}, \citenamefont
  {{Sevilla-Noarbe}}, \citenamefont {{Simard}}, \citenamefont {{Smith}},
  \citenamefont {{Soares-Santos}}, \citenamefont {{Sobreira}}, \citenamefont
  {{Suchyta}}, \citenamefont {{Swanson}}, \citenamefont {{Tarle}},
  \citenamefont {{Thomas}}, \citenamefont {{Wechsler}},\ and\ \citenamefont
  {{Weller}}}]{2016MNRAS.459...21K}%
  \BibitemOpen
  \bibfield  {author} {\bibinfo {author} {\bibfnamefont {D.}~\bibnamefont
  {{Kirk}}}, \bibinfo {author} {\bibfnamefont {Y.}~\bibnamefont {{Omori}}},
  \bibinfo {author} {\bibfnamefont {A.}~\bibnamefont {{Benoit-L{\'e}vy}}},
  \bibinfo {author} {\bibfnamefont {R.}~\bibnamefont {{Cawthon}}}, \bibinfo
  {author} {\bibfnamefont {C.}~\bibnamefont {{Chang}}}, \bibinfo {author}
  {\bibfnamefont {P.}~\bibnamefont {{Larsen}}}, \bibinfo {author}
  {\bibfnamefont {A.}~\bibnamefont {{Amara}}}, \bibinfo {author} {\bibfnamefont
  {D.}~\bibnamefont {{Bacon}}}, \bibinfo {author} {\bibfnamefont {T.~M.}\
  \bibnamefont {{Crawford}}}, \bibinfo {author} {\bibfnamefont
  {S.}~\bibnamefont {{Dodelson}}}, \bibinfo {author} {\bibfnamefont
  {P.}~\bibnamefont {{Fosalba}}}, \bibinfo {author} {\bibfnamefont
  {T.}~\bibnamefont {{Giannantonio}}}, \bibinfo {author} {\bibfnamefont
  {G.}~\bibnamefont {{Holder}}}, \bibinfo {author} {\bibfnamefont
  {B.}~\bibnamefont {{Jain}}}, \bibinfo {author} {\bibfnamefont
  {T.}~\bibnamefont {{Kacprzak}}}, \bibinfo {author} {\bibfnamefont
  {O.}~\bibnamefont {{Lahav}}}, \bibinfo {author} {\bibfnamefont
  {N.}~\bibnamefont {{MacCrann}}}, \bibinfo {author} {\bibfnamefont
  {A.}~\bibnamefont {{Nicola}}}, \bibinfo {author} {\bibfnamefont
  {A.}~\bibnamefont {{Refregier}}}, \bibinfo {author} {\bibfnamefont
  {E.}~\bibnamefont {{Sheldon}}}, \bibinfo {author} {\bibfnamefont {K.~T.}\
  \bibnamefont {{Story}}}, \bibinfo {author} {\bibfnamefont {M.~A.}\
  \bibnamefont {{Troxel}}}, \bibinfo {author} {\bibfnamefont {J.~D.}\
  \bibnamefont {{Vieira}}}, \bibinfo {author} {\bibfnamefont {V.}~\bibnamefont
  {{Vikram}}}, \bibinfo {author} {\bibfnamefont {J.}~\bibnamefont {{Zuntz}}},
  \bibinfo {author} {\bibfnamefont {T.~M.~C.}\ \bibnamefont {{Abbott}}},
  \bibinfo {author} {\bibfnamefont {F.~B.}\ \bibnamefont {{Abdalla}}}, \bibinfo
  {author} {\bibfnamefont {M.~R.}\ \bibnamefont {{Becker}}}, \bibinfo {author}
  {\bibfnamefont {B.~A.}\ \bibnamefont {{Benson}}}, \bibinfo {author}
  {\bibfnamefont {G.~M.}\ \bibnamefont {{Bernstein}}}, \bibinfo {author}
  {\bibfnamefont {R.~A.}\ \bibnamefont {{Bernstein}}}, \bibinfo {author}
  {\bibfnamefont {L.~E.}\ \bibnamefont {{Bleem}}}, \bibinfo {author}
  {\bibfnamefont {C.}~\bibnamefont {{Bonnett}}}, \bibinfo {author}
  {\bibfnamefont {S.~L.}\ \bibnamefont {{Bridle}}}, \bibinfo {author}
  {\bibfnamefont {D.}~\bibnamefont {{Brooks}}}, \bibinfo {author}
  {\bibfnamefont {E.}~\bibnamefont {{Buckley-Geer}}}, \bibinfo {author}
  {\bibfnamefont {D.~L.}\ \bibnamefont {{Burke}}}, \bibinfo {author}
  {\bibfnamefont {D.}~\bibnamefont {{Capozzi}}}, \bibinfo {author}
  {\bibfnamefont {J.~E.}\ \bibnamefont {{Carlstrom}}}, \bibinfo {author}
  {\bibfnamefont {A.~C.}\ \bibnamefont {{Rosell}}}, \bibinfo {author}
  {\bibfnamefont {M.~C.}\ \bibnamefont {{Kind}}}, \bibinfo {author}
  {\bibfnamefont {J.}~\bibnamefont {{Carretero}}}, \bibinfo {author}
  {\bibfnamefont {M.}~\bibnamefont {{Crocce}}}, \bibinfo {author}
  {\bibfnamefont {C.~E.}\ \bibnamefont {{Cunha}}}, \bibinfo {author}
  {\bibfnamefont {C.~B.}\ \bibnamefont {{D'Andrea}}}, \bibinfo {author}
  {\bibfnamefont {L.~N.}\ \bibnamefont {{da Costa}}}, \bibinfo {author}
  {\bibfnamefont {S.}~\bibnamefont {{Desai}}}, \bibinfo {author} {\bibfnamefont
  {H.~T.}\ \bibnamefont {{Diehl}}}, \bibinfo {author} {\bibfnamefont {J.~P.}\
  \bibnamefont {{Dietrich}}}, \bibinfo {author} {\bibfnamefont
  {P.}~\bibnamefont {{Doel}}}, \bibinfo {author} {\bibfnamefont {T.~F.}\
  \bibnamefont {{Eifler}}}, \bibinfo {author} {\bibfnamefont {A.~E.}\
  \bibnamefont {{Evrard}}}, \bibinfo {author} {\bibfnamefont {B.}~\bibnamefont
  {{Flaugher}}}, \bibinfo {author} {\bibfnamefont {J.}~\bibnamefont
  {{Frieman}}}, \bibinfo {author} {\bibfnamefont {D.~W.}\ \bibnamefont
  {{Gerdes}}}, \bibinfo {author} {\bibfnamefont {D.~A.}\ \bibnamefont
  {{Goldstein}}}, \bibinfo {author} {\bibfnamefont {D.}~\bibnamefont
  {{Gruen}}}, \bibinfo {author} {\bibfnamefont {R.~A.}\ \bibnamefont
  {{Gruendl}}}, \bibinfo {author} {\bibfnamefont {K.}~\bibnamefont
  {{Honscheid}}}, \bibinfo {author} {\bibfnamefont {D.~J.}\ \bibnamefont
  {{James}}}, \bibinfo {author} {\bibfnamefont {M.}~\bibnamefont {{Jarvis}}},
  \bibinfo {author} {\bibfnamefont {S.}~\bibnamefont {{Kent}}}, \bibinfo
  {author} {\bibfnamefont {K.}~\bibnamefont {{Kuehn}}}, \bibinfo {author}
  {\bibfnamefont {N.}~\bibnamefont {{Kuropatkin}}}, \bibinfo {author}
  {\bibfnamefont {M.}~\bibnamefont {{Lima}}}, \bibinfo {author} {\bibfnamefont
  {M.}~\bibnamefont {{March}}}, \bibinfo {author} {\bibfnamefont
  {P.}~\bibnamefont {{Martini}}}, \bibinfo {author} {\bibfnamefont
  {P.}~\bibnamefont {{Melchior}}}, \bibinfo {author} {\bibfnamefont {C.~J.}\
  \bibnamefont {{Miller}}}, \bibinfo {author} {\bibfnamefont {R.}~\bibnamefont
  {{Miquel}}}, \bibinfo {author} {\bibfnamefont {R.~C.}\ \bibnamefont
  {{Nichol}}}, \bibinfo {author} {\bibfnamefont {R.}~\bibnamefont {{Ogando}}},
  \bibinfo {author} {\bibfnamefont {A.~A.}\ \bibnamefont {{Plazas}}}, \bibinfo
  {author} {\bibfnamefont {C.~L.}\ \bibnamefont {{Reichardt}}}, \bibinfo
  {author} {\bibfnamefont {A.}~\bibnamefont {{Roodman}}}, \bibinfo {author}
  {\bibfnamefont {E.}~\bibnamefont {{Rozo}}}, \bibinfo {author} {\bibfnamefont
  {E.~S.}\ \bibnamefont {{Rykoff}}}, \bibinfo {author} {\bibfnamefont
  {M.}~\bibnamefont {{Sako}}}, \bibinfo {author} {\bibfnamefont
  {E.}~\bibnamefont {{Sanchez}}}, \bibinfo {author} {\bibfnamefont
  {V.}~\bibnamefont {{Scarpine}}}, \bibinfo {author} {\bibfnamefont
  {M.}~\bibnamefont {{Schubnell}}}, \bibinfo {author} {\bibfnamefont
  {I.}~\bibnamefont {{Sevilla-Noarbe}}}, \bibinfo {author} {\bibfnamefont
  {G.}~\bibnamefont {{Simard}}}, \bibinfo {author} {\bibfnamefont {R.~C.}\
  \bibnamefont {{Smith}}}, \bibinfo {author} {\bibfnamefont {M.}~\bibnamefont
  {{Soares-Santos}}}, \bibinfo {author} {\bibfnamefont {F.}~\bibnamefont
  {{Sobreira}}}, \bibinfo {author} {\bibfnamefont {E.}~\bibnamefont
  {{Suchyta}}}, \bibinfo {author} {\bibfnamefont {M.~E.~C.}\ \bibnamefont
  {{Swanson}}}, \bibinfo {author} {\bibfnamefont {G.}~\bibnamefont {{Tarle}}},
  \bibinfo {author} {\bibfnamefont {D.}~\bibnamefont {{Thomas}}}, \bibinfo
  {author} {\bibfnamefont {R.~H.}\ \bibnamefont {{Wechsler}}}, \ and\ \bibinfo
  {author} {\bibfnamefont {J.}~\bibnamefont {{Weller}}},\ }\href {\doibase
  10.1093/mnras/stw570} {\bibfield  {journal} {\bibinfo  {journal} {\mnras}\
  }\textbf {\bibinfo {volume} {459}},\ \bibinfo {pages} {21} (\bibinfo {year}
  {2016})},\ \Eprint {http://arxiv.org/abs/1512.04535} {arXiv:1512.04535}
  \BibitemShut {NoStop}%
\bibitem [{\citenamefont {{Kitching}}\ \emph {et~al.}(2015)\citenamefont
  {{Kitching}}, \citenamefont {{Heavens}},\ and\ \citenamefont
  {{Das}}}]{2015MNRAS.449.2205K}%
  \BibitemOpen
  \bibfield  {author} {\bibinfo {author} {\bibfnamefont {T.~D.}\ \bibnamefont
  {{Kitching}}}, \bibinfo {author} {\bibfnamefont {A.~F.}\ \bibnamefont
  {{Heavens}}}, \ and\ \bibinfo {author} {\bibfnamefont {S.}~\bibnamefont
  {{Das}}},\ }\href {\doibase 10.1093/mnras/stv193} {\bibfield  {journal}
  {\bibinfo  {journal} {\mnras}\ }\textbf {\bibinfo {volume} {449}},\ \bibinfo
  {pages} {2205} (\bibinfo {year} {2015})},\ \Eprint
  {http://arxiv.org/abs/1408.7052} {arXiv:1408.7052} \BibitemShut {NoStop}%
\bibitem [{\citenamefont {{Miyatake}}\ \emph {et~al.}(2016)\citenamefont
  {{Miyatake}}, \citenamefont {{Madhavacheril}}, \citenamefont {{Sehgal}},
  \citenamefont {{Slosar}}, \citenamefont {{Spergel}}, \citenamefont
  {{Sherwin}},\ and\ \citenamefont {{van Engelen}}}]{2016arXiv160505337M}%
  \BibitemOpen
  \bibfield  {author} {\bibinfo {author} {\bibfnamefont {H.}~\bibnamefont
  {{Miyatake}}}, \bibinfo {author} {\bibfnamefont {M.~S.}\ \bibnamefont
  {{Madhavacheril}}}, \bibinfo {author} {\bibfnamefont {N.}~\bibnamefont
  {{Sehgal}}}, \bibinfo {author} {\bibfnamefont {A.}~\bibnamefont {{Slosar}}},
  \bibinfo {author} {\bibfnamefont {D.~N.}\ \bibnamefont {{Spergel}}}, \bibinfo
  {author} {\bibfnamefont {B.}~\bibnamefont {{Sherwin}}}, \ and\ \bibinfo
  {author} {\bibfnamefont {A.}~\bibnamefont {{van Engelen}}},\ }\href@noop {}
  {\bibfield  {journal} {\bibinfo  {journal} {ArXiv e-prints}\ } (\bibinfo
  {year} {2016})},\ \Eprint {http://arxiv.org/abs/1605.05337}
  {arXiv:1605.05337} \BibitemShut {NoStop}%
\bibitem [{\citenamefont {{Singh}}\ \emph {et~al.}(2016)\citenamefont
  {{Singh}}, \citenamefont {{Mandelbaum}},\ and\ \citenamefont
  {{Brownstein}}}]{2016arXiv160608841S}%
  \BibitemOpen
  \bibfield  {author} {\bibinfo {author} {\bibfnamefont {S.}~\bibnamefont
  {{Singh}}}, \bibinfo {author} {\bibfnamefont {R.}~\bibnamefont
  {{Mandelbaum}}}, \ and\ \bibinfo {author} {\bibfnamefont {J.~R.}\
  \bibnamefont {{Brownstein}}},\ }\href@noop {} {\bibfield  {journal} {\bibinfo
   {journal} {ArXiv e-prints}\ } (\bibinfo {year} {2016})},\ \Eprint
  {http://arxiv.org/abs/1606.08841} {arXiv:1606.08841} \BibitemShut {NoStop}%
\bibitem [{\citenamefont {{Eifler}}\ \emph {et~al.}(2014)\citenamefont
  {{Eifler}}, \citenamefont {{Krause}}, \citenamefont {{Schneider}},\ and\
  \citenamefont {{Honscheid}}}]{2014MNRAS.440.1379E}%
  \BibitemOpen
  \bibfield  {author} {\bibinfo {author} {\bibfnamefont {T.}~\bibnamefont
  {{Eifler}}}, \bibinfo {author} {\bibfnamefont {E.}~\bibnamefont {{Krause}}},
  \bibinfo {author} {\bibfnamefont {P.}~\bibnamefont {{Schneider}}}, \ and\
  \bibinfo {author} {\bibfnamefont {K.}~\bibnamefont {{Honscheid}}},\ }\href
  {\doibase 10.1093/mnras/stu251} {\bibfield  {journal} {\bibinfo  {journal}
  {\mnras}\ }\textbf {\bibinfo {volume} {440}},\ \bibinfo {pages} {1379}
  (\bibinfo {year} {2014})},\ \Eprint {http://arxiv.org/abs/1302.2401}
  {arXiv:1302.2401} \BibitemShut {NoStop}%
\bibitem [{\citenamefont {{Hirata}}\ and\ \citenamefont
  {{Seljak}}(2003{\natexlab{a}})}]{2003PhRvD..67d3001H}%
  \BibitemOpen
  \bibfield  {author} {\bibinfo {author} {\bibfnamefont {C.~M.}\ \bibnamefont
  {{Hirata}}}\ and\ \bibinfo {author} {\bibfnamefont {U.}~\bibnamefont
  {{Seljak}}},\ }\href {\doibase 10.1103/PhysRevD.67.043001} {\bibfield
  {journal} {\bibinfo  {journal} {\prd}\ }\textbf {\bibinfo {volume} {67}},\
  \bibinfo {eid} {043001} (\bibinfo {year} {2003}{\natexlab{a}})},\ \Eprint
  {http://arxiv.org/abs/astro-ph/0209489} {astro-ph/0209489} \BibitemShut
  {NoStop}%
\bibitem [{\citenamefont {{Hirata}}\ and\ \citenamefont
  {{Seljak}}(2003{\natexlab{b}})}]{2003PhRvD..68h3002H}%
  \BibitemOpen
  \bibfield  {author} {\bibinfo {author} {\bibfnamefont {C.~M.}\ \bibnamefont
  {{Hirata}}}\ and\ \bibinfo {author} {\bibfnamefont {U.}~\bibnamefont
  {{Seljak}}},\ }\href {\doibase 10.1103/PhysRevD.68.083002} {\bibfield
  {journal} {\bibinfo  {journal} {\prd}\ }\textbf {\bibinfo {volume} {68}},\
  \bibinfo {eid} {083002} (\bibinfo {year} {2003}{\natexlab{b}})},\ \Eprint
  {http://arxiv.org/abs/astro-ph/0306354} {astro-ph/0306354} \BibitemShut
  {NoStop}%
\bibitem [{\citenamefont {{Krause}}\ and\ \citenamefont
  {{Eifler}}(2016)}]{2016arXiv160105779K}%
  \BibitemOpen
  \bibfield  {author} {\bibinfo {author} {\bibfnamefont {E.}~\bibnamefont
  {{Krause}}}\ and\ \bibinfo {author} {\bibfnamefont {T.}~\bibnamefont
  {{Eifler}}},\ }\href@noop {} {\bibfield  {journal} {\bibinfo  {journal}
  {ArXiv e-prints}\ } (\bibinfo {year} {2016})},\ \Eprint
  {http://arxiv.org/abs/1601.05779} {arXiv:1601.05779} \BibitemShut {NoStop}%
\bibitem [{\citenamefont {{Chang}}\ \emph {et~al.}(2013)\citenamefont
  {{Chang}}, \citenamefont {{Jarvis}}, \citenamefont {{Jain}}, \citenamefont
  {{Kahn}}, \citenamefont {{Kirkby}}, \citenamefont {{Connolly}}, \citenamefont
  {{Krughoff}}, \citenamefont {{Peng}},\ and\ \citenamefont
  {{Peterson}}}]{2013MNRAS.434.2121C}%
  \BibitemOpen
  \bibfield  {author} {\bibinfo {author} {\bibfnamefont {C.}~\bibnamefont
  {{Chang}}}, \bibinfo {author} {\bibfnamefont {M.}~\bibnamefont {{Jarvis}}},
  \bibinfo {author} {\bibfnamefont {B.}~\bibnamefont {{Jain}}}, \bibinfo
  {author} {\bibfnamefont {S.~M.}\ \bibnamefont {{Kahn}}}, \bibinfo {author}
  {\bibfnamefont {D.}~\bibnamefont {{Kirkby}}}, \bibinfo {author}
  {\bibfnamefont {A.}~\bibnamefont {{Connolly}}}, \bibinfo {author}
  {\bibfnamefont {S.}~\bibnamefont {{Krughoff}}}, \bibinfo {author}
  {\bibfnamefont {E.-H.}\ \bibnamefont {{Peng}}}, \ and\ \bibinfo {author}
  {\bibfnamefont {J.~R.}\ \bibnamefont {{Peterson}}},\ }\href {\doibase
  10.1093/mnras/stt1156} {\bibfield  {journal} {\bibinfo  {journal} {\mnras}\
  }\textbf {\bibinfo {volume} {434}},\ \bibinfo {pages} {2121} (\bibinfo {year}
  {2013})},\ \Eprint {http://arxiv.org/abs/1305.0793} {arXiv:1305.0793}
  \BibitemShut {NoStop}%
\bibitem [{\citenamefont {{Rozo}}\ \emph {et~al.}(2015)\citenamefont {{Rozo}},
  \citenamefont {{Rykoff}}, \citenamefont {{Abate}}, \citenamefont {{Bonnett}},
  \citenamefont {{Crocce}}, \citenamefont {{Davis}}, \citenamefont {{Hoyle}},
  \citenamefont {{Leistedt}}, \citenamefont {{Peiris}}, \citenamefont
  {{Wechsler}}, \citenamefont {{Abbott}}, \citenamefont {{Abdalla}},
  \citenamefont {{Banerji}}, \citenamefont {{Bauer}}, \citenamefont
  {{Benoit-L{\'e}vy}}, \citenamefont {{Bernstein}}, \citenamefont {{Bertin}},
  \citenamefont {{Brooks}}, \citenamefont {{Buckley-Geer}}, \citenamefont
  {{Burke}}, \citenamefont {{Capozzi}}, \citenamefont {{Carnero Rosell}},
  \citenamefont {{Carollo}}, \citenamefont {{Carrasco Kind}}, \citenamefont
  {{Carretero}}, \citenamefont {{Castander}}, \citenamefont {{Childress}},
  \citenamefont {{Cunha}}, \citenamefont {{D'Andrea}}, \citenamefont {{Davis}},
  \citenamefont {{DePoy}}, \citenamefont {{Desai}}, \citenamefont {{Diehl}},
  \citenamefont {{Dietrich}}, \citenamefont {{Doel}}, \citenamefont {{Eifler}},
  \citenamefont {{Evrard}}, \citenamefont {{Fausti Neto}}, \citenamefont
  {{Flaugher}}, \citenamefont {{Fosalba}}, \citenamefont {{Frieman}},
  \citenamefont {{Gaztanaga}}, \citenamefont {{Gerdes}}, \citenamefont
  {{Glazebrook}}, \citenamefont {{Gruen}}, \citenamefont {{Gruendl}},
  \citenamefont {{Honscheid}}, \citenamefont {{James}}, \citenamefont
  {{Jarvis}}, \citenamefont {{Kim}}, \citenamefont {{Kuehn}}, \citenamefont
  {{Kuropatkin}}, \citenamefont {{Lahav}}, \citenamefont {{Lidman}},
  \citenamefont {{Lima}}, \citenamefont {{Maia}}, \citenamefont {{March}},
  \citenamefont {{Martini}}, \citenamefont {{Melchior}}, \citenamefont
  {{Miller}}, \citenamefont {{Miquel}}, \citenamefont {{Mohr}}, \citenamefont
  {{Nichol}}, \citenamefont {{Nord}}, \citenamefont {{O'Neill}}, \citenamefont
  {{Ogando}}, \citenamefont {{Plazas}}, \citenamefont {{Romer}}, \citenamefont
  {{Roodman}}, \citenamefont {{Sako}}, \citenamefont {{Sanchez}}, \citenamefont
  {{Santiago}}, \citenamefont {{Schubnell}}, \citenamefont {{Sevilla-Noarbe}},
  \citenamefont {{Smith}}, \citenamefont {{Soares-Santos}}, \citenamefont
  {{Sobreira}}, \citenamefont {{Suchyta}}, \citenamefont {{Swanson}},
  \citenamefont {{Thaler}}, \citenamefont {{Thomas}}, \citenamefont {{Uddin}},
  \citenamefont {{Vikram}}, \citenamefont {{Walker}}, \citenamefont {{Wester}},
  \citenamefont {{Zhang}},\ and\ \citenamefont {{da
  Costa}}}]{2015arXiv150705460R}%
  \BibitemOpen
  \bibfield  {author} {\bibinfo {author} {\bibfnamefont {E.}~\bibnamefont
  {{Rozo}}}, \bibinfo {author} {\bibfnamefont {E.~S.}\ \bibnamefont
  {{Rykoff}}}, \bibinfo {author} {\bibfnamefont {A.}~\bibnamefont {{Abate}}},
  \bibinfo {author} {\bibfnamefont {C.}~\bibnamefont {{Bonnett}}}, \bibinfo
  {author} {\bibfnamefont {M.}~\bibnamefont {{Crocce}}}, \bibinfo {author}
  {\bibfnamefont {C.}~\bibnamefont {{Davis}}}, \bibinfo {author} {\bibfnamefont
  {B.}~\bibnamefont {{Hoyle}}}, \bibinfo {author} {\bibfnamefont
  {B.}~\bibnamefont {{Leistedt}}}, \bibinfo {author} {\bibfnamefont {H.~V.}\
  \bibnamefont {{Peiris}}}, \bibinfo {author} {\bibfnamefont {R.~H.}\
  \bibnamefont {{Wechsler}}}, \bibinfo {author} {\bibfnamefont
  {T.}~\bibnamefont {{Abbott}}}, \bibinfo {author} {\bibfnamefont {F.~B.}\
  \bibnamefont {{Abdalla}}}, \bibinfo {author} {\bibfnamefont {M.}~\bibnamefont
  {{Banerji}}}, \bibinfo {author} {\bibfnamefont {A.~H.}\ \bibnamefont
  {{Bauer}}}, \bibinfo {author} {\bibfnamefont {A.}~\bibnamefont
  {{Benoit-L{\'e}vy}}}, \bibinfo {author} {\bibfnamefont {G.~M.}\ \bibnamefont
  {{Bernstein}}}, \bibinfo {author} {\bibfnamefont {E.}~\bibnamefont
  {{Bertin}}}, \bibinfo {author} {\bibfnamefont {D.}~\bibnamefont {{Brooks}}},
  \bibinfo {author} {\bibfnamefont {E.}~\bibnamefont {{Buckley-Geer}}},
  \bibinfo {author} {\bibfnamefont {D.~L.}\ \bibnamefont {{Burke}}}, \bibinfo
  {author} {\bibfnamefont {D.}~\bibnamefont {{Capozzi}}}, \bibinfo {author}
  {\bibfnamefont {A.}~\bibnamefont {{Carnero Rosell}}}, \bibinfo {author}
  {\bibfnamefont {D.}~\bibnamefont {{Carollo}}}, \bibinfo {author}
  {\bibfnamefont {M.}~\bibnamefont {{Carrasco Kind}}}, \bibinfo {author}
  {\bibfnamefont {J.}~\bibnamefont {{Carretero}}}, \bibinfo {author}
  {\bibfnamefont {F.~J.}\ \bibnamefont {{Castander}}}, \bibinfo {author}
  {\bibfnamefont {M.~J.}\ \bibnamefont {{Childress}}}, \bibinfo {author}
  {\bibfnamefont {C.~E.}\ \bibnamefont {{Cunha}}}, \bibinfo {author}
  {\bibfnamefont {C.~B.}\ \bibnamefont {{D'Andrea}}}, \bibinfo {author}
  {\bibfnamefont {T.}~\bibnamefont {{Davis}}}, \bibinfo {author} {\bibfnamefont
  {D.~L.}\ \bibnamefont {{DePoy}}}, \bibinfo {author} {\bibfnamefont
  {S.}~\bibnamefont {{Desai}}}, \bibinfo {author} {\bibfnamefont {H.~T.}\
  \bibnamefont {{Diehl}}}, \bibinfo {author} {\bibfnamefont {J.~P.}\
  \bibnamefont {{Dietrich}}}, \bibinfo {author} {\bibfnamefont
  {P.}~\bibnamefont {{Doel}}}, \bibinfo {author} {\bibfnamefont {T.~F.}\
  \bibnamefont {{Eifler}}}, \bibinfo {author} {\bibfnamefont {A.~E.}\
  \bibnamefont {{Evrard}}}, \bibinfo {author} {\bibfnamefont {A.}~\bibnamefont
  {{Fausti Neto}}}, \bibinfo {author} {\bibfnamefont {B.}~\bibnamefont
  {{Flaugher}}}, \bibinfo {author} {\bibfnamefont {P.}~\bibnamefont
  {{Fosalba}}}, \bibinfo {author} {\bibfnamefont {J.}~\bibnamefont
  {{Frieman}}}, \bibinfo {author} {\bibfnamefont {E.}~\bibnamefont
  {{Gaztanaga}}}, \bibinfo {author} {\bibfnamefont {D.~W.}\ \bibnamefont
  {{Gerdes}}}, \bibinfo {author} {\bibfnamefont {K.}~\bibnamefont
  {{Glazebrook}}}, \bibinfo {author} {\bibfnamefont {D.}~\bibnamefont
  {{Gruen}}}, \bibinfo {author} {\bibfnamefont {R.~A.}\ \bibnamefont
  {{Gruendl}}}, \bibinfo {author} {\bibfnamefont {K.}~\bibnamefont
  {{Honscheid}}}, \bibinfo {author} {\bibfnamefont {D.~J.}\ \bibnamefont
  {{James}}}, \bibinfo {author} {\bibfnamefont {M.}~\bibnamefont {{Jarvis}}},
  \bibinfo {author} {\bibfnamefont {A.~G.}\ \bibnamefont {{Kim}}}, \bibinfo
  {author} {\bibfnamefont {K.}~\bibnamefont {{Kuehn}}}, \bibinfo {author}
  {\bibfnamefont {N.}~\bibnamefont {{Kuropatkin}}}, \bibinfo {author}
  {\bibfnamefont {O.}~\bibnamefont {{Lahav}}}, \bibinfo {author} {\bibfnamefont
  {C.}~\bibnamefont {{Lidman}}}, \bibinfo {author} {\bibfnamefont
  {M.}~\bibnamefont {{Lima}}}, \bibinfo {author} {\bibfnamefont {M.~A.~G.}\
  \bibnamefont {{Maia}}}, \bibinfo {author} {\bibfnamefont {M.}~\bibnamefont
  {{March}}}, \bibinfo {author} {\bibfnamefont {P.}~\bibnamefont {{Martini}}},
  \bibinfo {author} {\bibfnamefont {P.}~\bibnamefont {{Melchior}}}, \bibinfo
  {author} {\bibfnamefont {C.~J.}\ \bibnamefont {{Miller}}}, \bibinfo {author}
  {\bibfnamefont {R.}~\bibnamefont {{Miquel}}}, \bibinfo {author}
  {\bibfnamefont {J.~J.}\ \bibnamefont {{Mohr}}}, \bibinfo {author}
  {\bibfnamefont {R.~C.}\ \bibnamefont {{Nichol}}}, \bibinfo {author}
  {\bibfnamefont {B.}~\bibnamefont {{Nord}}}, \bibinfo {author} {\bibfnamefont
  {C.~R.}\ \bibnamefont {{O'Neill}}}, \bibinfo {author} {\bibfnamefont
  {R.}~\bibnamefont {{Ogando}}}, \bibinfo {author} {\bibfnamefont {A.~A.}\
  \bibnamefont {{Plazas}}}, \bibinfo {author} {\bibfnamefont {A.~K.}\
  \bibnamefont {{Romer}}}, \bibinfo {author} {\bibfnamefont {A.}~\bibnamefont
  {{Roodman}}}, \bibinfo {author} {\bibfnamefont {M.}~\bibnamefont {{Sako}}},
  \bibinfo {author} {\bibfnamefont {E.}~\bibnamefont {{Sanchez}}}, \bibinfo
  {author} {\bibfnamefont {B.}~\bibnamefont {{Santiago}}}, \bibinfo {author}
  {\bibfnamefont {M.}~\bibnamefont {{Schubnell}}}, \bibinfo {author}
  {\bibfnamefont {I.}~\bibnamefont {{Sevilla-Noarbe}}}, \bibinfo {author}
  {\bibfnamefont {R.~C.}\ \bibnamefont {{Smith}}}, \bibinfo {author}
  {\bibfnamefont {M.}~\bibnamefont {{Soares-Santos}}}, \bibinfo {author}
  {\bibfnamefont {F.}~\bibnamefont {{Sobreira}}}, \bibinfo {author}
  {\bibfnamefont {E.}~\bibnamefont {{Suchyta}}}, \bibinfo {author}
  {\bibfnamefont {M.~E.~C.}\ \bibnamefont {{Swanson}}}, \bibinfo {author}
  {\bibfnamefont {J.}~\bibnamefont {{Thaler}}}, \bibinfo {author}
  {\bibfnamefont {D.}~\bibnamefont {{Thomas}}}, \bibinfo {author}
  {\bibfnamefont {S.}~\bibnamefont {{Uddin}}}, \bibinfo {author} {\bibfnamefont
  {V.}~\bibnamefont {{Vikram}}}, \bibinfo {author} {\bibfnamefont {A.~R.}\
  \bibnamefont {{Walker}}}, \bibinfo {author} {\bibfnamefont {W.}~\bibnamefont
  {{Wester}}}, \bibinfo {author} {\bibfnamefont {Y.}~\bibnamefont {{Zhang}}}, \
  and\ \bibinfo {author} {\bibfnamefont {L.~N.}\ \bibnamefont {{da Costa}}},\
  }\href@noop {} {\bibfield  {journal} {\bibinfo  {journal} {ArXiv e-prints}\ }
  (\bibinfo {year} {2015})},\ \Eprint {http://arxiv.org/abs/1507.05460}
  {arXiv:1507.05460 [astro-ph.IM]} \BibitemShut {NoStop}%
\bibitem [{\citenamefont {{Bernstein}}\ and\ \citenamefont
  {{Huterer}}(2010)}]{2010MNRAS.401.1399B}%
  \BibitemOpen
  \bibfield  {author} {\bibinfo {author} {\bibfnamefont {G.}~\bibnamefont
  {{Bernstein}}}\ and\ \bibinfo {author} {\bibfnamefont {D.}~\bibnamefont
  {{Huterer}}},\ }\href {\doibase 10.1111/j.1365-2966.2009.15748.x} {\bibfield
  {journal} {\bibinfo  {journal} {\mnras}\ }\textbf {\bibinfo {volume} {401}},\
  \bibinfo {pages} {1399} (\bibinfo {year} {2010})},\ \Eprint
  {http://arxiv.org/abs/0902.2782} {arXiv:0902.2782 [astro-ph.CO]} \BibitemShut
  {NoStop}%
\bibitem [{\citenamefont {{Ma}}\ \emph {et~al.}(2006)\citenamefont {{Ma}},
  \citenamefont {{Hu}},\ and\ \citenamefont {{Huterer}}}]{2006ApJ...636...21M}%
  \BibitemOpen
  \bibfield  {author} {\bibinfo {author} {\bibfnamefont {Z.}~\bibnamefont
  {{Ma}}}, \bibinfo {author} {\bibfnamefont {W.}~\bibnamefont {{Hu}}}, \ and\
  \bibinfo {author} {\bibfnamefont {D.}~\bibnamefont {{Huterer}}},\ }\href
  {\doibase 10.1086/497068} {\bibfield  {journal} {\bibinfo  {journal} {\apj}\
  }\textbf {\bibinfo {volume} {636}},\ \bibinfo {pages} {21} (\bibinfo {year}
  {2006})},\ \Eprint {http://arxiv.org/abs/astro-ph/0506614} {astro-ph/0506614}
  \BibitemShut {NoStop}%
\bibitem [{\citenamefont {{Taylor}}\ and\ \citenamefont
  {{Kitching}}(2016)}]{2016arXiv160509130T}%
  \BibitemOpen
  \bibfield  {author} {\bibinfo {author} {\bibfnamefont {A.~N.}\ \bibnamefont
  {{Taylor}}}\ and\ \bibinfo {author} {\bibfnamefont {T.~D.}\ \bibnamefont
  {{Kitching}}},\ }\href@noop {} {\bibfield  {journal} {\bibinfo  {journal}
  {ArXiv e-prints}\ } (\bibinfo {year} {2016})},\ \Eprint
  {http://arxiv.org/abs/1605.09130} {arXiv:1605.09130} \BibitemShut {NoStop}%
\bibitem [{\citenamefont {{Hirata}}\ \emph {et~al.}(2004)\citenamefont
  {{Hirata}}, \citenamefont {{Padmanabhan}}, \citenamefont {{Seljak}},
  \citenamefont {{Schlegel}},\ and\ \citenamefont
  {{Brinkmann}}}]{HirataSeljak04}%
  \BibitemOpen
  \bibfield  {author} {\bibinfo {author} {\bibfnamefont {C.~M.}\ \bibnamefont
  {{Hirata}}}, \bibinfo {author} {\bibfnamefont {N.}~\bibnamefont
  {{Padmanabhan}}}, \bibinfo {author} {\bibfnamefont {U.}~\bibnamefont
  {{Seljak}}}, \bibinfo {author} {\bibfnamefont {D.}~\bibnamefont
  {{Schlegel}}}, \ and\ \bibinfo {author} {\bibfnamefont {J.}~\bibnamefont
  {{Brinkmann}}},\ }\href {\doibase 10.1103/PhysRevD.70.103501} {\bibfield
  {journal} {\bibinfo  {journal} {\prd}\ }\textbf {\bibinfo {volume} {70}},\
  \bibinfo {eid} {103501} (\bibinfo {year} {2004})},\ \Eprint
  {http://arxiv.org/abs/astro-ph/0406004} {astro-ph/0406004} \BibitemShut
  {NoStop}%
\bibitem [{\citenamefont {{Bridle}}\ and\ \citenamefont
  {{King}}(2007)}]{BridleKing07}%
  \BibitemOpen
  \bibfield  {author} {\bibinfo {author} {\bibfnamefont {S.}~\bibnamefont
  {{Bridle}}}\ and\ \bibinfo {author} {\bibfnamefont {L.}~\bibnamefont
  {{King}}},\ }\href {\doibase 10.1088/1367-2630/9/12/444} {\bibfield
  {journal} {\bibinfo  {journal} {New Journal of Physics}\ }\textbf {\bibinfo
  {volume} {9}},\ \bibinfo {pages} {444} (\bibinfo {year} {2007})},\ \Eprint
  {http://arxiv.org/abs/0705.0166} {arXiv:0705.0166} \BibitemShut {NoStop}%
\bibitem [{\citenamefont {{Singh}}\ \emph {et~al.}(2015)\citenamefont
  {{Singh}}, \citenamefont {{Mandelbaum}},\ and\ \citenamefont
  {{More}}}]{Singh15}%
  \BibitemOpen
  \bibfield  {author} {\bibinfo {author} {\bibfnamefont {S.}~\bibnamefont
  {{Singh}}}, \bibinfo {author} {\bibfnamefont {R.}~\bibnamefont
  {{Mandelbaum}}}, \ and\ \bibinfo {author} {\bibfnamefont {S.}~\bibnamefont
  {{More}}},\ }\href {\doibase 10.1093/mnras/stv778} {\bibfield  {journal}
  {\bibinfo  {journal} {\mnras}\ }\textbf {\bibinfo {volume} {450}},\ \bibinfo
  {pages} {2195} (\bibinfo {year} {2015})},\ \Eprint
  {http://arxiv.org/abs/1411.1755} {arXiv:1411.1755} \BibitemShut {NoStop}%
\bibitem [{\citenamefont {{Catelan}}\ \emph {et~al.}(2001)\citenamefont
  {{Catelan}}, \citenamefont {{Kamionkowski}},\ and\ \citenamefont
  {{Blandford}}}]{Catelan2001}%
  \BibitemOpen
  \bibfield  {author} {\bibinfo {author} {\bibfnamefont {P.}~\bibnamefont
  {{Catelan}}}, \bibinfo {author} {\bibfnamefont {M.}~\bibnamefont
  {{Kamionkowski}}}, \ and\ \bibinfo {author} {\bibfnamefont {R.~D.}\
  \bibnamefont {{Blandford}}},\ }\href {\doibase
  10.1046/j.1365-8711.2001.04105.x} {\bibfield  {journal} {\bibinfo  {journal}
  {\mnras}\ }\textbf {\bibinfo {volume} {320}},\ \bibinfo {pages} {L7}
  (\bibinfo {year} {2001})},\ \Eprint {http://arxiv.org/abs/astro-ph/0005470}
  {astro-ph/0005470} \BibitemShut {NoStop}%
\bibitem [{\citenamefont {{Hirata}}\ \emph {et~al.}(2007)\citenamefont
  {{Hirata}}, \citenamefont {{Mandelbaum}}, \citenamefont {{Ishak}},
  \citenamefont {{Seljak}}, \citenamefont {{Nichol}}, \citenamefont
  {{Pimbblet}}, \citenamefont {{Ross}},\ and\ \citenamefont
  {{Wake}}}]{Hirata07}%
  \BibitemOpen
  \bibfield  {author} {\bibinfo {author} {\bibfnamefont {C.~M.}\ \bibnamefont
  {{Hirata}}}, \bibinfo {author} {\bibfnamefont {R.}~\bibnamefont
  {{Mandelbaum}}}, \bibinfo {author} {\bibfnamefont {M.}~\bibnamefont
  {{Ishak}}}, \bibinfo {author} {\bibfnamefont {U.}~\bibnamefont {{Seljak}}},
  \bibinfo {author} {\bibfnamefont {R.}~\bibnamefont {{Nichol}}}, \bibinfo
  {author} {\bibfnamefont {K.~A.}\ \bibnamefont {{Pimbblet}}}, \bibinfo
  {author} {\bibfnamefont {N.~P.}\ \bibnamefont {{Ross}}}, \ and\ \bibinfo
  {author} {\bibfnamefont {D.}~\bibnamefont {{Wake}}},\ }\href {\doibase
  10.1111/j.1365-2966.2007.12312.x} {\bibfield  {journal} {\bibinfo  {journal}
  {\mnras}\ }\textbf {\bibinfo {volume} {381}},\ \bibinfo {pages} {1197}
  (\bibinfo {year} {2007})},\ \Eprint {http://arxiv.org/abs/astro-ph/0701671}
  {astro-ph/0701671} \BibitemShut {NoStop}%
\bibitem [{\citenamefont {{Mandelbaum}}\ \emph {et~al.}(2011)\citenamefont
  {{Mandelbaum}}, \citenamefont {{Blake}}, \citenamefont {{Bridle}},
  \citenamefont {{Abdalla}}, \citenamefont {{Brough}}, \citenamefont
  {{Colless}}, \citenamefont {{Couch}}, \citenamefont {{Croom}}, \citenamefont
  {{Davis}}, \citenamefont {{Drinkwater}}, \citenamefont {{Forster}},
  \citenamefont {{Glazebrook}}, \citenamefont {{Jelliffe}}, \citenamefont
  {{Jurek}}, \citenamefont {{Li}}, \citenamefont {{Madore}}, \citenamefont
  {{Martin}}, \citenamefont {{Pimbblet}}, \citenamefont {{Poole}},
  \citenamefont {{Pracy}}, \citenamefont {{Sharp}}, \citenamefont
  {{Wisnioski}}, \citenamefont {{Woods}},\ and\ \citenamefont
  {{Wyder}}}]{Mandelbaum11}%
  \BibitemOpen
  \bibfield  {author} {\bibinfo {author} {\bibfnamefont {R.}~\bibnamefont
  {{Mandelbaum}}}, \bibinfo {author} {\bibfnamefont {C.}~\bibnamefont
  {{Blake}}}, \bibinfo {author} {\bibfnamefont {S.}~\bibnamefont {{Bridle}}},
  \bibinfo {author} {\bibfnamefont {F.~B.}\ \bibnamefont {{Abdalla}}}, \bibinfo
  {author} {\bibfnamefont {S.}~\bibnamefont {{Brough}}}, \bibinfo {author}
  {\bibfnamefont {M.}~\bibnamefont {{Colless}}}, \bibinfo {author}
  {\bibfnamefont {W.}~\bibnamefont {{Couch}}}, \bibinfo {author} {\bibfnamefont
  {S.}~\bibnamefont {{Croom}}}, \bibinfo {author} {\bibfnamefont
  {T.}~\bibnamefont {{Davis}}}, \bibinfo {author} {\bibfnamefont {M.~J.}\
  \bibnamefont {{Drinkwater}}}, \bibinfo {author} {\bibfnamefont
  {K.}~\bibnamefont {{Forster}}}, \bibinfo {author} {\bibfnamefont
  {K.}~\bibnamefont {{Glazebrook}}}, \bibinfo {author} {\bibfnamefont
  {B.}~\bibnamefont {{Jelliffe}}}, \bibinfo {author} {\bibfnamefont {R.~J.}\
  \bibnamefont {{Jurek}}}, \bibinfo {author} {\bibfnamefont {I.-H.}\
  \bibnamefont {{Li}}}, \bibinfo {author} {\bibfnamefont {B.}~\bibnamefont
  {{Madore}}}, \bibinfo {author} {\bibfnamefont {C.}~\bibnamefont {{Martin}}},
  \bibinfo {author} {\bibfnamefont {K.}~\bibnamefont {{Pimbblet}}}, \bibinfo
  {author} {\bibfnamefont {G.~B.}\ \bibnamefont {{Poole}}}, \bibinfo {author}
  {\bibfnamefont {M.}~\bibnamefont {{Pracy}}}, \bibinfo {author} {\bibfnamefont
  {R.}~\bibnamefont {{Sharp}}}, \bibinfo {author} {\bibfnamefont
  {E.}~\bibnamefont {{Wisnioski}}}, \bibinfo {author} {\bibfnamefont
  {D.}~\bibnamefont {{Woods}}}, \ and\ \bibinfo {author} {\bibfnamefont
  {T.}~\bibnamefont {{Wyder}}},\ }\href {\doibase
  10.1111/j.1365-2966.2010.17485.x} {\bibfield  {journal} {\bibinfo  {journal}
  {\mnras}\ }\textbf {\bibinfo {volume} {410}},\ \bibinfo {pages} {844}
  (\bibinfo {year} {2011})},\ \Eprint {http://arxiv.org/abs/0911.5347}
  {arXiv:0911.5347} \BibitemShut {NoStop}%
\bibitem [{\citenamefont {{Joachimi}}\ \emph {et~al.}(2011)\citenamefont
  {{Joachimi}}, \citenamefont {{Mandelbaum}}, \citenamefont {{Abdalla}},\ and\
  \citenamefont {{Bridle}}}]{Joachimi11}%
  \BibitemOpen
  \bibfield  {author} {\bibinfo {author} {\bibfnamefont {B.}~\bibnamefont
  {{Joachimi}}}, \bibinfo {author} {\bibfnamefont {R.}~\bibnamefont
  {{Mandelbaum}}}, \bibinfo {author} {\bibfnamefont {F.~B.}\ \bibnamefont
  {{Abdalla}}}, \ and\ \bibinfo {author} {\bibfnamefont {S.~L.}\ \bibnamefont
  {{Bridle}}},\ }\href {\doibase 10.1051/0004-6361/201015621} {\bibfield
  {journal} {\bibinfo  {journal} {\aap}\ }\textbf {\bibinfo {volume} {527}},\
  \bibinfo {pages} {A26} (\bibinfo {year} {2011})},\ \Eprint
  {http://arxiv.org/abs/1008.3491} {arXiv:1008.3491 [astro-ph.CO]} \BibitemShut
  {NoStop}%
\bibitem [{\citenamefont {{Loveday}}\ \emph {et~al.}(2012)\citenamefont
  {{Loveday}}, \citenamefont {{Norberg}}, \citenamefont {{Baldry}},
  \citenamefont {{Driver}}, \citenamefont {{Hopkins}}, \citenamefont
  {{Peacock}}, \citenamefont {{Bamford}}, \citenamefont {{Liske}},
  \citenamefont {{Bland-Hawthorn}}, \citenamefont {{Brough}}, \citenamefont
  {{Brown}}, \citenamefont {{Cameron}}, \citenamefont {{Conselice}},
  \citenamefont {{Croom}}, \citenamefont {{Frenk}}, \citenamefont
  {{Gunawardhana}}, \citenamefont {{Hill}}, \citenamefont {{Jones}},
  \citenamefont {{Kelvin}}, \citenamefont {{Kuijken}}, \citenamefont
  {{Nichol}}, \citenamefont {{Parkinson}}, \citenamefont {{Phillipps}},
  \citenamefont {{Pimbblet}}, \citenamefont {{Popescu}}, \citenamefont
  {{Prescott}}, \citenamefont {{Robotham}}, \citenamefont {{Sharp}},
  \citenamefont {{Sutherland}}, \citenamefont {{Taylor}}, \citenamefont
  {{Thomas}}, \citenamefont {{Tuffs}}, \citenamefont {{van Kampen}},\ and\
  \citenamefont {{Wijesinghe}}}]{LFGAMA}%
  \BibitemOpen
  \bibfield  {author} {\bibinfo {author} {\bibfnamefont {J.}~\bibnamefont
  {{Loveday}}}, \bibinfo {author} {\bibfnamefont {P.}~\bibnamefont
  {{Norberg}}}, \bibinfo {author} {\bibfnamefont {I.~K.}\ \bibnamefont
  {{Baldry}}}, \bibinfo {author} {\bibfnamefont {S.~P.}\ \bibnamefont
  {{Driver}}}, \bibinfo {author} {\bibfnamefont {A.~M.}\ \bibnamefont
  {{Hopkins}}}, \bibinfo {author} {\bibfnamefont {J.~A.}\ \bibnamefont
  {{Peacock}}}, \bibinfo {author} {\bibfnamefont {S.~P.}\ \bibnamefont
  {{Bamford}}}, \bibinfo {author} {\bibfnamefont {J.}~\bibnamefont {{Liske}}},
  \bibinfo {author} {\bibfnamefont {J.}~\bibnamefont {{Bland-Hawthorn}}},
  \bibinfo {author} {\bibfnamefont {S.}~\bibnamefont {{Brough}}}, \bibinfo
  {author} {\bibfnamefont {M.~J.~I.}\ \bibnamefont {{Brown}}}, \bibinfo
  {author} {\bibfnamefont {E.}~\bibnamefont {{Cameron}}}, \bibinfo {author}
  {\bibfnamefont {C.~J.}\ \bibnamefont {{Conselice}}}, \bibinfo {author}
  {\bibfnamefont {S.~M.}\ \bibnamefont {{Croom}}}, \bibinfo {author}
  {\bibfnamefont {C.~S.}\ \bibnamefont {{Frenk}}}, \bibinfo {author}
  {\bibfnamefont {M.}~\bibnamefont {{Gunawardhana}}}, \bibinfo {author}
  {\bibfnamefont {D.~T.}\ \bibnamefont {{Hill}}}, \bibinfo {author}
  {\bibfnamefont {D.~H.}\ \bibnamefont {{Jones}}}, \bibinfo {author}
  {\bibfnamefont {L.~S.}\ \bibnamefont {{Kelvin}}}, \bibinfo {author}
  {\bibfnamefont {K.}~\bibnamefont {{Kuijken}}}, \bibinfo {author}
  {\bibfnamefont {R.~C.}\ \bibnamefont {{Nichol}}}, \bibinfo {author}
  {\bibfnamefont {H.~R.}\ \bibnamefont {{Parkinson}}}, \bibinfo {author}
  {\bibfnamefont {S.}~\bibnamefont {{Phillipps}}}, \bibinfo {author}
  {\bibfnamefont {K.~A.}\ \bibnamefont {{Pimbblet}}}, \bibinfo {author}
  {\bibfnamefont {C.~C.}\ \bibnamefont {{Popescu}}}, \bibinfo {author}
  {\bibfnamefont {M.}~\bibnamefont {{Prescott}}}, \bibinfo {author}
  {\bibfnamefont {A.~S.~G.}\ \bibnamefont {{Robotham}}}, \bibinfo {author}
  {\bibfnamefont {R.~G.}\ \bibnamefont {{Sharp}}}, \bibinfo {author}
  {\bibfnamefont {W.~J.}\ \bibnamefont {{Sutherland}}}, \bibinfo {author}
  {\bibfnamefont {E.~N.}\ \bibnamefont {{Taylor}}}, \bibinfo {author}
  {\bibfnamefont {D.}~\bibnamefont {{Thomas}}}, \bibinfo {author}
  {\bibfnamefont {R.~J.}\ \bibnamefont {{Tuffs}}}, \bibinfo {author}
  {\bibfnamefont {E.}~\bibnamefont {{van Kampen}}}, \ and\ \bibinfo {author}
  {\bibfnamefont {D.}~\bibnamefont {{Wijesinghe}}},\ }\href {\doibase
  10.1111/j.1365-2966.2011.20111.x} {\bibfield  {journal} {\bibinfo  {journal}
  {\mnras}\ }\textbf {\bibinfo {volume} {420}},\ \bibinfo {pages} {1239}
  (\bibinfo {year} {2012})},\ \Eprint {http://arxiv.org/abs/1111.0166}
  {arXiv:1111.0166 [astro-ph.CO]} \BibitemShut {NoStop}%
\bibitem [{\citenamefont {{Krause}}\ \emph {et~al.}(2016)\citenamefont
  {{Krause}}, \citenamefont {{Eifler}},\ and\ \citenamefont
  {{Blazek}}}]{keb16}%
  \BibitemOpen
  \bibfield  {author} {\bibinfo {author} {\bibfnamefont {E.}~\bibnamefont
  {{Krause}}}, \bibinfo {author} {\bibfnamefont {T.}~\bibnamefont {{Eifler}}},
  \ and\ \bibinfo {author} {\bibfnamefont {J.}~\bibnamefont {{Blazek}}},\
  }\href {\doibase 10.1093/mnras/stv2615} {\bibfield  {journal} {\bibinfo
  {journal} {\mnras}\ }\textbf {\bibinfo {volume} {456}},\ \bibinfo {pages}
  {207} (\bibinfo {year} {2016})},\ \Eprint {http://arxiv.org/abs/1506.08730}
  {arXiv:1506.08730} \BibitemShut {NoStop}%
\bibitem [{\citenamefont {{Joachimi}}\ and\ \citenamefont
  {{Bridle}}(2010)}]{JoachimiBridle10}%
  \BibitemOpen
  \bibfield  {author} {\bibinfo {author} {\bibfnamefont {B.}~\bibnamefont
  {{Joachimi}}}\ and\ \bibinfo {author} {\bibfnamefont {S.~L.}\ \bibnamefont
  {{Bridle}}},\ }\href {\doibase 10.1051/0004-6361/200913657} {\bibfield
  {journal} {\bibinfo  {journal} {\aap}\ }\textbf {\bibinfo {volume} {523}},\
  \bibinfo {eid} {A1} (\bibinfo {year} {2010})},\ \Eprint
  {http://arxiv.org/abs/0911.2454} {arXiv:0911.2454} \BibitemShut {NoStop}%
\bibitem [{\citenamefont {{Troxel}}\ and\ \citenamefont
  {{Ishak}}(2014{\natexlab{a}})}]{TroxelIshak14}%
  \BibitemOpen
  \bibfield  {author} {\bibinfo {author} {\bibfnamefont {M.~A.}\ \bibnamefont
  {{Troxel}}}\ and\ \bibinfo {author} {\bibfnamefont {M.}~\bibnamefont
  {{Ishak}}},\ }\href {\doibase 10.1103/PhysRevD.89.063528} {\bibfield
  {journal} {\bibinfo  {journal} {\prd}\ }\textbf {\bibinfo {volume} {89}},\
  \bibinfo {eid} {063528} (\bibinfo {year} {2014}{\natexlab{a}})},\ \Eprint
  {http://arxiv.org/abs/1401.7051} {arXiv:1401.7051} \BibitemShut {NoStop}%
\bibitem [{\citenamefont {{Hall}}\ and\ \citenamefont
  {{Taylor}}(2014{\natexlab{a}})}]{HallTaylor14}%
  \BibitemOpen
  \bibfield  {author} {\bibinfo {author} {\bibfnamefont {A.}~\bibnamefont
  {{Hall}}}\ and\ \bibinfo {author} {\bibfnamefont {A.}~\bibnamefont
  {{Taylor}}},\ }\href {\doibase 10.1093/mnrasl/slu094} {\bibfield  {journal}
  {\bibinfo  {journal} {\mnras}\ }\textbf {\bibinfo {volume} {443}},\ \bibinfo
  {pages} {L119} (\bibinfo {year} {2014}{\natexlab{a}})},\ \Eprint
  {http://arxiv.org/abs/1401.6018} {arXiv:1401.6018} \BibitemShut {NoStop}%
\bibitem [{\citenamefont {{Li}}\ \emph
  {et~al.}(2014{\natexlab{a}})\citenamefont {{Li}}, \citenamefont {{Hu}},\ and\
  \citenamefont {{Takada}}}]{2014PhRvD..89h3519L}%
  \BibitemOpen
  \bibfield  {author} {\bibinfo {author} {\bibfnamefont {Y.}~\bibnamefont
  {{Li}}}, \bibinfo {author} {\bibfnamefont {W.}~\bibnamefont {{Hu}}}, \ and\
  \bibinfo {author} {\bibfnamefont {M.}~\bibnamefont {{Takada}}},\ }\href
  {\doibase 10.1103/PhysRevD.89.083519} {\bibfield  {journal} {\bibinfo
  {journal} {\prd}\ }\textbf {\bibinfo {volume} {89}},\ \bibinfo {eid} {083519}
  (\bibinfo {year} {2014}{\natexlab{a}})},\ \Eprint
  {http://arxiv.org/abs/1401.0385} {arXiv:1401.0385} \BibitemShut {NoStop}%
\bibitem [{\citenamefont {{Chiang}}\ \emph {et~al.}(2014)\citenamefont
  {{Chiang}}, \citenamefont {{Wagner}}, \citenamefont {{Schmidt}},\ and\
  \citenamefont {{Komatsu}}}]{2014JCAP...05..048C}%
  \BibitemOpen
  \bibfield  {author} {\bibinfo {author} {\bibfnamefont {C.-T.}\ \bibnamefont
  {{Chiang}}}, \bibinfo {author} {\bibfnamefont {C.}~\bibnamefont {{Wagner}}},
  \bibinfo {author} {\bibfnamefont {F.}~\bibnamefont {{Schmidt}}}, \ and\
  \bibinfo {author} {\bibfnamefont {E.}~\bibnamefont {{Komatsu}}},\ }\href
  {\doibase 10.1088/1475-7516/2014/05/048} {\bibfield  {journal} {\bibinfo
  {journal} {\jcap}\ }\textbf {\bibinfo {volume} {5}},\ \bibinfo {eid} {048}
  (\bibinfo {year} {2014})},\ \Eprint {http://arxiv.org/abs/1403.3411}
  {arXiv:1403.3411} \BibitemShut {NoStop}%
\bibitem [{\citenamefont {{Li}}\ \emph
  {et~al.}(2014{\natexlab{b}})\citenamefont {{Li}}, \citenamefont {{Hu}},\ and\
  \citenamefont {{Takada}}}]{2014PhRvD..90j3530L}%
  \BibitemOpen
  \bibfield  {author} {\bibinfo {author} {\bibfnamefont {Y.}~\bibnamefont
  {{Li}}}, \bibinfo {author} {\bibfnamefont {W.}~\bibnamefont {{Hu}}}, \ and\
  \bibinfo {author} {\bibfnamefont {M.}~\bibnamefont {{Takada}}},\ }\href
  {\doibase 10.1103/PhysRevD.90.103530} {\bibfield  {journal} {\bibinfo
  {journal} {\prd}\ }\textbf {\bibinfo {volume} {90}},\ \bibinfo {eid} {103530}
  (\bibinfo {year} {2014}{\natexlab{b}})},\ \Eprint
  {http://arxiv.org/abs/1408.1081} {arXiv:1408.1081} \BibitemShut {NoStop}%
\bibitem [{\citenamefont {{Takada}}\ and\ \citenamefont
  {{Spergel}}(2014)}]{2014MNRAS.441.2456T}%
  \BibitemOpen
  \bibfield  {author} {\bibinfo {author} {\bibfnamefont {M.}~\bibnamefont
  {{Takada}}}\ and\ \bibinfo {author} {\bibfnamefont {D.~N.}\ \bibnamefont
  {{Spergel}}},\ }\href {\doibase 10.1093/mnras/stu759} {\bibfield  {journal}
  {\bibinfo  {journal} {\mnras}\ }\textbf {\bibinfo {volume} {441}},\ \bibinfo
  {pages} {2456} (\bibinfo {year} {2014})},\ \Eprint
  {http://arxiv.org/abs/1307.4399} {arXiv:1307.4399} \BibitemShut {NoStop}%
\bibitem [{\citenamefont {{Schaan}}\ \emph {et~al.}(2014)\citenamefont
  {{Schaan}}, \citenamefont {{Takada}},\ and\ \citenamefont
  {{Spergel}}}]{2014PhRvD..90l3523S}%
  \BibitemOpen
  \bibfield  {author} {\bibinfo {author} {\bibfnamefont {E.}~\bibnamefont
  {{Schaan}}}, \bibinfo {author} {\bibfnamefont {M.}~\bibnamefont {{Takada}}},
  \ and\ \bibinfo {author} {\bibfnamefont {D.~N.}\ \bibnamefont {{Spergel}}},\
  }\href {\doibase 10.1103/PhysRevD.90.123523} {\bibfield  {journal} {\bibinfo
  {journal} {\prd}\ }\textbf {\bibinfo {volume} {90}},\ \bibinfo {eid} {123523}
  (\bibinfo {year} {2014})},\ \Eprint {http://arxiv.org/abs/1406.3330}
  {arXiv:1406.3330} \BibitemShut {NoStop}%
\bibitem [{\citenamefont {{van Engelen}}\ \emph {et~al.}(2014)\citenamefont
  {{van Engelen}}, \citenamefont {{Bhattacharya}}, \citenamefont {{Sehgal}},
  \citenamefont {{Holder}}, \citenamefont {{Zahn}},\ and\ \citenamefont
  {{Nagai}}}]{2014ApJ...786...13V}%
  \BibitemOpen
  \bibfield  {author} {\bibinfo {author} {\bibfnamefont {A.}~\bibnamefont {{van
  Engelen}}}, \bibinfo {author} {\bibfnamefont {S.}~\bibnamefont
  {{Bhattacharya}}}, \bibinfo {author} {\bibfnamefont {N.}~\bibnamefont
  {{Sehgal}}}, \bibinfo {author} {\bibfnamefont {G.~P.}\ \bibnamefont
  {{Holder}}}, \bibinfo {author} {\bibfnamefont {O.}~\bibnamefont {{Zahn}}}, \
  and\ \bibinfo {author} {\bibfnamefont {D.}~\bibnamefont {{Nagai}}},\ }\href
  {\doibase 10.1088/0004-637X/786/1/13} {\bibfield  {journal} {\bibinfo
  {journal} {\apj}\ }\textbf {\bibinfo {volume} {786}},\ \bibinfo {eid} {13}
  (\bibinfo {year} {2014})},\ \Eprint {http://arxiv.org/abs/1310.7023}
  {arXiv:1310.7023} \BibitemShut {NoStop}%
\bibitem [{\citenamefont {{Osborne}}\ \emph {et~al.}(2014)\citenamefont
  {{Osborne}}, \citenamefont {{Hanson}},\ and\ \citenamefont
  {{Dor{\'e}}}}]{2014JCAP...03..024O}%
  \BibitemOpen
  \bibfield  {author} {\bibinfo {author} {\bibfnamefont {S.~J.}\ \bibnamefont
  {{Osborne}}}, \bibinfo {author} {\bibfnamefont {D.}~\bibnamefont {{Hanson}}},
  \ and\ \bibinfo {author} {\bibfnamefont {O.}~\bibnamefont {{Dor{\'e}}}},\
  }\href {\doibase 10.1088/1475-7516/2014/03/024} {\bibfield  {journal}
  {\bibinfo  {journal} {\jcap}\ }\textbf {\bibinfo {volume} {3}},\ \bibinfo
  {eid} {024} (\bibinfo {year} {2014})},\ \Eprint
  {http://arxiv.org/abs/1310.7547} {arXiv:1310.7547} \BibitemShut {NoStop}%
\bibitem [{\citenamefont {{Eifler}}\ \emph {et~al.}(2009)\citenamefont
  {{Eifler}}, \citenamefont {{Schneider}},\ and\ \citenamefont
  {{Hartlap}}}]{2009A&A...502..721E}%
  \BibitemOpen
  \bibfield  {author} {\bibinfo {author} {\bibfnamefont {T.}~\bibnamefont
  {{Eifler}}}, \bibinfo {author} {\bibfnamefont {P.}~\bibnamefont
  {{Schneider}}}, \ and\ \bibinfo {author} {\bibfnamefont {J.}~\bibnamefont
  {{Hartlap}}},\ }\href {\doibase 10.1051/0004-6361/200811276} {\bibfield
  {journal} {\bibinfo  {journal} {\aap}\ }\textbf {\bibinfo {volume} {502}},\
  \bibinfo {pages} {721} (\bibinfo {year} {2009})},\ \Eprint
  {http://arxiv.org/abs/0810.4254} {arXiv:0810.4254} \BibitemShut {NoStop}%
\bibitem [{\citenamefont {{Sellentin}}\ and\ \citenamefont
  {{Heavens}}(2016)}]{2016MNRAS.456L.132S}%
  \BibitemOpen
  \bibfield  {author} {\bibinfo {author} {\bibfnamefont {E.}~\bibnamefont
  {{Sellentin}}}\ and\ \bibinfo {author} {\bibfnamefont {A.~F.}\ \bibnamefont
  {{Heavens}}},\ }\href {\doibase 10.1093/mnrasl/slv190} {\bibfield  {journal}
  {\bibinfo  {journal} {\mnras}\ }\textbf {\bibinfo {volume} {456}},\ \bibinfo
  {pages} {L132} (\bibinfo {year} {2016})},\ \Eprint
  {http://arxiv.org/abs/1511.05969} {arXiv:1511.05969} \BibitemShut {NoStop}%
\bibitem [{\citenamefont {{White}}\ and\ \citenamefont
  {{Padmanabhan}}(2015)}]{2015JCAP...12..058W}%
  \BibitemOpen
  \bibfield  {author} {\bibinfo {author} {\bibfnamefont {M.}~\bibnamefont
  {{White}}}\ and\ \bibinfo {author} {\bibfnamefont {N.}~\bibnamefont
  {{Padmanabhan}}},\ }\href {\doibase 10.1088/1475-7516/2015/12/058} {\bibfield
   {journal} {\bibinfo  {journal} {\jcap}\ }\textbf {\bibinfo {volume} {12}},\
  \bibinfo {eid} {058} (\bibinfo {year} {2015})},\ \Eprint
  {http://arxiv.org/abs/1508.00566} {arXiv:1508.00566} \BibitemShut {NoStop}%
\bibitem [{\citenamefont {{Carron}}(2013)}]{2013A&A...551A..88C}%
  \BibitemOpen
  \bibfield  {author} {\bibinfo {author} {\bibfnamefont {J.}~\bibnamefont
  {{Carron}}},\ }\href {\doibase 10.1051/0004-6361/201220538} {\bibfield
  {journal} {\bibinfo  {journal} {\aap}\ }\textbf {\bibinfo {volume} {551}},\
  \bibinfo {eid} {A88} (\bibinfo {year} {2013})},\ \Eprint
  {http://arxiv.org/abs/1204.4724} {arXiv:1204.4724 [astro-ph.CO]} \BibitemShut
  {NoStop}%
\bibitem [{\citenamefont {{Foreman-Mackey}}\ \emph {et~al.}(2013)\citenamefont
  {{Foreman-Mackey}}, \citenamefont {{Hogg}}, \citenamefont {{Lang}},\ and\
  \citenamefont {{Goodman}}}]{2013PASP..125..306F}%
  \BibitemOpen
  \bibfield  {author} {\bibinfo {author} {\bibfnamefont {D.}~\bibnamefont
  {{Foreman-Mackey}}}, \bibinfo {author} {\bibfnamefont {D.~W.}\ \bibnamefont
  {{Hogg}}}, \bibinfo {author} {\bibfnamefont {D.}~\bibnamefont {{Lang}}}, \
  and\ \bibinfo {author} {\bibfnamefont {J.}~\bibnamefont {{Goodman}}},\ }\href
  {\doibase 10.1086/670067} {\bibfield  {journal} {\bibinfo  {journal} {\pasp}\
  }\textbf {\bibinfo {volume} {125}},\ \bibinfo {pages} {306} (\bibinfo {year}
  {2013})},\ \Eprint {http://arxiv.org/abs/1202.3665} {arXiv:1202.3665
  [astro-ph.IM]} \BibitemShut {NoStop}%
\bibitem [{\citenamefont {{Planck Collaboration}}\ \emph
  {et~al.}(2015{\natexlab{b}})\citenamefont {{Planck Collaboration}},
  \citenamefont {{Ade}}, \citenamefont {{Aghanim}}, \citenamefont {{Arnaud}},
  \citenamefont {{Ashdown}}, \citenamefont {{Aumont}}, \citenamefont
  {{Baccigalupi}}, \citenamefont {{Banday}}, \citenamefont {{Barreiro}},
  \citenamefont {{Bartlett}},\ and\ \citenamefont
  {et~al.}}]{2015arXiv150201589P}%
  \BibitemOpen
  \bibfield  {author} {\bibinfo {author} {\bibnamefont {{Planck
  Collaboration}}}, \bibinfo {author} {\bibfnamefont {P.~A.~R.}\ \bibnamefont
  {{Ade}}}, \bibinfo {author} {\bibfnamefont {N.}~\bibnamefont {{Aghanim}}},
  \bibinfo {author} {\bibfnamefont {M.}~\bibnamefont {{Arnaud}}}, \bibinfo
  {author} {\bibfnamefont {M.}~\bibnamefont {{Ashdown}}}, \bibinfo {author}
  {\bibfnamefont {J.}~\bibnamefont {{Aumont}}}, \bibinfo {author}
  {\bibfnamefont {C.}~\bibnamefont {{Baccigalupi}}}, \bibinfo {author}
  {\bibfnamefont {A.~J.}\ \bibnamefont {{Banday}}}, \bibinfo {author}
  {\bibfnamefont {R.~B.}\ \bibnamefont {{Barreiro}}}, \bibinfo {author}
  {\bibfnamefont {J.~G.}\ \bibnamefont {{Bartlett}}}, \ and\ \bibinfo {author}
  {\bibnamefont {et~al.}},\ }\href@noop {} {\bibfield  {journal} {\bibinfo
  {journal} {ArXiv e-prints}\ } (\bibinfo {year} {2015}{\natexlab{b}})},\
  \Eprint {http://arxiv.org/abs/1502.01589} {arXiv:1502.01589} \BibitemShut
  {NoStop}%
\bibitem [{\citenamefont {{Huterer}}\ \emph
  {et~al.}(2006{\natexlab{b}})\citenamefont {{Huterer}}, \citenamefont
  {{Takada}}, \citenamefont {{Bernstein}},\ and\ \citenamefont
  {{Jain}}}]{Hutereretal:06}%
  \BibitemOpen
  \bibfield  {author} {\bibinfo {author} {\bibfnamefont {D.}~\bibnamefont
  {{Huterer}}}, \bibinfo {author} {\bibfnamefont {M.}~\bibnamefont {{Takada}}},
  \bibinfo {author} {\bibfnamefont {G.}~\bibnamefont {{Bernstein}}}, \ and\
  \bibinfo {author} {\bibfnamefont {B.}~\bibnamefont {{Jain}}},\ }\href
  {\doibase 10.1111/j.1365-2966.2005.09782.x} {\bibfield  {journal} {\bibinfo
  {journal} {\mnras}\ }\textbf {\bibinfo {volume} {366}},\ \bibinfo {pages}
  {101} (\bibinfo {year} {2006}{\natexlab{b}})},\ \Eprint
  {http://arxiv.org/abs/arXiv:astro-ph/0506030} {arXiv:astro-ph/0506030}
  \BibitemShut {NoStop}%
\bibitem [{\citenamefont {{Hall}}\ and\ \citenamefont
  {{Taylor}}(2014{\natexlab{b}})}]{2014MNRAS.443L.119H}%
  \BibitemOpen
  \bibfield  {author} {\bibinfo {author} {\bibfnamefont {A.}~\bibnamefont
  {{Hall}}}\ and\ \bibinfo {author} {\bibfnamefont {A.}~\bibnamefont
  {{Taylor}}},\ }\href {\doibase 10.1093/mnrasl/slu094} {\bibfield  {journal}
  {\bibinfo  {journal} {\mnras}\ }\textbf {\bibinfo {volume} {443}},\ \bibinfo
  {pages} {L119} (\bibinfo {year} {2014}{\natexlab{b}})},\ \Eprint
  {http://arxiv.org/abs/1401.6018} {arXiv:1401.6018} \BibitemShut {NoStop}%
\bibitem [{\citenamefont {{Chisari}}\ \emph {et~al.}(2015)\citenamefont
  {{Chisari}}, \citenamefont {{Dunkley}}, \citenamefont {{Miller}},\ and\
  \citenamefont {{Allison}}}]{2015MNRAS.453..682C}%
  \BibitemOpen
  \bibfield  {author} {\bibinfo {author} {\bibfnamefont {N.~E.}\ \bibnamefont
  {{Chisari}}}, \bibinfo {author} {\bibfnamefont {J.}~\bibnamefont
  {{Dunkley}}}, \bibinfo {author} {\bibfnamefont {L.}~\bibnamefont {{Miller}}},
  \ and\ \bibinfo {author} {\bibfnamefont {R.}~\bibnamefont {{Allison}}},\
  }\href {\doibase 10.1093/mnras/stv1655} {\bibfield  {journal} {\bibinfo
  {journal} {\mnras}\ }\textbf {\bibinfo {volume} {453}},\ \bibinfo {pages}
  {682} (\bibinfo {year} {2015})},\ \Eprint {http://arxiv.org/abs/1507.03906}
  {arXiv:1507.03906} \BibitemShut {NoStop}%
\bibitem [{\citenamefont {{Troxel}}\ and\ \citenamefont
  {{Ishak}}(2014{\natexlab{b}})}]{2014PhRvD..89f3528T}%
  \BibitemOpen
  \bibfield  {author} {\bibinfo {author} {\bibfnamefont {M.~A.}\ \bibnamefont
  {{Troxel}}}\ and\ \bibinfo {author} {\bibfnamefont {M.}~\bibnamefont
  {{Ishak}}},\ }\href {\doibase 10.1103/PhysRevD.89.063528} {\bibfield
  {journal} {\bibinfo  {journal} {\prd}\ }\textbf {\bibinfo {volume} {89}},\
  \bibinfo {eid} {063528} (\bibinfo {year} {2014}{\natexlab{b}})},\ \Eprint
  {http://arxiv.org/abs/1401.7051} {arXiv:1401.7051} \BibitemShut {NoStop}%
\bibitem [{\citenamefont {{Henderson}}\ \emph {et~al.}(2016)\citenamefont
  {{Henderson}}, \citenamefont {{Allison}}, \citenamefont {{Austermann}},
  \citenamefont {{Baildon}}, \citenamefont {{Battaglia}}, \citenamefont
  {{Beall}}, \citenamefont {{Becker}}, \citenamefont {{De Bernardis}},
  \citenamefont {{Bond}}, \citenamefont {{Calabrese}}, \citenamefont {{Choi}},
  \citenamefont {{Coughlin}}, \citenamefont {{Crowley}}, \citenamefont
  {{Datta}}, \citenamefont {{Devlin}}, \citenamefont {{Duff}}, \citenamefont
  {{Dunkley}}, \citenamefont {{D{\"u}nner}}, \citenamefont {{van Engelen}},
  \citenamefont {{Gallardo}}, \citenamefont {{Grace}}, \citenamefont
  {{Hasselfield}}, \citenamefont {{Hills}}, \citenamefont {{Hilton}},
  \citenamefont {{Hincks}}, \citenamefont {{Hlo{\^z}ek}}, \citenamefont {{Ho}},
  \citenamefont {{Hubmayr}}, \citenamefont {{Huffenberger}}, \citenamefont
  {{Hughes}}, \citenamefont {{Irwin}}, \citenamefont {{Koopman}}, \citenamefont
  {{Kosowsky}}, \citenamefont {{Li}}, \citenamefont {{McMahon}}, \citenamefont
  {{Munson}}, \citenamefont {{Nati}}, \citenamefont {{Newburgh}}, \citenamefont
  {{Niemack}}, \citenamefont {{Niraula}}, \citenamefont {{Page}}, \citenamefont
  {{Pappas}}, \citenamefont {{Salatino}}, \citenamefont {{Schillaci}},
  \citenamefont {{Schmitt}}, \citenamefont {{Sehgal}}, \citenamefont
  {{Sherwin}}, \citenamefont {{Sievers}}, \citenamefont {{Simon}},
  \citenamefont {{Spergel}}, \citenamefont {{Staggs}}, \citenamefont
  {{Stevens}}, \citenamefont {{Thornton}}, \citenamefont {{Van Lanen}},
  \citenamefont {{Vavagiakis}}, \citenamefont {{Ward}},\ and\ \citenamefont
  {{Wollack}}}]{2016JLTP..tmp..144H}%
  \BibitemOpen
  \bibfield  {author} {\bibinfo {author} {\bibfnamefont {S.~W.}\ \bibnamefont
  {{Henderson}}}, \bibinfo {author} {\bibfnamefont {R.}~\bibnamefont
  {{Allison}}}, \bibinfo {author} {\bibfnamefont {J.}~\bibnamefont
  {{Austermann}}}, \bibinfo {author} {\bibfnamefont {T.}~\bibnamefont
  {{Baildon}}}, \bibinfo {author} {\bibfnamefont {N.}~\bibnamefont
  {{Battaglia}}}, \bibinfo {author} {\bibfnamefont {J.~A.}\ \bibnamefont
  {{Beall}}}, \bibinfo {author} {\bibfnamefont {D.}~\bibnamefont {{Becker}}},
  \bibinfo {author} {\bibfnamefont {F.}~\bibnamefont {{De Bernardis}}},
  \bibinfo {author} {\bibfnamefont {J.~R.}\ \bibnamefont {{Bond}}}, \bibinfo
  {author} {\bibfnamefont {E.}~\bibnamefont {{Calabrese}}}, \bibinfo {author}
  {\bibfnamefont {S.~K.}\ \bibnamefont {{Choi}}}, \bibinfo {author}
  {\bibfnamefont {K.~P.}\ \bibnamefont {{Coughlin}}}, \bibinfo {author}
  {\bibfnamefont {K.~T.}\ \bibnamefont {{Crowley}}}, \bibinfo {author}
  {\bibfnamefont {R.}~\bibnamefont {{Datta}}}, \bibinfo {author} {\bibfnamefont
  {M.~J.}\ \bibnamefont {{Devlin}}}, \bibinfo {author} {\bibfnamefont {S.~M.}\
  \bibnamefont {{Duff}}}, \bibinfo {author} {\bibfnamefont {J.}~\bibnamefont
  {{Dunkley}}}, \bibinfo {author} {\bibfnamefont {R.}~\bibnamefont
  {{D{\"u}nner}}}, \bibinfo {author} {\bibfnamefont {A.}~\bibnamefont {{van
  Engelen}}}, \bibinfo {author} {\bibfnamefont {P.~A.}\ \bibnamefont
  {{Gallardo}}}, \bibinfo {author} {\bibfnamefont {E.}~\bibnamefont {{Grace}}},
  \bibinfo {author} {\bibfnamefont {M.}~\bibnamefont {{Hasselfield}}}, \bibinfo
  {author} {\bibfnamefont {F.}~\bibnamefont {{Hills}}}, \bibinfo {author}
  {\bibfnamefont {G.~C.}\ \bibnamefont {{Hilton}}}, \bibinfo {author}
  {\bibfnamefont {A.~D.}\ \bibnamefont {{Hincks}}}, \bibinfo {author}
  {\bibfnamefont {R.}~\bibnamefont {{Hlo{\^z}ek}}}, \bibinfo {author}
  {\bibfnamefont {S.~P.}\ \bibnamefont {{Ho}}}, \bibinfo {author}
  {\bibfnamefont {J.}~\bibnamefont {{Hubmayr}}}, \bibinfo {author}
  {\bibfnamefont {K.}~\bibnamefont {{Huffenberger}}}, \bibinfo {author}
  {\bibfnamefont {J.~P.}\ \bibnamefont {{Hughes}}}, \bibinfo {author}
  {\bibfnamefont {K.~D.}\ \bibnamefont {{Irwin}}}, \bibinfo {author}
  {\bibfnamefont {B.~J.}\ \bibnamefont {{Koopman}}}, \bibinfo {author}
  {\bibfnamefont {A.~B.}\ \bibnamefont {{Kosowsky}}}, \bibinfo {author}
  {\bibfnamefont {D.}~\bibnamefont {{Li}}}, \bibinfo {author} {\bibfnamefont
  {J.}~\bibnamefont {{McMahon}}}, \bibinfo {author} {\bibfnamefont
  {C.}~\bibnamefont {{Munson}}}, \bibinfo {author} {\bibfnamefont
  {F.}~\bibnamefont {{Nati}}}, \bibinfo {author} {\bibfnamefont
  {L.}~\bibnamefont {{Newburgh}}}, \bibinfo {author} {\bibfnamefont {M.~D.}\
  \bibnamefont {{Niemack}}}, \bibinfo {author} {\bibfnamefont {P.}~\bibnamefont
  {{Niraula}}}, \bibinfo {author} {\bibfnamefont {L.~A.}\ \bibnamefont
  {{Page}}}, \bibinfo {author} {\bibfnamefont {C.~G.}\ \bibnamefont
  {{Pappas}}}, \bibinfo {author} {\bibfnamefont {M.}~\bibnamefont
  {{Salatino}}}, \bibinfo {author} {\bibfnamefont {A.}~\bibnamefont
  {{Schillaci}}}, \bibinfo {author} {\bibfnamefont {B.~L.}\ \bibnamefont
  {{Schmitt}}}, \bibinfo {author} {\bibfnamefont {N.}~\bibnamefont {{Sehgal}}},
  \bibinfo {author} {\bibfnamefont {B.~D.}\ \bibnamefont {{Sherwin}}}, \bibinfo
  {author} {\bibfnamefont {J.~L.}\ \bibnamefont {{Sievers}}}, \bibinfo {author}
  {\bibfnamefont {S.~M.}\ \bibnamefont {{Simon}}}, \bibinfo {author}
  {\bibfnamefont {D.~N.}\ \bibnamefont {{Spergel}}}, \bibinfo {author}
  {\bibfnamefont {S.~T.}\ \bibnamefont {{Staggs}}}, \bibinfo {author}
  {\bibfnamefont {J.~R.}\ \bibnamefont {{Stevens}}}, \bibinfo {author}
  {\bibfnamefont {R.}~\bibnamefont {{Thornton}}}, \bibinfo {author}
  {\bibfnamefont {J.}~\bibnamefont {{Van Lanen}}}, \bibinfo {author}
  {\bibfnamefont {E.~M.}\ \bibnamefont {{Vavagiakis}}}, \bibinfo {author}
  {\bibfnamefont {J.~T.}\ \bibnamefont {{Ward}}}, \ and\ \bibinfo {author}
  {\bibfnamefont {E.~J.}\ \bibnamefont {{Wollack}}},\ }\href {\doibase
  10.1007/s10909-016-1575-z} {\bibfield  {journal} {\bibinfo  {journal}
  {Journal of Low Temperature Physics}\ } (\bibinfo {year} {2016}),\
  10.1007/s10909-016-1575-z},\ \Eprint {http://arxiv.org/abs/1510.02809}
  {arXiv:1510.02809 [astro-ph.IM]} \BibitemShut {NoStop}%
\bibitem [{\citenamefont {{Benson}}\ \emph {et~al.}(2014)\citenamefont
  {{Benson}}, \citenamefont {{Ade}}, \citenamefont {{Ahmed}}, \citenamefont
  {{Allen}}, \citenamefont {{Arnold}}, \citenamefont {{Austermann}},
  \citenamefont {{Bender}}, \citenamefont {{Bleem}}, \citenamefont
  {{Carlstrom}}, \citenamefont {{Chang}}, \citenamefont {{Cho}}, \citenamefont
  {{Cliche}}, \citenamefont {{Crawford}}, \citenamefont {{Cukierman}},
  \citenamefont {{de Haan}}, \citenamefont {{Dobbs}}, \citenamefont
  {{Dutcher}}, \citenamefont {{Everett}}, \citenamefont {{Gilbert}},
  \citenamefont {{Halverson}}, \citenamefont {{Hanson}}, \citenamefont
  {{Harrington}}, \citenamefont {{Hattori}}, \citenamefont {{Henning}},
  \citenamefont {{Hilton}}, \citenamefont {{Holder}}, \citenamefont
  {{Holzapfel}}, \citenamefont {{Irwin}}, \citenamefont {{Keisler}},
  \citenamefont {{Knox}}, \citenamefont {{Kubik}}, \citenamefont {{Kuo}},
  \citenamefont {{Lee}}, \citenamefont {{Leitch}}, \citenamefont {{Li}},
  \citenamefont {{McDonald}}, \citenamefont {{Meyer}}, \citenamefont
  {{Montgomery}}, \citenamefont {{Myers}}, \citenamefont {{Natoli}},
  \citenamefont {{Nguyen}}, \citenamefont {{Novosad}}, \citenamefont {{Padin}},
  \citenamefont {{Pan}}, \citenamefont {{Pearson}}, \citenamefont
  {{Reichardt}}, \citenamefont {{Ruhl}}, \citenamefont {{Saliwanchik}},
  \citenamefont {{Simard}}, \citenamefont {{Smecher}}, \citenamefont {{Sayre}},
  \citenamefont {{Shirokoff}}, \citenamefont {{Stark}}, \citenamefont
  {{Story}}, \citenamefont {{Suzuki}}, \citenamefont {{Thompson}},
  \citenamefont {{Tucker}}, \citenamefont {{Vanderlinde}}, \citenamefont
  {{Vieira}}, \citenamefont {{Vikhlinin}}, \citenamefont {{Wang}},
  \citenamefont {{Yefremenko}},\ and\ \citenamefont
  {{Yoon}}}]{2014SPIE.9153E..1PB}%
  \BibitemOpen
  \bibfield  {author} {\bibinfo {author} {\bibfnamefont {B.~A.}\ \bibnamefont
  {{Benson}}}, \bibinfo {author} {\bibfnamefont {P.~A.~R.}\ \bibnamefont
  {{Ade}}}, \bibinfo {author} {\bibfnamefont {Z.}~\bibnamefont {{Ahmed}}},
  \bibinfo {author} {\bibfnamefont {S.~W.}\ \bibnamefont {{Allen}}}, \bibinfo
  {author} {\bibfnamefont {K.}~\bibnamefont {{Arnold}}}, \bibinfo {author}
  {\bibfnamefont {J.~E.}\ \bibnamefont {{Austermann}}}, \bibinfo {author}
  {\bibfnamefont {A.~N.}\ \bibnamefont {{Bender}}}, \bibinfo {author}
  {\bibfnamefont {L.~E.}\ \bibnamefont {{Bleem}}}, \bibinfo {author}
  {\bibfnamefont {J.~E.}\ \bibnamefont {{Carlstrom}}}, \bibinfo {author}
  {\bibfnamefont {C.~L.}\ \bibnamefont {{Chang}}}, \bibinfo {author}
  {\bibfnamefont {H.~M.}\ \bibnamefont {{Cho}}}, \bibinfo {author}
  {\bibfnamefont {J.~F.}\ \bibnamefont {{Cliche}}}, \bibinfo {author}
  {\bibfnamefont {T.~M.}\ \bibnamefont {{Crawford}}}, \bibinfo {author}
  {\bibfnamefont {A.}~\bibnamefont {{Cukierman}}}, \bibinfo {author}
  {\bibfnamefont {T.}~\bibnamefont {{de Haan}}}, \bibinfo {author}
  {\bibfnamefont {M.~A.}\ \bibnamefont {{Dobbs}}}, \bibinfo {author}
  {\bibfnamefont {D.}~\bibnamefont {{Dutcher}}}, \bibinfo {author}
  {\bibfnamefont {W.}~\bibnamefont {{Everett}}}, \bibinfo {author}
  {\bibfnamefont {A.}~\bibnamefont {{Gilbert}}}, \bibinfo {author}
  {\bibfnamefont {N.~W.}\ \bibnamefont {{Halverson}}}, \bibinfo {author}
  {\bibfnamefont {D.}~\bibnamefont {{Hanson}}}, \bibinfo {author}
  {\bibfnamefont {N.~L.}\ \bibnamefont {{Harrington}}}, \bibinfo {author}
  {\bibfnamefont {K.}~\bibnamefont {{Hattori}}}, \bibinfo {author}
  {\bibfnamefont {J.~W.}\ \bibnamefont {{Henning}}}, \bibinfo {author}
  {\bibfnamefont {G.~C.}\ \bibnamefont {{Hilton}}}, \bibinfo {author}
  {\bibfnamefont {G.~P.}\ \bibnamefont {{Holder}}}, \bibinfo {author}
  {\bibfnamefont {W.~L.}\ \bibnamefont {{Holzapfel}}}, \bibinfo {author}
  {\bibfnamefont {K.~D.}\ \bibnamefont {{Irwin}}}, \bibinfo {author}
  {\bibfnamefont {R.}~\bibnamefont {{Keisler}}}, \bibinfo {author}
  {\bibfnamefont {L.}~\bibnamefont {{Knox}}}, \bibinfo {author} {\bibfnamefont
  {D.}~\bibnamefont {{Kubik}}}, \bibinfo {author} {\bibfnamefont {C.~L.}\
  \bibnamefont {{Kuo}}}, \bibinfo {author} {\bibfnamefont {A.~T.}\ \bibnamefont
  {{Lee}}}, \bibinfo {author} {\bibfnamefont {E.~M.}\ \bibnamefont {{Leitch}}},
  \bibinfo {author} {\bibfnamefont {D.}~\bibnamefont {{Li}}}, \bibinfo {author}
  {\bibfnamefont {M.}~\bibnamefont {{McDonald}}}, \bibinfo {author}
  {\bibfnamefont {S.~S.}\ \bibnamefont {{Meyer}}}, \bibinfo {author}
  {\bibfnamefont {J.}~\bibnamefont {{Montgomery}}}, \bibinfo {author}
  {\bibfnamefont {M.}~\bibnamefont {{Myers}}}, \bibinfo {author} {\bibfnamefont
  {T.}~\bibnamefont {{Natoli}}}, \bibinfo {author} {\bibfnamefont
  {H.}~\bibnamefont {{Nguyen}}}, \bibinfo {author} {\bibfnamefont
  {V.}~\bibnamefont {{Novosad}}}, \bibinfo {author} {\bibfnamefont
  {S.}~\bibnamefont {{Padin}}}, \bibinfo {author} {\bibfnamefont
  {Z.}~\bibnamefont {{Pan}}}, \bibinfo {author} {\bibfnamefont
  {J.}~\bibnamefont {{Pearson}}}, \bibinfo {author} {\bibfnamefont
  {C.}~\bibnamefont {{Reichardt}}}, \bibinfo {author} {\bibfnamefont {J.~E.}\
  \bibnamefont {{Ruhl}}}, \bibinfo {author} {\bibfnamefont {B.~R.}\
  \bibnamefont {{Saliwanchik}}}, \bibinfo {author} {\bibfnamefont
  {G.}~\bibnamefont {{Simard}}}, \bibinfo {author} {\bibfnamefont
  {G.}~\bibnamefont {{Smecher}}}, \bibinfo {author} {\bibfnamefont {J.~T.}\
  \bibnamefont {{Sayre}}}, \bibinfo {author} {\bibfnamefont {E.}~\bibnamefont
  {{Shirokoff}}}, \bibinfo {author} {\bibfnamefont {A.~A.}\ \bibnamefont
  {{Stark}}}, \bibinfo {author} {\bibfnamefont {K.}~\bibnamefont {{Story}}},
  \bibinfo {author} {\bibfnamefont {A.}~\bibnamefont {{Suzuki}}}, \bibinfo
  {author} {\bibfnamefont {K.~L.}\ \bibnamefont {{Thompson}}}, \bibinfo
  {author} {\bibfnamefont {C.}~\bibnamefont {{Tucker}}}, \bibinfo {author}
  {\bibfnamefont {K.}~\bibnamefont {{Vanderlinde}}}, \bibinfo {author}
  {\bibfnamefont {J.~D.}\ \bibnamefont {{Vieira}}}, \bibinfo {author}
  {\bibfnamefont {A.}~\bibnamefont {{Vikhlinin}}}, \bibinfo {author}
  {\bibfnamefont {G.}~\bibnamefont {{Wang}}}, \bibinfo {author} {\bibfnamefont
  {V.}~\bibnamefont {{Yefremenko}}}, \ and\ \bibinfo {author} {\bibfnamefont
  {K.~W.}\ \bibnamefont {{Yoon}}},\ }in\ \href {\doibase 10.1117/12.2057305}
  {\emph {\bibinfo {booktitle} {Millimeter, Submillimeter, and Far-Infrared
  Detectors and Instrumentation for Astronomy VII}}},\ \bibinfo {series}
  {\procspie}, Vol.\ \bibinfo {volume} {9153}\ (\bibinfo {year} {2014})\ p.\
  \bibinfo {pages} {91531P},\ \Eprint {http://arxiv.org/abs/1407.2973}
  {arXiv:1407.2973 [astro-ph.IM]} \BibitemShut {NoStop}%
\bibitem [{\citenamefont {{Amendola}}\ \emph {et~al.}(2013)\citenamefont
  {{Amendola}}, \citenamefont {{Appleby}}, \citenamefont {{Bacon}},
  \citenamefont {{Baker}}, \citenamefont {{Baldi}}, \citenamefont {{Bartolo}},
  \citenamefont {{Blanchard}}, \citenamefont {{Bonvin}}, \citenamefont
  {{Borgani}}, \citenamefont {{Branchini}}, \citenamefont {{Burrage}},
  \citenamefont {{Camera}}, \citenamefont {{Carbone}}, \citenamefont
  {{Casarini}}, \citenamefont {{Cropper}}, \citenamefont {{de Rham}},
  \citenamefont {{Di Porto}}, \citenamefont {{Ealet}}, \citenamefont
  {{Ferreira}}, \citenamefont {{Finelli}}, \citenamefont
  {{Garc{\'{\i}}a-Bellido}}, \citenamefont {{Giannantonio}}, \citenamefont
  {{Guzzo}}, \citenamefont {{Heavens}}, \citenamefont {{Heisenberg}},
  \citenamefont {{Heymans}}, \citenamefont {{Hoekstra}}, \citenamefont
  {{Hollenstein}}, \citenamefont {{Holmes}}, \citenamefont {{Horst}},
  \citenamefont {{Jahnke}}, \citenamefont {{Kitching}}, \citenamefont
  {{Koivisto}}, \citenamefont {{Kunz}}, \citenamefont {{La Vacca}},
  \citenamefont {{March}}, \citenamefont {{Majerotto}}, \citenamefont
  {{Markovic}}, \citenamefont {{Marsh}}, \citenamefont {{Marulli}},
  \citenamefont {{Massey}}, \citenamefont {{Mellier}}, \citenamefont {{Mota}},
  \citenamefont {{Nunes}}, \citenamefont {{Percival}}, \citenamefont
  {{Pettorino}}, \citenamefont {{Porciani}}, \citenamefont {{Quercellini}},
  \citenamefont {{Read}}, \citenamefont {{Rinaldi}}, \citenamefont {{Sapone}},
  \citenamefont {{Scaramella}}, \citenamefont {{Skordis}}, \citenamefont
  {{Simpson}}, \citenamefont {{Taylor}}, \citenamefont {{Thomas}},
  \citenamefont {{Trotta}}, \citenamefont {{Verde}}, \citenamefont
  {{Vernizzi}}, \citenamefont {{Vollmer}}, \citenamefont {{Wang}},
  \citenamefont {{Weller}},\ and\ \citenamefont
  {{Zlosnik}}}]{2013LRR....16....6A}%
  \BibitemOpen
  \bibfield  {author} {\bibinfo {author} {\bibfnamefont {L.}~\bibnamefont
  {{Amendola}}}, \bibinfo {author} {\bibfnamefont {S.}~\bibnamefont
  {{Appleby}}}, \bibinfo {author} {\bibfnamefont {D.}~\bibnamefont {{Bacon}}},
  \bibinfo {author} {\bibfnamefont {T.}~\bibnamefont {{Baker}}}, \bibinfo
  {author} {\bibfnamefont {M.}~\bibnamefont {{Baldi}}}, \bibinfo {author}
  {\bibfnamefont {N.}~\bibnamefont {{Bartolo}}}, \bibinfo {author}
  {\bibfnamefont {A.}~\bibnamefont {{Blanchard}}}, \bibinfo {author}
  {\bibfnamefont {C.}~\bibnamefont {{Bonvin}}}, \bibinfo {author}
  {\bibfnamefont {S.}~\bibnamefont {{Borgani}}}, \bibinfo {author}
  {\bibfnamefont {E.}~\bibnamefont {{Branchini}}}, \bibinfo {author}
  {\bibfnamefont {C.}~\bibnamefont {{Burrage}}}, \bibinfo {author}
  {\bibfnamefont {S.}~\bibnamefont {{Camera}}}, \bibinfo {author}
  {\bibfnamefont {C.}~\bibnamefont {{Carbone}}}, \bibinfo {author}
  {\bibfnamefont {L.}~\bibnamefont {{Casarini}}}, \bibinfo {author}
  {\bibfnamefont {M.}~\bibnamefont {{Cropper}}}, \bibinfo {author}
  {\bibfnamefont {C.}~\bibnamefont {{de Rham}}}, \bibinfo {author}
  {\bibfnamefont {C.}~\bibnamefont {{Di Porto}}}, \bibinfo {author}
  {\bibfnamefont {A.}~\bibnamefont {{Ealet}}}, \bibinfo {author} {\bibfnamefont
  {P.~G.}\ \bibnamefont {{Ferreira}}}, \bibinfo {author} {\bibfnamefont
  {F.}~\bibnamefont {{Finelli}}}, \bibinfo {author} {\bibfnamefont
  {J.}~\bibnamefont {{Garc{\'{\i}}a-Bellido}}}, \bibinfo {author}
  {\bibfnamefont {T.}~\bibnamefont {{Giannantonio}}}, \bibinfo {author}
  {\bibfnamefont {L.}~\bibnamefont {{Guzzo}}}, \bibinfo {author} {\bibfnamefont
  {A.}~\bibnamefont {{Heavens}}}, \bibinfo {author} {\bibfnamefont
  {L.}~\bibnamefont {{Heisenberg}}}, \bibinfo {author} {\bibfnamefont
  {C.}~\bibnamefont {{Heymans}}}, \bibinfo {author} {\bibfnamefont
  {H.}~\bibnamefont {{Hoekstra}}}, \bibinfo {author} {\bibfnamefont
  {L.}~\bibnamefont {{Hollenstein}}}, \bibinfo {author} {\bibfnamefont
  {R.}~\bibnamefont {{Holmes}}}, \bibinfo {author} {\bibfnamefont
  {O.}~\bibnamefont {{Horst}}}, \bibinfo {author} {\bibfnamefont
  {K.}~\bibnamefont {{Jahnke}}}, \bibinfo {author} {\bibfnamefont {T.~D.}\
  \bibnamefont {{Kitching}}}, \bibinfo {author} {\bibfnamefont
  {T.}~\bibnamefont {{Koivisto}}}, \bibinfo {author} {\bibfnamefont
  {M.}~\bibnamefont {{Kunz}}}, \bibinfo {author} {\bibfnamefont
  {G.}~\bibnamefont {{La Vacca}}}, \bibinfo {author} {\bibfnamefont
  {M.}~\bibnamefont {{March}}}, \bibinfo {author} {\bibfnamefont
  {E.}~\bibnamefont {{Majerotto}}}, \bibinfo {author} {\bibfnamefont
  {K.}~\bibnamefont {{Markovic}}}, \bibinfo {author} {\bibfnamefont
  {D.}~\bibnamefont {{Marsh}}}, \bibinfo {author} {\bibfnamefont
  {F.}~\bibnamefont {{Marulli}}}, \bibinfo {author} {\bibfnamefont
  {R.}~\bibnamefont {{Massey}}}, \bibinfo {author} {\bibfnamefont
  {Y.}~\bibnamefont {{Mellier}}}, \bibinfo {author} {\bibfnamefont {D.~F.}\
  \bibnamefont {{Mota}}}, \bibinfo {author} {\bibfnamefont {N.}~\bibnamefont
  {{Nunes}}}, \bibinfo {author} {\bibfnamefont {W.}~\bibnamefont {{Percival}}},
  \bibinfo {author} {\bibfnamefont {V.}~\bibnamefont {{Pettorino}}}, \bibinfo
  {author} {\bibfnamefont {C.}~\bibnamefont {{Porciani}}}, \bibinfo {author}
  {\bibfnamefont {C.}~\bibnamefont {{Quercellini}}}, \bibinfo {author}
  {\bibfnamefont {J.}~\bibnamefont {{Read}}}, \bibinfo {author} {\bibfnamefont
  {M.}~\bibnamefont {{Rinaldi}}}, \bibinfo {author} {\bibfnamefont
  {D.}~\bibnamefont {{Sapone}}}, \bibinfo {author} {\bibfnamefont
  {R.}~\bibnamefont {{Scaramella}}}, \bibinfo {author} {\bibfnamefont
  {C.}~\bibnamefont {{Skordis}}}, \bibinfo {author} {\bibfnamefont
  {F.}~\bibnamefont {{Simpson}}}, \bibinfo {author} {\bibfnamefont
  {A.}~\bibnamefont {{Taylor}}}, \bibinfo {author} {\bibfnamefont
  {S.}~\bibnamefont {{Thomas}}}, \bibinfo {author} {\bibfnamefont
  {R.}~\bibnamefont {{Trotta}}}, \bibinfo {author} {\bibfnamefont
  {L.}~\bibnamefont {{Verde}}}, \bibinfo {author} {\bibfnamefont
  {F.}~\bibnamefont {{Vernizzi}}}, \bibinfo {author} {\bibfnamefont
  {A.}~\bibnamefont {{Vollmer}}}, \bibinfo {author} {\bibfnamefont
  {Y.}~\bibnamefont {{Wang}}}, \bibinfo {author} {\bibfnamefont
  {J.}~\bibnamefont {{Weller}}}, \ and\ \bibinfo {author} {\bibfnamefont
  {T.}~\bibnamefont {{Zlosnik}}},\ }\href {\doibase 10.12942/lrr-2013-6}
  {\bibfield  {journal} {\bibinfo  {journal} {Living Reviews in Relativity}\
  }\textbf {\bibinfo {volume} {16}} (\bibinfo {year} {2013}),\
  10.12942/lrr-2013-6},\ \Eprint {http://arxiv.org/abs/1206.1225}
  {arXiv:1206.1225} \BibitemShut {NoStop}%
\bibitem [{\citenamefont {{Wolz}}\ \emph {et~al.}(2012)\citenamefont {{Wolz}},
  \citenamefont {{Kilbinger}}, \citenamefont {{Weller}},\ and\ \citenamefont
  {{Giannantonio}}}]{2012JCAP...09..009W}%
  \BibitemOpen
  \bibfield  {author} {\bibinfo {author} {\bibfnamefont {L.}~\bibnamefont
  {{Wolz}}}, \bibinfo {author} {\bibfnamefont {M.}~\bibnamefont {{Kilbinger}}},
  \bibinfo {author} {\bibfnamefont {J.}~\bibnamefont {{Weller}}}, \ and\
  \bibinfo {author} {\bibfnamefont {T.}~\bibnamefont {{Giannantonio}}},\ }\href
  {\doibase 10.1088/1475-7516/2012/09/009} {\bibfield  {journal} {\bibinfo
  {journal} {\jcap}\ }\textbf {\bibinfo {volume} {9}},\ \bibinfo {eid} {009}
  (\bibinfo {year} {2012})},\ \Eprint {http://arxiv.org/abs/1205.3984}
  {arXiv:1205.3984} \BibitemShut {NoStop}%
\end{thebibliography}%

\appendix

\section{Cosmological parameter dependence}
\label{app:param_dpdce} 

\begin{figure}[H]
\centering
\includegraphics[width=0.42\columnwidth]{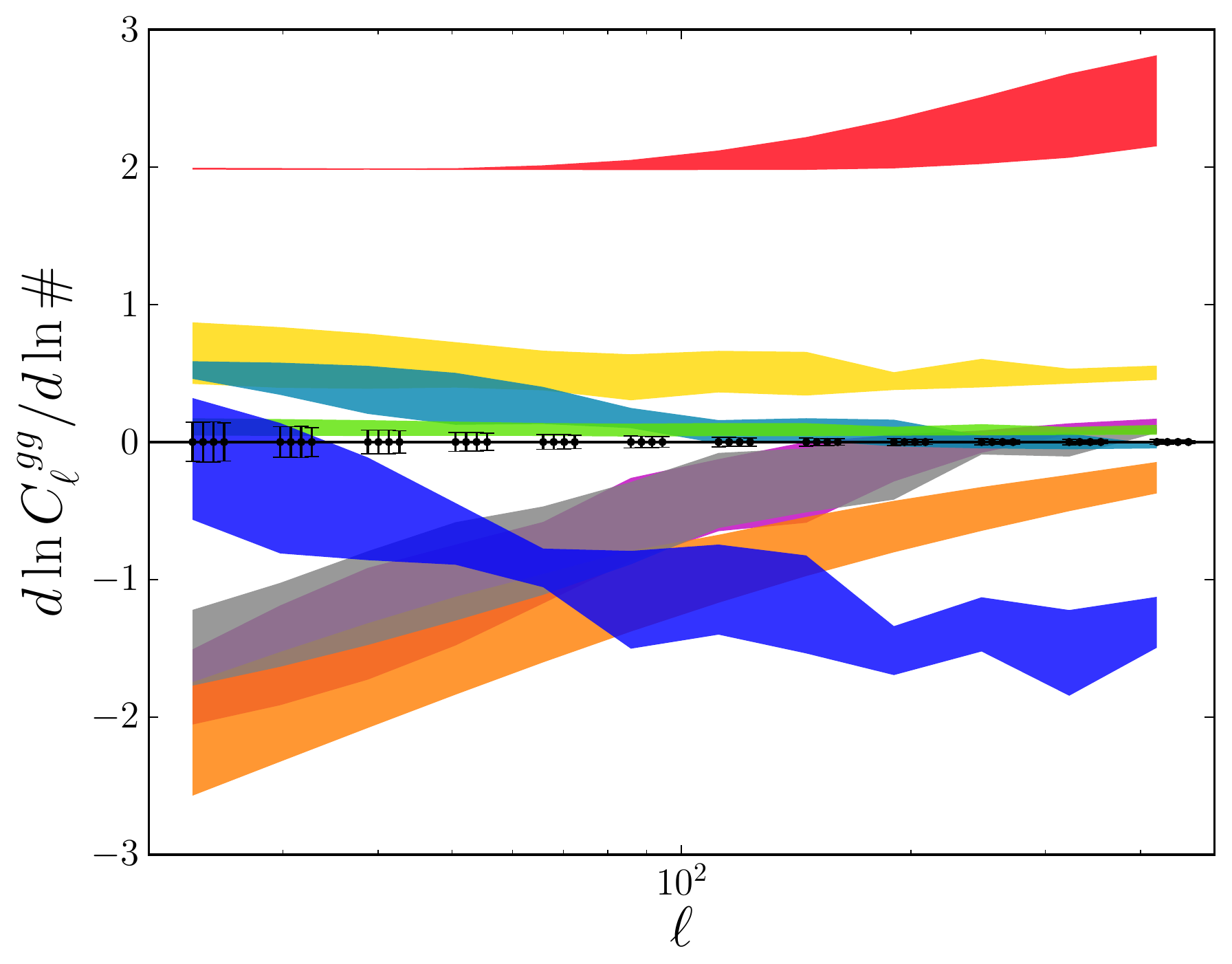}
\includegraphics[width=0.42\columnwidth]{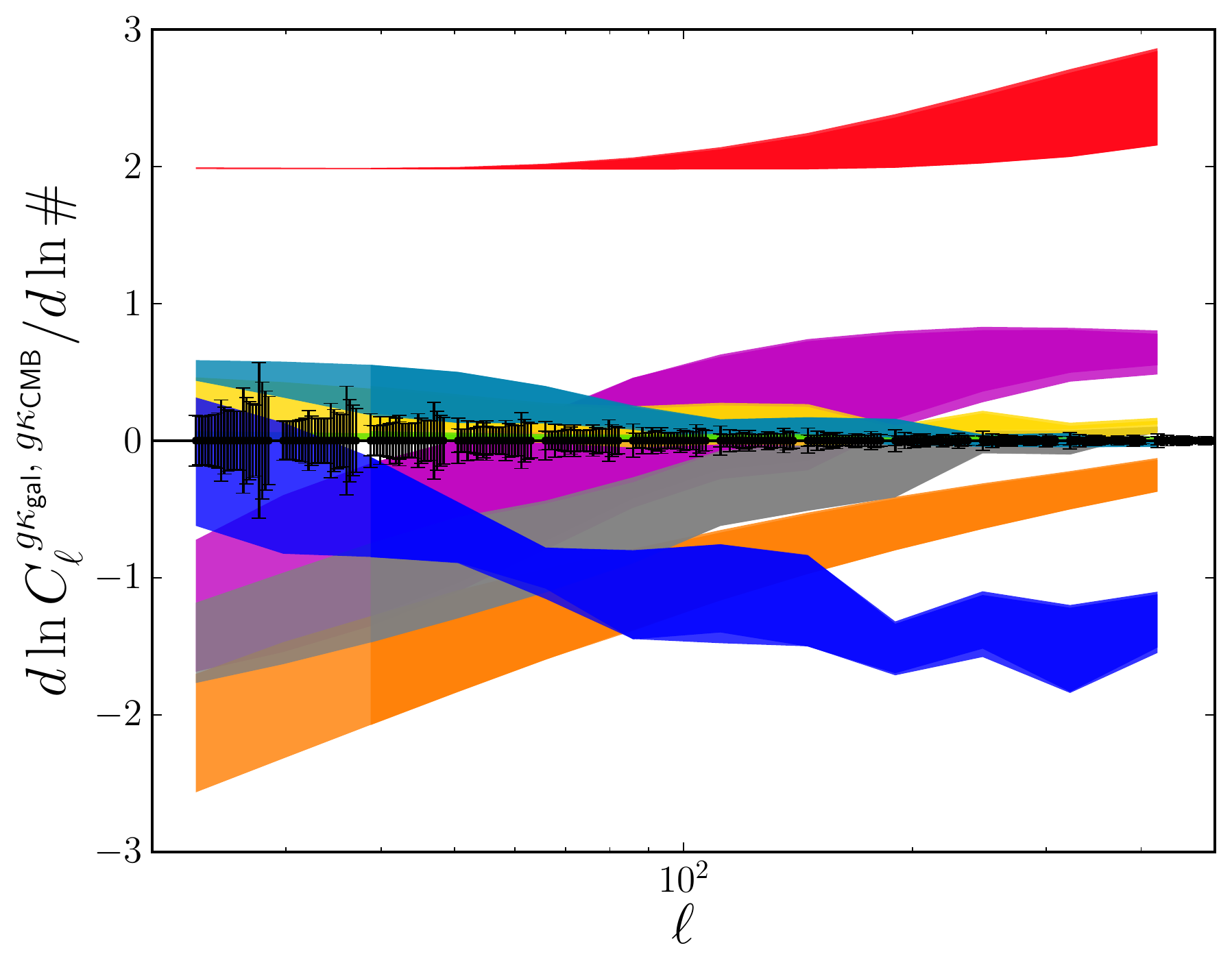}
\includegraphics[width=0.42\columnwidth]{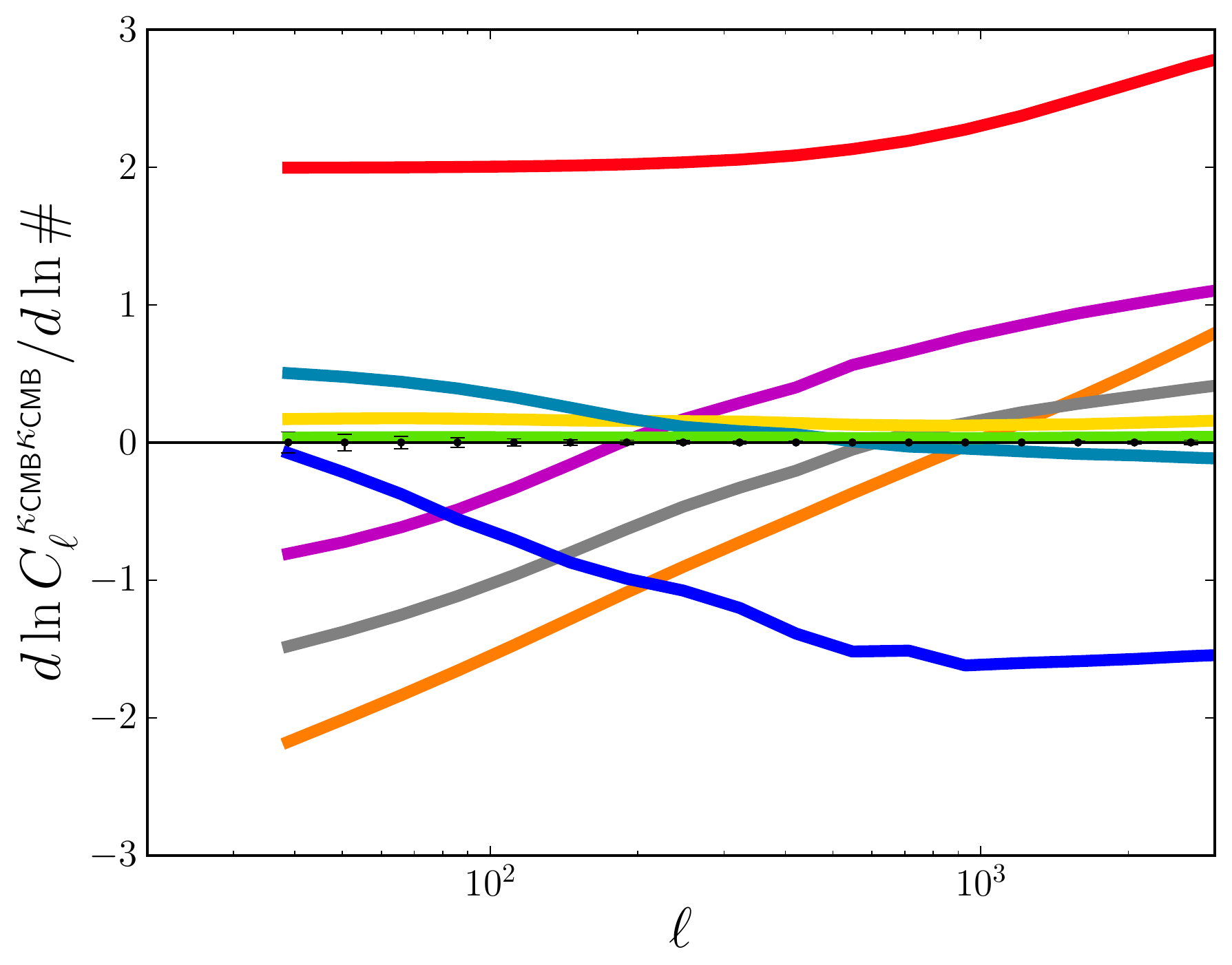}
\includegraphics[width=0.42\columnwidth]{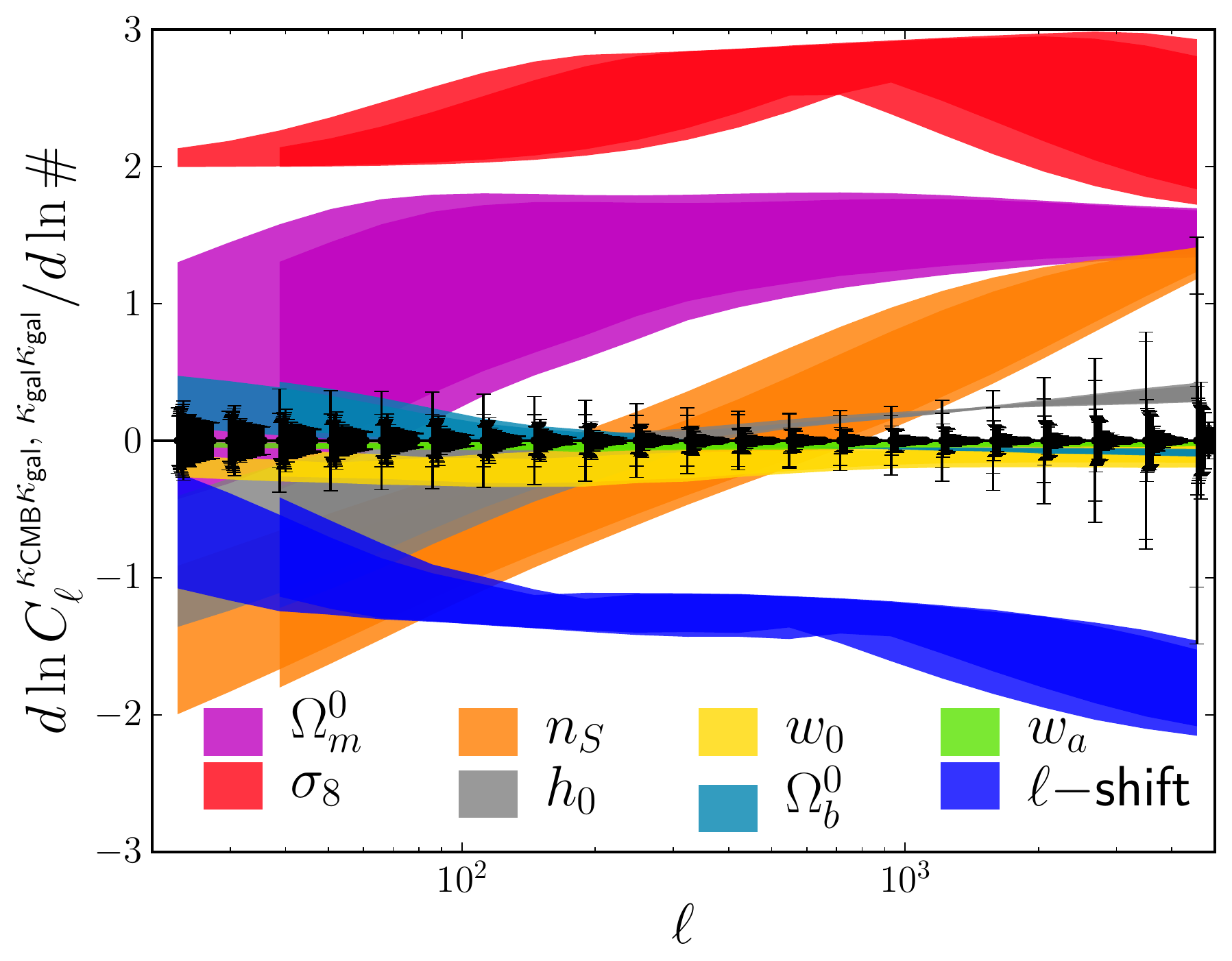}
\caption{
Logarithmic derivatives of observables with respect to cosmological parameters, showing the scalings of observables with parameters. Together with the covariance matrix in Fig.~\ref{fig:lsst_cor}, these plots visualize the information required for the Fisher forecast. They give intuition on the parameter dependences and degeneracies.
Top left: clustering; top right: galaxy-CMB lensing and galaxy-galaxy lensing; bottom left: CMB lensing auto-spectrum; bottom right: cosmic shear and galaxy lensing-CMB lensing.\\
The width of the lines or bands corresponds to the range of variation across tomographic bins. On these plots, a high absolute value corresponds to a strong parameter dependence. A positive value corresponds to an observable growing with the parameter. A horizontal curve corresponds to a multiplicative factor, and a slanted curve corresponds to a tilt in the observable, when the parameter is varied. Two curves identical modulo multiplicative factor correspond to a perfect degeneracy between parameters.\\
For example, all observables scale roughly as $\propto \sig^2$ in the linear regime and $\sig^3$ in the non-linear regime. The parameter $\Omega_m^0$ typically produces a tilt, and is strongly degenerate with $h_0$ for clustering (top left panel).
}
\label{fig:param_dpdces}
\end{figure}

In this appendix, we show the dependences of the various two-point correlation functions on the cosmological parameters (Fig.~\ref{fig:param_dpdces}).
These figures show the logarithmic derivative of the observables with respect to the parameters varied. 
Together with the covariance matrix in Fig.~\ref{fig:lsst_cor}, they are the key ingredients needed to perform a Fisher forecast. 
The width of the lines or bands displayed corresponds to the range of variation across tomographic bins.

These figures give intuition on the scaling of observables with parameters, and can be interpreted as follows. 

A $1\%$ change in parameter $p$ produces a $\frac{d \ln \mathcal{O}}{d \ln p} \%$ change in observable $\mathcal{O}$.
Therefore, observable $\mathcal{O}$ grows with parameter $p$ if $\frac{d \ln \mathcal{O}}{d \ln p}\geqslant 0$. Observable $\mathcal{O}$ is strongly dependent on parameter $p$ if $| \frac{d \ln \mathcal{O}}{d \ln p} |$ is large. In this case, a measurement of $\mathcal{O}$ will be very constraining for parameter $p$.

If $\frac{d \ln \mathcal{O}}{d \ln p}$ is independent of $\ell$, then the parameter $p$ simply acts as a multiplicative factor with $\mathcal{O} \propto p^{\frac{d \ln \mathcal{O}}{d \ln p}}$. For example, as seen in Fig.~\ref{fig:param_dpdces}, all observables roughly satisfy $\frac{d \ln \mathcal{O}}{d \ln \sig} = 2$, i.e. $\mathcal{O} \propto \sig^2$, in the linear regime and $\frac{d \ln \mathcal{O}}{d \ln \sig} = 3$, i.e. $\mathcal{O} \propto \sig^3$, in the non-linear regime. If the curve of $\frac{d \ln \mathcal{O}}{d \ln p}$ as a function of $\ell$ is slanted, it means that the parameter $p$ produces a tilt on $\mathcal{O}$. This is typically the case for $\Omega_m^0$.

This figure also allows to visualize parameter degeneracies and covariances. If two curves $\frac{d \ln \mathcal{O}}{d \ln p_1}(\ell)$ and $\frac{d \ln \mathcal{O}}{d \ln p_2} (\ell)$ are identical modulo a multiplicative factor, then the parameters $p_1$ and $p_2$ are perfectly degenerate. If these two curves are only similar modulo multiplicative factor (where ``similarity'' depends on the covariance matrix), then the parameters $p_1$ and $p_2$ are partially degenerate, and have a non-zero covariance.

\begin{figure}[H]
\centering
\includegraphics[width=0.7\columnwidth]{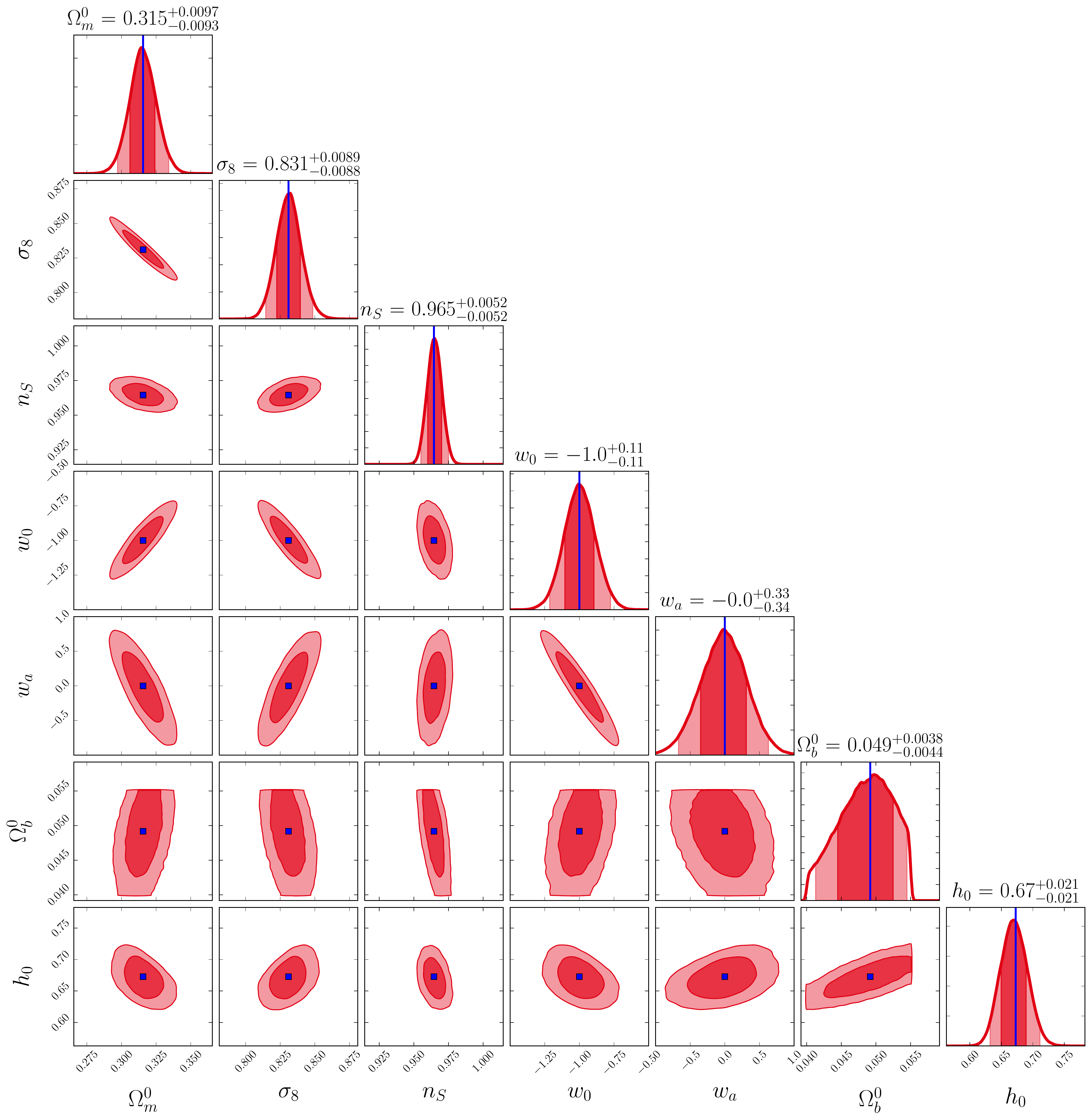}
\caption{
Cosmological constraints from LSST and CMB S4 lensing. We marginalize over all the nuisance parameters, relaxing completely the priors on shear biases. The red zones indicate the $68\%$ and $95\%$ confidence regions.
Striking features are the negative correlations for $\om - \sig$, $\sig - \w$, $\w - \wa$ and the positive correlations for $\om - \w$, $h_0 - \omb$, which can be understood in light of Fig.~\ref{fig:param_dpdces}. The irregular shape of the confidence region for $\omb$ is a consequence of the flat prior imposed and the poor constraining power of the data for this parameter.
}
\label{fig:corner_full_cosmo}
\end{figure}

Fig.~\ref{fig:corner_full_cosmo} shows the confidence regions for the 7 cosmological parameters varied in our analysis. The negative correlations for $\om - \sig$, $\sig - \w$, $\w - \wa$ and the positive correlations for $\om - \w$, $h_0 - \omb$, can be understood in light of Fig.~\ref{fig:param_dpdces}.

\section{Convergence of the MCMC chains; validation of the Fisher approximation}
\label{app:mcmc_fisher_convergence} 

\begin{figure}[H]
\centering
\includegraphics[width=0.49\columnwidth]{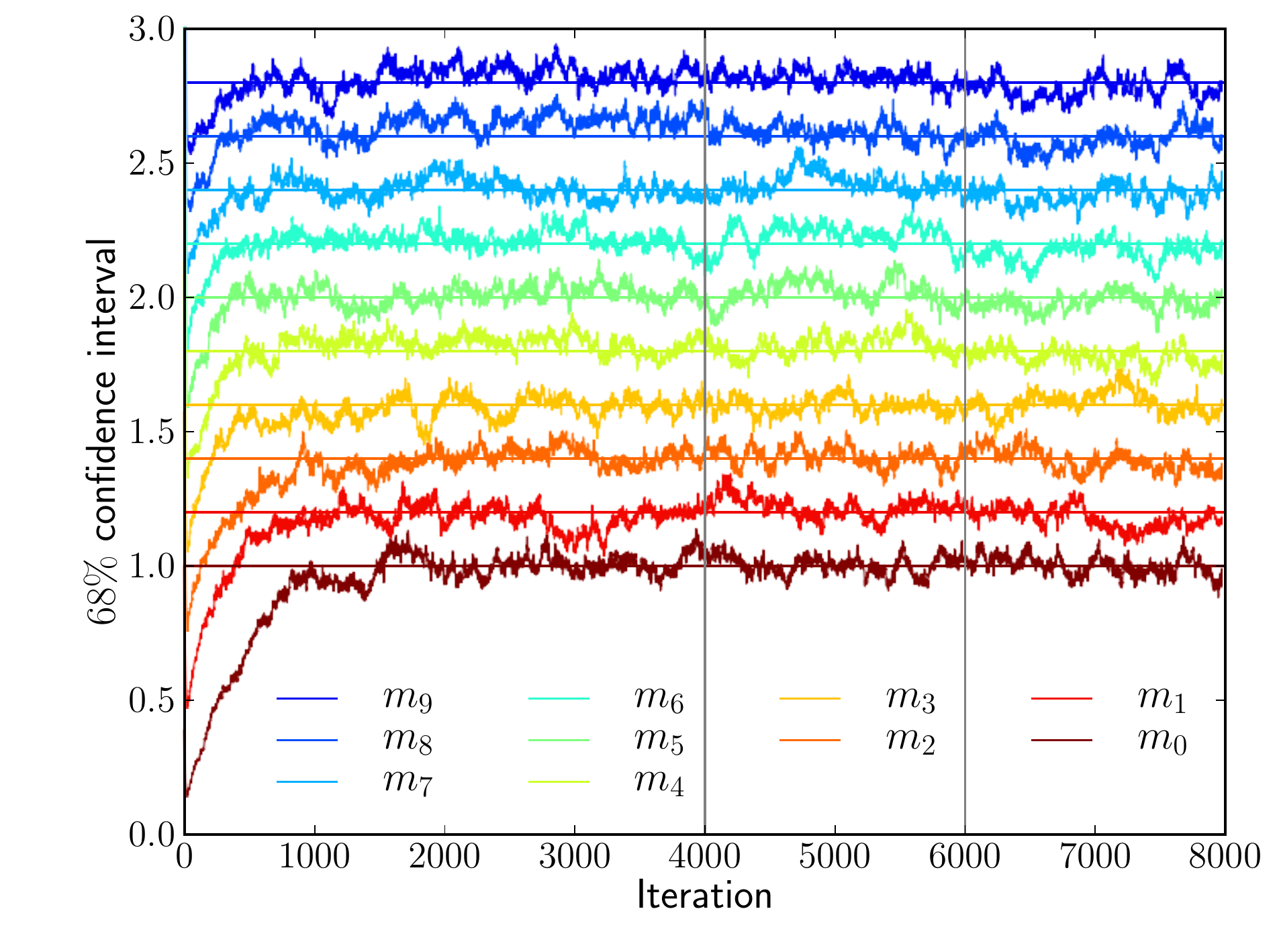}
\includegraphics[width=0.49\columnwidth]{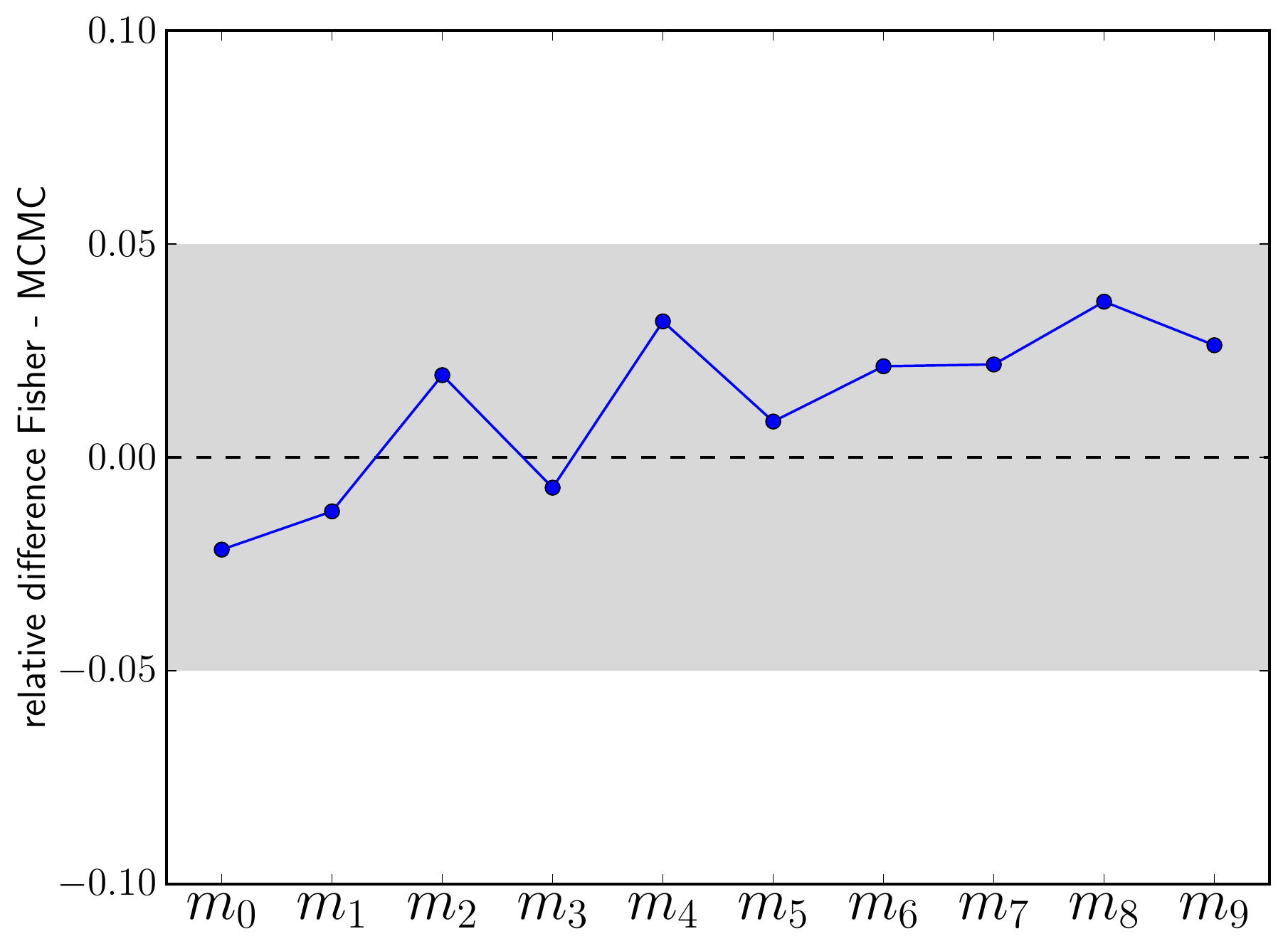}
\caption{
\textbf{Left panel:} At each of the 8000 iterations, the radius of the $68\%$ confidence interval for the shear biases is estimated from the $512$ walkers at that same iteration, and shown on this plot, normalized to the final quoted value and offset for clarity. The final quoted value is obtained by combining the last 4000 iterations. 
When comparing the estimated values from iterations 4000-6000 and 6000-8000, they differ by $\sqrt{2} \times (0.3\% \text{ to } 5\%)$.\\
\textbf{Right panel:} relative difference between Fisher approximation and MCMC result for the $68\%$ confidence radius of the shear biases. The agreement is better than $5\%$, which implies a corresponding convergence of the MCMC chains, and the near-Gaussianity of the posterior distribution for the shear biases.
}
\label{fig:convergence_mcmc}
\end{figure}

In this Appendix, we show the state of convergence of the MCMC chains and the agreement with the Fisher approximation. Fig.~\ref{fig:convergence_mcmc} shows that Fisher and MCMC forecasts agree to better than $5\%$ for all shear biases. 

\begin{figure}[H]
\centering
\includegraphics[width=0.7\columnwidth]{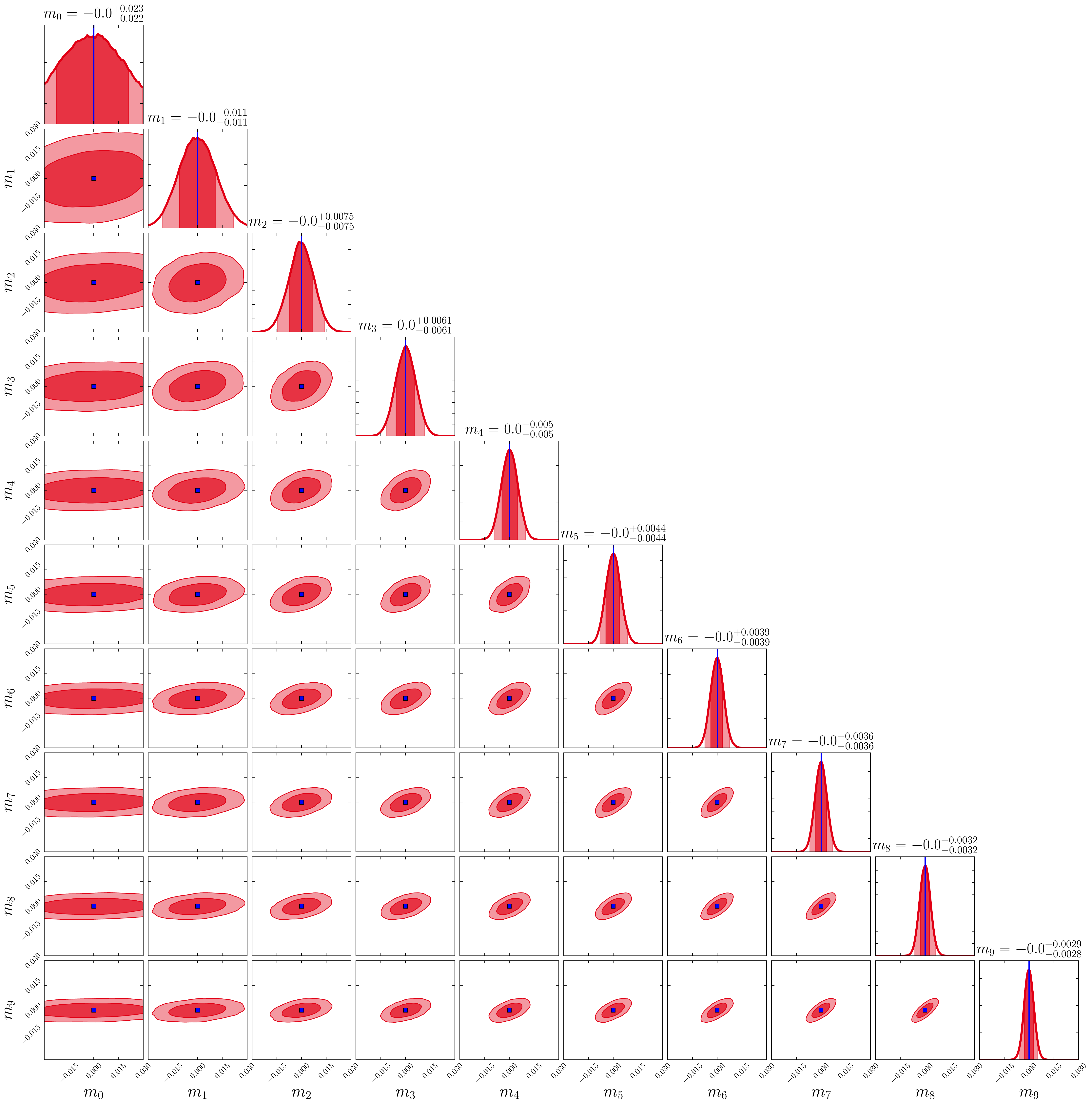}
\caption{
Self-calibration of the shear biases from LSST and CMB S4 lensing. We marginalize over all the nuisance parameters, relaxing completely the priors on shear biases. The red zones indicate the $68\%$ and $95\%$ confidence regions.
The posterior distributions are visually close to Gaussian, which explains the validity of the Fisher approximation shown in Fig.~\ref{fig:convergence_mcmc}.
}
\label{fig:corner_full_m}
\end{figure}

This good agreement is the result of two effects. First, it indicates convergence of the MCMC chains, which is not trivial given that we are jointly fitting 37 parameters. Second, it indicates that the posterior distribution for the shear biases is close to Gaussian. This was not obvious \textit{a priori}, and is in agreement with \cite{2012JCAP...09..009W}. We show in Fig.~\ref{fig:corner_full_m} that the posterior is indeed close to Gaussian for the shear biases.

\end{document}